\documentclass[aps,rmp,twocolumn,floats,nofootinbib]{revtex4} 
\usepackage{graphics,graphicx,epsfig} 
\usepackage{amssymb,color} 
\usepackage{hyperref,babel}
\usepackage{epsf,epstopdf,wrapfig} 
\usepackage {amsmath} 
 \usepackage{textgreek,bm}
\usepackage{sidecap} 
\usepackage{float}
\graphicspath{ {./figures/}}

\newcommand{\beq}{\begin{equation}} 
\newcommand{\eeq}{\end{equation}} 
\newcommand{\beqn}{\begin{eqnarray}} 
\newcommand{\eeqn}{\end{eqnarray}}

\begin{document} 

\title{Statistical mechanics for networks of real neurons}

\author{Leenoy Meshulam} \affiliation{Center for Computational Neuroscience and Department of Applied Mathematics\\ University of Washington, Seattle, Washington 98195 USA}
\author{William Bialek}
\affiliation{Joseph Henry Laboratories of Physics and Lewis--Sigler Institute for Integrative Genomics\\
Princeton University, Princeton, NJ 08544 USA}

\begin{abstract} 
Perceptions and actions, thoughts and memories result from  coordinated activity in hundreds or even thousands of neurons in the brain. It is an old dream of the physics community to provide a statistical mechanics description for these and other emergent phenomena of life.  These  aspirations appear in a new light because of developments in our ability to measure the electrical activity of the brain,  sampling thousands of individual neurons simultaneously over hours or days.  We review the  progress that has been made in bringing theory and experiment together, focusing on maximum entropy methods and a phenomenological renormalization group.  These approaches have uncovered new, quantitatively reproducible collective behaviors in networks of real neurons, and provide examples of rich parameter--free predictions that agree in detail with experiment.   
\end{abstract}

\date{\today}

\maketitle

\tableofcontents

\section{Introduction}

Neural networks have been an inspiring source of physics problems for generations.  The current revolution in artificial intelligence \cite{lecun2015deep,minae+al_24}  has roots in classical work that appeared a bit over sixty years ago in {\sl Reviews of Modern Physics} \cite{block1962,block+al1962,Rosenblatt1961}. The explicit effort to build network models grounded in statistical mechanics began in the 1970s \cite{cooper_73,little_74,little+shaw_75,little+shaw_78} and received important stimuli in the early 1980s \cite{hopfield1982,hopfield1984}, making connections to then new ideas about spin glasses \cite{mezard+al_87,amit_89}.  In the context of these models one could use statistical mechanics to address not just the dynamics of a network, but the way in which it learns from experience  \cite{levin+al_90,watkin+al_93}.  There is a path from this early work to current efforts of the theoretical physics community to understand the recent successes of  machine learning \cite{carleo+al_19,mehta+al_19,roberts_21,roberts+yaida_22}.  In \S\ref{sec:history} we provide a brief guide to this rich history, emphasizing points which seem especially relevant for recent developments connecting theory and experiment.

In the long history of physicists' engagement with neural networks, it must be admitted that the search for tractable models often loosened the connection of theory to experiments on real brains.   This problem became more urgent as methods became available to monitor, simultaneously, the electrical activity of tens, hundreds, and even thousands of neurons while animals engage in reasonably natural behaviors (\S\ref{sec-expts}).  If we imagine a statistical mechanics for neural networks, these tools give us access to something like a Monte Carlo simulation of the microscopic degrees of freedom.  This explosion of data calls out for new methods of analysis, and creates new opportunities for theory/experiment interaction.

Roughly twenty years ago it was suggested that maximum entropy methods could provide a very direct bridge from the new data on large numbers of neurons to explicit statistical physics models for these networks (\S\ref{sec-max_ent}).   In the simplest version of this approach, measurements on the mean activity and pairwise correlations among neurons result in an Ising spin glass model for patterns of activity in the network.  Importantly, all the couplings in the Ising model are determined by the measured correlations, and one can proceed to make {\em parameter free} predictions for higher order properties of the network.  The surprise was that these predictions, at least in some cases, are extraordinarily successful (\S\S\ref{sec-larger} and \ref{sec-subgroups}).  

The phenomenological success of the maximum entropy approach raised several questions.  Should we expect this success to generalize or was there something special about the first examples?  Does success tell us something about the underlying network?  If the models are so accurate, perhaps we should take them seriously as statistical physics problems: where are real networks in the phase diagram of possible networks (\S\ref{criticality})?  Can these models be given different interpretations, e.g. in terms of a smaller number of ``latent variables'' that are encoded by the network?

The relatively simple statistical physics models constructed via maximum entropy are in some cases are more successful than complex models motivated by biological details.   Why should simple models work?  In condensed matter physics we often describe macroscopic, emergent phenomena using models that are much simpler than the underlying microscopic mechanisms.  This works not because we are lucky but because the renormalization group (RG) tells us that in many cases there is only a small number of relevant operators, so that models  simplify as we restrict our attention to longer length scales.  Inspired by these ideas, there have been efforts to explicitly coarse--grain the patterns of activity in very large networks (\S\ref{chapter-RG}).  The very first such efforts revealed surprisingly precise scaling behaviors, in some cases with exponents that are reproducible in the second decimal place.  These initial results now have been confirmed in other systems.

As with maximum entropy methods, the success of coarse–graining in uncovering interesting collective behaviors of real neurons raises several questions. The observation of scaling suggests that the dynamics of these networks is controlled by some nontrivial fixed point of the RG.  But are these phenomenological analyses sufficient to identify fixed point behaviors in cases that we understand?  Could the observed scaling behaviors emerge in some other way?   Are these behaviors universal?

When physicists first wrote down statistical mechanics models for neural networks, it was not clear if these models should be taken as metaphors or if they should be taken seriously as theories of real brains.\footnote{One can trace the metaphorical description of coordinated activity in the brain as being like collective effects in a magnet back even further, at least to \citet{cragg+temperley_54}.}  If forced to choose, most people would have voted for metaphors, since real brains surely are too complicated to be captured in the physicists' drive for simplification.  While it emphatically is too soon to claim that we have a theory of the brain, progress that we review here makes clear that we can have the  precise quantitative connections between theory and experiment that we have in the rest of physics.  As experiments on the physics of living systems improve, we should ask more of our theories.  

Finally, in case thinking about the brain is not sufficient motivation, networks of neurons provide a prototype of living systems with many degrees of freedom (Appendix \ref{sec-sequences+}).  Even a single protein molecule typically is composed of more than one hundred amino acids, and the structures and functions of these molecules emerge from interactions among these many more microscopic elements.   At the next scale up, membrane patches and protein droplets self--organize in ways that most likely reflect phase separation.  The identities and internal states of cells are determined by the expression levels of large numbers of genes that form an interacting regulatory network.  In developing embryos and tissues more generally the movements of individual cells organize into macroscopic flows.  In populations of bacteria, swarms of insects, schools of fish, and flocks of birds we see collective movements and decision making.  In all these examples---and, of course, in networks of neurons---what we recognize as the functional behavior of living systems is a macroscopic behavior that emerges from interactions of many components on a smaller scale.  

In the inanimate world, statistical mechanics provides a powerful and predictive framework within which to understand emergent phenomena.  It is an old dream of the physics community that we could have a statistical mechanics of emergent phenomena in the living world as well.   We encourage the reader to think of what we review here as progress toward realizing this dream.

\section{Some history}
\label{sec:history}

Today, neural network models are known to many different communities: physicists and applied mathematicians, computer scientists and engineers, neurobiologists and cognitive scientists.  Neural networks are at the heart of an ongoing revolution in artificial intelligence, and are making their way into many aspects of scientific data analysis, from cell biology to CERN.  Here we provide  a brief (and perhaps idiosyncratic) reminder of how some of these ideas developed.  

\subsection{Prehistoric times}
\label{sec-prehistory}

The engagement of physicists with neurons and the brain has a long and fascinating history.  Our modern understanding of electricity has its roots in the 1700s with observations  on nerves and muscles.  The understanding of optics and acoustics that emerged in the 1800s was continuous with the exploration of vision and hearing.   This involved thinking not just about the optics of the eye or the mechanics of the inner ear, but about the inferences that our brains can derive from the data collected by these physical instruments.

The idea that the brain is made out of discrete cells, connected by synapses, dates from late 1800s  \cite{Cajal1894}.   The electrical signals from individual nerve cells (neurons) were first recorded in the 1920s, starting with the cells in sense organs that provide the input to the brain \cite{Adrian1928}.   Observing these small signals required instruments no less sensitive than those in contemporary physics laboratories.  The crucial observation is that neurons communicate by generating discrete, identical pulses of voltage across their membranes; these pulses are called action potentials or, more colloquially, spikes.  

By the 1950s there was a clear mathematical description of the dynamics underlying the generation and propagation of spikes \cite{Hodgkin+Huxley_1952}.  Perhaps surprisingly, the terms in these equations could be taken literally as representing the action of real physical components---ion channel proteins that allow the flow of specific ions across the cell membrane, and which open and close (or ``gate'') in response to the transmembrane voltage.  The progress from macroscopic phenomenology to the dynamics of individual channels is a beautiful chapter in the interaction of physics and biology.  The classic textbook account is \citet{Aidley1998}; \citet{Dayan+Abbott_2001} discuss phenomenological models for spiking activity; and a broader biological context is provided by \citet{Kandel+al_2001}.   \citet{spikesbook} describe the way in which sequences of spikes represent information about the sensory world, and \citet{bialek_12} connects channels and spikes to other problems in the physics of biological systems.

Even before the mechanisms were clear, people began to think about how  the quasi--digital character of spiking could be harnessed to do computations \cite{mcculloch+pitts_43}.   This work comes after the foundational work of \citet{turing_37} on universal computation, but before any practical modern computers. The goal of this work was to show that the basic facts known about neurons were sufficient to support computing essentially anything.  On the one hand this is a very positive theoretical development: the brain could be a computer, in a deep sense.  On the other hand it is disappointing, since if the brain is a universal computer there is not much more that one can say about the dynamics..

The way in which computation emerges from neurons in this early work clearly involves interactions among large numbers of cells in a network.  Although single neurons can have remarkably precise dynamics in relation to sensory inputs and motor outputs \cite{spikesbook,nemenman+al_08,hires+al_15,srivastava+al_17}, there are many indications that  our perceptions and actions, thoughts and memories typically are connected to the activity in many hundreds, perhaps even hundreds of thousands of neurons.  Relevant activity in these large networks must be coordinated or collective.  

The idea that collective neural activity in the brain might be described with statistical mechanics was very much influenced by observations on the electroencephalogram or EEG  \cite{wiener_58}.   The EEG is a macroscopic measure of activity, traditionally done simply by placing electrodes on the scalp, and the existence of the EEG is prima facie evidence that the electrical activity of many, many neurons must be correlated. There is also the remarkable story of a demonstration by Adrian, in which he sat quietly with his eyes closed with electrodes attached to his head.  The signals, sent to an oscilloscope, showed the characteristic ``alpha rhythm'' that occurs in resting states, roughly an oscillation at $\sim 10\,{\rm Hz}$.  When asked to add two numbers in his head, the rhythm disappeared, replaced by less easily described patterns of activity \cite{adrian+matthews_34}.  This should dispel any lingering doubts that your mental life is related to the electrical activity of your brain.

In the simplest models for neural dynamics,  we describe the state of each neuron  ${\rm i}$ at time $t$ by a binary or Ising variable $\sigma_{\rm i}(t)$; $\sigma_{\rm i}(t) = +1$ means that the neuron is active, and $\sigma_{\rm i}(t) = 0$ means that the neuron is silent.\footnote{For the moment ``active'' is a deliberately vague term.  We could mean that the cell is in some relatively sustained state, perhaps steadily firing action potentials over a reasonable fraction of a second.  Alternatively, we might be looking in very small time windows and asking about the presence or absence of single spikes.  Resolving this vagueness will be essential in connecting theory with experiment, below.}  We imagine the dynamics proceeding in discrete time steps $\Delta \tau$.  Each neuron sums inputs from other neurons, weighted by the strength $J_{\rm ij}$ of the synapse  or connection from cell ${\rm j}\rightarrow {\rm i}$, and neurons switch into the active state if the total input is above a threshold:
\begin{equation}
\sigma_{\rm i}(t+\Delta\tau ) = \Theta\left[ \sum_{\rm j} J_{\rm ij}\sigma_{\rm j}(t) - \theta_{\rm i}\right].
\label{JJH1}
\end{equation}
The nature of the dynamics is encoded in the matrix $J_{\rm ij}$ of synaptic strengths.  If we think about arbitrary matrices, then the dynamics can be arbitrarily complex; progress depends on simplifying assumptions.  It is useful to organize our discussion around two extreme simplifications.  But keep in mind as we follow these threads that many of the developments occurred in parallel, and that there was considerable crosstalk.

\subsection{From perceptrons to deep networks}

\begin{figure}[t]
\includegraphics[width = \linewidth]{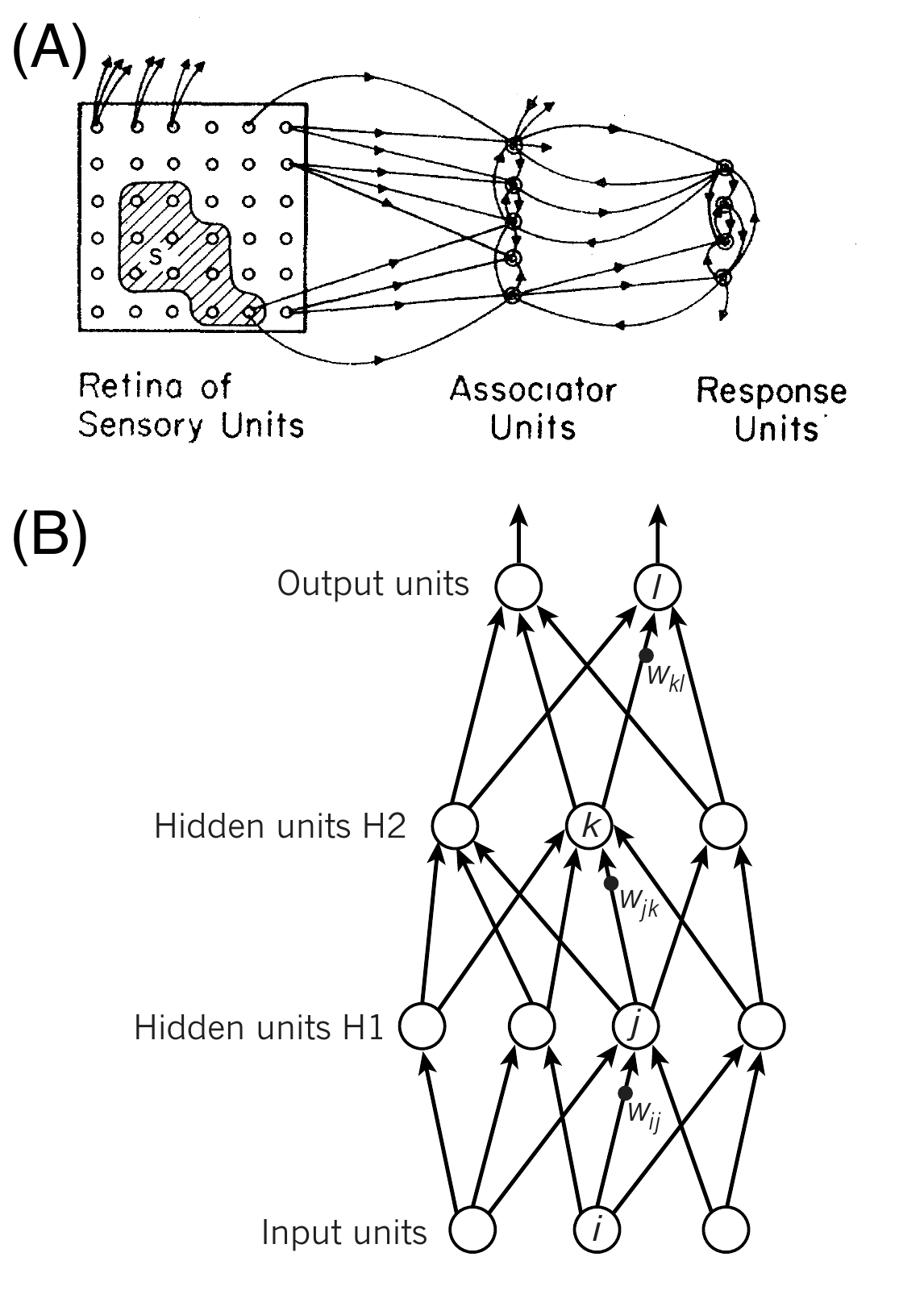}
\caption{Neural networks with a feed--forward architecture, or ``perceptrons.'' (A) An early version of the idea, from \citet{block1962}.  (B) A modern version, with additional hidden layers.  The first steps in the modern AI revolution involved similar networks, with many hidden layers, that achieved human--level performance on image classification and other tasks \cite{lecun2015deep}. . \label{perceptrons}}
\end{figure}

One popular simplification is to assume that $J_{\rm ij}$ has a feed--forward, layered structure.  This is the ``perceptron'' architecture \cite{block1962,block+al1962,Rosenblatt1961}, illustrated in Fig \ref{perceptrons}A, which is simpler to analyze precisely because there are no feedback loops.  It is convenient to label the neurons also by the layer $\ell$ in which they reside, and to generalize from binary variables to continuous ones, so that
\begin{equation}
x_{\rm i}^{(\ell +1 )} = g\left[ \sum_{\rm j} W^{(\ell +1)}_{\rm ij} x_{\rm j}^{(\ell )} -\theta_{\rm i}^{(\ell +1 )}\right] ,
\end{equation}
where the propagation through layers replaces propagation through time and  $g[\cdot ]$ is a monotonic nonlinear function. Thus each neuron computes a single projection of its possible inputs from the previous layer, and then outputs a nonlinear function of this projection.  

In the limit that $g[\cdot ]$ becomes a step function we recover binary variables and neuron $\rm i$ in layer $\ell +1$, can be thought of a dividing the space of its inputs in half, with a hyperplane perpendicular to the vector 
\begin{equation}
\bm{V} = \{ V_{\rm j}\} = \{W^{(\ell +1)}_{\rm ij} \}.
\end{equation}
Thus the elementary computation is a binary classification of inputs,
\begin{equation}
{\bm x} \rightarrow y = \Theta (\bm{V\cdot x} - \theta)  
\end{equation}
We could imagine having access to many examples  of the input vector $\bm x$  labelled by the correct classification $y$, and thereby learning the optimal vector $\bm V$.  This picture of learning to classify was present already $\sim$1960, although it would  take the full power of modern statistical physics to say that we really understand it.  Crucially, if we think of the the $\{x_{\rm i}\}$ or $\{\sigma_{\rm i}\}$ as being the microscopic variables in the system and the $J_{\rm ij}$ as being the interactions among these variables, then learning is statistical mechanics in the space of interactions \cite{gardner_88,gardner+derrida_88,levin+al_90,watkin+al_93}.

Although many of the computations done by the brain can be framed as classification problems, such as attaching names or words to images, very few can be solved by a single step of linear separation.   Again this was clear at the start, but development of these ideas took decades.   Enthusiasm was dampened by an emphasis on what  two layer networks could {\em not} do \cite{MinksyPapert1969}, but eventually it became clear that multilayer perceptrons are much more powerful \cite{LeCun1987,LapedesFarber1988}, and theorems were proven to show that these systems can approximate any function \cite{hornik+al_89}.   As with the simple perceptron,  optimal weights $W$ can be learned by fitting to many examples of input/output pairs.  Importantly this doesn't require access to the ``correct'' answers at every layer; instead if we work with continuous variables then the goodness of fit across many layers can be differentiated using the chain rule, and errors propagated back through the network to adjust the weights \cite{rumelhart+al_1986}.

Fast forward from the late 1980s to the mid 2010s.  The few layers of early perceptrons became the many layers of ``deep networks,'' in the spirit of Fig \ref{perceptrons}B;  comparing the two panels of Fig \ref{perceptrons} emphasizes the continuity of ideas across the decades. Advances in computing power and storage made it possible not just to simulate these models efficiently, but to solve the problem of finding optimal synaptic weights by comparing against millions or even billions of examples.  These explorations led to networks so large that  the number of weights needed to specify the network vastly exceeded the number of examples.   Contrary to well established intuitions these ``over parameterized'' models worked, generalizing to new examples rather than over--fitting to the training data.  Although we don't fully understand them, these developments have fueled a revolution in artificial intelligence (AI).

\subsection{Symmetric networks}
\label{sec-hopfield1}

Feed--forward networks have the property that if $J_{\rm ij}$ is nonzero, then $J_{\rm ji} = 0$.   \citet{hopfield1982,hopfield1984}  considered the opposite simplification:   if neuron $\rm i$ is connected to neuron $\rm j$, then neuron $\rm j$ is connected to neuron $\rm i$, and the strength of the connection is the same, so that $J_{\rm ij} = J_{\rm ji}$. In this case the dynamics in Eq (\ref{JJH1}) have a Lyapunov function: at each time step the ``energy''
\begin{equation}
E = -{1\over 2}\sum_{\rm ij}  \sigma_{\rm i} J_{\rm ij} \sigma_{\rm j} +\sum_{\rm i} \theta_{\rm i}\sigma_{\rm i}
\label{JJHE}
\end{equation}
either decreases or stays constant.  The evolution of the network state stops at local minima of the energy $E$, and only at these local minima.
We recognize this energy function as an Ising model with pairwise interactions among the spins (neurons).  This very explicit connection of neural dynamics to statistical physics triggered an avalanche of work, and  textbook accounts of these ideas appeared quickly \cite{amit_89,hertz+al_91}. 

\begin{figure}
\includegraphics[width = \linewidth]{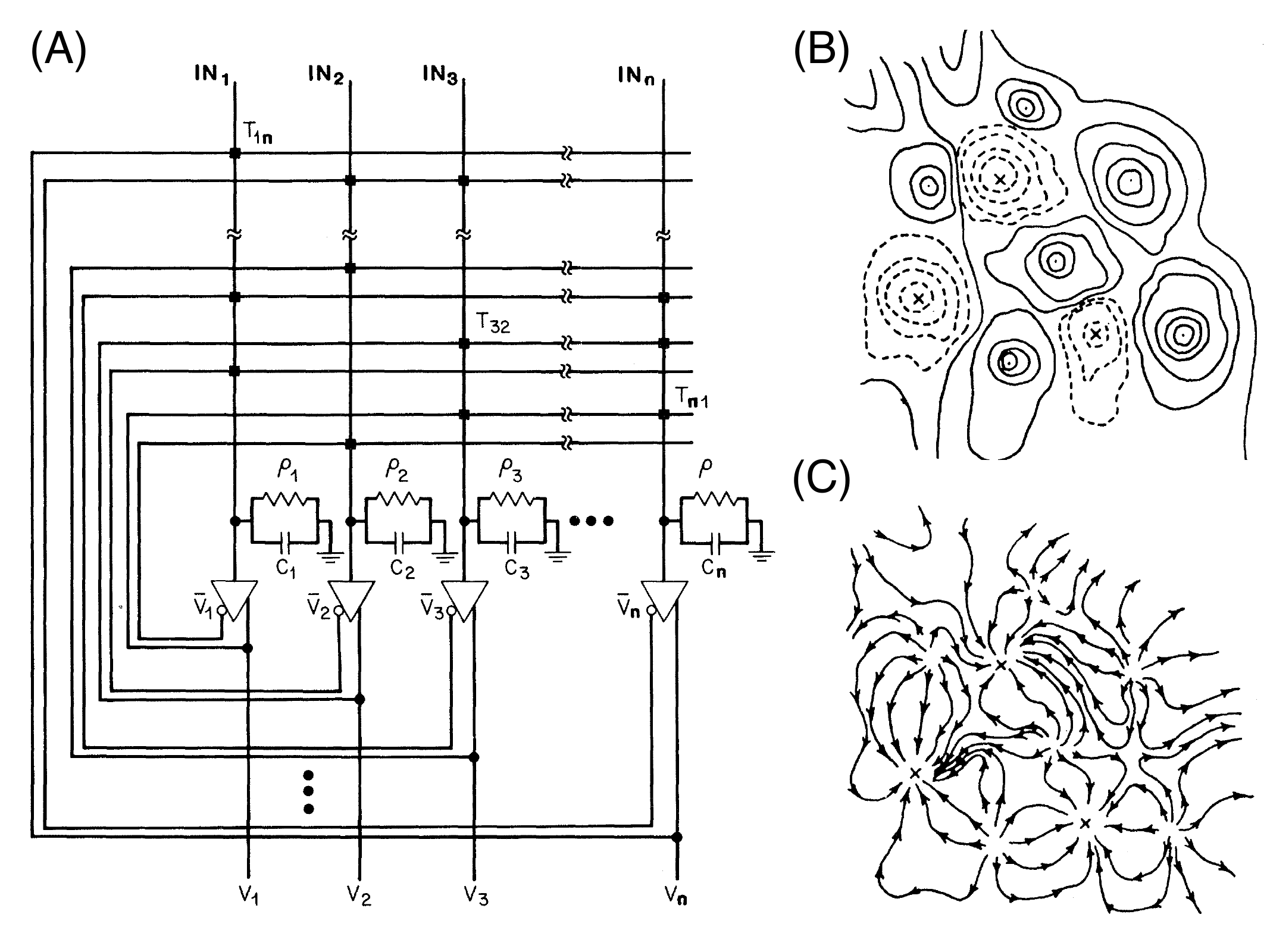}
\caption{Equivalent circuit and dynamics in a symmetric network \cite{hopfield+tank1986}.  (A) ... (B) Schematic energy function for the circuit in (A); solid contours are above a mean level and dashed contours below, with X marking fixed points at the bottoms of energy valleys.  (C) Corresponding dynamics, shown as a flow field.  
\label{fig-hopfield}}
\end{figure}

It was useful in visualizing the dynamics of symmetric networks that they can be realized by simple circuit components, using amplifiers with saturating outputs in place of neurons, as in Fig \ref{fig-hopfield}.  As with perceptrons one generalize to soft spins, now in continuous time; one version of these dynamics is
\begin{equation}
\tau{{dx_{\rm i}}\over{dt}} = - x_{\rm i} + \sum_{\rm j} J_{\rm ij} g(x_{\rm j}) .
\label{dyn_net1}
\end{equation}
These models have the same collective behaviors as Ising spins \cite{hopfield1984}.

A crucial point is that one can ``program''  symmetric networks to place local minima at desired states.  Since the dynamics will flow spontaneously toward these minima and stop, we can think of this programming as storing memories in the network, which then can be recovered by initializing the state anywhere in the relevant basin of attraction.  Taking the mapping of the Lyapunov function to an energy more seriously, this memory storage represents a sculpting of the energy landscape, which is a more general idea.  As an example, we can think about the evolution of amino acid sequences in proteins sculpting the energy landscape for folding.

To illustrate the idea of memory storage, consider the case where the thresholds $\theta_{\rm i} = 0$.  Suppose we can construct a matrix of synaptic weights such that
\begin{equation}
J_{\rm ij} = J\xi_{\rm i}\xi_{\rm j} ,
\end{equation}
where the $\xi_{\rm i} = \pm 1$ are again a set of (now fixed) binary or Ising variables.  Then the energy function becomes
\begin{equation}
E = -{J\over 2} \sum_{\rm ij} \sigma_{\rm i} \xi_{\rm i}\xi_{\rm j} \sigma_{\rm j} = -{J\over 2} \left( \sum_{\rm i} \sigma_{\rm i} \xi_{\rm i}\right)^2 = 
-{J\over 2} \left( \vec\sigma {\mathbf \cdot} \vec\xi \right)^2 .
\label{1memory}
\end{equation}
Because both $\vec\sigma$ and $\vec\xi$ are binary vectors the energy is minimized when these vectors are equal.\footnote{Because we set the thresholds to zero, the globally sign--flipped solution $\vec\sigma = - \vec\xi$ also is allowed.} If want to be a bit fancier we can transform $\sigma_{\rm i} \rightarrow \tilde\sigma_{\rm i} = \sigma_{\rm i} \xi_{\rm i}$, and we then realize that Eq (\ref{1memory}) is gauge equivalent to the mean--field ferromagnet.  

Crucially, we can generalize this construction,
\begin{equation}
J_{\rm ij} = J\left( \xi_{\rm i}^{(1)} \xi_{\rm j}^{(1)}  + \xi_{\rm i}^{(2)} \xi_{\rm j}^{(2)}  + \cdots + \xi_{\rm i}^{(K)} \xi_{\rm j}^{(K)}\right) .
\label{JK}
\end{equation}
If network has $N$ neurons, and the number of these terms $K\ll N$, then typically the vectors $\vec\xi^{(\mu)}$ are orthogonal, and the energy function will have multiple minima at $\vec\sigma = \vec\xi^{(\mu)}$: we have a model that stores $K$ memories.  

To make this more rigorous let's imagine that the states of the network are not just the minima of the energy function, but are drawn from a Boltzmann distribution at some inverse temperature $\beta$; it is plausible that this emerges from a noisy version of the dynamics in Eq (\ref{JJH1}).  Then we have
\begin{eqnarray}
P(\vec\sigma ) &=& {1\over Z} \exp\left[ -\beta E(\vec \sigma )\right]\\
E(\vec \sigma ) &=& - {{J_0}\over N}\sum_{{\rm ij} =1}^N \sum_{\mu =1}^K \sigma_{\rm i} \xi_{\rm i}^\mu \xi_{\rm j}^\mu\sigma_{\rm j} ,
\end{eqnarray}
where we use the usual normalization of interactions by a factor $N$ to insure a thermodynamic limit.  Because the stored patterns are fixed, this is a statistical mechanics problem with quenched disorder, a special kind of mean--field spin glass.  As a first try we can take the stored patterns to be random vectors, which might make sense if we are describing a region of the brain where the mapping between the features of what we remember and the identities of neurons is very abstract.  We can measure the success of recalling memories by measuring the order parameters
\begin{equation}
m_\mu = \overline{\langle {\vec\xi}^\mu {\mathbf \cdot} \vec\sigma \rangle} ,
\end{equation}
where $\langle \cdots \rangle$ denotes an average over the ``thermal'' fluctuations in the neural state $\vec\sigma$ and $\overline{\cdots}$ denotes an average over the random choice of the patterns ${\vec\xi}^\mu$.

Shortly before the introduction of these models, there had been dramatic developments in the statistical mechanics of disordered systems, including the solution of the fully mean--field Sherrington--Kirkpatrick spin glass model \cite{mezard+al_87}.  These tools could be applied to neural networks, resulting in a phase diagram mapping the order parameters $\{m_\mu\}$ as function of the fictitious temperature and the storage density $\alpha = K/N$, all in the thermodynamic limit $N\rightarrow\infty$ \cite{amit+al1985,amit+al1987}.   In the limit of zero temperature, below a critical $\alpha_c = 0.138$ only one of the $m_\mu$ will be nonzero, and it takes values close to one; this survives to finite temperatures.  Thus there is a whole phase in which this model provides effective even if not quite perfect recall.  By now we think of neural network models not as an application of statistical mechanics, but as a source of problems.

An important feature of the dynamics is that it is ``associative.''  Many initial states will relax to the same local minimum of the energy, which is equivalent to saying the same memory can be recalled from many different cues.  In particular, we can imagine that the many bits represented by the state $\{\sigma_{\rm i}\}$ can be grouped into features, e.g. parts of the image of a face, the sound of the person's voice, ... .  Under many conditions if one set of features is given and the others randomized, the nearest local minimum will have all the features correctly aligned \cite{hopfield1982}.  The fact that our mind conjures an image in response to a sound or a fragrance had once seemed mysterious, and this provides a path to demystification, built on the idea that stored and recalled memories are collective states of the network.

The synaptic matrix in Eq (\ref{JK}) has an important feature.  Suppose that the network is currently in some state $\vec\sigma$ and we would like to add this state to the list of stored memories---i.e. we would like the network to learn the current state.  Following Eq (\ref{JK}) we should change the synaptic weights
\begin{equation}
J_{\rm ij} \rightarrow J_{\rm ij}  + J\sigma_{\rm i}\sigma_{\rm j} .
\end{equation}
First we note that the connection between neurons $\rm i$ and $\rm j$ changes in a way that depends only on these two neurons.  This locality of the learning rule is in a way remarkable, since we might have thought that sculpting the energy landscape would require more global manipulations. Second, the change in synaptic strength depends on the correlation between the pre--synaptic neuron $\rm j$ and the post--synaptic neuron $\rm i$: if the cells are active together, the synapse should be strengthened.  This simple rule sometimes is summarized by saying that neurons that ``fire together wire together,'' and there is considerable evidence that real synapses change in this way.  Indeed, although this idea has its origins in classical discussions  \cite{Hebb1949,James1904}, more direct measurements demonstrating that correlated activity leads to long lasting increases of synaptic strength came only in the decade before Hopfield's work \cite{bliss+lomo1973}.

In the first examples, the goal of computation was to recover a stored pattern from partial information (associative memory). Beyond memory,  \citet{hopfield+tank1985} soon showed that one could construct networks that  solve classical optimization problems, and that many biologically relevant problems could be cast in this form \cite{hopfield+tank1986}.   At the same time, the idea of simulated annealing \cite{kirkpatrick+al_83} led people to take much more seriously the mapping between ``computational'' problems of optimization and the ``physical'' problems of finding minimum energy states of many--body systems.  This led, for example, to connections between statistical mechanics and computational complexity \cite{kirkpatrick+selman_94,monasson+al_99}.
 From an engineering point of view, models for neural networks connected immediately to the possibility of using modern chip design methods to build analog, rather than digital circuits \cite{Mead1989}.   Taken together, these simple symmetric models of neural networks  formed a nexus among statistical physics, computer science, neurobiology, and engineering.

\subsection{Perspectives}

Our emphasis in this review is on networks of real neurons.  But it would be foolish to ignore what is happening in the world of engineered, artificial networks, which proceeds at a terrifying pace, realizing many of the old dreams for artificial intelligence (AI).   Not so long ago we would have emphasized the tremendous progress being made on problems such as image recognition or game playing, where deep networks achieved something that approximates human level performance.  Today, popular discussion is focused on generative AI, with networks that produces text and images that have a striking realism.  Our theoretical understanding of why these things work remains quite weak.  There are  engineering questions about what practical problems can be solved with confidence by such systems, and ethical questions about how humanity will interact with these machines.   The successes of AI even have led to some to suggest that the physicists' notions of understanding might themselves be superseded.   In opposition to this,  many physicists are hopeful that ideas from statistical mechanics will help us build a better understanding of modern AI \cite{mehta+al_19,carleo+al_19,roberts+yaida_22}.

In a different direction, many physicists have been interested in more explicitly dynamical models of neural networks   \cite{vogels+al_05}, as in Eq (\ref{dyn_net1}).  Guided by the statistical physics of disordered systems, one can study networks in which the matrix of synaptic connections is drawn at random, perhaps from an ensemble that captures some established features of real connectivity patterns. These same ideas can be used for probabilistic models of binary neurons;  notable developments include the development of a dynamical mean--field theory for these systems \cite{vreeswijk+sompolinsky_98}.

Against the background of these theoretical developments, there has been a revolution in the experimental exploration of the brain, driven by techniques that combine methods from physics, chemistry and biology.  We believe that this provides an unprecedented opportunity to connect statistical physics ideas to quantitative measurements on network dynamics in real brains.   We turn first to an overview of the experimental state of the art.

\section{New experimental methods}
\label{sec-expts}

Much of what we know about the brain has been learned by recording the electrical activity of one neuron at a time with metal microelectrodes.  If we have a theoretical framework in which interesting things happen through collective activity in the network, however, it is difficult to see how we could make progress without experimental methods for recording from many neurons simultaneously.\footnote{It's important that the ``order parameters'' in these theories are not simply the summed activity of all the neurons in the network, and hence don't correspond simply to something like the EEG.  If we take the Hopfield model as an example, then near its capacity the patterns of activity live in a space with dimensionality proportional to the size of the network itself (there are many order parameters), so there shouldn't really be any simple path to dimensionality reduction.}  Several groups were recording from pairs of neurons already in the 1960s, but systematic efforts to record from many neurons took until the 1980s.  

\begin{figure}
\includegraphics[width=\linewidth]{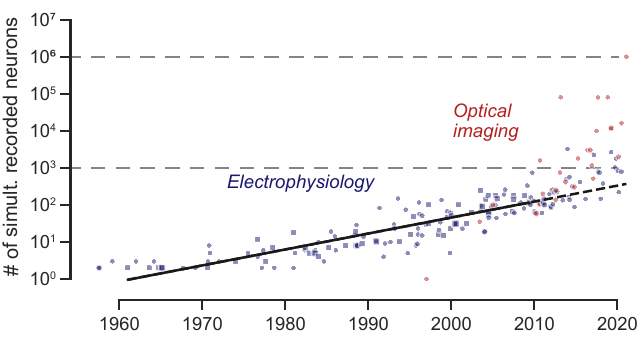}
\caption{Summarizing the growth in scale of neural recordings, adapted from \citet{urai+al_22}.   Number of neurons recorded simultaneously with electrodes and electrode arrays (blue);  squares from the earlier survey by \citet{stevenson+kording2011}.  Number of neurons recorded simultaneously with optical imaging methods (red).  Exponential growth for electrode recordings (black line), with a doubling time of $7.4\pm0.4\,{\rm yr}$.
\label{NvsYear}}
\end{figure}

For five decades we saw exponential growth in the number of neurons that can be monitored simultaneously with arrays of electrodes (Fig \ref{NvsYear}), with a doubling time of $7.4\pm0.4\,{\rm yr}$ \cite{stevenson+kording2011}.  Impressively, progress followed essentially the same pace over the last decade, so that $\sim10^3$ cells now are accessible almost routinely in many different brain areas and many different organisms; these developments are described in \S\S\ref{sec-arrays} and \ref{sec-neuropixels}.  This century also brought a fundamentally new technique, with animals genetically engineered so that their neurons produce fluorescent proteins with fluorescence intensity modulated by electrical activity (\S\ref{sec-imagingmethods});  these methods are approaching $\sim10^6$ neurons \cite{demas+al2021,manley2024simultaneous}. This progress creates new challenges for data analysis, but more deeply new opportunities for testing once speculative theories.  These developments also have a beauty of their own that we hope to capture here.  

Before we begin, note that as methods diversified, ``recording from $N$ neurons'' came to mean different things, so a simple plot of $N_{\rm max}$ vs time doesn't capture everything that is going on in these experiments.  These features of the experiments matter for theory, so we try to provide a guide.
We caution that we are theorists reviewing experimental developments, and references are meant to be illustrative rather than exhaustive.

\subsection{Electrode arrays}
\label{sec-arrays}

\begin{figure}[b]
\includegraphics[width = \linewidth]{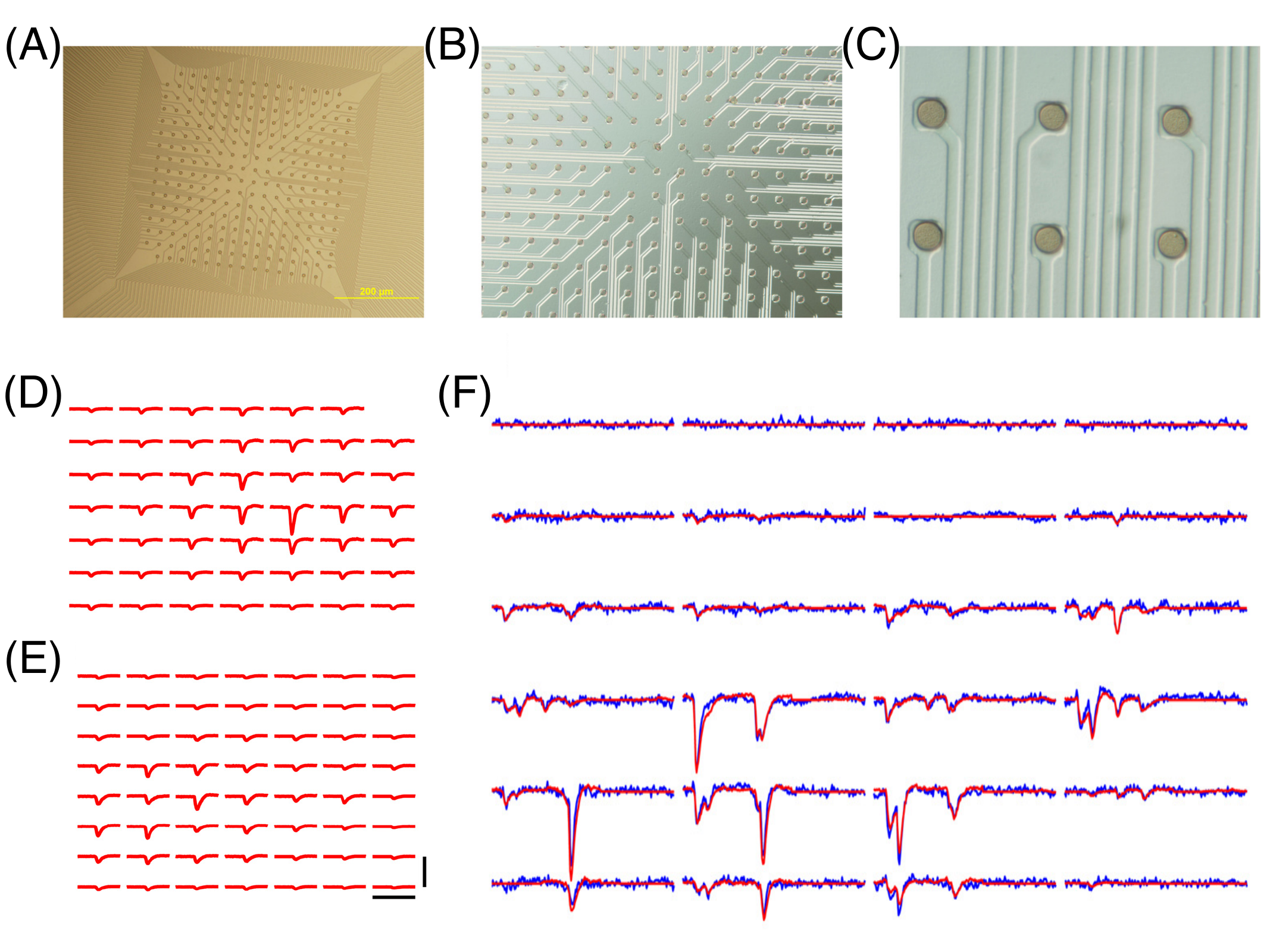}
\caption{Array of 252 electrodes for recording from the retina \cite{marre+al2012}. (A, B, C) Views of the electrode array at increasing magnification.  Distance between electrodes in $30\,\mu{\rm m}$.  (D, E) Examples of the stereotyped voltage traces---the templates $T_{n\alpha}(\tau )$ in Eq (\ref{spikesort1})---associated with two different cells.  Scale bars are $6.5\,{\rm ms}$ and $200\,\mu{\rm V}$.  (F)  Raw voltage traces (blue) and reconstruction by superposing templates as in Eq (\ref{spikesort1}).  Snippets are $20\,{\rm ms}$ in duration. \label{retina_array}}
\end{figure}

Rather studying neurons in an intact brain, one can culture the cells in a dish, allowing them to connect into a network.  In $\sim$1980, it was appreciated that the culture dish could be instrumented with an array of electrodes, giving access to the electrical activity of many if not all of the neurons in such artificial networks \cite{pine+gilbert1982}.  Most of the brain is 3D, so this doesn't generalize, but the retina can be quite flat, at least locally.  Placing a patch of a dissected retina onto an array of electrodes gives access to the ``ganglion cells'' that carry information from the eye to the brain and come together to form the optic nerve \cite{meister+al1994}.  Techniques progressed from recording a handful of cells simultaneously to arrays that can capture tens and eventually hundreds, as in  Fig \ref{retina_array}A--C \cite{segev+al2004,litke+al2004,marre+al2012}.  In some cases it is possible to achieve electrode densities high enough to record not just from large numbers ($100+$)  of ganglion cells but from a large fraction of the ganglion cells in a small patch of the retina, so we can access everything that the brain ``sees'' about a small patch of the visual world.  These sorts of experiments have become routine, in retinas from salamanders, from mice, and from primates whose visual systems are much like our own.  There are efforts to scale up to recording from $1000+$ cells in this way \cite{tsai2017very}

In electrode arrays, each electrode picks up signals from multiple cells and each cell appears on multiple electrodes.  Thus there is a deconvolution problem, referred to as ``spike sorting.'' This can be solved because the spikes generated by individual neurons are stereotyped.  Concretely this means that we can write the voltage $v_{n}(t)$ on the  $n^{\rm th}$ electrode as a sum of terms contributed by action potentials from cell $\alpha$ at times $t_{\rm i}^\alpha$,
\begin{equation}
v_{n}(t) = \sum_\alpha\sum_{\rm i} T_{n\alpha}(t-t_{\rm i}^\alpha) + \eta_n(t) ,
\label{spikesort1}
\end{equation}
where the $T_{n\alpha} (\tau )$ are ``templates'' that express how cells appear at electrodes and $\eta_n(t)$ is residual noise (Fig~\ref{retina_array}D--F).   In outline, one can learn these templates by finding candidate spike events that stand well above the background noise, clustering these, using the cluster centers as matched filters to identify more candidate spikes, and iterating.  There are many challenges in turning this outline in a working algorithm; for the multi--electrode arrays used in recording from the retina, see the discussions by \citet{prentice+al_11} and \citet{marre+al2012}.    An important test of spike sorting is that spikes from a single neuron should never come closer in time than a refractory period of $\sim 1\,{\rm msec}$.

Before searching for collective behaviors in the population of neurons, experiments with multi--electrode arrays provide an efficient way of exploring the properties of many individual cells.  Neurons throughout the brain can be divided into cell types, with different types exhibiting, for example, different responses to sensory inputs, different three--dimensional structures, and more recently different patterns of gene expression.  The retina is a classic example, with classification based on structure dating back to the classic work of \citet{Cajal1893}.  Electrode arrays provide a direct view of how cells of a particular type tile the retina in a lattice, and how the lattices of different cell types interdigitate \cite{field+chichilnisky_07,roy+al_21}.\footnote{Although generally forming a lattice, the regions of the visual world to which individual cells respond (``receptive fields'') can be quite irregular.  Experiments using the electrode arrays also show that the irregularities in the receptive fields of neighboring cells are coordinated, so that they interlock and provide more uniform coverage of the visual world \cite{gauthier+al_09}.   For a theoretical discussion see \citet{liu+al_09}.}   In addition to classification based on their responses to visual inputs, the templates $T_{n\alpha }(\tau)$ derived from spike sorting can be thought of as  ``electrical images'' of each cell, and these images also aid in the classification of neural cell types \cite{wu+al_23}.

\subsection{Multiple electrodes in 3D}
\label{sec-neuropixels}

A different approach is to insert multiple electrodes deep into brain tissue, which also has a long history.  Where classical experiments brought a metal tip as close as possible to a single neuron, it was appreciated that multiple closely spaced tips, e.g. with wires twisted into a stereotrode or tetrode, could resolve multiple neurons from a small volume \cite{mcnaughton+al_83,wilson+mcnaughton_93}. The introduction of methods from semiconductor fabrication made it possible to build arrays of 100 silicon needles that could be inserted into the cortex \cite{jones+al_92}.  

\begin{figure}[t]
\includegraphics[width = \linewidth]{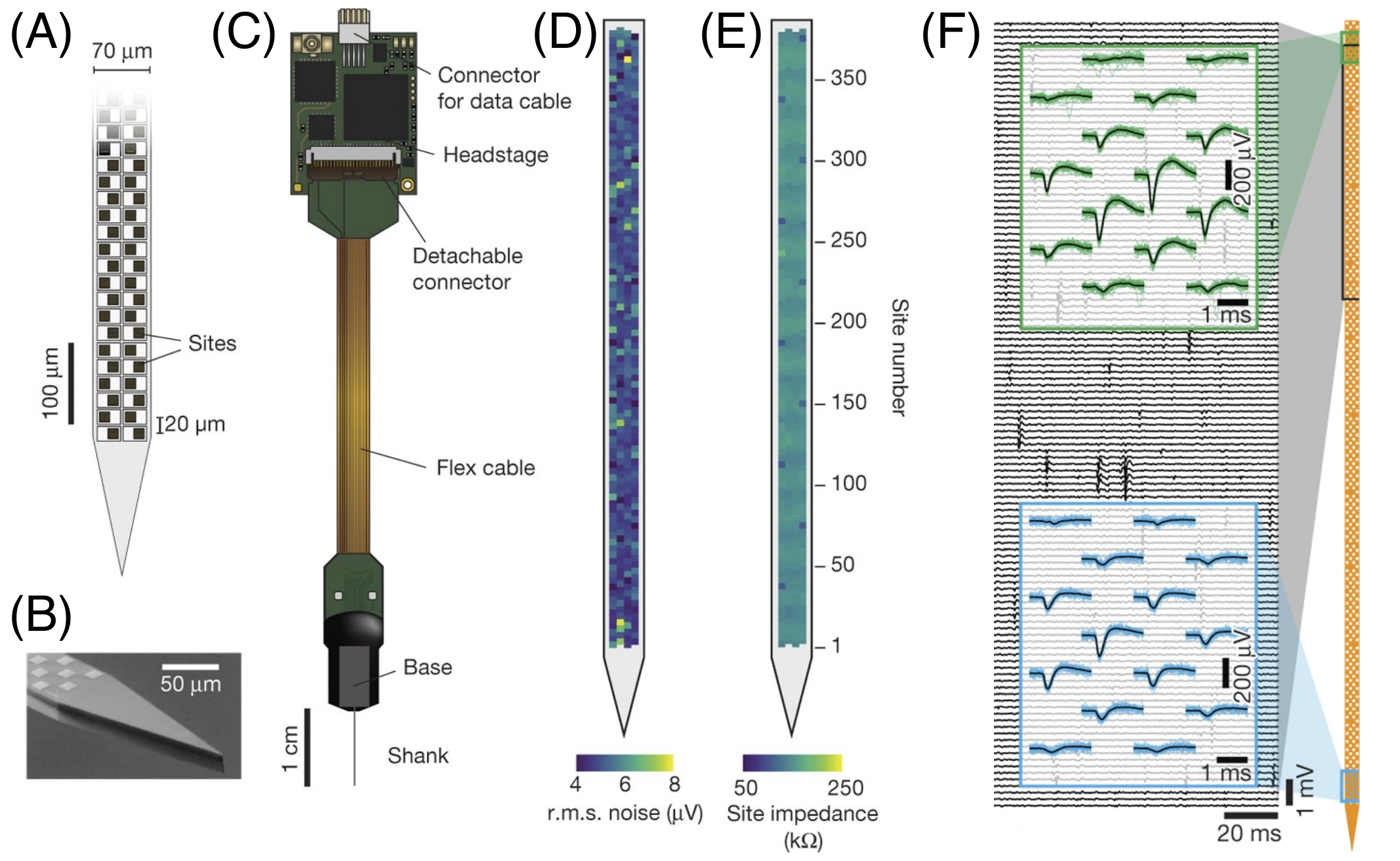}
\caption{The ``neuropixel'' probe, with 384 electrodes arrayed along a single  shank \cite{jun+al_17}. (A) Schematic of probe tip, showing checkerboard layout of active electrode sites. (B)   Scanning electron microscope image of probe tip. (C) Probe packaging, including flexible cable and headstage electronics for  data transmission. (D)  Example of root--mean--square voltage noise levels in a bandwidth that captures the action potentials; 
$\delta V_{\rm rms} = 5.1 \pm 0.6 \, \mu{\rm V}$. (E) Typical site impedance in saline,   measured for each site with sinusoidal $1\,{\rm nA}$ injected currents at $1\,{\rm  kHz}$; $Z =  149\pm 6\,{\rm k}\Omega$.
(F) A short segment of raw voltage recordings in the mouse brain.  Insets show the short snippets from multiple nearby electrodes that are identified as spikes from the same neuron, with $30$ waveforms superposed to illustrate the stereotypy of these signals.  The angle with which the shank penetrated the brain was chosen to sample many different ares; upper electrodes are in the motor cortex, lower electrodes in the dorsal tenia tecta.
\label{neuropixels}}
\end{figure}

Jumping ahead two decades, further miniaturization has led to integrated arrays of multiple electrodes along a single shaft coupled with pre--processing electronics as illustrated in Fig \ref{neuropixels} \cite{jun+al_17}.  The most recent such devices have 1000+ sensors along a single probe, capable of resolving hundreds of individual neurons \cite{steinmetz2021neuropixels}.  Although it is most common to deploy these arrays in studies on rodent brains, they can also be adapted to primates, where comparisons to the human brain are easier \cite{trautmann+al2023}. Alternative methods make use of polymer materials for flexible electrodes \cite{chung2019high}.  In particular these allow very long term recordings, monitoring the same neurons over weeks or months, e.g. as the animal learns \cite{zhao+al2023}.   Importantly all these methods, as with classical single neuron recordings, provide access to the full stream of action potentials generated by each neuron, down to millisecond precision.  

Although our emphasis here is on basic scientific questions, an important stimulus for continued development of these techniques is their potential for clinical applications.  In particular there is the program of constructing ``brain computer interfaces,'' where electrode arrays monitor the activity of many neurons in motor cortex and these signals are decoded to generate commands e.g. for a robot arm or cursor \cite{serruya+al_02,carmena+al_03,taylor+al_02,musallam+al_04}.  More recently these techniques have emerged from the laboratory to experimental treatments of humans   \cite{hochberg+al_12,willett+al_21}.  This is a rapidly developing field, in which not only experimental methods but also our theoretical understanding of neural dynamics and coding is contributing to practical medical goals.

Versions of these tools have been commercialized, leading to an explosive increase is large scale experiments across a wide range of brain regions in many different animals; a snapshot of this activity can be found in \citet{steinmetz2018challenges}.  As with the electrode arrays in \S\ref{sec-arrays}, signals from individual neurons appear at multiple electrodes and individual electrodes pick up multiple neurons, so there is a problem of spike sorting.  With thousands of neurons, this problem is on a much larger scale than before, and there is a particular drive to have fully automated methods \cite{chung+al_17}. Progress continues, but the problem is not fully solved.  We would add that different analyses are sensitive to different systematic errors in the sorting process.

\subsection{Imaging methods}
\label{sec-imagingmethods}

It is an old idea that we might be able to see the electrical activity of neurons, literally.  The first implementation was with voltage sensitive dyes that insert into the cell membrane and have optical properties (absorption or fluorescence) that shift in response to the large electric fields associated with the action potential  \cite{cohen+salzberg_78}.   The exploration of the brain (and living systems more generally) was revolutionized by the discovery that there are proteins which are intrinsically fluorescent, without the need for cofactors \cite{shinomura+al_62,johnson+al_62}.  These proteins were then tuned, by changing their amino acid sequences, to have different colors as well as fluorescence that responds to environmental signals \cite{tsien_09}.  Decades after their initial discovery, genetic engineering allowed the insertion of these sequences into the genome \cite{prasher+al_92,chalfie+al_94}, placing them under the control of regulatory elements that are active in neurons or even in restricted classes of neurons.

\begin{figure}[b]
	\includegraphics[width=\linewidth]{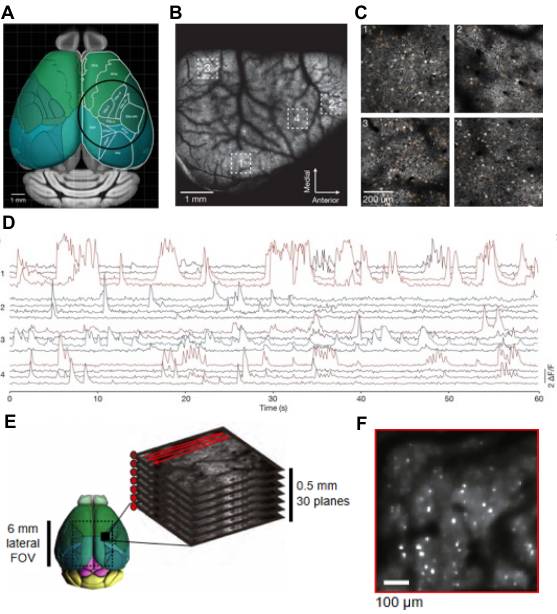}
	\caption{Large scale imaging of neural activity. (A) Schematic of the dorsal surface of the mouse cortex. The overlaid circle corresponds to the field of view  when imaging with via the ``mesoscope" setup ($5\,{\rm mm}$ radius). (B) Mesoscope image from a mouse brain expressing the fluorescent calcium indicator protein GCaMP6f. (C) Four fields of view (indicated in B), with regions of interest (orange) drawn manually around individual neurons. (D) Normalized fluorescence activity traces for 16 neurons extracted from the four regions in (B). Sampling rate $9.6\,{\rm Hz}$. (E) An example field of view  for volumetric imaging via Light Beads Microscopy,  which recently enabled monitoring $\sim10^6$ neurons simultaneously.   The fast pulse structure of the laser source is used to produce ``beads'' (red dots) at different depths across $0.5\,{\rm mm}$, and this then is scanned laterally, enabling volumetric recording. (F) The standard deviation of fluorescence in an imaging plane $183\, \mu{\rm m}$  below the cortical surface in a mouse brain expressing GCaMP6f. Panels (A--D) adapted from \citet{sofroniew2016large}, (E, F) from \citet{manley2024simultaneous}.  \label{imaging}}
\end{figure}

Taking inspiration from voltage--sensitive dyes, the ideal would be to have a genetically encoded, fluorescent membrane protein that responds directly to the voltage across the membrane. There is continuing progress toward this goal  \cite{Jin+al2012,Abdelfattah+al2019, villette2019, platisa+al2023},  but current indicator molecules are not quite sufficient for long term recordings from large populations of neurons.
What we do have are fluorescent proteins that respond to changes in intracellular calcium concentration, which provides a slightly indirect, low--pass filtered trace of electrical activity; these now are widely used \cite{tian2012neural, Chen+al_2013,ZhangLooger2023GCamp8}.  To make the most of these signals requires sophisticated microscopy, such as scanning two--photon methods \cite{helmchen2005deep}. With these tools we can observed reasonably large areas of the brain with single cell resolution, as in Fig \ref{imaging}A--C. In addition, there now are engineered  proteins that insert into the membrane and act as light--gated channels, making it possible to inject controlled pulses of current it individual neurons both to excite and inhibit these cells through optical control \cite{rickgauer2014simultaneous,packer+al_15}.
 
 In many regions of the brain, we do not see the full dynamical behavior of neuronal networks unless the animal is engaged in behavior.  Evidently having the ``sample'' moving and behaving is in tension with high--resolution microscopy.  One solution is to miniaturize the microscope so that the animal can carry the instrument as it moves through its environment \cite{ziv+al2013, zong2017fast}.  Alternatively one can hold the animal's head fixed under a stationary microscope but allow it to run on rotating ball, using the movement of the ball to compute how the animal would have moved through the environment. This computed trajectory is then used to generate virtual reality \cite{harvey2009intracellular, dombeck2010functional};  an example of a virtual reality setup is shown in Fig \ref{setup}B below. It is possible to simulate not just the animal's visual experience of running through the world, but even its olfactory experience \cite{radvansky2018olfactory}.
 
Imaging methods allow flexible tradeoffs among spatial resolution, temporal resolution, the area over which one records, and the signal--to--noise ratio for each individual cell.  Importantly, as seen in Figs~\ref{imaging}D and \ref{setup}C, there is a regime in which the transient periods of neural activity stand out well above the background noise of the measurements from individual cells. If the aim is to record simultaneously from as many neurons as possible, one can reach ``every neuron in the brain'' of smaller animals, such as larval zebrafish, at the expense of visiting each neuron rather infrequently \cite{ahrens2013whole}. More generally it is possible to combine methods, providing single cell recordings at high time resolution while monitoring a much larger area of the brain at lower resolution \cite{barson2020simultaneous}.

A special case is the small worm {\em Caenorhabditis elegans}, which has only 302 neurons in total; as in many invertebrates these neurons have names and numbers and thus are identifiable across individuals.  {\em C.~elegans} was the first organism in which the pattern of synaptic connectivity was traced at electron microscope resolution \cite{white+al_86},  and this ``connectome'' has been revisited with modern methods \cite{varshney+al_11,cook+al_19}.  The worm is largely transparent, so that optical methods can be used directly to monitor and drive neural activity without dissection, even in freely moving worms \cite{leifer+al_11}.   Recordings from 100+ neurons in this system reflect a macroscopic fraction of all the neurons, so that we are approaching ``whole brain'' imaging with single cell resolution \cite{nguyen+al2016}. The neurons in {\em C.~elegans} do not generate the discrete, stereotyped action potentials that are familiar in other organisms, so the graded fluorescence signals in imaging experiments are a more direct correlate of slower, continuous electrical dynamics. Advances in experimental technique make it possible to identify neurons as their activity is monitored, placing them in the context of the known connectivity, and the combination of recording and stimulation has resulted ``pump--probe'' measurements that map the functional connections between 10,000+ pairs of cells \cite{randi+al_23}.   These data provide the opportunity to formulate and test more global theoretical ideas about network dynamics.
 
If we want to visit each neuron often enough to make full use of the time resolution  allowed by the calcium response of the fluorescent proteins, then there will be limits on the number of neurons that can be monitored.  Scanning in two dimensions one can now reach 1000+ neurons, as in the example discussed at length in \S\S\ref{sec-subgroups} and \ref{chapter-RG}.  Scanning in depth poses additional challenges \cite{weisenburger2019volumetric,Zhang+al2021}, but new ``light bead'' methods make use of the very short time scale of laser pulses to  collect from multiple depths almost simultaneously, as shown in Figs \ref{imaging}E and F \cite{demas+al2021, manley2024simultaneous}.    These methods are pushing toward monitoring one million cells.

The raw data from an imaging experiment is a movie: fluorescence intensity vs time in each of $\sim 10^6$ pixels.  What we want are signals labelled by the cells that generate them, not by pixels.  This involves two essential steps: discarding all changes in light intensity that result from sources other than electrical activity (primarily motion of the brain), and grouping together the pixels that belong to each cell.  In many cases these steps need to be done in three dimensions, combining signals from a ``z--stack'' in which the microscope's plane of focus has been stepped through the thickness of the brain region under study.  These are challenging problems in data analysis, and a wide range of mathematical and algorithmic ideas have been brought to bear:  local correlations \cite{SmithHäusser2010}, dictionary learning \cite{Pachitariu2013}, graph-cut related algorithms \cite{Kaifosh2014}, independent component analysis \cite{Mukamel2009}, and non--negative  matrix factorization  \cite{Maruyama2014}.

The fact that neurons generate discrete action potentials means that if we look in small time bins the natural variables are binary, inviting a connection to Ising models.  Calcium--sensitive indicators do not give us direct access to the time resolution that is needed for this binary description.  There are several efforts to reconstruct the $\sim{\rm msec}$ spikes that underlie the $\sim100\,{\rm msec}$ calcium signals, but we suspect that these will be  overtaken by advances in engineering directly voltage--sensitive proteins.  An alternative, which we use below, is to discretize the calcium signals, admitting that the resulting binary variables necessarily refer to ``active'' and ``inactive'' states of the cell rather than to the presence or absence of action potentials (Fig \ref{setup}C).

\subsection{Perspectives}

Experimental methods for monitoring the electrical activity of neurons continue to evolve rapidly.  It is interesting to look ahead, and make some predictions about where the methods will be in five or ten years.  Again we caution that we are theorists surveying the state of experiments.

In recordings based on electrodes and electrode arrays we can expect two major trends. The first is better coverage and higher sampling density.  It is tempting to focus on the largest scale experiments as these are perhaps the most tantalizing opportunities to test the applicability of statistical physics ideas. In practice, however, more neurons often come at the expense of lower sampling density, which matters deeply for comparison with theory (e.g. \S\ref{sec-density}), so one would like to be careful. We expect that the push for ``whole brain'' coverage soon will by complemented by a push for denser sampling:  instead of choosing between high density sampling in a small region, often in 2D, or sparse sampling of much larger areas in 3D, experiments will get much closer to recording every neuron in progressively larger volumes.   The second trend is toward longer duration recordings, with chronic presence of electrodes in the animal brain. Recent efforts have provided proof of concept for recordings that last for weeks; we expect this to become more routine, reaching toward  experiments that last months or even years.   The central challenge is verifying that we are monitoring the exact same set of cells throughout the entire recording. The big advantage, of course, is that the animal can be monitored in its home cage, in different environments, at different times of the day, as it engages in a fuller range of behaviors.   The longest time scale recordings will give a unique view of neural dynamics during learning.

On the optical front, the growth in number of neurons that we can (literally) see simultaneously has recently accelerated dramatically, as seen clearly in  Fig \ref{NvsYear}.  Faster, more selective scanning is in the works, which should allow more imaging techniques to reach the realm of $\sim 10^6$ neurons, with improved signal--to--noise ratio. Currently, when imaging $10^5-10^6$ neurons, the loss of temporal resolution is significant, with a drop to acquisition rates below $10\,{\rm Hz}$.
As with electrodes where sampling density in space matters, here it is the sampling density in time that can be problematic.  There are  tradeoffs among speed, number of neurons, the signal--to--noise ratio in each neuron, and total amount of optical power delivered to the brain, but these are specific to each imaging modality and we can hope for progress. Another intriguing direction is selective acquisition; following methods used in astrophysics, if we can concentrate on the exact locations of the neurons, we can scan more quickly and use the same number of photons more efficiently. Additionally, there is steady improvement in methods to express both indicators and light--gated channels in the same cells, often targeting specific classes of cells.   This will bring to larger animals the kind of complete survey of functional connectivity that currently is possible only in {\em C.~elegans}  \cite{randi+al_23}, as well as making it possible to probe causal connections between neural activity and motor output.

Finally, a significant breakthrough would be if voltage-sensitive fluorescent proteins become fully viable.
The demands are severe: proteins must respond on a millisecond time scale, with large amplitude changes in fluorescence, and cells must be programmed to insert these proteins into the membrane. When this happens, it will become possible to monitor thousands to millions of neurons with a resolution where we see every individual action potential, giving us the precision of electrodes and the survey capacity of optical imaging.

\section{Maximum entropy as a path to connect theory and experiment}
\label{sec-max_ent}

New experimental methods create new opportunities to test our theories.  For neural networks, monitoring the electrical activity of tens, hundreds, or thousands of neurons simultaneously should allow us to test statistical approaches to these systems in detail. Doing this requires taking much more seriously the connection between our models and real neurons, a connection that sometimes has been tenuous.  Can we really take the spins $\sigma_{\rm i}$ in Eq (\ref{JJHE}) to represent the presence or absence of an action potential in cell $\rm i$?
We will indeed make this identification, and our goal  will be an accurate description of the probability distribution out of which the ``microscopic'' states of a large network are drawn.  Note that, as in equilibrium statistical mechanics, this would be the beginning and not the end of our understanding. 

We will see that maximum entropy models provide a path that starts with data and constructs models that  have a very direct connection to statistical physics.   Our focus here is on networks of neurons, but it is important that the same concepts and methods are being used to study a much wider range of living systems, and there are important lessons to be drawn from seeing all these problems as part of the same project (Appendix \ref{sec-sequences+}).

\subsection{Basics of maximum entropy}
\label{sec-basics}

Consider a network of neurons, labelled by ${\rm i} =1,\, 2,\, \cdots ,\, N$, each with a state $\sigma_{\rm i}$.  In the simplest case where these states of individual neurons are binary---active/inactive, or spiking/silent---then the network as a whole has access to $\Omega = 2^N$ possible states 
\begin{equation}
\bm{\sigma} \equiv \{\sigma_1,\, \sigma_2,\, \cdots ,\, \sigma_N \}.
\end{equation}
These states mean something to the organism:  they may represent sensory inputs, inferred features of the surrounding world, plans, motor commands, recalled memories, or internal thoughts.  But before we can build a dictionary for these meanings we need a lexicon, describing which of the possible states actually occur, and how often.    More formally, we would like to understand the probability distribution $P(\bm{\sigma})$.  We might also be interested in sequences of states over time, $P[\{\bm{\sigma}(t_1),\, \bm{\sigma}(t_2),\,\cdots \}]$, but for simplicity we focus first on states at a single moment in time.

The distribution $P(\bm{\sigma})$ is a list of $\Omega$ numbers that sum to one.  Even for modest size networks this is a very long list, $\Omega \sim 10^{30}$ for $N= 100$.   To be clear, there is no way that we can measure all these numbers in any realistic experiment.  More deeply, large networks could not visit all of their possible states in the age of the universe, let alone the lifetime of a single organism.  This shouldn't bother us, since one can make similar observations about the states of molecules in the air around us, or the states of all the atoms in a tiny grain of sand.  The fact that the number of possible states $\Omega$ is (beyond) astronomically large does not stop us from asking questions about the distribution from which these states are drawn.

The enormous value of $\Omega$ does mean, however, that answering  questions about the distribution from which the states are drawn requires the answer to be, in some sense, simpler than it could be.  If $P(\bm{\sigma})$ really were just a list of $\Omega$ numbers with no underlying structure, we could never make a meaningful experimental prediction.  Progress in the description of many--body systems depends on the discovery of some regularity or simplicity, and without such simplifying hypotheses nothing can be inferred from any reasonable amount of data.   The maximum entropy method is a way of being explicit about our simplifying hypotheses.

We can imagine mapping each microscopic state $\bm{\sigma} $ into some perhaps more macroscopic observable $f(\bm{\sigma})$, and from reasonable experiments we should be able to estimate the average of this observable $\langle f(\bm{\sigma})\rangle_{\rm expt}$.  If we think this observable is an important and meaningful quantity, it makes sense to insist that any theory we write down for the distribution $P(\bm{\sigma})$ should predict this expectation value correctly, 
\begin{equation}
\langle f(\bm{\sigma} )\rangle_{P} \equiv \sum_{\bm{\sigma}} P(\bm{\sigma})f(\bm{\sigma} ) = \langle f(\bm{\sigma} )\rangle_{\rm expt} .
\end{equation}
There might be several such meaningful observables, so we should have
\begin{equation}
\langle f_\mu (\bm{\sigma} )\rangle_{P} \equiv \sum_{\bm{\sigma}} P(\bm{\sigma})f_\mu(\bm{\sigma} ) = \langle f_\mu(\bm{\sigma} )\rangle_{\rm expt} 
\label{constraints}
\end{equation}
for $\mu = 1,\, 2,\, \cdots ,\, K$.  These are strong constraints, but so long as the number of these observables $K \ll \Omega$ there are infinitely many distributions consistent with Eq (\ref{constraints}).  How do we choose among them?

There are many ways of saying, in words, how we would like to make our choice among the $P(\bm{\sigma}) $ that are consistent with the measured expectation values of observables.  We would like to pick the simplest or  least structured model.  We would like not to inject into our model any information beyond what is given to us by the measurements $\{\langle f_\mu(\bm{\sigma} )\rangle_{\rm expt} \}$.  From a different point of view, we would like drawing states out of the distribution $P(\bm{\sigma}) $ to generate samples that are as random as possible while still obeying the constraints in Eq (\ref{constraints}).  It might seem that each choice of words generates a new discussion---what do we mean, mathematically,  by ``least structured,''  or ``as random as possible''?

Introductory courses in statistical mechanics make some remarks about entropy as a measure of our ignorance about the microscopic state of a system, but this connection often is left quite vague.  In laying the foundations of information theory, Shannon made this connection precise  \cite{shannon1948mathematical}. If we ask a question, we have the intuition that we ``gain information'' when we hear the answer.  If we want to attach a number to this information gain, then the {\em unique} measure that is consistent with natural constraints is the entropy of the distribution out of which the answers are drawn.  Thus, if we ask for the microscopic state of a system, the information we gain on hearing the answer is (on average) the entropy of the distribution over these microscopic states.  Conversely, if the entropy is less than its maximum possible value, this reduction in entropy measures how much we already know about the microscopic state even before we see it.  As a result, for states to be as random as possible---to be sure that we do not  inject extra  information about these states---we need to find the distribution that has the maximum entropy.

Maximizing the entropy subject to constraints defines a variational problem, maximizing
\begin{widetext}
\begin{equation}
\tilde S = -\sum_{\bm{\sigma}} P(\bm{\sigma})\ln P(\bm{\sigma}) -\sum_{\mu =1}^K  \lambda_\mu \left[\sum_{\bm{\sigma}}  P(\bm{\sigma})f_\mu(\bm{\sigma} ) -\langle f_\mu(\bm{\sigma} )\rangle_{\rm expt} \right] - \lambda_0 \left[ \sum_{\bm{\sigma}} P(\bm{\sigma})  -1 \right] ,
\label{maxentproblem}
\end{equation}
\end{widetext}
where the $\lambda_\mu$ are Lagrange multipliers. We include an additional term ($\propto \lambda_0$) to constrain the normalization, so we can treat each entry in the distribution as an independent variable.  Then 
\begin{eqnarray}
{{\delta \tilde S }\over{\delta P(\bm{\sigma})}} &=& 0 \\
\Rightarrow P(\bm{\sigma}) &=& {1\over {Z(\{\lambda_\mu\})}} \exp\left[ - E(\bm{\sigma})\right]
\label{maxent1}\\
E(\bm{\sigma}) &=& \sum_{\mu =1}^K  \lambda_\mu f_\mu(\bm{\sigma} ) .\label{maxent2}
\end{eqnarray}
Thus the model we are looking for is equivalent to an equilibrium statistical mechanics problem in which the ``energy'' is a sum of terms, one for each of the observables whose expectation values we constrain; the Lagrange multipliers become coupling constants in the effective energy. To finish the construction we need to adjust these couplings $\{\lambda_\mu\}$ to satisfy Eq (\ref{constraints}), and in general this is a hard problem; see Appendix \ref{sec-inference}.   Importantly, if we have some set of expectation values that we are matching, and we want to add one more, this just adds one more term to the {\em form} of the energy function, but in general implementing this extra constraint requires adjusting all the coupling constants.

To make the connections explicit, recall that we can define thermodynamic equilibrium as the state of maximum entropy given the constraint of fixed mean energy.  This optimization problem is solved by the Boltzmann distribution.  In this view the (inverse) temperature is a Lagrange multiplier that enforces the energy constraint, opposite to usual view of controlling the temperature and predicting the energy.  The Boltzmann distribution generalizes if other expectation values are constrained \cite{landau+lifshitz}.

The maximum entropy argument gives us the form of the probability distribution, but we also need the coupling constants.  We can think of this as being an ``inverse statistical mechanics'' problem, since we are given expectation values or correlation functions and need to find the couplings, rather than the other way around.   Different formulations of this problem have a long history in the mathematical physics community \cite{keller+zumino1959,kunkin+frisch_69,chayes+al_84}.  An early application to living systems involved reconstructing the forces that hold together the array of gap junction proteins which bridge the membranes of two cells in contact \cite{braun+al1984}.  As attention focused on networks of neurons, finding the relevant coupling constants came to be described as the ``inverse Ising'' problem, as will become clear below.

In statistical physics there is in some sense a force driving systems toward equilibrium, as encapsulated in the H--theorem.  In many cases this force triumphs, and what we see is a state with maximal entropy subject only to a very few constraints.  In the networks of neurons that we study here, there is no H--theorem, and the list of constraints will be quite long compared to what we are used to in thermodynamics.  This means that the probability distributions we write down will be mathematically equivalent to some equilibrium statistical mechanics problem, but they do not describe an equilibrium state of the system we are actually studying.  This somewhat subtle relationship between maximum entropy as a description of thermal equilibrium and maximum entropy as a tool for inference was outlined long ago by  \citet{jaynes1957information,jaynes1982rationale}.

If we don't have any constraints then the maximum entropy distribution is uniform over all $\Omega$ states.  Each observable whose expectation value we constrain lowers the maximum allowed value of the entropy, and if we add enough constraints we eventually reach the true entropy and hence the true distribution.  Often it make sense to group the observables  into one--body, two--body, three--body terms, etc..  Having constrained all the $k$--body observables for $k\leq K$, the maximum entropy model makes parameter--free predictions for correlations among groups of $k>K$ variables.  This provides a powerful path to testing the model, and defines a natural generalization of connected correlations \cite{schneidman+al2003}.

The connection of maximum entropy models to the Boltzmann distribution gives us intuition and practical computational tools.  It can also leave the impression that we are describing a system in equilibrium, which would be a disaster.  In fact the maximum entropy distribution describes thermal equilibrium {\em only} if the observable that we constrain is  the energy in the mechanical sense.  There is no obstacle to building maximum entropy models for the distribution of states in a non--equilibrium system.

Although we can usefully think of states distributed over an energy landscape, as we have formulated the maximum entropy construction this description works for  states at one moment in time.  Thus we cannot conclude that the dynamics by which the system moves from one state to another are analogous to Brownian motion on the effective energy surface.  There are infinitely many models for the dynamics that are consistent with this description, and most of these will not obey detailed balance.  Recent work shows how to explore a large family of dynamical models consistent with the maximum entropy distribution, and applies these ideas to  collective animal behavior \cite{chen+al_23}.  There also are generalizations of the maximum entropy method to describe distributions of trajectories, as we discuss below (\S\ref{sec-more+less}); maximum entropy models for trajectories sometimes are called maximum caliber \cite{presse+al2013,ghosh+al_20}.  Finally we note that, for better or worse, the symmetries that are central to many problems in statistical physics in general are absent from the systems we will be studying;  flocks and swarms are an exception, as discussed in \S\ref{app-flocks}.

 To conclude this introduction, we emphasize that maximum entropy is unlike usual theories.  We don't start with a theoretical principle or even a model.  Rather, we start with some features of the data and test the hypothesis that these features alone encode everything we need to describe the system. 
 Whenever we use this approach we are referring back to the basic structure of the optimization problem defined in Eq (\ref{maxentproblem}), and its formal solution in Eqs (\ref{maxent1}, \ref{maxent2}),  but  there is no single maximum entropy model, and each time we need to be explicit:  Which are the observables $f_\mu$ whose measured expectation values we want our model to reproduce?  Can we find the corresponding Lagrange mutlipliers $\lambda_\mu$?  Do these parameters have a natural interpretation?  Once we answer these questions, we can ask whether these relatively simple statistical physics descriptions make predictions that agree with experiment.  There is an unusually clean separation between learning the model (matching observed expectation values) and testing the model (predicting new expectation values).  In this sense we can think of maximum entropy as predicting a set of {\em parameter free} relations among different aspects of the data.  Finally,  we will have to think  carefully about what it means for models to ``work.''  We begin with early explorations at relatively small $N$ (\S\ref{sec-neuron1}), then turn to a wide variety of larger networks (\S\ref{sec-larger}), and finally address how these analyses can catch up to the experimental frontier (\S\ref{sec-more+less}).

\subsection{First connections to neurons}
\label{sec-neuron1}

Suppose we observe three neurons, and measure their mean activity as well as their pairwise correlations.  Given these measurements, should we be surprised by how often the three neurons are active together?  Maximum entropy provides a way of answering this question, generating a ``null model'' prediction assuming all the correlation structure is captured in the pairs, and this was appreciated $\sim$2000 \cite{martignon+al2000}.  Over the next several years a more ambitious idea emerged:  could we build maximum entropy models for patterns of activity in larger populations of neurons?  The first target for this analysis was a population of neurons in the salamander retina, as it responds to naturalistic visual inputs \cite{schneidman2006weak}. 

\begin{figure}[t]
\includegraphics[width=\linewidth]{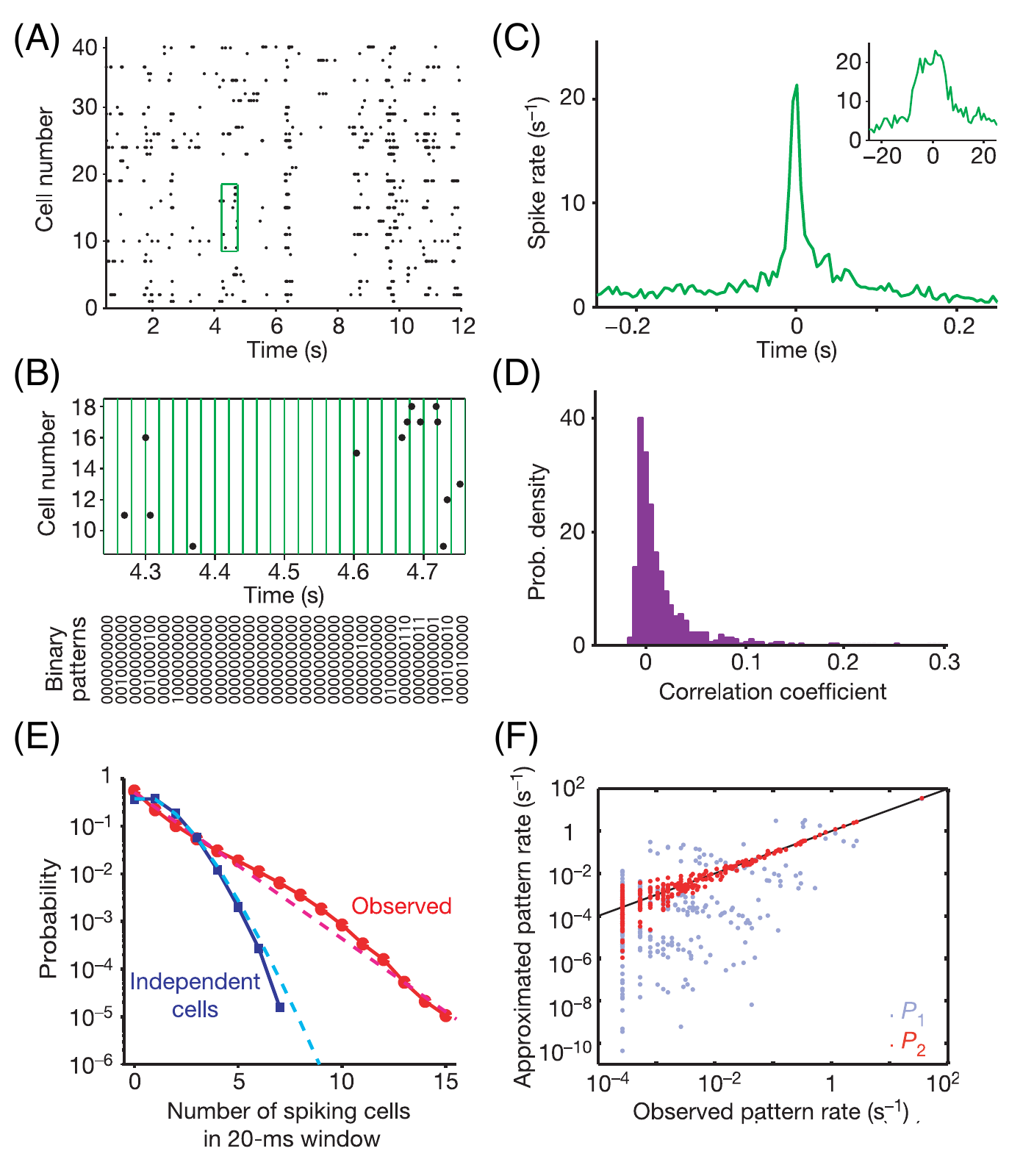}
\caption{Responses of the salamander retina to naturalistic movies \cite{schneidman2006weak}. (A) Raster plot of the action potentials from $N=40$ neurons.  Each dot represents a spike from one cell. (B) Expanded view of the green box in (A), showing the discretization of time into bins of width $\Delta\tau = 20\,{\rm ms}$.  The result (bottom) is that the state of the network is a binary word $\{\sigma_{\rm i}\}$. (C) Correlations between two neurons.  Results are shown as the probability per unit time of a spike in cell $\rm j$ (spike rate) given that there is a spike in cell $\rm i$ at time $t=0$; the plateau at long times should be the mean rate $r_{\rm j} = \langle\sigma_{\rm j}\rangle/\Delta\tau$.  There a peak with a width $\sim 100\,{\rm ms}$, related to time scales in the visual input, and a peak with width $\sim 20\,{\rm ms}$ emphasizes in the inset; this motivates the choice of bins size. (D) Distribution of (off--diagonal) correlation coefficients, from Eq (\ref{corr-def}), across the population of $N=40$ neurons. (E) Probability that $K$ out of the $N=40$ neurons are active in the same time bin (red) compared with expectations if activity of each neuron were independent of all the others (blue).  Dashed lines are exponential (red) and Poisson (blue), to guide the eye.  (F) Predicted occurrence rates of different binary patterns vs the observed rates, for the  independent model $P_1$ [Eqs (\ref{eq-E1}, \ref{eq-P1}), blue] and the pairwise maximum entropy model $P_2$ [Eqs (\ref{eq-E2}, \ref{eq-P2}),  red].
\label{salamander1}}
\end{figure}

In response to natural movies, the output neurons of the  retina---the ``ganglion cells'' that carry visual signals from eye to brain, and which as a group form the optic nerve---are sparsely activated, generating an average of just a few spikes per second each (Fig \ref{salamander1}A, B).  Those initial experiments monitored populations of up to forty neurons in a small patch of the retina, with recordings of up to one hour.  Pairs of neurons have temporal correlations with a relatively sharp peak or trough on a broad background that tracks longer timescales in the visual input (Fig \ref{salamander1}C).  If we discretize time into bins of $\Delta\tau = 20\,{\rm ms}$ then we capture most of the short time correlations but still have a very low probability of seeing two spikes in the same bin, so that responses of neuron $\rm i$ become binary,\footnote{The literature is mixed in sometimes choosing $\sigma_{\rm i} = \pm 1$ and sometimes  $\sigma_{\rm i} = \{0,1\}$; this choice  is arbitrary.  Here we use the $\sigma_{\rm i} = \{0,1\}$ representation, which makes some things easier.  In real neurons active and inactive states emphatically are not symmetric, so the elegance of the familiar  $\sigma_{\rm i} = \pm 1$ is lost.}
 $\sigma_{\rm i} = \{0,1\}$.  

If we define as usual the fluctuations around the mean,
\begin{equation}
\delta\sigma_{\rm i} = \sigma_{\rm i} - \langle \sigma_{\rm i}\rangle ,
\end{equation}
 then the data sets were large enough to get good estimates of the covariance
 \begin{equation}
C_{\rm ij} = \langle \delta\sigma_{\rm i}\delta\sigma_{\rm j}\rangle  = \langle \sigma_{\rm i} \sigma_{\rm j}\rangle_c,
\label{cov-def}
\end{equation}
wheer $\langle\cdots\rangle_c$ denotes the connected part of the correlations; in many cases we have more intuition about   the correlation matrix
 \begin{equation}
\tilde C_{\rm ij} = {{C_{\rm ij}}\over\sqrt{C_{\rm ii} C_{\rm jj}}} .
\label{corr-def}
\end{equation}
Importantly, these pairwise correlations are weak:  almost all of the $|\tilde C_{{\rm i}\neq{\rm j}}| < 0.1$, and the bulk of these correlations are just a few percent  (Fig \ref{salamander1}D).  The recordings are long enough that these weak correlations are statistically significant, and almost none of the matrix elements are zero within errors.  Correlations thus are weak and widespread, which seems to be common across many different regions of the brain.

If we look just at two neurons, the approximation that they are independent of one another is very good, because the correlations are so weak.  But if we look more globally then the widespread correlations combine to have qualitative effects.  As an example, we can ask for the probability that $K$ out of $N=40$ neurons are active in the same time bin, $P_N(K)$, and we find that this has a much longer tail than expected if the cells were independent (Fig \ref{salamander1}E); simultaneous activity of $K=10$ neurons already is $\sim 10^3\times$ more likely than in the independent model.  

If we focus on $N=10$ neurons then the experiments are long enough to sample all $\Omega \sim 10^3$ states, and  the probabilities of these different binary words depart dramatically from the predictions of an independent model (Fig \ref{salamander1}F).  If we group the different binary words by the total number of active neurons, then the predictions of the independent model actually are {\em anti}--correlated with the real data.  We emphasize that these failures occur despite the fact that pairwise correlations are weak, and that they are visible at a relatively modest $N=10$.

If we want to build a model for the patterns of activity in networks of neurons it certainly makes sense to insist that we match the mean activity of each cell.   At the risk of being pedantic, what this means explicitly is that we are looking for a probability distribution over network states, $P_1\left(\bm{\sigma}\right)$ that has the maximum entropy while correctly predicting the expectation values
\begin{equation}
m_{\rm i}\equiv \langle \sigma_{\rm i}\rangle_{\rm expt} = \langle \sigma_{\rm i}\rangle_{P_1} .
\label{matchmean1}
\end{equation}
Referring back to Eq (\ref{constraints}), the observables that we constrain become 
\begin{equation}
\{f_\mu^{(1)} \} \rightarrow\{ \sigma_{\rm i}\};
\end{equation}
note that ${\rm i} = 1,\, 2,\, \cdots ,\, N$, where $N$ is the number of neurons.
To implement these constraints we need one Lagrange multiplier  for each neuron,
 and it is convenient to write this multiplier as an ``effective field'' $h_{\rm i}$, so that the general Eqs (\ref{maxent1}, \ref{maxent2}) become
\begin{eqnarray}
P_1\left(\bm{\sigma}\right) &=& {1\over {Z_1}}\exp\left[ - E_1 (\bm{ \vec\sigma})\right]\\
E_1 (\bm{ \vec\sigma}) &=& \sum_\mu \lambda_\mu^{(1)} f_\mu^{(1)}\\
&=& \sum_{{\rm i}=1}^N h_{\rm i} \sigma_{\rm i}.
\label{eq-E1}
\end{eqnarray}
We notice that $E_1$ is the energy function for independent spins in local fields, and so the probability distribution over states factorizes, 
\begin{equation}
P_1\left(\bm{\sigma}\right) \propto  \prod_{{\rm i}=1}^N e^{-h_{\rm i}\sigma_{\rm i}} .
\label{eq-P1}
\end{equation}
Thus a maximum entropy model which matches only the mean activities of individual neurons is a model in which the activity of each cell is independent of all the others.  We have seen that this model is in dramatic disagreement with the data.

A natural first step in trying to capture the non--independence of neurons is to build a maximum entropy model that matches pairwise correlations.  Thus, we are looking for a distribution $P_2\left(\bm{\sigma}\right)$ that has maximum entropy while matching the mean activities as in Eq (\ref{matchmean1}) and also the covariance of activity
\begin{equation}
C_{\rm ij} \equiv \langle \delta\sigma_{\rm i}\delta\sigma_{\rm j}\rangle_{\rm expt} = \langle \delta\sigma_{\rm i}\delta\sigma_{\rm j}\rangle_{P_2} .
\label{matchC1}
\end{equation}
In the language of Eq (\ref{constraints}) this means that we have a second set of relevant observables 
\begin{equation}
\{f_\nu^{(2)} \} \rightarrow \{\sigma_{\rm i}\sigma_{\rm j}\}.
\end{equation}
As before we need one Lagrange multiplier for each constrained observable, and  it is useful to think of the Lagrange multiplier that constrains  
$\sigma_{\rm i}\sigma_{\rm j}$ as being a ``spin--spin'' coupling $\lambda_{\rm ij} = J_{\rm ij}$.  Recalling that each extra constraint adds a term to the effective energy function, Eqs (\ref{maxent1}, \ref{maxent2}) become
\begin{eqnarray}
P_2(\bm{\sigma}) &=& {1\over {Z_2(\{h_{\rm i}; J_{\rm ij}\})}} e^{-E_2 (\bm{\sigma})} .
\label{eq-P2}\\
E_2 (\bm{ \vec\sigma}) &=& \sum_\mu \lambda_\mu^{(1)} f_\mu^{(1)} + \sum_\mu \lambda_\mu^{(2)} f_\mu^{(2)}\\
&=& \sum_{{\rm i}=1}^N h_{\rm i} \sigma_{\rm i} + {1\over 2}\sum_{{\rm i}\neq {\rm j}}J_{\rm ij} \sigma_{\rm i}\sigma_{\rm j}.
\label{eq-E2}
\end{eqnarray}
This is exactly an Ising model with pairwise interactions among the spins---not an analogy but a mathematical equivalence.

Ising models for networks of neurons have a long history, as described in \S\ref{sec-hopfield1}.  In their earliest appearance, these models emerged from a hypothetical, simplified model of the underlying dynamics.  Here they emerge as the least structured models consistent with {\em measured} properties of the network. As a result, we arrive not at some arbitrary Ising model, where we are free to choose the fields and couplings, but at a particular model that describes the actual network of neurons we are observing.   To complete this construction we have to adjust the fields and couplings to match the observed mean activities and correlations.  Concretely we have to solve  Eqs (\ref{matchmean1}, \ref{matchC1}), which can be rewritten as
\begin{eqnarray}
\langle \sigma_{\rm i}\rangle_{\rm expt}
 &=& \langle \sigma_{\rm i}\rangle_{P_2} = {{\partial\ln Z_2(\{h_{\rm i}; J_{\rm ij}\})}\over{\partial h_{\rm i}}} \label{pairwise_means}\\
\langle \sigma_{\rm i}\sigma_{\rm j} \rangle_{\rm expt} &=& \langle \sigma_{\rm i}\sigma_{\rm j} \rangle_{P_2} = {{\partial\ln Z_2(\{h_{\rm i}; J_{\rm ij}\})}\over{\partial J_{\rm ij}}} .
\label{pairwise_corrs}
\end{eqnarray}
With $N=10$ neurons this is challenging but can be done exactly, since the partition function is a sum over just $\Omega\sim 1000$ terms.   Once we are done, the model is specified completely.  Anything that we compute is a prediction, and there is no room to adjust parameters in search of better agreement with the data.

As noted above, with $N=10$ neurons the experiments are long enough to get a reasonably full sampling of the probability distribution over $\bm{\sigma}$.  This provides the most detailed possible test of the model $P_2$, and in Fig \ref{salamander1}F we see that the agreement between theory and experiment is excellent, except for very rare patterns where errors in the estimate of the probability are larger.  Similar results are obtained for other groups of $N=10$ cells drawn out of the full population of $N=40$.  Quantitatively we can measure the Jensen--Shannon divergence between the estimated distribution $P_{\rm data}(\bm{\sigma})$ and the model $P_2(\bm{\sigma})$; across multiple choices of ten cells this fluctuates by a factor of two around $D_{JS} = 0.001\,{\rm bits}$, which means that it takes thousands of independent observations to distinguish the model from the data.

The architecture of the retina is such that many individual output neurons can be driven or inhibited by a single common neuron that is internal to the circuitry.  This is one of many reasons that one might expect significant combinatorial regulation in the patterns of activity, and there were serious efforts to search for these effects \cite{schnitzer+meister_03}.  The success of a pairwise model thus came as a considerable surprise.  

The results in the salamander retina, with natural inputs, were quickly confirmed in the primate retina using simpler inputs \cite{shlens+al2006}.  Those experiments covered a larger area and thus could focus on sub--populations of neurons belonging to a single class, which are arrayed in a relatively regular lattice.  In this case not only did the pairwise model work very well, but the effective interactions $J_{\rm ij}$ were confined largely to nearest neighbors on this lattice.  

Pairwise maximum entropy models also were reasonably successful in describing patterns of activity across $N\leq 10$ neurons sampled from a cluster of cortical neurons kept alive in a dish \cite{tang+al2008}.  This work also pointed to the fact that dynamics did not correspond to Brownian motion on the energy surface.

These early successes with small numbers of neurons raised many questions.  For example, the interaction matrix $J_{\rm ij}$ contained a mix of positive and negative terms, suggesting that frustration could lead to many local minima of the energy function or equivalently local maxima of the probability $P(\bm{\sigma})$, as in the Hopfield model (\S\ref{sec-hopfield1}); could these ``attractors'' have a function in representing the visual world?  Relatedly, an important consequence of the collective behavior in the Ising model is that if we know that state of all neurons in the network but one, then we have a parameter--free prediction for the probability that this last neuron will be active; does this allow for error correction?  To address these and other issues one must go beyond $N\sim 10$ cells, which was already possible experimentally.  But at larger $N$ one needs more powerful methods for solving the inverse problem that is at the heart of the maximum entropy construction, as described in Appendix \ref{sec-inference}.

The equivalence to equilibrium models entices us to describe the couplings $J_{\rm ij}$ as ``interactions,'' but there is no reason to think that these correspond to genuine connections between cells.  In particular, $J_{\rm ij}$ is symmetric because it is an effective interaction driving the equal--time correlations of activity in cells $\rm i$ and $\rm j$, and these correlations are symmetric by definition.  If we go beyond single time slices to describe  trajectories of activity over time, then with multiple cells the effective interactions can become asymmetric and break time--reversal invariance.  

Before leaving the early work, it is useful to step back and ask about the goals and hopes from that time.  As reviewed above, the use of statistical physics models for neural networks has a deep history.  Saying that the brain is described by an Ising model captured both the optimism and (one must admit) the na\"ivet\'e of the physics community in approaching the phenomena of life.  One could balance optimism and na\"ivet\'e by retreating to the position that these models are metaphors, illustrating what could happen rather than being theories of what actually happens.  The success of maximum entropy models in the retina gave an example of how statistical physics ideas could provide a quantitative theory for networks of real neurons.

\subsection{Larger networks of neurons}
\label{sec-larger}

The use of maximum entropy for networks of real neurons quickly triggered almost all  possible reactions:  (a) It should never work, because systems are not in equilibrium, have combinational interactions, ... .  (b) It could work, but only under uninteresting conditions. (c) It should always work, since these models are very expressive.  (d) It works at small $N$, but this is a poor guide to what will happen at large $N$. (e) Sure, but why not use [favorite alternative], for which we have efficient algorithms? 

Perhaps the most concrete response to these issues is just to see what happens as we move to more examples, especially in larger networks.  But we should do this with several questions in mind, some of which were very explicit in the early literature \cite{roudi2009pairwise,macke+al_11}.  First, finding the maximum entropy model that matches the desired constraints---that is, solving Eqs (\ref{constraints})---becomes more difficult at larger $N$.  Can we be sure that we are testing the maximum entropy idea, and our choice of constraints, rather than the efficacy of our algorithms for solving this problem?

Second, as $N$ increases the maximum entropy construction becomes very data hungry. This concern often is phrased as the usual problem of ``over--fitting,'' when the number of parameters in our model is too large to fully constrained by the data.  But in the maximum entropy formulation the problem is even more fundamental.  The maximum entropy construction builds the least structured model consistent with a set of known expectation values.  With a finite amount of data, if our list of expectation values is too long then the claim that we ``know'' these features of the system just isn't true, and this problem arises even before we try to build the maximum entropy model. 

Third, because correlations are spread widely in these networks, if one develops a perturbation theory around the limit of independent neurons then factors of $N$ appear in the series, e.g. for the entropy per neuron.   Success at modest $N$ might thus mean that we are in a perturbative regime, which would be much less interesting.  The question of whether success is perturbative is subtle, since at finite $N$ all properties of the maximum entropy model are analytic functions of the correlations, and hence if we carry perturbation theory far enough we will get the right answer \cite{sessak+monasson_09}. 

Finally, in statistical mechanics we are used to the idea of a large $N$, thermodynamic limit. Although this carries over to model networks \cite{amit_89}, it is not obvious how to use this idea in thinking about networks of real neurons.  Naive extrapolation of results from maximum entropy models of $N=10-20$ neurons in the retina indicated that something special had to happen by $N\sim 200$, or else the entropy would vanish; this was interesting because $N\sim 200$ is the number cells that are ``looking'' at overlapping regions of the visual world \cite{schneidman2006weak}.   A more sophisticated extrapolation imagines a large population of neurons in which mean activities and pairwise correlations are drawn at random from the same distribution as found in recordings from smaller numbers of neurons \cite{tkacik2006ising,tkacik2009spin}. This sort of extrapolation is motivated in part by the observation that ``thermodynamic'' properties of the maximum entropy models learned for $N=20$ or $N=40$ retinal neurons match the behavior of such random models at the same $N$.  If we now extrapolate to $N=120$ there are striking collective behaviors, and we will ask if these are seen in real data from $N>100$ cells.

Early experiments in the retina already were monitoring $N=40$ cells, and the development of numerical methods described in  Appendix \ref{sec-inference} quickly allowed analysis of these larger data sets \cite{tkacik2006ising,tkacik2009spin}.  With $N=40$ cells one cannot check the predictions for probabilities of individual patterns $P(\bm{\sigma})$, but one can check the probability that $K$ out of $N$ cells are active in the same small time bin, as in Fig.~\ref{salamander1}E, or the correlations among triplets of neurons.   At $N=40$ we see the first hints that constraining pairwise correlations is not quite enough to capture the full structure of the network.  There are disagreements between theory and experiment in the tails of the distribution $P_N(K)$, and more importantly a few percent disagreement at $K=0$.  This may not seem like much, but since the network is completely silent in roughly half of the $\Delta \tau = 20\,{\rm ms}$ time bins, the data determine $P_N(K=0)$ very precisely, and a one percent discrepancy is hugely significant.

A new generation of electrode arrays made it possible to record $N = 100 - 200$ cells, densely sampling a small patch of the retina (\S\ref{sec-arrays}).  As an example, these experiments could capture the signals from $N_{\rm max} = 160$ ganglion cells in a $(450\,\mu{\rm m})^2$ area of the salamander retina that contains a total of $N\sim 200$ cells, and these recordings are stable for $\sim 1.5\,{\rm hr}$.

\begin{figure}
\includegraphics[width=\linewidth]{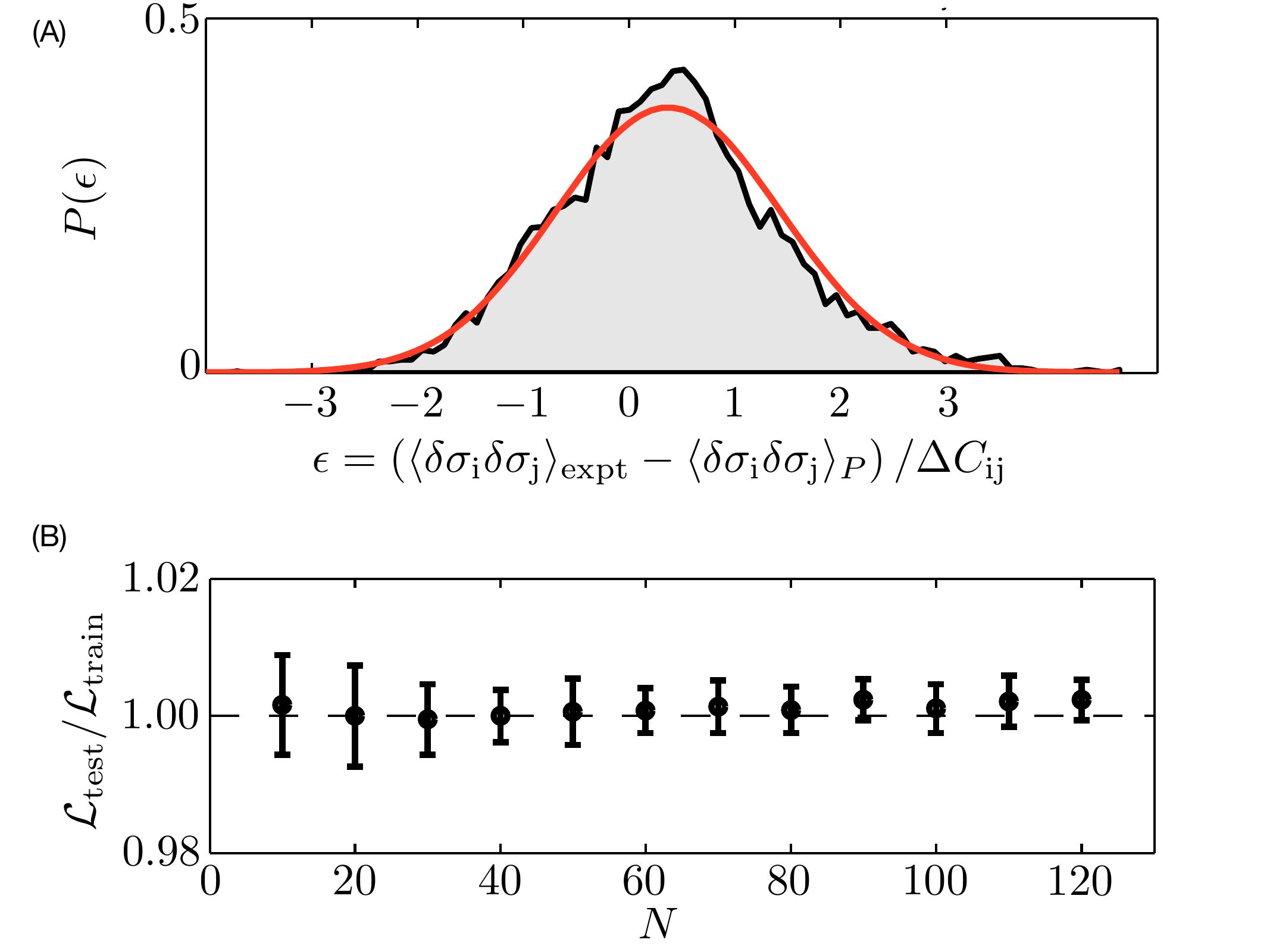}
\caption{Fitting, but not over--fitting, with $N\sim 100$ neurons \cite{tkavcik2014searching}. (A) Distribution of errors in the prediction of pairwise correlations, after adjusting the parameters $\{h_{\rm i}; J_{\rm ij}\}$, for $N=100$.  Prediction errors are in units of the measurement error $\Delta C_{\rm ij}$ for each element of the covariance matrix.  Red line shows a Gaussian with zero mean and unit variance.  (B) Log--likelihood [Eq (\ref{LL-def})] of test data not used in constructing the maximum entropy model, in units of the result for the training data.  At $N = 10$ it is not surprising that these agree, since the number of parameters $\{h_{\rm i}; J_{\rm ij}\}$ is small.  But we see this agreement persists at the $\sim 1\%$ level out to $N= 120$, showing that even models for relatively large networks are not overfit.  
\label{NoOverFit}}
\end{figure}

As explained in Appendix \ref{sec-inference}, we can build maximum entropy models at larger $N$ by using Monte Carlo simulation to estimate expectation values in the model, comparing with the measured expectation values, and then adjusting the coupling constants to improve the agreement.  Necessarily this doesn't yield an exact solution to the constraint Eqs (\ref{constraints}), but this seems acceptable since we are trying to match expectation values that are estimated from experiment and these have errors.  Figure \ref{NoOverFit}A shows that with $N=100$ we can match the observed pairwise correlations within experimental error \cite{tkavcik2014searching}.  More precisely the errors in predicting the elements of the covariance matrix $C_{\rm ij}$ [Eq (\ref{cov-def})] are nearly Gaussian, with a variance equal to the variance of the measurement errors. This suggests, strongly, that one can successfully fit, but not over--fit, a maximum entropy model to these data.

The test for fitting vs over--fitting in Fig \ref{NoOverFit}A looks at each pair of cells individually, but part of the worry is that at large $N$ we can have accurate estimates of individual elements $C_{\rm ij}$ while under--determining the global properties of the matrix. We can  take a familiar empirical approach, measuring the means $\langle\sigma_{\rm i}\rangle$ and covariances $\langle\delta \sigma_{\rm i}\delta \sigma_{\rm j}\rangle_c$ in $90\%$ of the data, using these to infer the parameters  $\{h_{\rm i}; J_{\rm ij}\}$ in a maximum entropy model, and then testing the predictions of the model [Eqs (\ref{eq-E2}, \ref{eq-P2})] on the remaining $10\%$. The fundamental measure of model quality is the log--likelihood of the data, which we can normalize per sample and per neuron
\begin{equation}
{\cal L} = {1\over N}\langle \log P\left(\bm{\sigma}\right)\rangle_{\rm expt} .
\label{LL-def}
\end{equation}
Figure \ref{NoOverFit}B shows that $\cal L$ is the same, to better than one percent, whether we evaluate it over the training data or over the test data.  This is true at $N=10$, where surely there can be no question that we have enough samples, and it is true at $N=120$.

Different networks of neurons, in different organisms and different regions of the brain, have different correlation structures.  One should thus be wary of generalizations such as ``an hour is enough data for one hundred neurons."   But at least in the context of experiments on the retina, there is no question that maximum entropy models can be learned reliably from the available data, and that there is no over--fitting.  Said another way, the models really are the solutions to the mathematical problem that we set out to solve (\S\ref{sec-basics}): What is the minimal model consistent with a set of expectation values measured in experiment?  These models do not carry signatures of the algorithm that we used to find them, nor are they systematically perturbed by the finiteness of the data on which they are based.  This answers the first two questions formulated above.

Given that we can construct the maximum entropy models reliably, what do we learn?  To begin, the small discrepancies in predicting the probability that $K$ out $N$ neurons are active simultaneously, $P_N(K)$, become larger as $N$ increases.  The simplest solution to this problem is to add one more constraint, insisting that the maximum entropy model match the observed $P_N(K)$ exactly.  This adds only $\sim N$ constraints to a problem in which we already have $N(N+1)/2$, so the resulting ``K--pairwise'' models are not significantly more complex.  

Again, at the risk of being pedantic let's formulate matching of the observed $P_N(K)$ as constraining expectation values.  If we introduce the Kronecker delta for integers $n$ and $m$,
\begin{eqnarray}
\delta (n,m) &=& 1 \,\,\,\,\, n=m\\
&=& 0 \,\,\,\,\, n\neq m ,
\end{eqnarray}
then
\begin{equation}
P_N(K) = {\bigg\langle} \delta \left(K, \sum_{\rm i}^N \sigma_{\rm i} \right) {\bigg\rangle} .
\label{PKofN_expval}
\end{equation}
Thus to match $P_N(K)$ we want to enlarge our set of observables to include
\begin{equation}
\{f_\mu^{\rm (counts)}\} \rightarrow \bigg{\{} \delta \left(K, \sum_{\rm i}^N \sigma_{\rm i} \right)\bigg{\}} .
\end{equation}
As before, each new constraint adds a term to the effective energy,
\begin{equation}
E\left(\bm{\sigma}\right) = \sum_\mu \lambda_\mu^{\rm (counts)} f_\mu^{\rm (counts)} =
\sum_{K=0}^N \lambda_K \delta \left(K, \sum_{\rm i}^N \sigma_{\rm i} \right) .
\end{equation}
It is useful to think of this as an effective potential that acts on the summed activity,
\begin{equation}
\sum_{K=0}^N \lambda_K \delta \left(K, \sum_{\rm i}^N \sigma_{\rm i} \right) =  V \left( \sum_{\rm i}^N \sigma_{\rm i} \right) .
\end{equation}

Putting the pieces together, the maximum entropy model that matches the mean activity of individual neurons, the correlations between pairs of neurons, and the probability that $K$ out of $N$ are active simultaneously 
takes the form
\begin{eqnarray}
\label{K_pairwise}
P_{2k}(\bm{\sigma}) &=& {1\over {Z_{2k}}} e^{-E_{2k} (\bm{\sigma})} \\
E_{2k} (\bm{\sigma}) &=& \sum_{{\rm i}=1}^N h_{\rm i} \sigma_{\rm i} + {1\over 2}\sum_{{\rm i}\neq {\rm j}}J_{\rm ij} \sigma_{\rm i}\sigma_{\rm j} + V\left( \sum_{{\rm i}=1}^N \sigma_{\rm i}\right) .
\label{K_pairwiseB}
\end{eqnarray}
We refer to this as the ``K--pairwise'' model \cite{tkavcik2014searching}.

\begin{figure}[b]
\includegraphics[width=\linewidth]{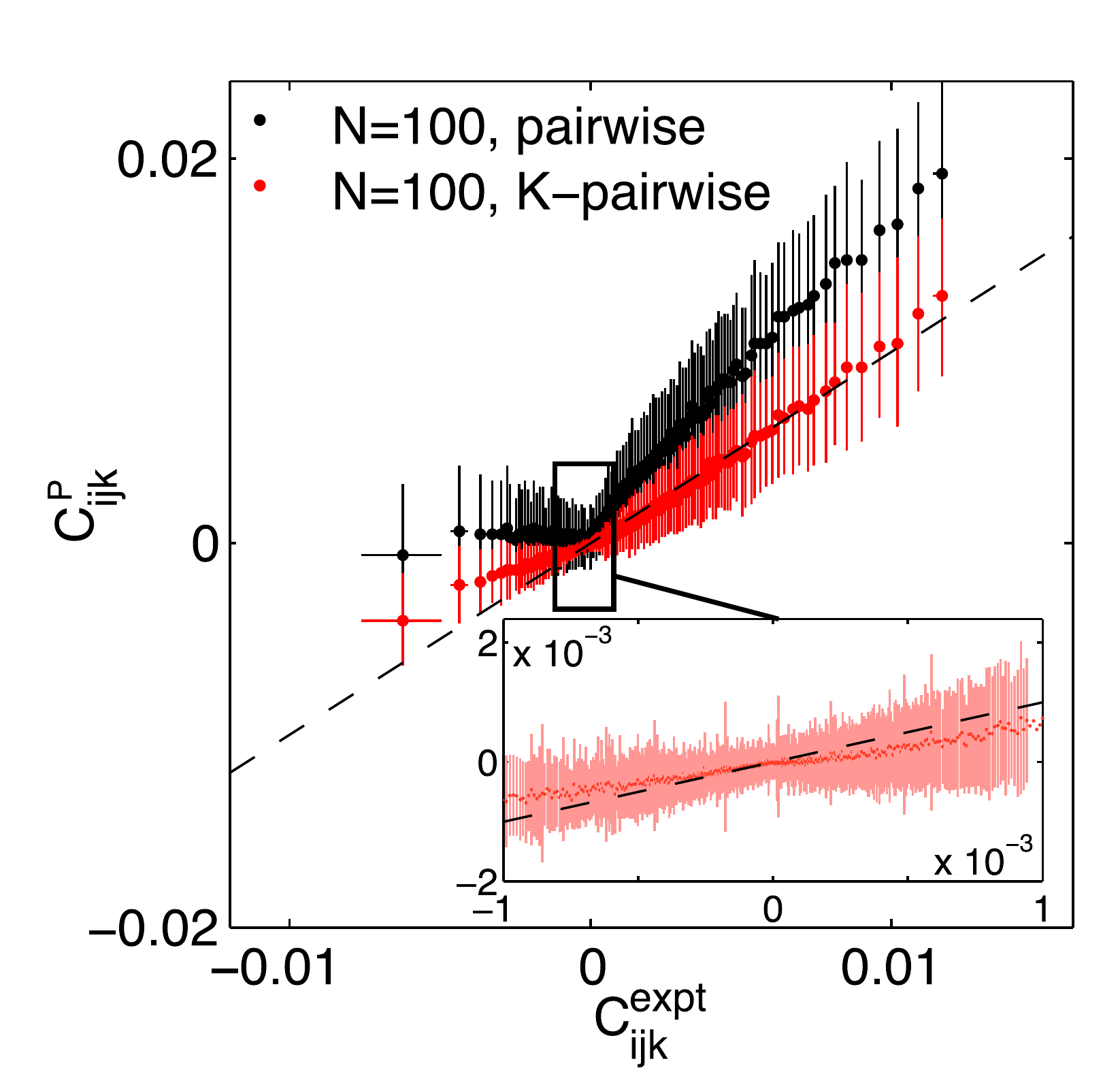}
\caption{Triplet correlations for $N=100$ cells in the retina \cite{tkavcik2014searching}. Measured $C_{\rm ijk}$ (x-axis) vs predicted by the model (y-axis), shown for a single subgroup. The $\sim 1.6\times 10^5$ distinct triplets are grouped into 1000 equally populated bins; error bars in x are s.d. across the bin. The corresponding values for the predictions are grouped together, yielding the mean and the s.d. of the prediction (y- axis). Inset zooms in on the bulk of the predictions at small correlation, for the K--pairwise model.  The original reference used $\sigma_{\rm i} = \pm 1$, so that all the $C_{\rm ijk}$ shown here are $8\times$ larger than they would be in the $\sigma_{\rm i} = \{0, 1\}$ representation.  \label{retina_triplets}}
\end{figure}

We can test this model immediately by estimating the correlations among triplets of neurons,
\begin{equation}
C_{\rm ijk} = \langle\left(\sigma_{\rm i} - \langle\sigma_{\rm i}\rangle\right)\left(\sigma_{\rm j} - \langle\sigma_{\rm j}\rangle\right)
\left(\sigma_{\rm k} - \langle\sigma_{\rm k}\rangle\right)\rangle .
\label{triplet1}
\end{equation}
Figure \ref{retina_triplets} shows the results with averages computed in both the pairwise and K--pairwise models, plotted vs.~the experimental values.  The discrepancies are very small, although still roughly three times larger than the experimental errors in the estimates of the correlations themselves \cite{tkavcik2014searching}; we will see that one can sometimes get even better agreement (\S\ref{sec-subgroups}). Note that the  potential $V$ which we add to match the constraint on $P_N(K)$ does not carry any information about the identities of the individual neurons.  It thus is interesting that including this term improves the prediction of all the triplet correlations, which do depend on neural identity.

\begin{figure}[t]
\includegraphics[width=\linewidth]{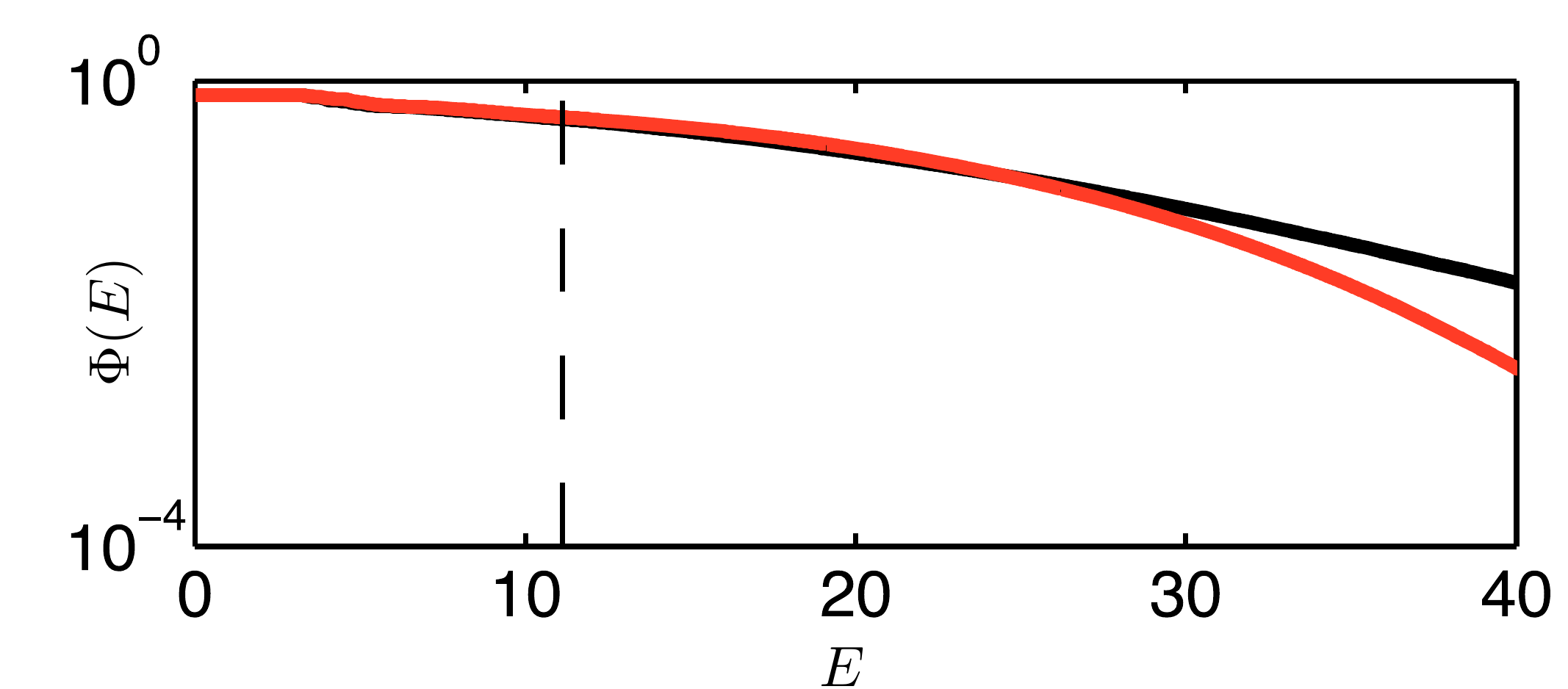}
\caption{The cumulative distribution of energies for $N=120$ neurons \cite{tkavcik2014searching}. $\Phi(E)$ is defined in Eq (\ref{eq-PhiE}), and averages are over data (black) or the theory (red).  Dashed vertical line denotes an energy $E_{2k}(\bm{\sigma})$ such that the particular state $\bm{\sigma}$ should occur on average once during the duration of the experiment. 
\label{fig-PhiE}}
\end{figure}

With $N=100$ cells we cannot check, as in Fig \ref{salamander1}F, the probability of every state of the network.  But the model assigns to every state an energy $E_{2k} (\bm{\sigma})$, and we can ask about the distribution of this energy over the states that we see in the experiment vs.~the expectation if states are drawn out of the model.  To emphasize the extremes we look at the high energy tail,
\begin{equation}
\Phi(E) = \langle \Theta\left[E - E_{2k} (\bm{\sigma}) \right]\rangle ,
\label{eq-PhiE}
\end{equation}
where $\Theta(x)$ is the unit step function and the expectation value can be taken over the data or the theory.  Figure \ref{fig-PhiE} shows the comparison between theory and experiment.  Note that the plot extends far past the point where individual states are predicted to occur once over the duration of the experiment, but we can make meaningful statements in this regime because there are (exponentially) many such states.  Close agreement between theory and experiment extends out to $E \sim 25$, corresponding to states that are predicted to occur roughly once per fifty years.

\begin{figure}[b]
\includegraphics[width=\linewidth]{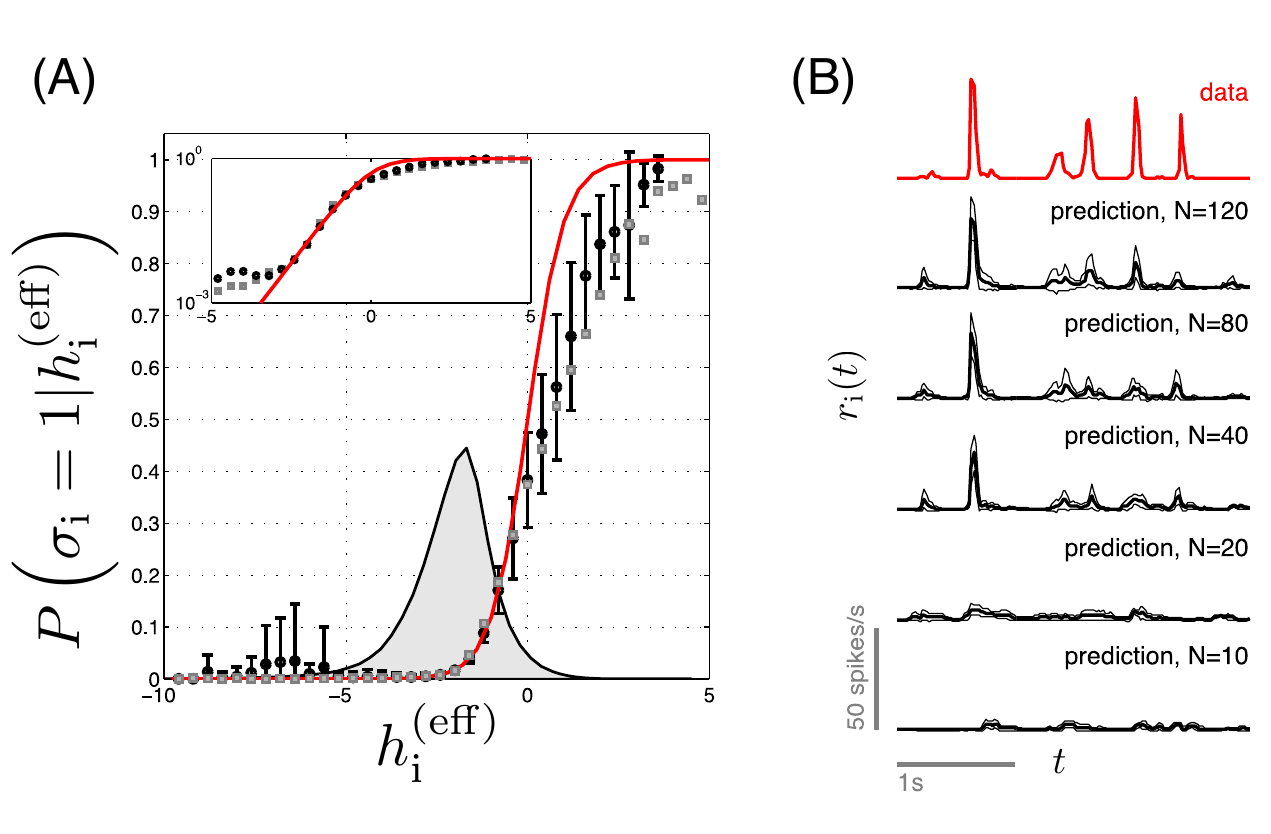}
\caption{Effective fields and the collective character of neural activity in the retina \cite{tkavcik2014searching}.  (A) The probability that a single neuron is active given the state of the rest of the network, with  $N=120$.  Points with error bars are the data, with the effective field computed from the model as in Eq (\ref{heff2}).  Red line is the prediction from Eq (\ref{heff_main}), and grey points are results with the purely pairwise rather than ``K--pairwise'' model.  Shaded grey region shows the distribution of fields across the experiment, emphasizing that the errors at large positive field are in the tail of the distribution.  Inset shows the same results on a logarithmic scale for probability. (B) Probability of a single neuron being active as a function of time in a repeated naturalistic movie, normalized as the probability per unit time of an action potential (spikes/s). Top, in red, experimental data. Lower traces, in black, predictions based on states of other neurons in an $N$--cell group, based on Eqs (\ref{heff1}, \ref{heff2}). Solid lines are the mean prediction across all repetitions of the movie, and thin lines are the envelope $\pm$ one standard deviation.
\label{retina_heff}}
\end{figure}

This class of models predicts that neural activity is collective.  Thus in a population of $N$ cells, if we know the state of $N-1$ we can make a prediction of the probability that the last cell will be active, 
\begin{equation}
P(\sigma_{\rm i} = 1 | \{\sigma_{{\rm i}\neq {\rm j}}\}) = {1\over{1 + \exp\left[- h_{\rm i}^{\rm eff}(\{\sigma_{{\rm i}\neq {\rm j}}\})\right]}},
\label{heff_main}
\end{equation}
where we can think of the other neurons as applying an effective field to the one neuron that we focus on,\footnote{The original presentation used $\sigma_{\rm i} = \pm1$, leading to a  factor of two in the definition of the effective field; see  Eq (25) in \citet{tkavcik2014searching}.}
\begin{eqnarray}
h_{\rm i}^{\rm eff}(\{\sigma_{{\rm i}\neq {\rm j}}\}) &=& E\left( \sigma_1,\, \sigma_2,\, \cdots ,\, \sigma_{\rm i}=1,\, \cdots ,\, \sigma_N\right) \nonumber\\
&&\, -  E\left( \sigma_1,\, \sigma_2,\, \cdots ,\, \sigma_{\rm i}= 0,\, \cdots ,\, \sigma_N\right) .
\nonumber\\
&&
\label{heff2}
\end{eqnarray}
For each neuron and for each moment in time we can calculate the effective field predicted by the theory, with no free parameters, and we can group together all instances in which this field is in some narrow range and ask if the probability of the cell being active agrees with Eq (\ref{heff1}).  Results are shown in Fig.~\ref{retina_heff}A.

We see that the predictions of Eqs (\ref{heff1}) and (\ref{heff2}) in the K--pairwise model agree well with experiment throughout the bulk of the distribution of effective fields, but that discrepancies arise in the tails.  These deviations are $\sim 1.5\times$ the error bars of the measurement, but have some systematic structure, suggesting that we are capturing much but not quite all of the collective behavior under conditions where neurons are driven most strongly.  

The results in Fig.~\ref{retina_heff}A combine data across all times  to estimate the probability of activity in one cell given the state of the rest of the network.  It is interesting to unfold these results in time.  In particular, the structure of the experiment was such that the retina saw the same movie many times, and so we can condition on a particular moment in the movie, as shown for one neuron in Fig \ref{retina_heff}B.  It is conventional to plot not the probability of being active in a small bin but the corresponding  ``rate'' \cite{spikesbook}
\begin{equation}
r_{\rm i}(t) = \langle \sigma_{\rm i} (t)\rangle /\Delta\tau ,
\end{equation}
where $\sigma_{\rm i}(t)$ denotes the state of neuron $\rm i$ at time $t$ relative to (in this case) the visual inputs.  We see in the top trace of Fig.~\ref{retina_heff}B that single neurons are active very rarely, with essentially zero probability of spiking between brief transients that  generate on average one or a few spikes. This pattern is common in response to naturalistic stimuli, and very difficult to reproduce in models  \cite{maheswaranathan+al_23}.

The maximum entropy models provide an extreme opposite point of view, making no reference to the visual inputs;  instead activity is determined by the state of the rest of the network.  We see that this approach correctly predicts  sparse activity, with near zero rate between transients that are  timed correctly relative to the input.   Although here we see just one cell, the average neuron exhibits an $r_{\rm i}(t)$ that has $\sim 80\%$ correlation with the theoretical predictions at $N=120$.  There is no sign of saturation, and it seems likely we would make even more precise predictions from models based on all $N\sim 200$ cells in this small patch of the retina.  The possibility of predicting activity without reference to the visual input suggests that the ``vocabulary'' of the retina's output is restricted, and that as with spelling rules this should allow for error--correction \cite{loback+al_17}.

\begin{figure}[t]
\includegraphics[width=\linewidth]{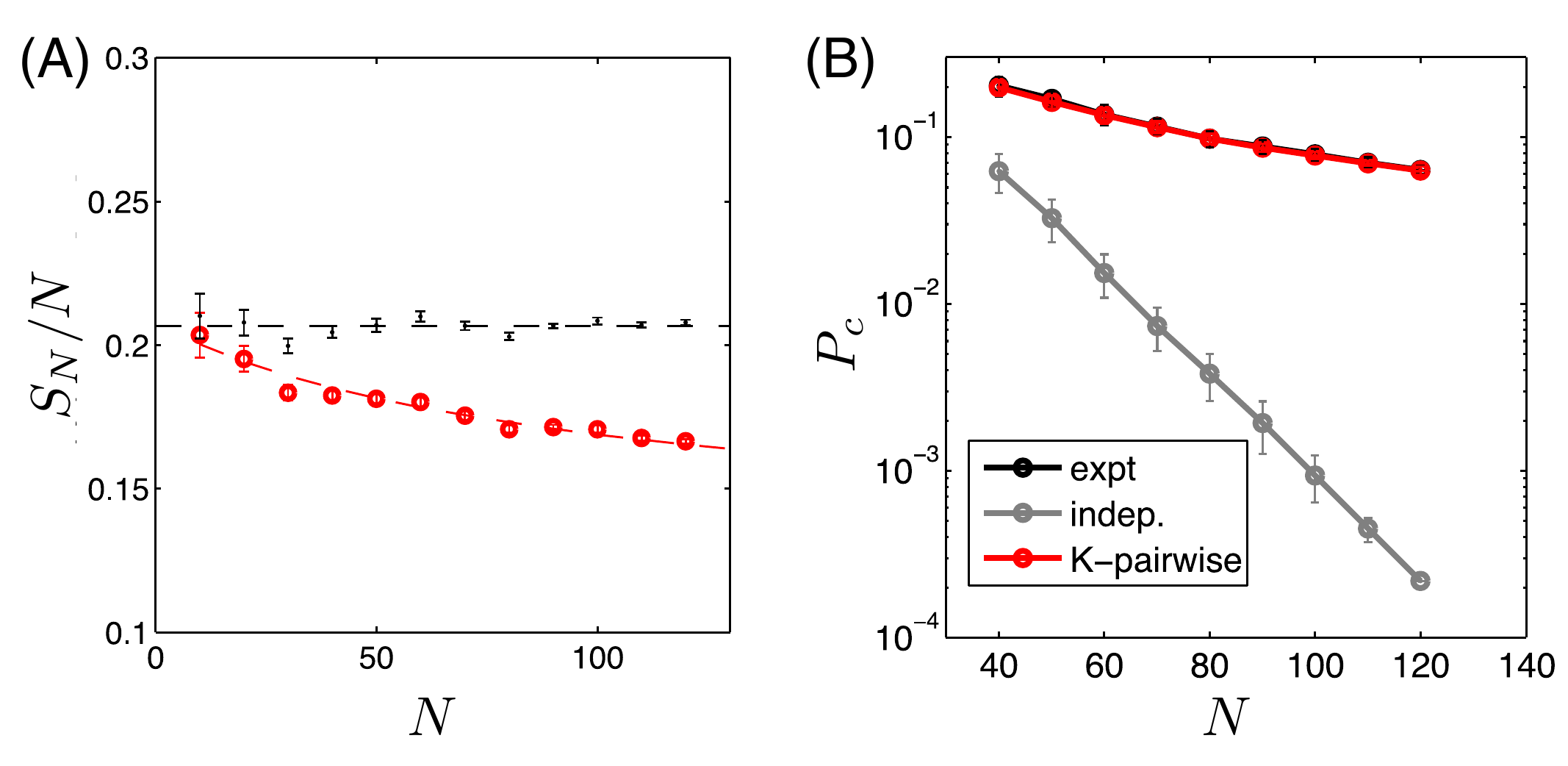}
\caption{Entropy and coincidences in the activity of the retinal network \cite{tkavcik2014searching}. (A) Entropy predicted in K--pairwise models (red) and in the approximation that all neurons are independent (grey).  Models are constructed independently for many subgroups of size $N$ chosen out of the total population $N_{\rm max} = 160$, and error bars include the variance across these groups.  (B) Probability that two randomly chosen states of the network are the same, again for many subgroups of size $N$.  Results for real data (black), shuffled data (grey), and the K--pairwise models (red).   
\label{entropy_retina}}
\end{figure}

Perhaps the most basic prediction from maximum entropy models is the entropy itself.    There are several ways that we can estimate the entropy.  First, in the K--pairwise model we can see that the effective energy of the completely silent state, from Eq (\ref{K_pairwiseB}), is zero, which means that the probability of this state is just the inverse of the partition function.  Further, in this model, the probability of complete silence matches what we observe experimentally.  Thus we can estimate the free energy of the model from the data, and then we can estimate the mean energy of the model from Monte Carlo, giving us an estimate of the entropy.  An alternative is to generalize the model by introducing a fictitious temperature, as will be discussed in \S\ref{sec-thermo}.  Then at $T=0$ the entropy must be zero and at $T\rightarrow\infty$ the entropy must be $N\log 2$, while the derivative of the entropy is related as always to the heat capacity.  Thus  the entropy of our model for the real system at $T=1$ becomes\footnote{We write $C_v$ not because we are worried about whether the volume is constant, but to avoid confusion with the covariance matrix $C_{\rm ij}$.}
\begin{equation}
S_N(T=1) = \int_0^1 dT\,{{C_v(T)}\over T} = N\log 2 - \int_1^\infty dT\,{{C_v(T)}\over T} ,
\label{S_estCv}
\end{equation}
where the heat capacity is related as usual to the variance of the energy, $C_v = \langle (\delta E)^2\rangle/T^2$, that we can estimate from Monte Carlo simulations at each $T$.
There is also a check that the two estimates in Eq (\ref{S_estCv}) should agree.  All of these methods agree with one another at the percent level, with results shown in Fig.~\ref{entropy_retina}A.

The $\sim 25\%$ reduction in entropy is significant, but more dramatic (and testable) is the prediction that the distribution over states is extremely inhomogeneous.  Recall that if the distribution is uniform over some effective number of states $\Omega_{\rm eff}$ then the entropy is $S = \log\Omega_{\rm eff}$ and the probability that two states chosen at random will be the same is $P_c = 1/\Omega_{\rm eff}$; for non--uniform distributions we have $S \geq -\log(P_c)$.  If neurons were independent then with $N$ cells we would have $P_c\propto e^{-\alpha N}$, and this is what we see in the data once they are shuffled to remove correlations (Fig.~\ref{entropy_retina}B).  But the real data show a much more gradual decay with $N$, and this is captured perfectly by the K--pairwise maximum entropy models.

At $N = 120$ the {\em logarithm} of the coincidence probability (both measured and predicted) is an order of magnitude smaller than the entropy predicted by the model.  Perhaps related is that the free energy per neuron---which, as discussed above, can be obtained directly from the probability of the fully silent state---also decreases dramatically as $N$ increases.  At $N=120$ the free energy is just a few percent of the either the entropy or the mean energy, reflecting near perfect cancelation between these terms; one can see this also in a much simpler model that only matches $P_N(K)$ and not the individual means or pairwise correlations \cite{tkavcik2013simplest}.  Importantly, these behaviors are captured by the K--pairwise model smoothly from $N<40$ through $N> 100$, indicating that what we learned at more modest $N$ really does extrapolate up a scale comparable to the whole population of cells in a patch of the retina.  We will have to work harder to decide if we can see the emergence of a true thermodynamic limit.

Finally, we should address the question of whether these results can be recovered as perturbations to a model of independent neurons.  At lowest order in perturbation theory, there is a simple relationship between the observed correlations and the inferred interactions $J_{\rm ij}$ in the pairwise model \cite{sessak+monasson_09},  and we can check this relationship against the values of $J_{\rm ij}$ inferred from correctly matching the observed correlations.  In the retina, large deviations from lowest order perturbation theory are visible already at $N=15$, and correspondingly models built from the perturbative estimates of $J_{\rm ij}$ are orders of magnitude further away from the data than the full model \cite{tkavcik2014searching}. Higher order perturbative contributions to the entropy would be comparable to one another for $N=20$ retinal neurons even in a hypothetical network where all correlations were scaled down by a factor of two from the real data \cite{azhar+bialek_10}.  We conclude that the success of maximum entropy models in describing networks of real neurons is not something we can understand in low order perturbation theory.  Interestingly, simulations of  models with pure 3-- and 4--spin interactions at $N\sim 20$ show that pairwise maximum entropy models typically are good approximations to the real distribution both in the weak correlation limit and in the limit of strong, dense interactions \cite{Merchan+Nemenman_2016}.

The retina is a very special part of the brain, and one might worry that the success of maximum entropy models is somehow tied to these special features.  It thus is important that the same methods work in capturing the collective behavior of neurons in very distant parts of the brain.  An example is in prefrontal cortex, which is involved in a wide range of higher cognitive functions.

\begin{figure}
\includegraphics[width=\linewidth]{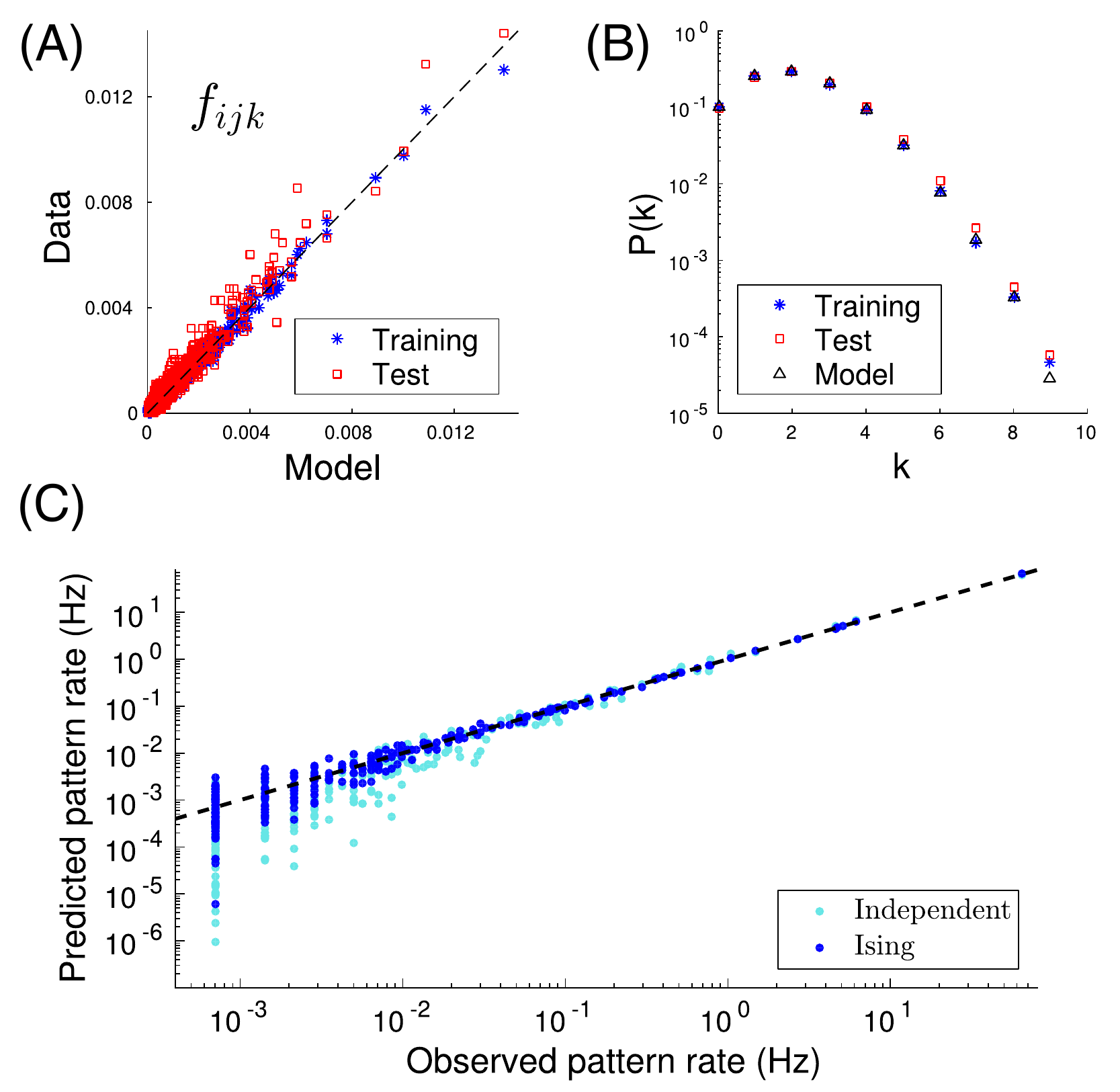}
\caption{Pairwise maximum entropy models describe collective behavior of $N=37$ neurons in  prefrontal cortex \cite{tavoni2017functional}. (A) Observed vs predicted triplet correlations among all neurons.  Training results (blue) are predictions from the same segment of the experiment where the pairwise correlations were measured; test results (red)  are in a different segment of the experiment. (B) Probability that $K$ out of $N$ neurons are active simultaneously, comparing predictions of the model with data in training and test segments.  (C) Rate at which patterns of spiking and silence appear in a subset of ten neurons, comparing predicted vs observed rates in an independent model (cyan) and in the pairwise model (blue).
\label{fig-tavoni}}
\end{figure}

Experiments recording simultaneous activity from several tens of neurons in prefrontal cortex were analyzed with maximum entropy methods, and an example of the results is shown in Fig.~\ref{fig-tavoni} \cite{tavoni2017functional}.  We see that these models pass the same tests as in the retina, correctly predicting triplet correlations, the probability of $K$ out of $N$ cells being active simultaneously, and the probabilities for particular patterns of activity in subgroups of $N=10$ cells.  Extending this analysis across multiple experimental sessions it was possible to detect changes in the coupling matrix $J_{\rm ij}$ as the animal learned to engage in different tasks.  These changes were concentrated in subsets of cells which also were preferentially re--activated during sleep between sessions.  One should be careful about giving too mechanistic an interpretation of the Ising models that emerge from these analyses, but it is exciting to see the structure of the models connect to independently measurable functional dynamics in the network.  This is true even in the farthest reaches of the cortex, the regions of the brain that we use for thinking, planning, and deciding.

The Ising model also gives us a way of exploring how the network would respond to hypothetical perturbations \cite{tavoni+al_16}.  If we increase the magnetic field uniformly across all the cells in the population of prefrontal neurons, the predicted changes in activity are far from uniform.  For some cells the response and the derivative of the response (susceptibility) are on a scale expected if neurons respond independently to applied fields, but there are groups of cells that co--activate much more, with susceptibilities peaking at intermediate fields.  It is tempting to think that these groups of cells have some functional significance, and this is supported by the fact that in the real data (with no fictitious fields) the groups of cells identified in this way remain co--activated over relatively long periods of time.  

At the opposite extreme of organismal complexity, the worm {\em C.~elegans} is an attractive target for these analyses because one can record not just from a large number of cells but from a large fraction of the entire brain at single cell resolution (\S\ref{sec-imagingmethods}).  A major challenge is that these neurons do not generate discrete action potentials or bursts, so the signal are not naturally binary.  A first step was to discretize the continuous fluorescence signals into three levels, and construct a Potts--like model that matched the population of each state and the probabilities that pairs of neurons are in the same state \cite{chen+al_19}.  Although these early data sets were limited, this simple model succeeded in predicting off--diagonal elements of the correlation matrix that were unconstrained, the probability that $K$ of $N$ neurons are in the same state, and the relative probabilities of different states in relation to the effective fields generated by the rest of the network.   The fact that the same statistical physics approaches work in worms and in mammalian cortex is encouraging, though we should see more compelling tests with the next generation of experiments.

A very different approach is to study networks of neurons that have been removed from the animal and kept alive in a dish.  There is a long history of work on these ``cultured networks,'' and as noted above (\S\ref{sec-arrays}) some of the earliest experiments recording from many neurons were done with networks that had been grown onto an array of electrodes \cite{pine+gilbert1982}.  Considerable interest was generated by the observation that patterns of activity in cultured networks of cortical neurons consist of ``avalanches'' that exhibit at least some degree of scale invariance (\S\ref{sec-avalanches}).  Recent work returns to these data and shows that detailed patterns of spiking and silence are well described by pairwise maximum entropy models, reproducing triplet correlations and the probability that $K$ out of $N=60$ neurons are active simultaneously \cite{SampaioFilho+al2024}.

\begin{figure*}
\includegraphics[width=\linewidth]{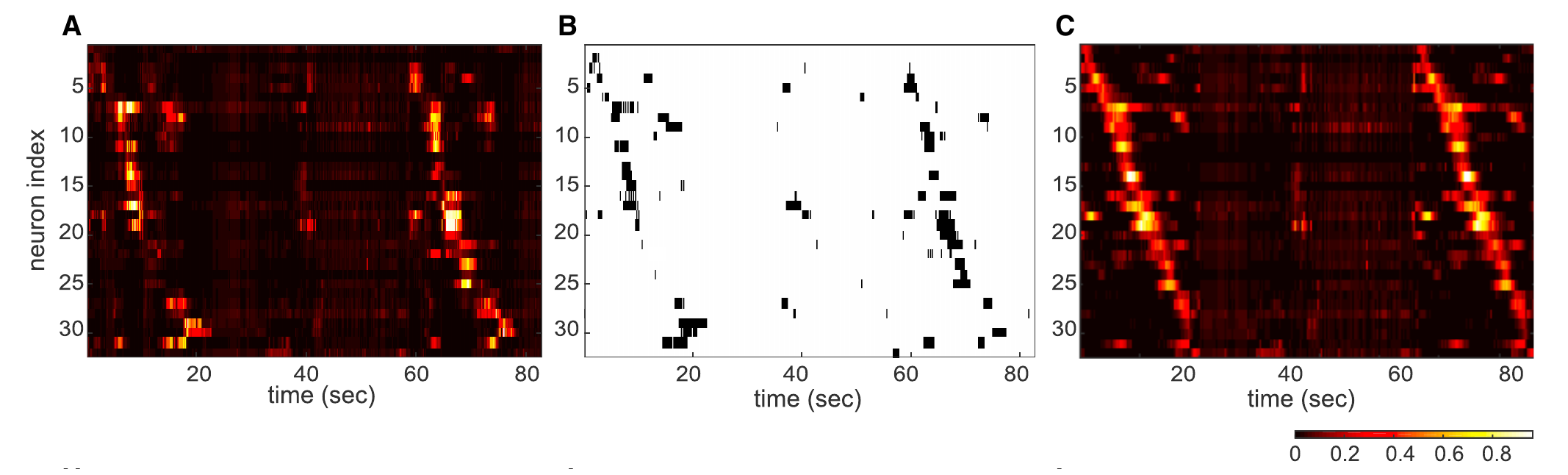}
\caption{Collective behavior in the mouse hippocampus \cite{meshulam2017collective}. (A) Predicted probability of activity for single neurons, computed from the effective field in the pairwise maximum entropy model.  Focus is on 32 place cells that should be active in sequence as the mouse runs along a virtual track.  During the first run, cells 21--25 are predicted to ``miss'' their place fields, but all cells are predicted to be active in the second run.
(B) Real data of place cell activity during two runs down the linear track, in the same time window as  (A) and (C); note the missed events for cells 21--25 in the first run. (C) Predicted probability  from the independent place cell model. There is no indication of when fields should be missed.
\label{missedfields}}
\end{figure*}

As a final example we consider populations of $N\sim 100$ neurons in the mouse hippocampus \cite{meshulam2017collective}.  The hippocampus plays a central role in navigation and episodic memory, and is perhaps best known for its population of ``place cells,'' neurons that are active only when the animal moves to a particular position in its environment.  First discovered in rodents \cite{OKeefe1971}, it is thought that the whole population of these cells together provides the animal with a cognitive map \cite{okeefe1978hippocampus}.  More recent work shows how this structure extends to three dimensions, and across hundreds of meters in bats \cite{tsoar+al_11,yartsev+ulanovsky_13}.

As the animal explores its environment, or runs along a virtual track, the mean activity of individual neurons is quite small, as in the examples above.  Most pairs of neuron have negative correlations, as expected if activity is tied to the position---if each cell is active in a different place, then on average one cell being active means that other cells must be silent, generating anti--correlations.  Indeed it is tempting to make a model of the hippocampus in which some positional signal is computed by the brain, with inputs from many regions, and each cell in the hippocampus is active or silent depending on the value of this positional signal.  This model is specified by the ``place fields'' of each cell, the probability that a cell is active as a function of position, and these can be estimated directly from the data; given the place fields all other properties of the network are determined with no adjustable parameters.

The place field of cell $\rm i$ is defined by the average activity conditional on the position $x$ along a track,
\begin{equation}
\langle\sigma_{\rm i}\rangle_{ x} = F_{\rm i}( x ) .
\label{placefield}
\end{equation}
If activity in each cell depends independently on position, then the pairwise correlations are driven by the fact that all cells experience the same $x$, drawn from some distribution $P(x)$ across the experiment.  The quantitative prediction is that
\begin{eqnarray}
C_{\rm ij} &\equiv& \langle \sigma_{\rm i} \sigma_{\rm j}\rangle - \langle \sigma_{\rm i} \rangle \langle \sigma_{\rm j}\rangle \nonumber \\
&=& \int dx \,P( x ) F_{\rm i}( x )F_{\rm j}( x ) \nonumber\\
&&\,\,\,   - \left[ \int dx \,P( x ) F_{\rm i}( x )\right] \left[\int dx \,P( x ) F_{\rm j}( x )\right] .
\end{eqnarray} 
The covariance matrix elements $C_{\rm ij}$ have a pattern that is qualitatively similar to the real data, but quantitatively very far off.  In particular the eigenvalue spectrum of the matrix predicted in this way falls very rapidly, while the real spectrum has a slow, nearly power--law decay \cite{meshulam2017collective}.  This is a first hint that the neurons in the hippocampal network share information, and hence exhibit collective behavior, beyond just place.

A new generation of experiments monitoring 1000+ neurons in the hippocampus provides unique opportunities for theory, as discussed in \S\S\ref{sec-subgroups} and \ref{chapter-RG} below.  Here we want to emphasize the way in which collective dynamics emerge from maximum entropy models of $N\sim 100$ cells. 
Equations (\ref{heff_main}, \ref{heff2}) and Figure \ref{retina_heff} remind us that models for the joint distribution of activity in a neural population also predict the probability for one neuron to be active given the state of the rest of the network.  We can go through the same exercise for a population of cells in the hippocampus:  construct the pairwise maximum entropy model, and for each neuron at each moment compute the probability that it will be active given the state of all the other neurons; results are shown in Fig \ref{missedfields}A.  

We see in Figure \ref{missedfields}A that, roughly speaking, cells are predicted to be active in sequence. This makes sense since these are place cells, and the mouse is running at nearly constant speed along a virtual track, so cells with place fields arrayed along the track should be activated one after the other. Interestingly the calculation leading to this prediction makes no reference to the (virtual) position of the mouse, or even to the idea of place fields, but only to the dependence of activity in one cell on the rest of the network.  In this window of time the mouse actually makes two trips along the track, and perhaps surprisingly the predictions for the two trips are different.    On the first trip it is predicted that several of the cells will ``miss'' their place fields, while all cells should be active in sequence on the second trip.  This is exactly what we see in the data (Fig \ref{missedfields}B).  If  neurons were driven only by the animal's position this wouldn't happen  (Fig \ref{missedfields}C).  Thus what might have seemed like unpredictable variation really reflects the collective behavior of the network, and is captured very well by the Ising model, with no additional parameters.  We return to Ising models for the hippocampus in \S\ref{sec-subgroups} below.

\subsection{Doing more and doing less}
\label{sec-more+less}

Is there any sense in which maximum entropy models are ``better'' than alternative models?  The pairwise maximum entropy models are singled out because they have the minimal structure needed to match the mean activity and two--point correlations in the network.  But how different are they from other models that would also match these data?  We could imagine, for example, that once we specify the full matrix of correlations then the set of allowed models is very tightly clustered in its predictions about higher order structure in the patterns of activity, in which case saying that these models ``work'' doesn't say much about the underlying physics.

One can build a statistical mechanics on the space of probability distributions $p(\bm{\sigma})$, defining a ``version space'' by all the models that match a given set of pairwise correlations within some tolerance $\epsilon$.  We can construct a Boltzmann weight over this space in which the entropy of the underlying distribution plays the role of the (negative) energy,
\begin{eqnarray}
Q\left[ p(\bm{\sigma}) \right] &\propto& \delta\left[ 1 - \sum_{\bm{\sigma}} p( \bm{\sigma} )\right] {\mathbf U}_\epsilon\left[ p(\bm{\sigma}); \{m_{\rm i}, C_{\rm ij}\}\right] \nonumber\\
&&\,\,\,\,\,\times\exp\left[ -\beta\sum_{\bm{\sigma}} p(\bm{\sigma}) \ln p(\bm{\sigma})\right] ,
\end{eqnarray}
where the first term in the product enforces normalization, the second term selects distributions that match expectation values within $\epsilon$, and the last term is the Boltzmann weight \cite{Obuchi+al_2015}.  Note that this is the  maximum entropy distribution of distributions (!) consistent with a particular mean value of the entropy and the measured expectation values.  As $\beta \rightarrow \infty$,  $Q$ condenses around the maximum entropy distribution, while as $\beta\rightarrow 0$ all distributions consistent with the expectation values are given equal weight.  

If  the matrix $C_{\rm ij}$ is chosen at random then one can use methods from the statistical mechanics of disordered systems to develop an analytic theory that compares the similarity of the true distribution to those drawn at random from the ensembles of distributions at varying $\beta$.  In this random setting the maximum entropy models are not special, and in a rough sense all models that match the low--order correlations are equally good approximations \cite{Obuchi+al_2015}.  Importantly this is {\em not} true for the real data on retinal neurons, where the maximum entropy model gives a better description than the typical model that matches the pairwise correlations, and this advantage grows with $N$:  ``for large networks it is better to pick the most random model than to pick a model at random''  \cite{ferrari+al_17}.

One way that we could do more in describing the patterns of neural activity is to address their time dependence more explicitly.  In particular for the retina we know that the network is being driven by visual inputs. We can repeat the movie many times and ask about the mean activity of each cell at a given moment in the movie, $\langle \sigma_{\rm i}(t)\rangle$.  In addition, as before, we can measure the correlations between neurons at the same moment in time, $\langle \sigma_{\rm i}(t)\sigma_{\rm j}(t)\rangle$.    Thus we want to find a model for the distribution over sequences or {\em trajetcories} of network states $P_{\rm traj} \left[ \bm{\sigma}(t) \right]$ that has maximal entropy and matches the time--dependent mean activity
\begin{eqnarray}
m_{\rm i}(t) &\equiv& \langle \sigma_{\rm i}(t)\rangle_{\rm ext} \\
&=& \langle \sigma_{\rm i}(t)\rangle_{P_{\rm traj}}
\end{eqnarray}
as well as the time averaged equal time correlations
\begin{eqnarray}
C_{\rm ij} &\equiv& {1\over {\tilde T}}\sum_t \langle \delta \sigma_{\rm i}(t) \delta \sigma_{\rm j}(t)\rangle_{\rm expt}\\
&=&{1\over {\tilde T}}\sum_t \langle \delta \sigma_{\rm i}(t) \delta \sigma_{\rm j}(t)\rangle_{P_{\rm traj}},
\end{eqnarray}
where $\tilde T$ is the duration of our observations in units of the time bin width $\Delta\tau$.

This is an instance of the general structure presented in \S\ref{sec-basics}, where the first set of observables is of the form 
\begin{equation}
\{f_\mu\} \rightarrow \{f_{{\rm i}, t}\} = \{ \sigma_{\rm i}(t)\} .
\end{equation}
To constrain the expectation value of each of these terms we need a separate Lagrange multiplier, and as before we think of these as local field that now depend on time, $\lambda_{{\rm i}, t} = h_{\rm i} (t)$.  In addition we have observables of the form
\begin{equation}
\{ f_\mu \} \rightarrow \{ f_{\rm ij} \} = \bigg{\{}  \sum_t    \sigma_{\rm i}(t)   \sigma_{\rm j}(t) \bigg{\}} ,
\end{equation}
and for each of these we again have a separate Lagrange multiplier that we think of as a spin--spin coupling, $\lambda_{\rm ij} = J_{\rm ij}$.  The general Eqs (\ref{maxent1}, \ref{maxent2}) now take the form
\begin{eqnarray}
P_{\rm traj} \left[  \bm{\sigma}(t) \right] &=& {1\over Z_{\rm traj}}\exp\left( - E_{\rm traj}\left[ \bm{\sigma}(t) \right]\right)
\label{stimdep1}\\
E_{\rm traj}\left[ \bm{\sigma}(t) \right] &=& \sum_{t}\sum_{\rm i} h_{\rm i}(t)\sigma_{\rm i}(t) \nonumber\\
&&\,\,\,\,\,\,\,\,\,\,+ {1\over 2}\sum_t \sum_{\rm ij} J_{\rm ij} \sigma_{\rm i}(t)\sigma_{\rm j}(t).
\label{stimdep2}
\end{eqnarray}
In this class of models, correlations arise both because different neurons may be subject to correlated time--dependent fields and because of effects intrinsic to the network.  If all of the correlations were driven by visual inputs then matching the correlations would lead to $J_{\rm ij} = 0$, but this never happens with real data.

One can go further and assume some form for the relation between the time dependent field $h_{\rm i}(t)$ and the visual inputs.  The simplest possibility is that the field is a spatiallly and temporally filtered version of the light intensity pattern shown to the retina \cite{granot-atedgi+al2013}, 
\begin{equation}
h_{\rm i}(t) = h_{\rm i}^0 + \int d^2 x \int d\tau \,K_{\rm i} (\vec{\mathbf x},\tau) I(\vec{\mathbf x},t- \tau),
\end{equation}
so that these stimulus--dependent maximum entropy models can be seen as a generalization of the widely used ``linear/nonlinear'' models for single neurons \cite{Dayan+Abbott_2001}.  These also are  the maximum entropy models consistent with the correlation between single neuron and the movie istelf,  $ \langle \sigma_{\rm i}(t) I(\vec{\mathbf x},t- \tau)\rangle$, which can be estimated without having to repeat the movie.

An alternative is to determine the time--dependent fields from experiments with a repeated movie,  and then fit a separate model to the dependence of the field on the input movie \cite{ferrari+al2018}.  This two step procedure has the advantage that incompleteness of  the model for the stimulus dependence of the field does not influence the estimates of the interactions $J_{\rm ij}$.  Indeed, the fact that we can predict the activity of single neurons in the retina from other neurons, without reference to the visual input (Fig \ref{retina_heff}), means that the problem of disentangling stimulus dependence of the fields from true interactions in non--trivial.

One of the interesting questions is how the decomposition into field and interactions connects to the distribution of sensory inputs.  We know that single neurons adapt their (apparent) input/output relationships to the input statistics, perhaps in ways that maximize the magnitude or efficiency of the sensory information that is conveyed by the resulting sequence of action potentials,\footnote{For a recent review see \citet{Bialek_24}.} and there are generalizations of this idea to populations of neurons \cite{tkacik+al_10}. In the language of the stimulus--dependent maximum entropy models, this suggests that the mapping from sensory inputs to the fields $h_{\rm i}(t)$ will change when we change the distribution of inputs.  But what happens to the interactions $J_{\rm ij}$?  Recent work in the retina suggests that the interaction matrix may be largely invariant across different ensembles of input movies \cite{hoshal+al_23}.  Despite the fact that the interactions themselves don't vary with the input ensemble, their presence enhances the reliability with which brief segments of the neural response can be used to make choices among a set of possible ensembles.  

Our discussion began with models that match the mean activity of each neuron and their pairwise correlations, with these correlations measured at equal times, resulting in Eqs (\ref{eq-P2}, \ref{eq-E2}).  The stimulus dependent models capture time dependent mean activity, where time is measured relative to the sensory inputs, but still match only the equal--time correlations.  A natural alternative is to capture time dependent correlations but simplify by matching only the global mean activity.  Concretely this means that we want to find a maximum entropy model for sequences or trajectories of states, 
\begin{equation}
P_{\rm traj2} \left[  \bm{\sigma}(t) \right] = {1\over {Z_{\rm traj2}}} \exp\left( - E_{\rm traj2}\left[  \bm{\sigma}(t) \right] \right) ,
\end{equation}
where the subscript reminds us that this is a (second) model for trajectories.  We want to match experimental observations of the mean activity, averaging over its time dependence
\begin{equation}
\bar m_{\rm i} \equiv  {1\over {\tilde T}} \sum_t \langle \sigma_{\rm i}(t)\rangle_{\rm expt}
=    \langle \sigma_{\rm i}(t) \rangle_{P_{\rm traj2}} .
\label{match_trajB1}
\end{equation}
In addition we want to match the pairwise correlations across time,
\begin{eqnarray}
C_{\rm ij}(\tau) &\equiv&  {1\over{\tilde T}}\sum_t \langle \delta \sigma_{\rm i}(t) \delta \sigma_{\rm j}(t+ \tau )\rangle_{\rm expt},\nonumber\\
&=& \langle \delta \sigma_{\rm i}(t) \delta \sigma_{\rm j}(t+ \tau )\rangle_{P_{\rm traj2}} 
\label{match_trajB2}
\end{eqnarray}
where as before $\delta \sigma_{\rm i}(t) = \sigma_{\rm i}(t) - m_{\rm i}$; we assume that the system is statistically stationary, so that correlations depend only on time differences.  

This is another instance of the general structure presented in \S\ref{sec-basics}, where one set of observables is of the form
\begin{equation}
\{ f_\mu^{\rm (means)}\}  \rightarrow  {\bigg\{}  \sum_t \sigma_{\rm i}(t)  {\bigg\}} ,
\end{equation}
and a second set of observables is of the form
\begin{equation}
\{ f_\mu^{\rm (pairs,  \tau )} \}\rightarrow  {\bigg\{} \sum_t \sigma_{\rm i}(t) \sigma_{\rm j}(t+\tau )  {\bigg\}}.
\end{equation}
As before we identify an effective field $h_{\rm i}= \lambda_{\rm i}$ as the Lagrange multiplier constraining the means of individual neurons, and now the Lagrange multipliers that  pairs of neurons separated by a time $\tau$ can be identified as  time dependent couplings $J_{\rm ij}(\tau )$.
The general Eq (\ref{maxent2}) thus becomes
\begin{eqnarray}
E_{\rm traj2}\left[  \bm{\sigma}(t) \right]  &=& \sum_t\sum_{\rm i} h_{\rm i}  \sigma_{\rm i}(t) 
\nonumber\\
&&\,\,\,\,\,\,\,\,\,\, + {1\over 2}\sum_{t\tau} \sum_{\rm ij} \sigma_{\rm i} (t)J_{\rm ij} (\tau)\sigma_{\rm j} (t+\tau ) .\nonumber\\
&&\label{Etraj2}
\end{eqnarray}
While this is a natural counterpoint to the model of Eq (\ref{stimdep2}), it has been less widely explored.

Perhaps the most important features of this class of models is that it gives us a chance to explore the breakdown of time reversal invariance in neural activity.  Note that in Eq (\ref{Etraj2}) we can swap indices on neurons and time, together, so that ${\rm i}, t \leftrightarrow {\rm j}, t'$, and this leaves the energy $E_{\rm traj2}$ unchanged.  This means that $J_{\rm ij}(\tau ) = J_{\rm ji}(-\tau)$.  But time reversal invariance would require $J_{\rm ij}(\tau ) = J_{\rm ij}(-\tau)$, which not be the result of solving the matching conditions in Eq (\ref{match_trajB2}).  This emphasizes the conceptual point that we can have maximum entropy models for systems whose dynamics violate detailed balance.  It also opens a path to investigating more concretely how patterns of neural activity represent the arrow of time \cite{lynn+al_22a,lynn+al_22b}.

In some cases pairwise maximum entropy models are not enough to capture the full structure of the network.  A simple idea is that we need more constraints, and we see how this worked in the retina where fixing the distribution $P_N(K)$ allowed for much closer agreement with experiment (\S\ref{sec-larger}).  An alternative is that we don't need more terms, just different terms.  Are there different paths to simplification, or at least to understanding why simplification is possible?

A key feature of pairwise models is that they involve matching $\sim N^2$ features of the data, many fewer than the $\sim 2^N$ parameters that would be required to describe an arbitrary probability distribution.   But as the experimentally accessible $N$ increases, eventually even $\sim N^2$ becomes too big, and we can't reliably determine the entire matrix of correlations.  If we want to keep to the maximum entropy strategy we have to find a way of working with fewer constraints, ideally $\sim N$.

An early idea was that instead of matching the full matrix of correlations we could match the distribution of these correlations across all pairs, which can be estimated more reliably \cite{Castellana+Bialek_2014}.  This problem really is the inverse of the usual spin glass:  rather than choosing interactions $J_{\rm ij}$ from a distribution and computing correlations, we choose the correlations from a distribution and infer the interactions.  If we fix only the first two moments of the distribution of correlations then the interactions develop a block structure, reminiscent of the hierarchy of correlations in replica symmetry breaking \cite{mezard+al_87}.  We can find a phase diagram in the space of these moments, which depends crucially on how they scale with $N$.  As experiments progress from $N\sim 100$ to $N\sim 10^4$ and even $N\sim 10^6$ (\S\ref{sec-imagingmethods}), it seems likely that this approach of constraining distributions will become more useful.

Another approach is to go back to the basic maximum entropy formulation but choose only a limited set of quantities whose expectation values we should constrain.  These quantities need not be simple objects such as $\sigma_{\rm i}$ or $\sigma_{\rm i}\sigma_{\rm j}$.  As an example, we could imagine a neuron somewhere else in the brain that takes inputs from the network we are studying, sums these inputs and compares with a threshold, as in the \citet{mcculloch+pitts_43} model described in \S\ref{sec-prehistory}.  Concretely we can consider the activity of these hypothetical neurons
\begin{equation}
y_\mu = \Theta\left(\sum_{{\rm i}=1}^N W_{\mu{\rm i}} \sigma_{\rm i} - \theta_\mu \right),
\label{ydef}
\end{equation}
and ask for the maximum entropy distribution that matches the expectation values of $\{y_\mu\}$ that we compute from the data.  Following the arguments above, this distribution has the form
\begin{equation}
P_{\rm out} \left( \bm{\sigma}  \right) = {1\over{Z_{\rm out}\left(\{g_{\mu}\}\right)}}
\exp\left[ - \sum_{\mu =1}^K g_\mu y_\mu \right],
\label{RPmodel}
\end{equation}
where the subscript reminds us that this model focuses on a (possible) output of the network rather than directly on the network state itself; we can set $\theta_\mu = 1$ by choosing units for the weights. Notice that the number of parameters $\{g_\mu; W_{\mu{\rm i}}\}$ is then $(N+1)K$, which is much less than $N^2$ if the number of output neurons $K\ll N$.   If we could view the weights and thresholds $\{W_{\mu{\rm i}}\}$ as given then we would have only $K$ parameters, set by the expectation values $\{y_\mu \}$, and we could even let $K \gg N$ without concerns about undersampling in experiments of reasonable length.  Surprisingly, one can make progress by choosing weights  at random  \cite{maoz+al_21}.

\begin{figure}[t]
\includegraphics[width = \linewidth]{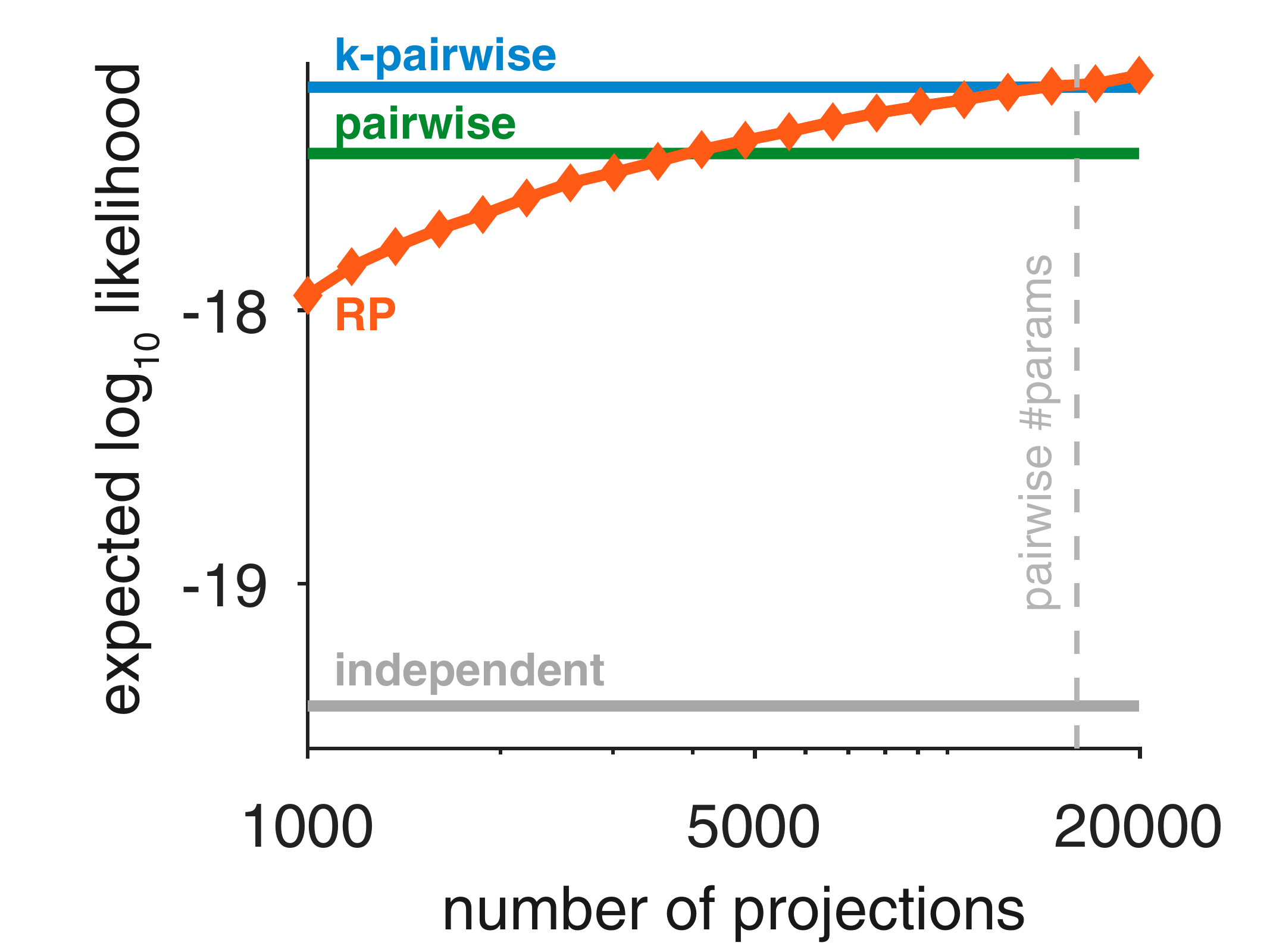}
\caption{The surprising success of random projections in describing a population of $N=178$ neurons in visual cortex \cite{maoz+al_21}.  RP are the models defined by  Eqs (\ref{ydef}, \ref{RPmodel});  independent ($P_1$), pairwise ($P_2$), and K--pairwise ($P_{2k}$) models are as described in \S\S\ref{sec-neuron1} and \ref{sec-larger}. In each case models are learned from random halves of the data, and likelihoods are computed from the held out data; plot shows averages over these random splits.  For the RP models there is also an average over many random choices for projections $\{W_{\mu{\rm i}}\}$ in Eq (\ref{ydef}); variations are small. Dashed line marks the point where the complexity of the RP models matches that of the pairwise models. \label{randproj}}
\end{figure}

Figure \ref{randproj} shows the behavior of models with random  weights or projections $\{W_{\mu{\rm i}}\}$ as applied to a population of $N=178$ neurons in the visual cortex of a macaque monkey as the animal was shown relatively simple images. Experiments were long enough that one could construct, reliably, the pairwise and K--pairwise maximum entropy models, with more than $15,000$ parameters.  The performance of these models can be measured, as usual, by estimating the mean log--likelihood of the data in the model, and we see that the models that match correlations generated the data with $\sim 100\times$ higher probability than a model of independent neurons.  More than half of this gain is achieved in the random projection models with only $K=1000$ projections, and the full performance is recovered if we allow $K\sim 15,000$ projections.  This is not, by itself, an improvement on the K--pairwise models, but here the projections have been chosen at random.  

By iteratively pruning the variables $y_\mu$ which make weak contributions to the performance of the model and replacing them with new random projections, one arrives at models with the same performance but $10\times$ fewer parameters.  One can make random choices from distributions in which different numbers of the $W_{\mu{\rm i}}$ are allowed to be nonzero, and it is suggestive that performance is best when the ``in degree'' of the connections $\{\sigma_{\rm i}\} \rightarrow y_\mu$ is small, less than ten.  These results indicate that relatively simple maximum entropy models that matched a set of strongly nonlinear functions of the network state can be very effective, although more work will be needed to understand their scaling with $N$.  It is especially attractive that these functions can be interpreted as the activity of downstream neurons.

The strategy of choosing random projections and then editing these choices is surprisingly successful.
This leaves the question, however, of whether there is a best choice of constraints given a set of possibilities.  Intuitively we measure the quality of a model by the probability that it generates the data.  More formally, a model for the distribution $P(\bm{\sigma} )$ defines a code in which each state is $\bm{\sigma}$ assigned a code word with length
\begin{equation}
\ell (\bm{\sigma}) = - \log_2 P(\bm{\sigma})\,{\rm bits} ,
\end{equation}
and hence the average amount of space needed to describe the data is \cite{shannon1948mathematical,cover+thomas_91} 
\begin{equation}
\langle \ell \rangle  = - \langle \log_2 P(\bm{\sigma})\rangle_{\rm expt} \,{\rm bits} .
\end{equation}
But maximum entropy distributions are special because, substituting from Eqs (\ref{maxent1}, \ref{maxent2}),
\begin{eqnarray}
 - \langle \ln P(\bm{\sigma})  \rangle_{\rm expt} &=& \ln Z + \sum_{\mu =1}^K \lambda_\mu \langle f_\mu (\bm{\sigma})\rangle_{\rm expt}\\
 &=& \ln Z + \sum_{\mu =1}^K \lambda_\mu \langle f_\mu (\bm{\sigma})\rangle_{P}\\
&=& -\langle  \ln P(\bm{\sigma})  \rangle_{P} = S[P] .
\end{eqnarray}
Thus---for maximum entropy models---the space into which we compress the real data is equal to the entropy of the distribution that we construct.  This means that we will achieve the greatest compression by choosing constraints that minimize the entropy of the corresponding maximum entropy model, so we arrive at a ``minimax entropy'' principle \cite{zhu+al_97}.

The minimax entropy principle is compelling but intractable in general.  If we want to constrain a subset of the pairwise correlations, then the problem simplifies enormously if we insist that the pairs we constrain form a tree with no loops.  On a tree the forward statistical mechanics problem is exactly solvable, the entropy reduction can written as a sum over the pairwise mutual informations, and there is a greedy algorithm to find the pairs which maximize the sum; the result is that we can construct the optimal tree--like model  with minimal computational effort \cite{lynn+al_23a}. The surprise is that at least in some cases the optimal tree model captures some though not all features in the collective behavior of $1000+$ neurons \cite{lynn+al_24}.  While such restricted models may not achieve the full accuracy that we hope for, they may provide a literal backbone for constructing more precise models.  It remains to be seen if there are other limits in which the minimax entropy principle becomes tractable.

Finally, when should we expect that simplified models are possible at all?  Again, the distribution $P(\bm{\sigma} )$ is a list of $2^N$ positive numbers, constrained only adding up to one; in principle these numbers could be arbitrary, as in the random energy model \cite{derrida_81}.  But a single variable $\sigma_{\rm i}$ can share only a limited amount of information with the rest of the network $\{\sigma_{{\rm j}\neq{\rm i}}\}$.  Since variables are binary, knowing the exact state of the rest of the network can provide at most one bit of information about $\sigma_{\rm i}$, although ``knowing the exact state of the rest of the network'' involves specifying $O(N-1)$ bits.  We can simplify our models if this knowledge can be compressed without losing information  \cite{bialek+al_20}.

In an Ising model where variables live on a regular lattice and interact  with their neighbors, the influence of the entire network on a single variable can be summarized by knowing the state of the neighboring variables, or $\sim z \,{\rm bits}$, where $z$ is the coordination number of the lattice.  So long as interactions are short--ranged, this number of {\em relevant} bits stays fixed even as $N\rightarrow\infty$.  In models with long--ranged interactions, including mean--field models, the averaging over many variables reduces the variance of the effective field acting on a single spin, and this again allows for compression in our description of the interactions.  Compressibility in the influence of the whole network on one variable is related to the sub--extensive behavior of mutual information between halves of the system, and this is violated  in the random energy model and in cryptographic systems \cite{ngampruetikorn+schwab_23}.  The information shared among neurons is compressible, even in cases where simple pairwise models fail  \cite{ramirez+bialek_21}.   While compressibility seems to be a requirement for simplification, it is not clear how to use compression to construct explicit simplified models.

\begin{figure*}[t]
	\includegraphics[width=\textwidth]{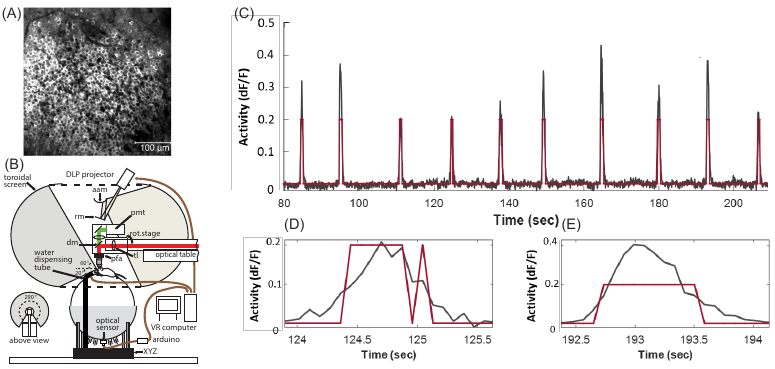}
	\caption{Imaging electrical activity in the mouse hippocampus. (A) $500\times500\,\mu{\rm m}^2$ region in hippocampal area CA1. Image is constructed in $1/30\,{\rm s}$ frames using a scanning two photon microscope.  Here the fluorescence intensity is integrated over time, so that each identifiable neuron appears as bright.  (B) Virtual reality setup where the mouse's  head is fixed while it runs on a ball; the rotation of the ball is used to compute the effective trajectory through space, driving a movie appropriate to these movements.  (C) Continuous fluorescence signal from a single neuron (black), emphasizing the high signal--to--noise ratio and the ease of defining a binary on/off version of the cell's activity (red). (D) On a finer time scale, we understand enough about the dynamics of the indicator molecule to identify slow decay of a fluorescence transient as an on/off flickering of the underlying activity. (E) A simpler case where the cell is ``on'' when the fluorescence signal is above threshold. 
	Panels (A, B) adapted from \cite{meshulam2019RG}, panels (C, D, E) adapted from \citet{meshulam2017collective}, with thanks to JL Gauthier, CD Brody, and DW Tank. 
		\label{setup}}
\end{figure*}

\section{A unique test}
\label{sec-subgroups}

As we started to see successful maximum entropy analyses of $\sim 100$ neurons (\S\ref{sec-larger}), the experimental frontier moved to recording 1000+ neurons from a single region of the brain.  Among other things (see \S\ref{chapter-RG}), these larger populations offer the chance to construct many different groups of $N=100$ cells from the same region of the brain, and to ask how the success or failure of models varies across these groups of neurons.  We can do more, and choose groups of cells such that the distribution of mean activities and pairwise correlations are essentially the same---different populations that look the same in the low order statistics that are the inputs to the maximum entropy construction.  Is the success of maximum entropy somehow guaranteed by the form of these low order data?  We will see that this is not the case, and that success therefore points to underlying structure in the network.  More deeply, we will see that when these models succeed, the match between theory and experiment is surprisingly precise, which may provide a more general lesson about the opportunities for theory in the physics of complex biological systems.

\begin{figure*}[t]
	\includegraphics[width = \textwidth]{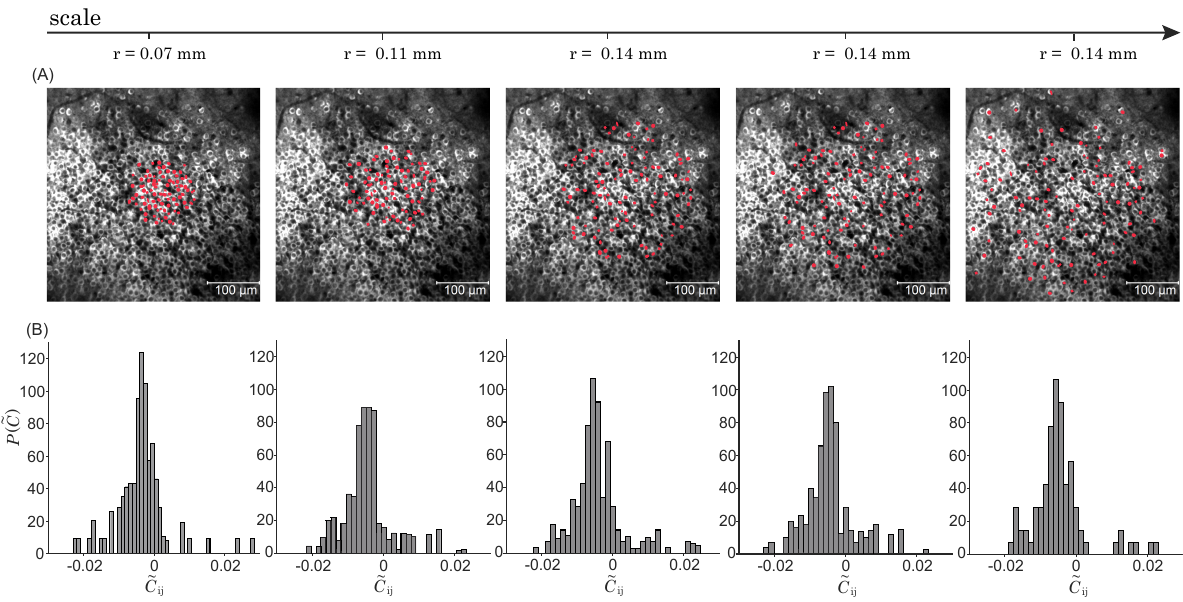}
	\caption{Subgroups of $N=100$ cells with different sampling density \cite{Meshulam+al_2021}. (A) Image of the CA1 region in mouse hippocampus, showing fluorescence signals from 1000+ neurons, as in Fig~\ref{setup}A.  Red dots indicate cells chosen, as described in the text, from five circles of increasing radius (top).  (B) Distribution of correlation coefficients, from Eq (\ref{corrcoef-V}),  across each population.
		\label{lower_order_stats}}
\end{figure*}

\subsection{Many groups of $N=100$ neurons}

 The initial application of this approach was  in the mouse hippocampus, specifically the CA1 region where cells are largely in a single plane \cite{Meshulam+al_2021}. A typical field of view is shown in Fig \ref{setup}a. Scanning two--photon microscopy covers 500$\rm {\mu m}$ at a frame rate of $30\,{\rm Hz}$ to monitor the calcium--modulated fluorescence of 1000+ cells as the animal runs repeatedly along a four meter long virtual track, as detailed in Fig \ref{setup}b. These experiments have been done on multiple animals, in each case collecting $\sim 30\,{\rm min}$ of data.   Roughly half of the cells that one sees in these recordings are ``place cells'' that are consistently active only when the mouse is in a small region of the track.

The raw data in these experiments are movies, as discussed in \S\ref{sec-imagingmethods}. There is a conventional pipeline to associate groups of pixels with individual cells, so that we have time series of fluoresence in response to electrical activity as in Fig \ref{setup}c.  We want to go one step further in and reduce the signal from each cell $\rm i$ to a binary variable $\sigma_{\rm i}$.  The simplest approach is to set a threshold, and because baselines are stable and noise levels are low, this is unambiguous.   We can do a little better, however, since if we see a pulse of fluorescence that falls from its peak and then recovers, we can use  our understanding of the dynamics of calcium unbinding from the indicator molecule to identify a flicker between on and off states. Two examples of the binarization process can be seen in Fig \ref{setup}d-e. A fully detailed description can be found in  \cite{meshulam2017collective}. As explained in \S\ref{sec-neuron1}, we can compute from these binary variables the mean activity\footnote{For the sake of clarity we repeat in this section some of the definitions from above.}
\begin{equation}
m_{\rm i} = \langle \sigma_{\rm i}\rangle  ,
\end{equation}
the covariance matrix
\begin{equation}
C_{\rm ij} = \langle (\sigma_{\rm i} - m_{\rm i}) (\sigma_{\rm j} - m_{\rm j}) \rangle  ,
\end{equation}
and the correlation matrix
\begin{equation}
{\tilde C}_{\rm ij} = {{C_{\rm ij}}\over\sqrt{C_{\rm ii}C_{\rm jj}} }.
\label{corrcoef-V}
\end{equation}

If we point randomly to one cell in the experiment and draw a circle of radius $r = 0.07\,{\rm mm}$ then we have a dense sampling of $N=100$ cells, as shown in Fig \ref{lower_order_stats}A.  If we increase the size of the circle until we enclose roughly twice as cells, we could choose randomly and create a new population of $N=100$ cells.  But this network would be noticeably different; in particular, the distribution of correlations between pairs of neurons,  $\tilde C_{\rm ij}$, would be different because there is a tendency for cells that are closer together to be more strongly correlated.

Rather than choosing completely at random, we can swap cells from the initial dense sampling with cells in the larger area,  and for each swap with check the distribution of  $\tilde C_{\rm ij}$ in the new population.  Formally, in the $k^{\rm th}$ circle we have some distribution $P_k({\tilde C})$, and in the $k+1^{\rm st}$ circle after each swap we have a new distribution $P_{k+1}({\tilde C})$.  We test the similarity of these distribution by estimating their Kullback--Leibler divergence,
\begin{equation}
	D_{KL}\left[ P_k({\tilde C}) || P_{k+1}({\tilde C}) \right] 
	= \int d{\tilde C}  P_k({\tilde C}) \log_2
	\left[ \frac{ P_k({\tilde C}) }{ P_{k+1}({\tilde C}) } \right].
	\label{DKL_def}
\end{equation}
As we make successive swaps we check that $D_{KL}$ remains small, until we have swapped half of the cells, at which point we enlarge the circle yet again.  The result, as shown in Fig \ref{lower_order_stats}, is a set of five subgroups of $N=100$ cells chosen from increasingly large areas and hence lower sampling density, but with distributions of correlations that are almost all indistinguishable.  

From the formal perspective we want to hold the distribution of correlations fixed as we look at different subgroups so that we are solving essentially similar problems.  From the functional perspective holding this distribution fixed also insures that a nearly constant fraction of the cells in each subgroup are place cells.

\subsection{Maximum entropy models for subgroups}

Starting the construction of Fig \ref{lower_order_stats} from ten randomly chosen cells in each of three animals, we have 150 distinct examples of $N=100$ cell subgroups, all with very similar low order statistics.  For each of these subgroups we can construct the maximum entropy model that matches the mean activity of each cell and the matrix of pairwise correlations.  As a reminder, from Eqs (\ref{eq-P2}) and (\ref{eq-E2}), the result is a model of the form
\begin{eqnarray}
	P_2(\bm{\sigma}) &=& {1\over {Z_2(\{h_{\rm i}; J_{\rm ij}\})}} e^{-E_2 (\bm{\sigma})} .
\\
	E_2 (\bm{ \sigma}) &=& \sum_{{\rm i}=1}^N h_{\rm i} \sigma_{\rm i} + {1\over 2}\sum_{{\rm i}\neq {\rm j}}J_{\rm ij} \sigma_{\rm i}\sigma_{\rm j} .
	\label{E2-V}
\end{eqnarray}
Note this is the original form of the model  discussed in \S\S \ref{sec-neuron1} and \ref{sec-larger}, without the additional constraint added in Eq (\ref{K_pairwise}) to give a better description of the retina.  Methods for choosing the parameters $\{h_{\rm i} ; J_{\rm ij}\}$ to match the  experimentally measured expectation values $\{m_{\rm i}; C_{\rm ij}\}$ are summarized in  Appendix \ref{sec-inference}.

Drawing from the discussion above, we can subject the predictions of these models to multiple tests:
\begin{itemize}
\item The probability that $K$ out of $N$ neurons are active simultaneously, $P_N(K)$ from Eq (\ref{PKofN_expval}).
\item The distribution of the effective energy, $E = E_2(\bm{\sigma})$ from Eq (\ref{E2-V}).
\item The correlations among triplets of neurons, $C_{\rm ijk}$ from Eq (\ref{triplet1}).
\item The fine--grained structure of triplet correlations, comparing the model's prediction errors with the experimental errors in estimating these correlations.
\item The probability that a single neuron $\rm i$ is active given the state of the rest of the network, as summarized by the effective field, $h_{\rm i}^{\rm eff}(\{\sigma_{{\rm j}\neq{\rm i}}\})$ from Eqs (\ref{heff_main}, \ref{heff2}).
\item The distribution of effective fields given the state of a single neuron.
\end{itemize}
We emphasize that each of these tests looks not just at a single number.  Thus $N=100$ cells have $\sim 1.6\times 10^5$ distinct triplet correlations, and the distribution of energies has a perhaps surprisingly rich structure.

We also note, once more, that there is no room for fitting in any of the tests.  All of the parameters of our description are determined by matching the means and pairwise correlations, so that everything else is a parameter free prediction.  The maximum entropy construction is the hypothesis that all signatures of collective activity in the network can be found in the low order statistics, and thus can be viewed as providing a set of predicted relations between these aspects of the data and the higher order statistics. 

\begin{figure*}[t]
	\includegraphics[width=\textwidth]{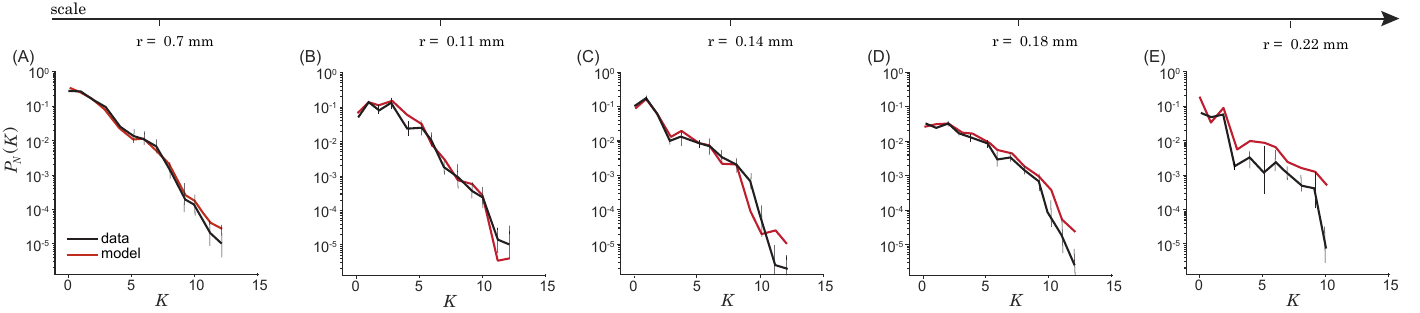}
	\caption{Distribution of summed activity in $N=100$ hippocampal neurons sampled at different densities  \cite{Meshulam+al_2021}. (A) The probability $P_N(K)$ that $K$ out of the $N=100$ neurons in the population are active simultaneously, for cells chosen from the smallest selection radius, $r = 0.07\,{\rm mm}$ at left in Fig \ref{lower_order_stats}. Model predictions (red) compared with data (black); error bars are standard deviations across random halves of the experiment.  (B--E) As in (A), but for populations chosen from larger areas, with $r=0.11, 0.14, 0.18,$ and $0.22\,{\rm mm}$ (top), moving toward the right in Fig \ref{lower_order_stats}.
	\label{P_K}}
\end{figure*}

\subsection{Success depends on sampling density}
\label{sec-density}

\begin{figure*}
	\includegraphics[width=\textwidth]{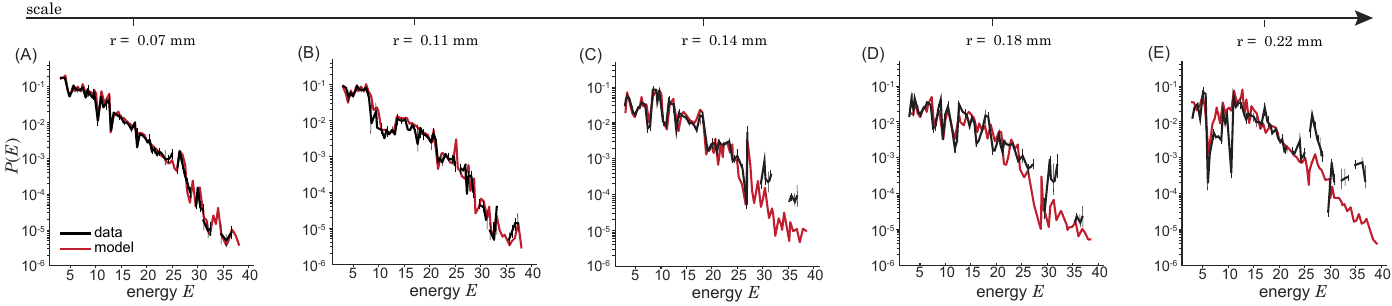}
	\caption{Distribution of effective energy, or surprise, in $N=100$ hippocampal neurons sampled at different densities \cite{Meshulam+al_2021}.
	(A) The distribution, $P(E)$, of effective energies or log probabilities, that the model assigns to every possible state in the network,  for cells in the smallest selection radius, $r = 0.07\,{\rm mm}$ at left in Fig \ref{lower_order_stats}. The distribution over states predicted by the model (red) is compared with the distribution over states as they occur in the experiment (black); both computed with a bin size bin size $\Delta E =  0.75$. Error bars are  standard deviations across random halves of the experiment. (B--E) As in (A), but for populations chosen from larger areas, with $r=0.11, 0.14, 0.18,$ and $0.22\,{\rm mm}$ (top), moving toward the right in Fig \ref{lower_order_stats}.
	\label{P_E}}
\end{figure*}

\begin{figure*}
	\includegraphics[width=\textwidth]{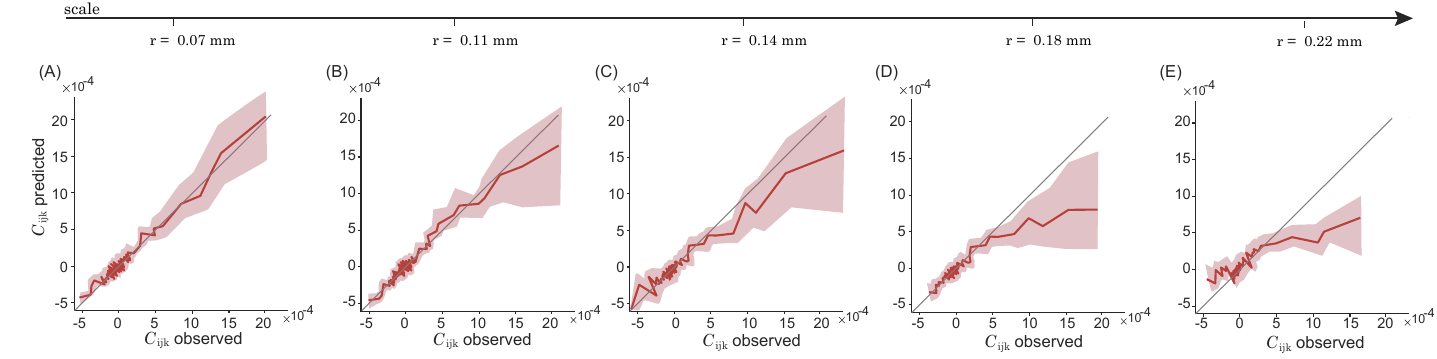}
	\caption{Trends in triplet correlations, in $N=100$ hippocampal neurons sampled at different densities  \cite{Meshulam+al_2021}. (A)  Predicted vs observed  triplet correlations $C_{\rm ijk}$,  for cells in the smallest selection radius, $r = 0.07\,{\rm mm}$ at left in Fig \ref{lower_order_stats}.  On the x-axis, the $\sim 1.6 \times 10^5$ distinct triplet correlations are grouped together into 100 adaptive bins. The corresponding values computed from the model are binned in the same way, shown on the y-axis. Shaded area indicates standard deviation of predictions within each bin. (B--E) As in (A), but for populations chosen from larger areas, with $r=0.11, 0.14, 0.18,$ and $0.22\,{\rm mm}$ (top), moving toward the right in Fig \ref{lower_order_stats}.
	\label{3pt_bin}}
\end{figure*}

\begin{figure*}
	\includegraphics[width=\textwidth]{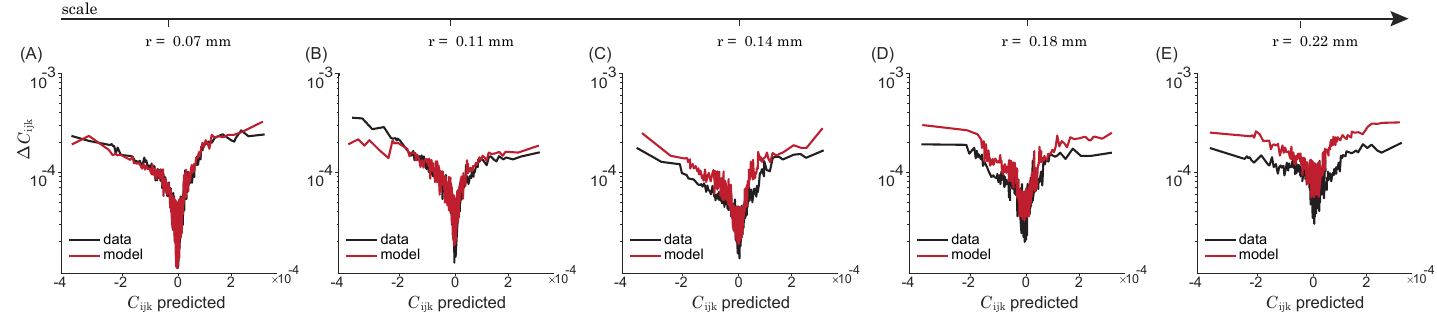} 
	\caption{Triplet correlations, in detail, for $N=100$ hippocampal neurons sampled at different densities  \cite{Meshulam+al_2021}. (A)  Comparison of the maximum entropy model prediction errors (red) for individual triplet correlations, $C_{\rm ijk}$  in Eq (\ref{C3}), with the measurement errors (black) from the data itself; for cells in the smallest selection radius, $r = 0.07\,{\rm mm}$ at left in Fig \ref{lower_order_stats}. Values on the x-axis are grouped together into 500 adaptive bins. (B--E) As in (A), but for populations chosen from larger areas, with $r=0.11, 0.14, 0.18,$ and $0.22\,{\rm mm}$ (top), moving toward the right in Fig \ref{lower_order_stats}.
	\label{3pt_ind}}
\end{figure*}

\begin{figure*}
	\includegraphics[width=\textwidth]{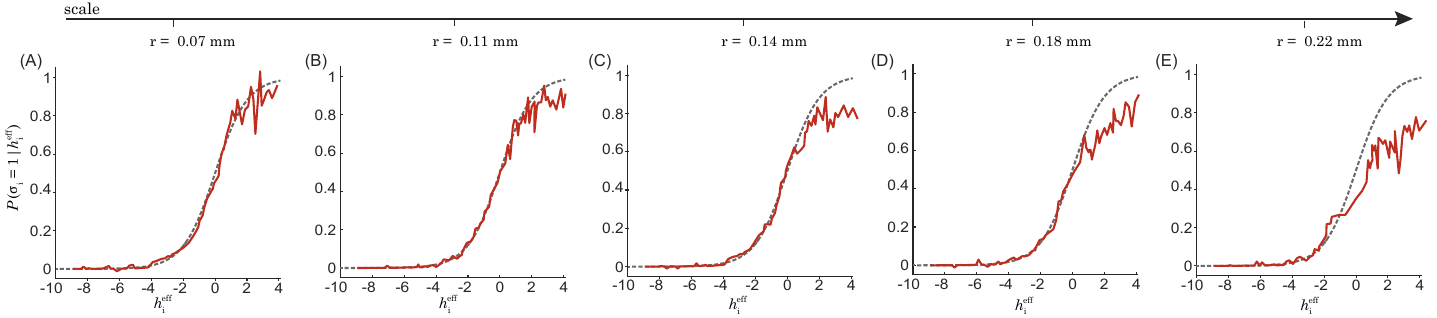}
	\caption{Collective behavior and effective fields in $N=100$ hippocampal neurons sampled at different densities \cite{Meshulam+al_2021}. (A) Probability of individual neurons to be active given the state of the network, summarized by the effective field from  Eq (\ref{heff_main});   for cells in the smallest selection radius, $r = 0.07\,{\rm mm}$ at left in Fig \ref{lower_order_stats}.  Data in red, prediction of Eq (\ref{heff2}) in dashed black line.   (B--E) As in (A), but for populations chosen from larger areas, with $r=0.11, 0.14, 0.18,$ and $0.22\,{\rm mm}$ (top), moving toward the right in Fig \ref{lower_order_stats}.	
	\label{eff}}
\end{figure*}

\begin{figure*}
	\includegraphics[width=\textwidth]{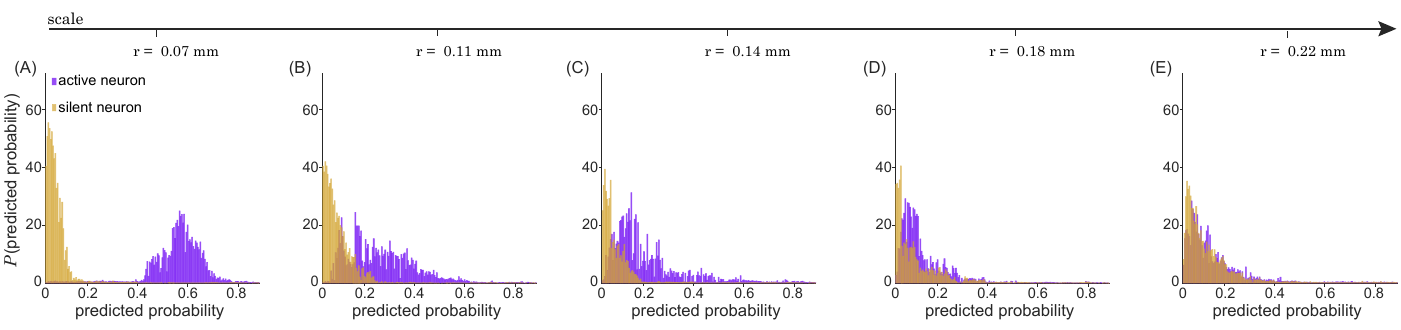} 
	\caption{Inferring the effective fields in $N=100$ hippocampal neurons sampled at different densities  \cite{Meshulam+al_2021}. (A) Distribution of effective fields $h^{\rm eff}$ given that a neuron is active (purple) or silent (gold), with effective field measured by the predicted probability of activity  from  Eq (\ref{logit});  cells here are in the smallest selection radius, $r = 0.07\,{\rm mm}$ at left in Fig \ref{lower_order_stats}.  (B--E) As in (A), but for populations chosen from larger areas, with $r=0.11, 0.14, 0.18,$ and $0.22\,{\rm mm}$ (top), moving toward the right in Fig \ref{lower_order_stats}.
	\label{pp}}
\end{figure*}

Following the agenda outlined above, we want to test the predictions of maximum entropy models against six distinct features of the data.  We do this in populations of $N=100$ cells drawn from regions of different size (Fig \ref{lower_order_stats}), so we can see how the quality of predictions depends on sampling density.  We will see that there is a systematic decay in the quality of predictions as density goes down, and that some features of the data are ``easier'' to get right than others.  Here we focus on describing the results from one example shown in Figs \ref{P_K}--\ref{pp}; Additional examples from more animals are shown in Fig \ref{predictions_all_mice}. We summarize the results across all examples and provide perspective in \S\ref{sec-precision}.

{\em Distribution of summed activity.}  Starting with the first applications of maximum entropy ideas to neurons, it has been appreciated that an important signature of collective behavior is the probability $P_N(K)$ that $K$ out of the $N$ neurons in the network are active in the same small time bins. Figure \ref{P_K} shows the $P_N(K)$ for the five groups of $N=100$ cells shown in Fig \ref{lower_order_stats}.
We see that the most  spatially contiguous group has the best quantitative agreement with the data, even down to very small probabilities, e.~g.  $P_N(K=12) \sim 10^{-5}$ (Fig \ref{P_K} A).  The observed $P_N(K)$ changes as we samples cells less densely from larger areas, but the corresponding maximum entropy models predict these changes reasonably well out to a sampling radius $r = 0.18\,{\rm mm}$ (Figs \ref{P_K}B--D).  Finally, when we sample from the largest area, predictions fail completely, with disagreements larger than experimental errors already at $K=3$ (Figs \ref{P_K}E).

{\em Distribution of effective energy, or surprise.}  Maximum entropy models predict the probability of every pattern of activity and silence in the network, or equivalently how surprised we should be by each of these microscopic states.  The negative logarithm of this probability defines an effective energy, and we can compare the distribution of this energy across the states that occur in the data with the distribution predicted by the model, as in Fig \ref{P_E}. Overall, model predictions are in excellent quantitative agreement with experimental observations for the subgroups selected from the two smallest radii (Fig \ref{P_E}A, B). We start seeing disagreements at $r=0.14\,{\rm mm}$, but even then only in the tails of very rare events $E>24$ (Fig \ref{P_E}C).  For the largest two radii  disagreements are  more notable (Fig \ref{P_E}D, E); for networks built by sparse sampling from the largest radii, significant prediction errors are visible already at  $E\sim 7$ (Fig \ref{P_E}E).  As we will emphasize below, the experimental distribution $P(E)$ has fine scale features that might be mistaken for noise, but are not, and these are reproduced by the maximum entropy model for the most densely sampled networks.

{\em Trends in triplet correlations.}  The maximum entropy models we consider here match the two--neuron correlations in the network, so a natural test is to ask about three--neuron or triplet correlations, as in Eq (\ref{triplet1}),
\begin{equation}
	C_{\rm ijk}\equiv \langle (\sigma_{\rm i} -m_{\rm i} ) (\sigma_{\rm j}- m_{\rm j} )(\sigma_{\rm k}-m_{\rm k} )\rangle .
	\label{C3}
\end{equation}
For $N=100$ cells, there are $\sim 1.6 \times 10^5$ distinct ways to choose a triplet. In Figure \ref{3pt_bin} we group the observed triplet correlations into bins, and show the mean and standard deviation of predicted correlations in each bin; perfect predictions would fall on a line of unit slope. For the three subgroups selected from the most compact regions (Fig \ref{3pt_bin}A--C), and hence with the most dense sampling, predictions are close to the line across the full dynamic range of the data. For $r=0.18\,{\rm mm}$ the model begins to underestimate the  larger, less common correlations, $|C_{\rm ijk}| = \gtrsim 3 \times 10^{-4}$ (Fig \ref{3pt_bin}D). Finally, with neurons chosen sparsely from the largest area, there is limited success with the  smallest  $|C_{\rm ijk}|$ and systematic underestimates of the (absolute) correlations over most of the dynamic range (Fig \ref{3pt_bin}E).

{\em Triplet correlations, in detail.}   Figure \ref{3pt_bin} tests the ability of the maximum entropy models to capture the trends in triplet correlations, but doesn't quite tell us whether  the individual elements of the correlation tensor $C_{\rm ijk}$ are correct  in detail.  To get at this we want to compare the errors in the model's predictions with the errors in measurement.  Once again we collect the observed correlations into small bins, and within each bin we compute the root-mean-square error in the model predictions and estimate the root-mean-square errors in measurement of the correlation itself from the data; we focus in particular on the bulk of the triplets with  $|C_{\rm ijk}| < 4\times 10^{-4}$.  Figure \ref{3pt_ind} compares these predictions and measurement errors across groups of $N=100$ cells drawn from increasingly large areas.  We see that the two measures of error are essentially identical in the smallest, most densely sampled network;  without overfitting, it is hard to perform any better than this (Fig \ref{3pt_ind}A).  As we sample cells from larger regions, the two error measures gradually separate (Fig \ref{3pt_ind}B--D), until the prediction errors are consistently larger than experimental errors across the full range of  correlations that we probe here (Fig \ref{3pt_ind}E).

{\em Collective behavior and effective fields.} One of the characteristics of a population whose behavior is collective is that the  activity in the network as a whole can be strongly predictive of individual member's activity. Using the equivalence between our maximum entropy model and an Ising model with competing interactions, this predictive power is summarized by an ``effective field,'' $h^{\rm eff}$, acting on each neuron, as in Eq (\ref{heff2}); in the model used here [Eq (\ref{E2-V})] this becomes
\begin{eqnarray}
	h^{\rm eff}_{\rm i} &=& E ( \sigma_1, \, \cdots ,\,\sigma_{\rm i}=0, \,\cdots , \,{\sigma_N} ) 
	\nonumber \\
	&& 
	\,\,\,\,\,\,\,\,\,\, 
	- E ( \sigma_1, \, \cdots ,\,\sigma_{\rm i}=1, \,\cdots , \,{\sigma_N} ) \nonumber\\
	&=& h_{\rm i} + \sum_{{\rm j}\neq{\rm i}} J_{\rm ij}\sigma_{\rm j} .
	\label{heff_def}
\end{eqnarray}
The effective field predicts the probability for any single neuron to be active at a single moment in time, given the active/silent state of all the other neurons in the population at the same time point; from Eq (\ref{heff_main}),
\begin{equation}
	P(\sigma_{\rm i} =1| h^{\rm eff}_{\rm i} ) = \frac{1}{1+\exp(-h^{\rm eff}_{\rm i})} .
	\label{logit}
\end{equation}
In Figure \ref{eff} we examine the quality of these predictions as a function of sampling density, as with previous tests.  For each cell ${\rm i}$, and for every moment in time, we can compute the effective field from the state of all the other neurons, and then we can estimate the probability that cell ${\rm is}$ is active given that the field falls into some small bin.  We see that the agreement between theory and experiment is very good for network built from the most dense sampling (Fig \ref{eff}A), even at the extremes of the effective field.   The quality of predictions falls gradually as we sample with lower density from larger areas (Figs \ref{eff}B--E).  In particular with dense sampling there are moments when the effective field is large  enough that we predict a cell to be active with near certainty, and these predictions are correct.  This strong (if rare) prediction fails as we look at lower density populations, and this error spreads to lower and lower probabilities until the models even fail at negative fields.

{\em Inferring the effective field.}  If the effective field acting on a neuron is large and positive, our models predict that the neuron should be active; quantitatively the model predicts the probability of activity as in Fig \ref{eff}.  Can we turn this around and use the activity or silence of one cell to predict the state of the rest of the network, as summarize by the effective field?  These questions are related by Bayes' rule,
\begin{equation}
P(h_{\rm i}^{\rm eff} | \sigma_{\rm i} = 1) = {1\over{P(\sigma_{\rm i} = 1)}} P(\sigma_{\rm i} = 1 | h_{\rm i}^{\rm eff}  ) P(h_{\rm i}^{\rm eff}) ,
\end{equation}
and similarly for $\sigma_{\rm i} = 0$.
Because the distribution of effective fields has a non--trivial form, it is not easy to guess how these distributions will look.  In particular we would like to see that active neurons point to a state of the network that generates large positive fields, and conversely for inactive neurons, so that $P(h_{\rm i}^{\rm eff} | \sigma_{\rm i} = 1)$ and $P(h_{\rm i}^{\rm eff} | \sigma_{\rm i} = 0)$ are distinguishable.  We test this distinguishability in Fig \ref{pp}, expressing the effective field as the predicted probability of activity through Eq (\ref{logit}). We see that when we build networks by dense sampling from a small region, the two distributions are almost non--overlapping (Fig \ref{pp}A), so that the activity or silence of a single cell is maximally informative about state of the network as whole.  Overlap is visible as soon as we sample from $r=0.11\,{\rm mm}$ (Fig \ref{pp}B), and continues to grow (Fig \ref{pp}C, D) until at the sparsest sampling from the largest area the two distributions overlap almost completely (Fig \ref{pp}E). It is interesting to note that as quality of predictions falls off with the increased sampling radius, the distribution of fields conditional on an active neuron  (purple) moves toward the distribution  conditional an a silent neuron (yellow), rather than both changing towards each other.  This is consistent with the errors in Fig (\ref{eff}) starting at large positive field and spreading toward lower values.  Taken together these results show  that as the radius increases the model predicts more false negatives, i.e.  ``misses'' predicting the activity of a neuron that was active; more precisely the model fails to connect active neurons with the associated states of the network.  It also suggests that false negatives are more difficult to avoid than false positives.

\subsection{Precision matters}
\label{sec-precision}

\begin{figure*}[t]
	\includegraphics[width=\textwidth]{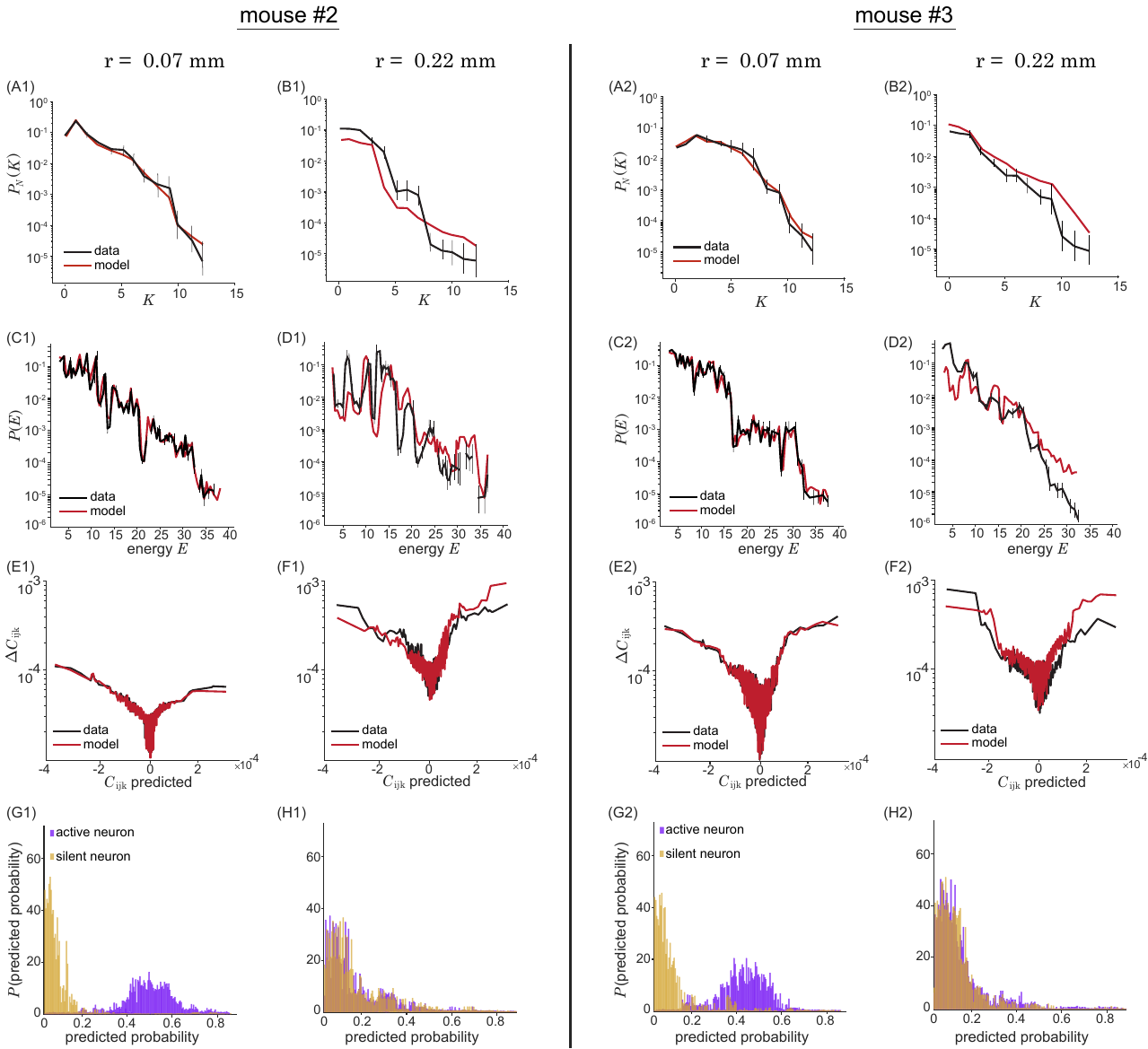}
	\caption{Model predictions examples for two more mice. Predictions shown for the subgroups constructed from the two extreme radii, with the same starting point.  Panels (A1-H1) from experiments in mouse \#2, (A2-H2) from mouse \#3. (A, C, E, G) predictions for the subgroup sampled from the smallest radius (neighboring cells). (B, D, F, H) predictions for the subgroup sampled from the largest radius (entire field of view). (A, B) Distributions of summed activity as in Fig \ref{P_K} for mouse 1. (C, D)  Distributions of effective energy as in Fig \ref{P_E}. (E, F)  Detailed triplet correlations as in Fig \ref{3pt_ind}. (G, H) Inferences of the effective field from the activity of a single neuron as in Fig \ref{pp}.
	\label{predictions_all_mice}}
\end{figure*}

The first thing we notice is that theory and experiment really can agree {\em very} well.  We see this especially in the panels of Figs~\ref{P_K} through \ref{pp} that refer to sampling $N=100$ neurons from the smallest radius, and in the left columns of the examples from two additional mice in Fig ~\ref{predictions_all_mice}.   It is particularly striking that maximum entropy models can reproduce bumps and wiggles in the energy distribution that one might have dismissed as noise, though this would be inconsistent with the measured error bars, and that the bulk of the $\sim 10^5$ individual triplet correlations are reproduced within experimental error.  We saw similar results in the first such analysis of the hippocampus \cite{meshulam2017collective}, but it is reassuring to see that this detailed quantitative success is reproducible across different populations of neurons in independent experiments on multiple animals.  While we expect this sort of reproducibility in physics, we should not take it for granted in the complex context of a functioning brain.

It is equally important that success is not automatic.  If we look only at $N=100$ cells from the smallest radius, where we have essentially complete sampling of a local network, everything ``works'' and it is hard to assess the significance of this result.  It could be that the models are so expressive that they can explain anything.  It could also be that while we see the multiple tests of the model as being different, the model or the real network ties these different quantities together so strongly that all succeed or fail together.  Looking at networks built by sampling at lower density from larger regions from larger regions we see that neither of these ideas are correct.  Different networks can be more or less well described by pairwise maximum entropy models, even though they have similar low--order statistics,\footnote{In particular, this lays to rest the early speculation that success or failure of these models could be predicted from the mean activity of the neurons alone \cite{roudi2009pairwise}.} and different features of the data can be captured or not by these models, suggesting that there is a hierarchy of difficulty to the different experimental tests.  More deeply, success or failure of the model must be telling us something about the underlying network beyond what we see in the pairwise correlations.

It is useful to look at the performance of the model at a middle level of sampling density ($r = 0.14\,{\rm mm}$), corresponding to panels (C) in Figs~\ref{P_K}--\ref{pp}.  We see that the predictions for the probability of $K$ neurons being active simultaneously (Fig~\ref{P_K}C), for the distribution of effective energies (Fig~\ref{P_E}C), and for the trends in triplet correlations (Fig~\ref{3pt_bin}) are not bad, and if this were the best we had seen we might think it was a success.  But when we look at the triplets in detail (Fig~\ref{3pt_ind}C), the activity of neurons as a function of the effective field (Fig~\ref{eff}), and the ability to infer the effective field from the activity of individual neurons (Fig~\ref{pp}C), the agreement between theory and experiment is noticeably worse.  This trend continues as we build networks by sampling the same number of cells from larger areas.

While the quality of predictions generally goes down at lower sampling densities, these failures happen in a well defined order.  Good predictions of $P_N(K)$ survive longest (Fig~\ref{P_K}), followed by the distribution of effective energies (Fig \ref{P_E}) and trends in triplet correlations (Fig \ref{P_E}),  which are roughly equivalent in performance.   The three more challenging properties to predict also follow an internal hierarchy, with the inference of the effective fields being the most difficult, with significant disagreements arising even at $r=0.11\,{\rm mm}$ (Fig \ref{pp}).

We also identify two particularly intriguing examples of model success and failure. To begin, it is interesting that we can have models which capture the trends in triplet correlations (e.g. Fig~\ref{3pt_bin}C) while the prediction errors for individual triplet correlations are outside the experimental errors (Fig~\ref{3pt_ind}C).  The failure to predict individual triplet correlations is a hint that something more serious is going wrong, and again this gets worse at lower sampling density.  Another observation is that a model can give a decent description of how the activity of a single cell depends on the network state through the effective field (e.g. Fig~\ref{eff}C) while it is almost impossible to distinguish the distributions of fields consistent with that cell being active or silent (Fig~\ref{pp}C).  This is not so much a disagreement with data as a breakdown in the interpretability of the model:  we would like to be able to say that, because behavior is collective, an active cell is responding to a positive field imposed the rest of the network, but this proves to be the most fragile of predictions.

The ability of these statistical physics models to reproduce all $N^3/3! \sim 10^5$ triplet correlations within the errors of the measurements gives a sense for the power of this theoretical approach.  Not so long ago it seemed sensible to consider alternative models that capture qualitative features of the triplet correlations \cite{macke2011common}, but now we see that is possible for pairwise models to predict all the triplet correlations within errors.  Importantly this success depends on the density of sampling.

Even if a large network of neurons is described exactly by a pairwise model, the distribution over states in a subnetwork will be described only approximately by such models.    The approximation gets better if the interactions in the whole network are largely within the subnetwork.  The success of pairwise models when applied to dense sampling from restricted areas suggests, strongly, that interactions and inputs are spatially local.  Although still somewhat controversial, this locality is consistent with more direct measurements over many years \cite{wiener+al_89,hampson+al_99,rickgauer2014simultaneous}.

The hippocampus is especially interesting because of the well--studied ``place cells'' that play a role in navigation, as discussed in \S\ref{sec-larger}.  We have seen how a model in which cells respond independently to the animal's position fails to capture the variability of responses in repeated movements through the same region, while the maximum entropy models predict this behavior as a response of individual neurons to the state of whole network without reference to position (Fig~\ref{missedfields}).  In looking more generally at alternative models (\S\ref{sec-alternatives}) we will see that the independent place cell model also fails to account for the triplet correlations (Fig~\ref{place_fail}).  This failure is quite dramatic, but only because we have seen the detailed quantitative success of the maximum entropy approach.  The conclusion is that cells in the hippocampus share information about more than just place, and this information is captured by the couplings in the Ising model.   This information can be quantified in bits \cite{meshulam2017collective}, and this picture is consistent with models in which place selectivity itself is an emergent property of the network \cite{treves+al_92}.

\begin{figure}
	\includegraphics[width=\linewidth]{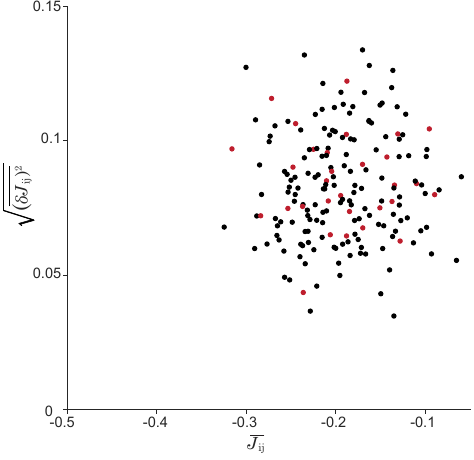}
	\caption{Mean vs standard deviation of the coupling constants in maximum entropy models constructed for each of the 150 populations of all sampling radii in three animals. The populations associated with the smallest radii are highlighted in red (10 from each animal).
		\label{Jscatter}}
\end{figure}

The success of pairwise maximum entropy models may be more surprising because the different examples of $N=100$ neurons really are different, even though the distributions of low--order statistics are quite similar.   We can see this directly from the data by comparing examples of $P_N(K)$  from three different animals (Figs~\ref{P_K}A and \ref{predictions_all_mice}A1, A2).  We can make the same point looking at the data through the lens of the models by comparing examples of $P(E)$ (Figs~\ref{P_E}A and \ref{predictions_all_mice}C1, C2). If we look at the models themselves, they all are spin--glass--like, with fluctuations in the couplings $J_{\rm ij}$ from link to link that are comparable to the mean coupling, as shown in Fig \ref{Jscatter}.  All of the models are in a regime of reasonably strong coupling, with $N\overline{J} \sim N\overline{(\delta J)^2} \sim 1$, but all the models are different in the precise form of the matrix $J_{\rm ij}$;\footnote{As usual we write $\overline{\cdots}$ to denote an average over ``disorder'' in the parameters of a model, to be distinguished from $\langle \cdots \rangle$ which denotes an average over variables drawn from the model at fixed parameters.  Here the ``disorder'' is the variation of couplings across all $N(N-1)/2 \sim 5000$ distinct pairs in each population of $N=100$ neurons.} 
these differences from subgroup to subgroup are larger than expected if the matrix elements were being drawn independently from a fixed underlying distribution.  Importantly, these differences are present even in the populations drawn from the smallest radii, where the models are most successful.  On the one hand these observations indicate  that relatively simple statistical physics models of real living systems are succeeding in capturing how particular systems behave, in detail.  On the other hand, this leaves open the question of whether there is something more universal in this behavior,  to which we return in \S\ref{sec-RG}.

Finally, there is a more general lesson to be drawn from this larger scale survey:  as the quality of measurements on biological systems improves we should aspire to the kind of detailed, quantitative theory/experiment comparison that we expect in other areas of physics.

\section{Criticality}
\label{criticality}

Correlations between two neurons in a network typically are weak but widespread.  This is reminiscent of what happens in mean--field models. As an example, for a mean--field ferromagnet all the pairwise correlations are equal and $C \sim 1/N$ \cite{sethna2021statistical,kivelson+al_24}.  If we take this analogy seriously, then correlations in network with $N\sim 100$ cells should be $C \sim 0.01$, which is in fact smaller than what we see.  More seriously, while we can observe a varying number of neurons the actual size of the network is fixed by the patterns of connectivity, and the ``real'' values of $N$ are even larger.  The familiar statistical physics models thus make it difficult to understand how the correlations, averaged over all pairs of cells, can reach $N\bar C \gg 1$.  There are two broad possibilities: such large correlations could be driven by fluctuating external fields, or could emerge from tuning of the system close to a critical point.

Living systems are not random combinations of their components, and it is a challenge to define what is special.  If the number of interacting components is large, we might expect that behaviors can be organized into a phase diagram.  Critical points in the phase diagram are special in many ways:  collective coordinates can be infinitely sensitive to variations in external parameters; correlations can extend throughout the system, far beyond the range of direct interactions; fluctuations and responses can occur over a wide range of time scales, with the longest time scale growing with the size of the system.  For all these reasons, and more, many groups have suggested that biological systems might be tuned, or self--tuned, to a critical point \cite{bak_96,mora+bialek_11,munoz2018}.

\begin{figure}[t]
\includegraphics[width=\linewidth]{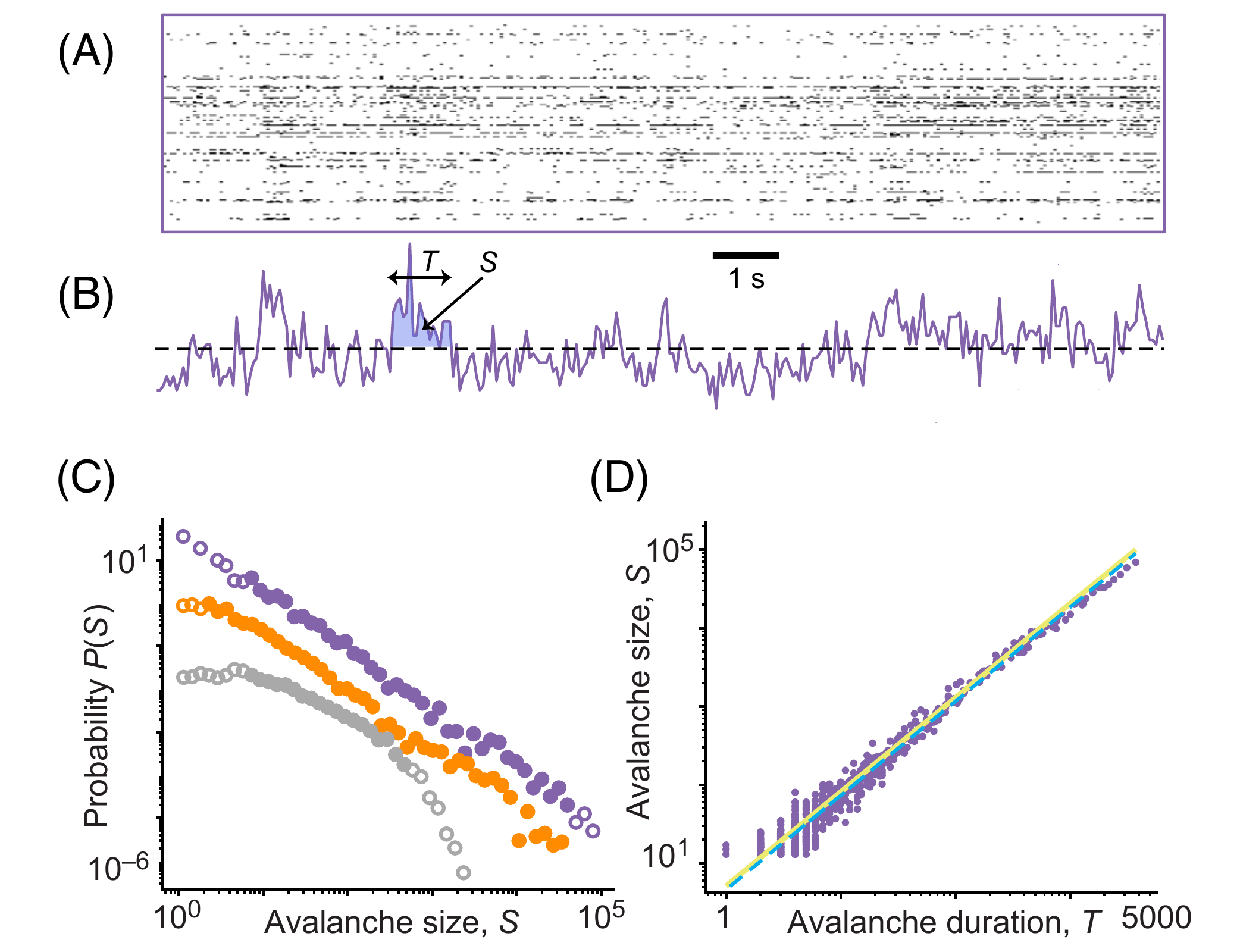}
\caption{Avalanches in a population of $N=208$ neurons in the motor cortex of a mouse as it runs along a track \cite{fontenele+al_24}.  (A) Spike rasters from all $N$ cells, where a dot represents the occurence of a spike. (B) Total number of spikes in $\Delta\tau = 50\,{\rm ms}$ bins.  Dashed line marks a threshold to define an avalanche; the area $S$ defines the avalanche size and $T$ the duration.  (C) Distribution of avalanche sizes $P(S)$, where the threshold has been set to the eighth percentile in the distribution of spike count.  Full data (purple);  projection onto the largest $5$ principal components of the activity (orange), which preserves much though not all of the scaling behavior; projection onto the remaining $N-5$ components (grey).  Distributions are shifted vertically for clarity.  (D) Avalanche size $S$ vs duration $T$, measured in units of $\Delta\tau$; each point is a single avalanche event. Lines are power laws from different model predictions.
\label{avalanches}}
\end{figure}

\subsection{Avalanches and dynamics}
\label{sec-avalanches}

The idea of criticality in networks of neurons was given considerable stimulus by the emergence of models for self--organized criticality \cite{tang+al_87,bak+al_87} in which, as the name suggests, dynamical systems can ``tune themselves'' to criticality rather than requiring precise adjustment of some underlying parameters.  The simplest models of self--organized criticality describe a (stylized) sandpile, with sand dropping randomly onto the surface, and criticality is the statement that the avalanches which collapse the high peaks in the pile occur in all sizes, with a power--law distribution.  The first suggestion that criticality might be relevant to the brain was the observation of ``neural avalanches'' in the activity of neural networks in a dish with an array of electrodes on its bottom surface  \cite{beggs+plenz_03}.  Activity in these systems consists of long periods of quiet punctuated by bursts, and these bursts are avalanche--like in the sense that the random occurrence of activity in one or a few cells triggers activity in other cells, spreading through the network.    Power laws are seen not just in the amplitude of the avalanches but also in their duration and in the mean amplitude as a function of duration; the averaged trajectories of avalanches with different duration can be rescaled to a universal form \cite{friedman+al_12}. 

In the earliest experiments,  activity was defined by the signal at a single electrode exceeding some threshold, and scaling often was confined to a narrow range. More recent experiments resolve the spikes from single neurons in intact brains \cite{fontenele+al_19} and resolve scaling over three decades  \cite{fontenele+al_24}.  An example, from neurons in the motor cortex of a behaving mouse, is shown in Fig \ref{avalanches}.

A central feature of criticality is critical slowing down.  In a low--dimensional dynamical system we expect to see one slow mode appear as the system parameters approach a bifurcation, but in a system with many degrees of freedom we can see a macroscopic density of modes with decay rates approaching zero; in many cases this is understandable as a result of dynamic scaling, as discussed in \S\ref{sec-RG}.  \citet{solovey+al_15} took a more phenomenological approach, analyzing electrocorticographic recordings (ECoG) in primates.

\begin{figure}[t]
\includegraphics[width=\linewidth]{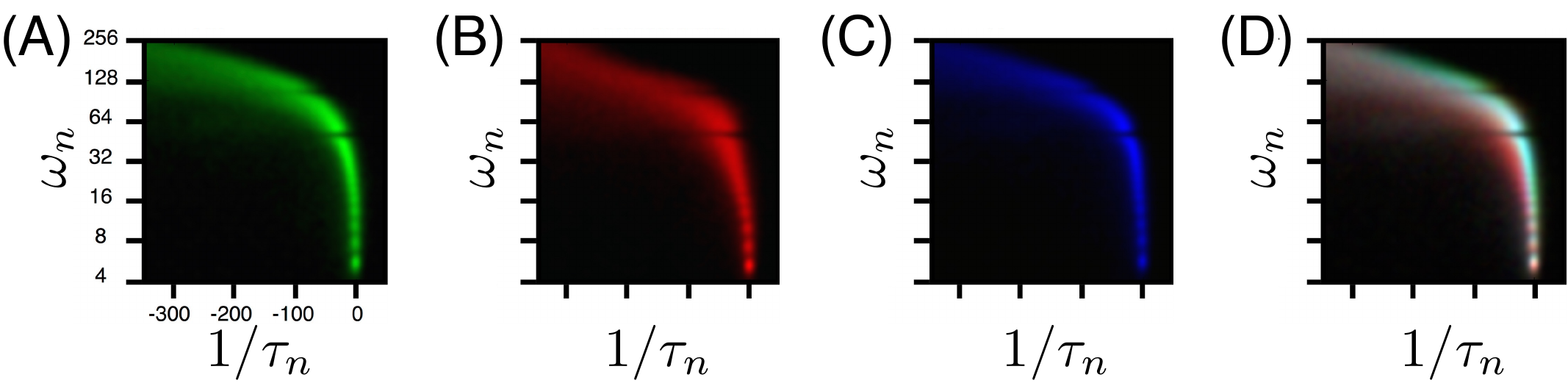}
\caption{Spectra of ECoG fluctuations in primate cortex \cite{solovey+al_15}.  Frequencies $\omega_n$  and decay rates $1/tau_n$ are extracted from the eigenvalues of the local dynamics matrix, as in Eq (\ref{ecog_eig}), and the density of these points is mapped across a long recording.  (A)  In an awake animal (green).  (B) Under anesthesia (red). (C) After recovery from anesthesia (blue).  (D) The three spectra are superposed.  The similarity of results before and after anesthesia is shown by the dominance of cyan rather than separate blue and green regions.  The faster decays under anesthesia are shown by the leftward displacement of the red density.  
\label{ecog}}
\end{figure}

ECoG is done by placing an array of electrodes on the surface of the brain; this is similar to electroencephalography (EEG), which uses an electrode array on the surface of the skull.  ECoG cannot resolve individual neurons, but offers higher spatial resolution than EEG; it often is used in neurosurgery to map brain areas in humans.  The dynamics of the voltage signals $\{V_\mu (t)\}$ are nonlinear, but one can make progress in a locally linear approximation. Concretely, the linear approximation is
\begin{equation}
V_\mu (t) = \sum_\nu A_{\mu\nu} V_\nu (t-\Delta\tau) + \epsilon_\mu (t),
\end{equation}
where $\mu = 1,\, 2,\, \cdots ,\, 128$, the time resolution $\Delta\tau = 1\,{\rm ms}$, $\epsilon_\mu$ is a noise term that we try to minimize by adjusting the dynamical matrix $A_{\mu\nu}$.  Because we expect linearity to work only locally, the dynamical matrix is fit to short ($500\,{\rm ms}$) segments of the data.  In each segment we can find the spectrum of the dynamical matrix,
\begin{eqnarray}
\sum_\nu A_{\mu\nu}\phi_\nu^{n} &=& \Lambda_n \phi_\mu^n\\
\Lambda_n &=& e^{ -(i\omega_n + 1/\tau_n)\Delta\tau },
\label{ecog_eig}
\end{eqnarray}
which defines a collection of modes with frequencies $\omega_n$ and relaxation times $\tau_n$.  Combining data across many segments we find a density in the $(\omega,1/\tau )$ plane, as shown  in Fig \ref{ecog}.  We see that there is a substantial concentration of modes with large values of $\tau$, almost touching the stability line $1/\tau = 0$ (Fig \ref{ecog}A).  Perhaps even more remarkably, the density shifts away from the stability line, toward shorter relaxation times, as the animal is anesthetized (Fig \ref{ecog}B) and  then the slow modes reappear as the animal wakes up (Fig \ref{ecog}C).  Not only do we see signs of critical slowing down, but these are associated with consciousness as opposed to sleep.

Analyses of ECoG and avalanches share the need for making choices.  Avalanches need to be defined, at least by a threshold, time is discretized, and care sometimes needs to be taken in marking the ends of these events.  The dynamics of ECoG signals surely are nonlinear, and the locally linear approximation uncovers interesting structure but one might worry that eigenvalue spectra have a clear meaning only in this approximation.  Given that there are choices to be made, one view is that there is a correct version of these choices, and the other view is that (within reason) these choices shouldn't matter for our conclusions.  In the case of avalanches, power--law size distributions are visible across many choices of parameters, but from the start it has been noticed that scaling exponents vary \cite{beggs+plenz_03}.  Given this sensitivity to choices in the analysis, it is not clear whether we should expect universality of exponents across different networks of neurons.  As we will emphasize in \S\ref{sec-RG}, in the examples that we understand scaling is very precise, and in a sense becomes clearer the more closely we look.  In addition, criticality is more than power laws.  Recent work has drawn attention to universality in the temporal form of avalanches \cite{friedman+al_12} and pushed for more precise tests of scaling over larger dynamic range \cite{fontenele+al_24}, but this has been challenging.  We hope that improved experimental methods will make it possible to address these issues more fully.

A much simpler notion of dynamical criticality arises in thinking about how the brain integrates signals over time.  As an example, we (and other animals) move our eyes to compensate for the rotation of our head.  This requires that our eye muscles apply a force related to the rotational displacement, else the eyes would relax back to their resting position.  But we measure the rotation of our head using our vestibular system, and this is an inertial sensor; viscosity of the fluid inside the semicircular canal converts the acceleration signal into a velocity signal, but it is left to our brains to do one more integral, converting velocity into displacement.  This ``oculomotor integrator'' has been studied for decades, in many different organisms.  A large class of models for the underlying circuits can be approximated as dynamical systems that have a line attractor, with position along the line corresponding to position of the eye \cite{seung_96}.  The linearized dynamics of such a system has a true zero mode, so that the network is poised at a bifurcation or critical point between stable and unstable behavior.  In practice the integration is leaky, but on times scales orders of magnitude longer than the relaxation times of individual neurons in the network \cite{aksay+al_01}, so these systems must be very close to critical.  How this relates to the underlying connections among neurons is an active topic of investigation, in model organisms ranging from zebrafish to primates and exemplifying current efforts to map synaptic connectivity completely \cite{joshua+lisberger_15,vishwanathan+al_24}.

Finally, a short discussion of dynamical criticality in relation to learning.  Many of these considerations are common to a broader class of dynamical systems, so let's think about continuous variables $x_{\rm i}(t)$ that obey quite general equations of motion
\begin{equation}
{{dx_{\rm i}}\over{dt}} = F_{\rm i}\left({\bm x}, t; {\bm \theta} \right) ,
\end{equation}
where ${\bm\theta} = \{\theta_\alpha\}$ are the adjustable parameters that we imagine can be learned by assessing the performance of the network, e.~g. following the gradient of some cost function.  We assume that this cost can be measured locally in time, and that the total cost $\bf C$ is an integral over time, so that
\begin{equation}
{\bf C} = \int dt\, {\cal C}[{\bm x}(t), t] .
\end{equation}
As an example, some of the variables $\bm x$ could be motor outputs, with $\cal C$ measuring the distance between these outputs and some desired trajectory.  The cost depends implicitly on the parameters $\mathbf \theta$ through the equations of motion, which makes it difficult to compute how the cost changes when we change parameters.

One strategy to make the dependence on parameters more explicit is to attach  extra terms to the local cost $\cal C$ that acts Lagrange multipliers to enforce the equations of motion: rather than finding the minimum of $\bf C$ with respect to parameters, we minimize the action
\begin{eqnarray}
{\cal S} &=& \int dt,{\cal L}\left[ {\bm x}(t), {\bm \lambda}(t),t ; {\bm \theta} \right]\\
{\cal L} &=& {\cal C} [ {\bm x}(t), t] - \sum_{\rm i} \lambda_{\rm i} \left[ {{dx_{\rm i}}\over{dt}} 
- F_{\rm i} \left( {\bm x}, t; {\bm \theta} \right)  \right].
\end{eqnarray}
Notice that along a trajectory that obeys the equations of motion we have
\begin{equation}
{{d{\bf C}}\over{d\theta_\alpha}} = {{d {\cal S}}\over{d\theta_\alpha}} .
\label{dCdtheta1}
\end{equation}
The use of Lagrange multipliers as auxiliary dynamical variables goes back to Pontryagin,\footnote{For an accessible source see \citet{pontrjagin}.} and has found wide application in control theory.  It reappears in the use of field theoretic methods for classical stochastic dynamics \cite{martin+al_73}, and its relevance to connecting learning and dynamics in networks of neurons was emphasized by \citet{krishnamurthy+al_20}.

Taking the derivatives in Eq (\ref{dCdtheta1}), we find
\begin{eqnarray}
{{d{\bf C}}\over{d\theta_\alpha}} 
  &=&\sum_{\rm i}\int dt\, \left[  { {\partial{\cal C}} \over {\partial x_{\rm i}}} {{dx_{\rm i}}\over {d\theta_\alpha} } 
 - \lambda_{\rm i}(t) {d\over{dt}} {{dx_{\rm i}}\over {d\theta_\alpha} } 
 \right] \nonumber\\
 &&
+\sum_{\rm i} \int dt\, \lambda_{\rm i}(t) \left[ {{\partial F_{\rm i}}\over{\partial\theta_\alpha}}
 +{{\partial F_{\rm i}}\over{\partial x_{\rm j}}}{{dx_{\rm j}}\over {d\theta_\alpha} } \right] .
 \label{dCdtheta2}
\end{eqnarray}
Further, if we are careful about the boundary conditions the extremum with respect of ${\mathbf x}(t)$ can be written as an equation for the dynamics of the Lagrange multipliers,
\begin{eqnarray}
{ {\delta {\cal S}} \over {\delta x_{\rm i}(t)} } &=& 0\\
\Rightarrow { {d\lambda_{\rm i}(t)} \over {dt} } &=& 
-\sum_{\rm j} { {\partial F_{\rm j}} \over {\partial x_{\rm i}} } \lambda_{\rm j} (t) 
- { {\partial{\cal C}} \over {\partial x_{\rm i}}}
\label{dlambdadt}
\end{eqnarray}
Substituting into Eq (\ref{dCdtheta2}) and again integrating by parts, we find that all the terms with $dx_{\rm i}/d\theta_\alpha$ cancel, leaving 
\begin{equation}
{{d{\bf C}}\over{d\theta_\alpha}} = \sum_{\rm i} \int dt\, \lambda_{\rm i}(t)  {{\partial F_{\rm i}}\over{\partial\theta_\alpha}} .
\label{gradCresult}
\end{equation}

We see from Eq (\ref{dlambdadt}) that the auxiliary variables $\bm \lambda$ (locally) grow or shrink exponentially, and this is determined by the eigenvalues of the matrix $ { {\partial F_{\rm j}} / {\partial x_{\rm i}} }$ evaluated along the trajectory.  Importantly, this is the transpose of the  dynamical matrix that determines, through the equations of motion, whether two nearby trajectories ${\bm x}(t)$ and ${\bm x}(t) +\delta{\bm x}(t)$ separate or converge with time.  Thus if the network dynamics is fully stable, with negative Lyapunov exponents, then $\bm\lambda$ will decay exponentially, and through Eq (\ref{gradCresult}) the gradient of the cost with respect to parameters also will be exponentially small, making it difficult to learn.  Conversely, if the network dynamics are strongly chaotic, with positive Lyapunov exponents, then $\bm \lambda$ will grow exponentially and so will the gradient, again making it difficult to learn.  The only way to insure that the gradient of the cost function has ${\mathcal O}(1)$ contributions from all along the trajectory is for the network dynamics to be characterized by Lyapunov exponents near zero---the regime of dynamical criticality.  Recurrent networks near criticality may also be more effective because they have access to a wider range of time scales.  These observations are broadly in agreement with empirical results \cite{vorontsov+al_17,pascanu+al_13,bertschinger+natschlager_04}.

\subsection{An effective thermodynamics}
\label{sec-thermo}

Now that we can construct accurate models for the statistical mechanics of real neural networks, it becomes natural to ask if there is a thermodynamics that emerges as $N\rightarrow \infty$.
While heat and temperature don't have any meaning in these systems, thermodynamics is about the interplay of energy and entropy \cite{sethna2021statistical,kivelson+al_24}, and these have clear significance for networks of neurons.  
We have written the probability distribution over patterns of activity as
\begin{equation}
P\left(\bm{\sigma}\right) = {1\over Z} e^{-E(\bm{\sigma})} ,
\label{P6C1}
\end{equation}
so that the effective energy $E(\bm{\sigma})$ is just the negative log probability of a state.  The negative logarithm of the probability, in turn, has an information theoretic meaning as the length of the ideal codeword for describing each pattern of activity, or more simply as the natural measure of how surprised we should be when we observe that pattern \cite{shannon1948mathematical,cover+thomas_91,Mezard+Montanari_2009}. 

As a reminder, when we compute the partition function in Eq (\ref{P6C1}) we have
\begin{eqnarray}
Z &=& \sum_{\bm{\sigma}}\exp\left[ - E\left(\bm{\sigma}\right) \right]\\
&=& \int dE\, \rho(E) e^{-E} ,
\end{eqnarray}
where the density of states
\begin{equation}
\rho (E) = \sum_{\bm{\sigma}}\delta\left[ E - E\left(\bm{\sigma}\right)\right]
\end{equation}
becomes smooth at large $N$, so that
\begin{equation}
\rho (E) \approx  {1\over{\Delta E}} e^{S(E)} ,
\end{equation}
where $S(E)$ is the microcanonical entropy and $\Delta E$ is a scale to get the units right.  We expect, as usual, that both energy and entropy will be proportional to the number of degrees of freedom $N$, so that
\begin{eqnarray}
E &=& N \epsilon\\
 \lim_{N \rightarrow \infty} {{S(E)}\over {N  }} &=&  s(\epsilon = E/N ),
\end{eqnarray}
and hence
\begin{equation}
Z \rightarrow {{N  }\over{\Delta E}} \int d\epsilon\, \exp\left[ -N    f (\epsilon )\right]
\label{ZA}
\end{equation}
where the free energy density $f(\epsilon ) = \epsilon - s(\epsilon )$.  At large $N $ the dominant states are those with energy per degree of freedom $\epsilon_*$ such that $\partial s(\epsilon )/\partial\epsilon = 1$, and
\begin{equation}
Z \approx  {{N  }\over{\Delta E}}e^{-N  f(\epsilon_*)} \int d\epsilon\, \exp\left[ N  s''(\epsilon_* ) (\epsilon - \epsilon_*)^2 + \cdots \right] .
\label{ZB}
\end{equation}
Thus the ``stiffness'' that holds the log probability of states near its typical value is the (negative) second derivative of the entropy, and the resulting variance in the energy density is the specific heat $c = 1/[- s''(\epsilon_* )]$.   

Equation (\ref{ZB}) makes clear that something special happens if $s''(\epsilon_* )\rightarrow 0$, so that the (linear) stiffness holding the energy near its typical value vanishes.  Formally the variance of the energy, and hence the specific heat, diverges as $N\rightarrow\infty$.  This is a critical point.

In statistical mechanics we have the equivalence of ensembles, telling us that what we compute at fixed temperature is essentially the same as what we compute with fixed energy, if the number of degrees of freedom is large \cite{sethna2021statistical}.  In information theory the corresponding idea is ``typicality,''  that almost all the states that we actually see have the same log probability \cite{cover+thomas_91,Mezard+Montanari_2009}.  When the specific heat diverges  the fluctuations in log probability become very large so that the approach to typicality at large $N$ becomes anomalously slow.

The fact that the microcanonical entropy is an increasing function of the energy means that states which are individually less likely are more numerous.  For neurons there is a useful intuition based on the fact that spikes are less likely than silences.  Thus, particular states in which more neurons are active are less probable than those in which fewer neurons are active.  But there are more ways of arranging $K$ spikes among $N$ neurons when $K$ is larger (until $K = N/2$, which essentially never happens).  This tradeoff between the frequency and multiplicity of states is exactly the tradeoff between energy and entropy.

The typical states that we observe have an energy such that $dS/dE = 1$, which means that the tradeoff between the frequency  and multiplicity is balanced.  Usually this balancing is local, but at a critical point it extends over a wider range of energies or frequencies.

In a finite population of neurons can of course never see a true divergence in the specific heat.  What we can do is to ask whether the specific heat or variance in log probability is large when compared with hypothetical networks that have similar but slightly different properties.  One way to construct such networks is to introduce a fictitious temperature, 
\begin{equation}
P(\bm{\sigma}) = {1\over {Z}} e^{-E (\bm{\sigma})} \rightarrow {1\over {Z(T)}} e^{-E (\bm{\sigma})/T} .
\label{Teff}
\end{equation} 
Varying $T$ gives us one slice through the space of possible networks:  at large $T$ we finds models where neurons are more active and less correlated than in the real network, and conversely at small $T$. 

The initial exploration of thermodynamics for $N=40$ cells in the retinal network showed that the specific heat 
\begin{equation}
c(T) = {{\langle (\delta E)^2\rangle}\over{NT^2}}
\end{equation}
was large, and further that there is a peak in $c(T)$ close to the model of the real network at $T=1$ \cite{tkacik2006ising,tkacik2009spin}.  This means that real networks have an unusually large dynamic range for the surprise carried by individual patterns of activity, and that this is a property not shared by plausible but slightly different networks.  We also can construct hypothetical networks in which individual elements of the correlation matrix are chosen at random from the observed distribution of matrix elements, and maximum entropy models for these randomized networks show almost identical thermodynamic behavior.  But we can build random networks in this way at larger $N$, with the prediction that the peak of the specific heat should be even larger and closer to $T=1$ for $N \sim 100$.  This prediction was confirmed in analysis of next generation experiments with $N = 100 - 160$ 
 \cite{tkacik2015signatures}.

\begin{figure}[t]
\includegraphics[width=\linewidth]{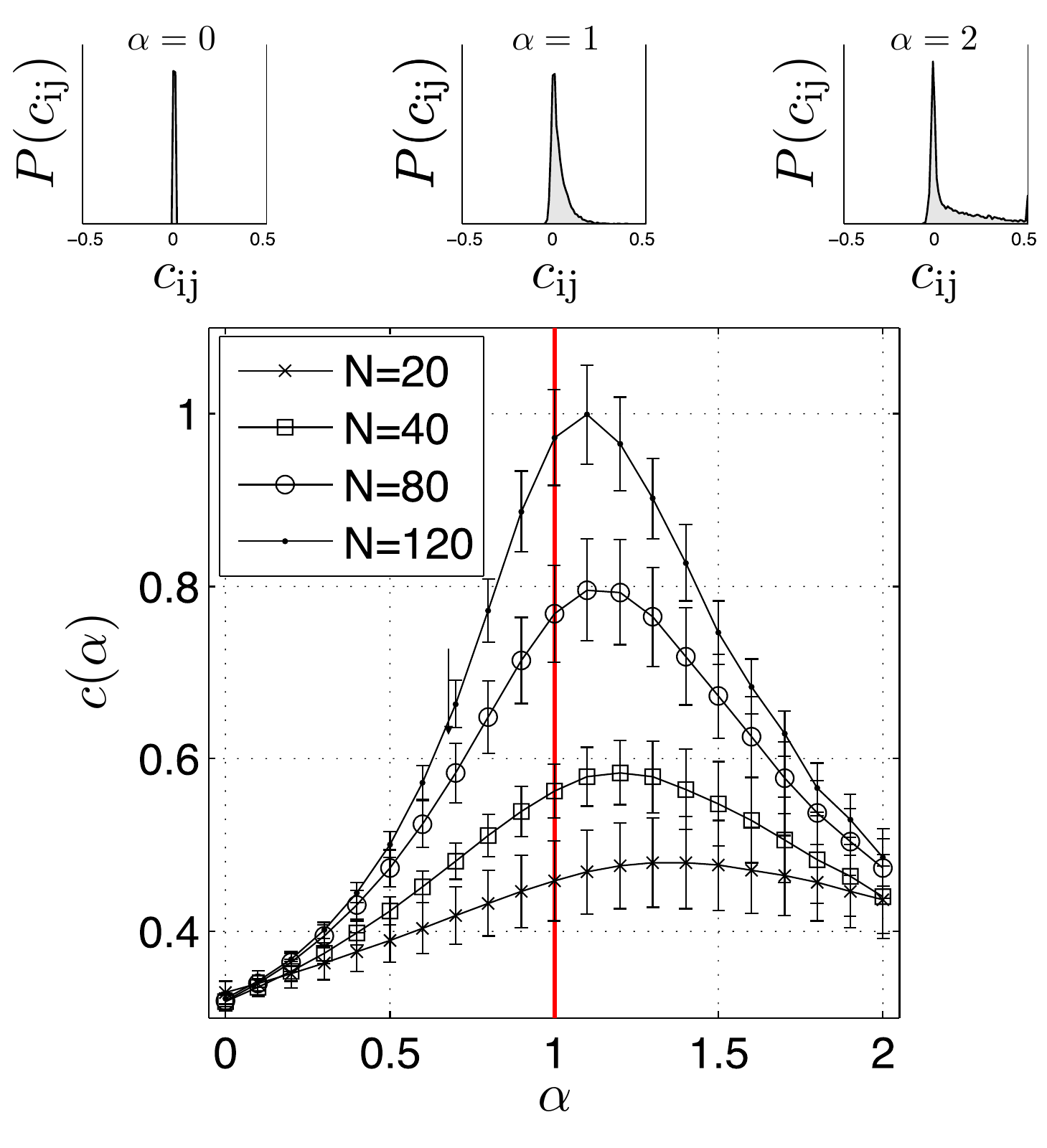}
\caption{Specific heat vs strength of interactions \cite{tkacik2015signatures}.  We construct a series of maximum entropy models for networks such that all neurons have the same mean activity as in the real network, but varying interactions and hence correlations, Eqs (\ref{Ealpha1}, \ref{Ealpha2}). Main panel shows the specific heat $c(\alpha )$ vs interaction strength $\alpha$, for different populations of $N$ cells chosen out of an experiment on $N=160$ cells in the retina. Error bars are SDs over 10 networks at each $N$ and $\alpha$.  Upper panels show the distribution of correlation coefficients for all pairs of neurons at three values of $\alpha$; $\alpha = 1$ is the real network. \label{CvsAlpha}}
\end{figure}

One may reasonably object that temperature is an artificial construct. Perhaps more reasonable is to divide the effective energy function into one piece that controls the activity of individual neurons and one that controls their interaction, then ask what happens as we change the strength of interactions while keeping the mean activity of each neuron fixed.  As an example, we can generalize the K--pairwise model of Eq (\ref{K_pairwise}) to write
\begin{eqnarray}
E_{2k\alpha} (\bm{\sigma}; \alpha ) &=& E_{\rm ind} (\bm{\sigma}) + \alpha E_{\rm int} (\bm{\sigma})
\label{Ealpha0}\\
E_{\rm ind} (\bm{\sigma}) &=& \sum_{{\rm i}=1}^N h_{\rm i}(\alpha )\sigma_{\rm i}
\label{Ealpha1}\\
E_{\rm int} (\bm{\sigma}) &=&  {1\over 2}\sum_{{\rm i}\neq {\rm j}}J_{\rm ij} \sigma_{\rm i}\sigma_{\rm j} + V\left( \sum_{{\rm i}=1}^N \sigma_{\rm i}\right) .
\label{Ealpha2}
\end{eqnarray}
Note that to fix the mean activity of each neuron we must adjust the local field $h_{\rm i}$ as we change the interaction strength $\alpha$.  If we start with the parameters that describe a population of $N = 120$ neurons in the retina, we obtain the results in Fig \ref{CvsAlpha} \cite{tkacik2015signatures}.

As we change $\alpha$ we produce models of possible networks that in many ways are quite plausible.  The extreme $\alpha = 0$ describes  neurons that turn on and off independently, which is extreme.  But even $\alpha =2$ describes a network in which pairwise correlation are still reasonable, with a sharper peak at $c_{\rm ij} = 0$ and a longer tail.   The strength of correlations varies monotonically with $\alpha$, but the specific heat does not---there is a peak within $\sim 10\%$ of $\alpha =1$.  This peak is higher and closer to $\alpha =1$ at larger $N$.  If we compare the K--pairwise model to the pure pairwise model, the peak is higher,  sharper, and closer to the real system in the more accurate model; these effects also are clearer when the retina is responding to more naturalistic stimuli, even though the pattern of correlations is not simply inherited from the visual inputs (\S\ref{sec-alternatives}).

\begin{figure}[b]
\includegraphics[width=\linewidth]{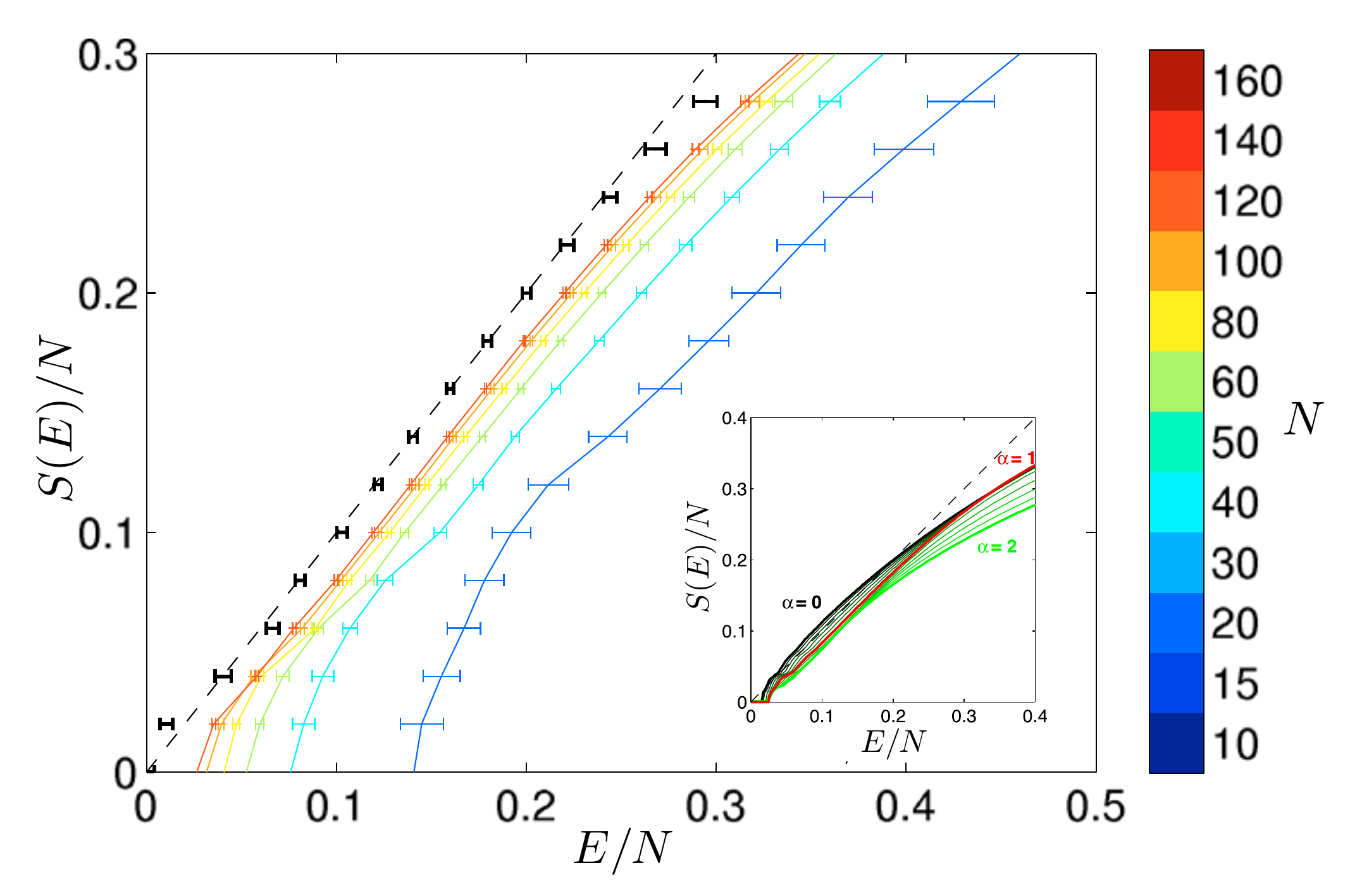}
\caption{Entropy vs energy in maximum entropy models for neural activity in the retina \cite{tkacik2015signatures}.  Main panel shows results for models at varying $N$, with black points based on extrapolation $N\rightarrow\infty$.  Error bars are standard deviations  across multiple networks at each $N$, and dashed line is $S = E$.  Inset shows results at $N=120$ with varyin g$\alpha$, as in Fig \ref{CvsAlpha}, showing that the near linearity of entropy vs energy breaks down as we move away from $\alpha = 1$.  
\label{SvsE}}
\end{figure}

We have emphasized that typical states in a Boltzmann--like distribution are those in which the tradeoff between the frequency and multiplicity of states is balanced, locally; at a critical point  this balance extends over a broader range of probabilities.  A striking feature of the maximum entropy models learned from the retina, for example, is that the frequency/multiplicity balance extends almost perfectly over a finite range of probabilities, so that the entropy is a nearly linear function of energy.  This can be seen over a limited dynamic range by directly counting states in the raw data, but the models make this prediction across ten orders of magnitude in probability (Fig \ref{SvsE}).   Importantly these models make correct predictions over this full range, as seen in Fig \ref{fig-PhiE}.  Linearity of entropy vs energy is equivalent to Zipf's law for the rank ordered probabilities of individual states \cite{mora+bialek_11}, and breaks if we move away from $\alpha = 1$.  The near linearity of entropy vs energy is seen also in much simpler maximum entropy models which capture the probability that $K$ out of $N$ neurons are active but discard information about the identity of the cells \cite{tkavcik2013simplest}.

The results in this section point strongly to the idea that real networks of neurons are poised at non--generic values of their underlying parameter, generating phenomenology that we associate with critical behavior in simpler systems.  Importantly, we can construct models which are close to the real system but different, quantitively, and aspects of this behavior fall away.  In the (admittedly coarser) observations on the human brain it even seems that one can drive the system away from critical behavior through anesthesia.  

\subsection{Bridges between dynamics and theromdynamics}

Notions of criticality in dynamics and thermodynamics seem very different.
But one can also build maximum entropy models for temporal sequences of states, e.g. matching pairwise temporal correlations; as noted above  this is sometimes called ``maximum caliber'' \cite{presse+al2013,ghosh+al_20}.  Among other things this dispels the idea that maximum entropy describes only equilibrium systems.  For networks of neurons we could be interested either in an autonomous description of the dynamics or a description that is locked to external signals, for example the visual inputs to the retina.   There also have been dynamical maximum entropy models for flocks \cite{Cavagna+al_2014b}. 

If we try to match pairwise correlations not just at equal time but also at unequal times, we are asking quite a lot of the data and arrive at a very complicated model.  As a first try one can build models for the summed activity of the network, that is for the number of neurons $K_t$ that are active in a small window of size $\Delta\tau$ surrounding the time $t$ \cite{mora+al2015}.  As noted above, applied to single time points this model focuses attention on  the surprising tradeoff between the probability and numerousity of network states with different numbers of spikes \cite{tkavcik2013simplest}.  

Concretely we can ask for the maximum entropy model that matches to distribution of the number of active neurons at one moment in time $P_N(K)$ from Eq (\ref{PKofN_expval}), and the joint distribution at two times, $P(K_t, K_{t+\tau})$ for  $\tau = 1,\, 2,\, \cdots ,\, v$.  The resulting model is of the form
\begin{eqnarray}
P_{\rm dyn} \left(\{K_t\}\right) &=& {1\over {Z_{\rm dyn}}}\exp\left[ - E_{\rm dyn} \left(\{K_t\}\right) \right]
\label{PKt}\\
E_{\rm dyn}\left(\{K_t\}\right) &=& \sum_{t=1}^{\tilde T} V_N(K_t ) -  \sum_{t=1}^{\tilde T}  \sum_{\tau =1}^v J_\tau (K_t , K_{t+\tau}) .\nonumber\\
&&
\label{EKt}
\end{eqnarray}
where $\tilde T$ is the (large) window of our observations, not to be confused with the temperature $T = 1/\beta$.
We have to adjust the ``potential'' $V_N (K)$ to match $P_N(K)$, and we adjust the ``interactions''  $J_\tau (K , K')$ to match the joint probabilities  $P(K_t, K_{t+\tau})$ that $K$ and $K'$ neurons are active in bins separated by $\tau$.  Because this is a one--dimensional model with local interactions it can be solved exactly by transfer matrix methods, avoiding Monte Carlo simulation.  

\begin{figure}[t]
\includegraphics[width=\linewidth]{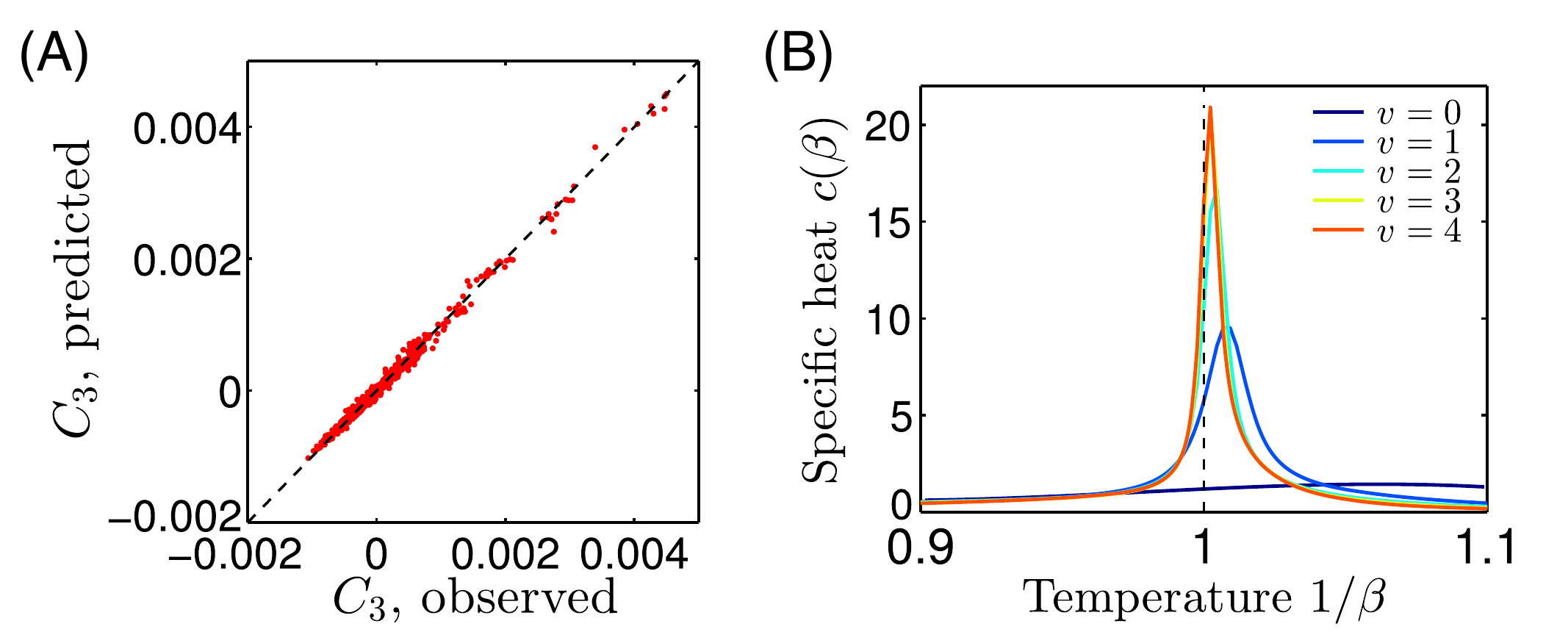}
\caption{Maximum entropy models for the dynamics of summed activity across $N = 185$ neurons in the retina \cite{mora+al2015}.  (A) Correlations among three successive time bins, from Eq (\ref{C3_def}), computed for a subgroup of $N=61$ cells.  (B) Specific heat as function of a fictitious temperature, for the full population of $N=185$ cells.  Different curves are for models that match pairwise temporal correlations in different numbers of time bins. Note the higher, sharper peak close to the real system at $\beta=1$ as the model matches more of the data. \label{fig-Kdynamic}}
\end{figure}

This approach was applied to experiments on a population of $N=185$ neurons in the rat retina, responding to videos of randomly moving bars; the binary variables $\sigma_{\rm i}(t)$ mark the spiking vs silence of neuron $\rm i$ in a bin of width $\Delta\tau = 10\,{\rm ms}$ surrounding the time $t$, $K_t = \sum_{\rm i} \sigma_{\rm i} (t)$ \cite{marre+al2012}.    As above, since we are matching correlations between pairs of times we can test the model by looking at triplets.  Specifically we can consider
\begin{eqnarray}
C_3 (K,\, K',\, K'') &=& P(K_t = K,\, K_{t+1} = K', K_{t+2} = K'')\nonumber\\
&&\,\,\,\,\,  - P_N(K)P_N(K')P_N(K'') ,
\label{C3_def}
\end{eqnarray}
which measures (connected) correlations among the numbers of active neurons in three successive time bins.   The number of distinct triplets becomes quite large, so this was tested on a subgroup of $N = 61$ cells, as shown Fig \ref{fig-Kdynamic}A; agreement betwen theory and experiment is excellent \cite{mora+al2015}.

Models that capture temporal correlation also give us a chance to look more deeply at the tradeoff between probability and numerosity of network states.   Again we generalize to vary the inverse temperature $\beta$,
\begin{equation}
P_{{\rm dyn} \beta} \left(\{K_t\}\right) = {1\over {Z_{\rm dyn}(\beta)}}\exp\left[ - \beta E\left(\{K_t\}\right) \right],
\end{equation}
with $E_{\rm dyn} \left(\{K_t\}\right)$ the same function as in Eq (\ref{EKt}).  We expect the mean energy will be proportional both to the number of neurons $N$ and to the duration of our observations $\tilde T$, so we can define a specific heat 
\begin{equation}
c_{\rm dyn}(\beta ) = {1\over{N\tilde T}} \beta^2 \langle \left( \delta E_{\rm dyn}\right)^2\rangle ,
\end{equation}
with results shown in Fig \ref{fig-Kdynamic}B.
The large variance in log probability occurs only when $\beta$ is within a few percent of the value $\beta = 1$ that describes the real network. This is becomes clearer as we move to more accurate models, increasing the range $v$ over which we match the temporal correlations.  Quantitatively, the specific heat  is $\sim 50\times$ larger than if neurons were uncorrelated.  We can think of different values of $\beta$ as describing possible networks with different levels of correlation, and the sharp peak in specific heat at $\beta =1$ means that the real network has collective behavior that is very different from other possible networks, even those that differ very subtly.   

The analysis of neural avalanches focuses on the summed activity of the network,  the same collective variable $K$ considered here.  In simple branching models  \cite{beggs+plenz_03} one can again estimate the specific heat, and it diverges exactly at the critical value of the branching parameter that allows for a power--law distribution of avalanche sizes and durations \cite{mora+al2015}.   This suggests that the thermodynamics of trajectories is capturing the same critical behavior as the dynamical analyses, but without adjustable parameters in the definition of avalanche events. 

\subsection{Alternatives}
\label{sec-alternatives}

For physicists, criticality is an evocative concept.  The rich phenomenology of critical points inspired the deep ideas of scaling, culminating the modern formulation of the renormalization group.  It is very exciting that something of this flavor arises in the complex context of living systems, whether in networks of neurons or swarms of midges \cite{attanasi+al_14b}.  For biologists, in contrast,  it can seem that invoking criticality is an example of imposing physics concepts onto a biological system, and we should worry about this too.  

An essential tool in the experimental investigation of critical phenomena is the ability to tune the control parameters, pushing the system toward or away from the critical point and exploring the whole critical region.  In addition, we usually have experimental probes that couple directly to the order parameter, whether it is the magnetization in a ferromagnet, the density of a fluid, or the degree of molecular alignment in a liquid crystal.  For networks of neurons these tools largely are absent.

An interesting exception is provided by culturing networks of neurons in a dish, where one can  manipulate the microscopic parameters.  By changing the mix of excitatory and inhibitory neurons one can see transitions in the behavior of the network: changes of just a few percent in the relative populations of the two cell types produce dramatic qualitative effects, reminiscent of phase transitions \cite{chen+dzakpasu_10}.

More generally,   modern experiments provide us with data analogous to the record of a Monte Carlo simulation, the simultaneous trajectories of the all the microscopic elements (neurons) over time.  The challenge is to draw inferences from these data  about where the real network is poised in the phase diagram of possible networks.  As we have explained, the construction of maximum entropy models provides us with one way of doing this.

The maximum entropy model consistent with pairwise correlations is an Ising model, with the activity of neurons in the role of spins, and it thus is tempting to think of the coefficients $J_{\rm ij}$ as ``interactions'' between the neurons.  This language seems natural for physicists, but we should be careful. Even in magnets we know that these are effective interactions, often mediated by fluctuations in additional degrees of freedom that we do not account for directly.  In the extreme,  a  magnetic dipole interaction can be thought of as arising from each spin interacting independently only with the local magnetic field, rather than directly with other spins.

To make these connections explicit it is useful to change from $\sigma_{\rm i} = \{0, 1\}$ to the more familiar Ising variable $s_{\rm i} = 2\sigma_{\rm i} -1 = \pm 1$, and to change   sign conventions for the fields and couplings.  Then the  conventional Ising model with pairwise interactions,
\begin{equation}
P\left( \bm{s}\right) = {1\over Z}\exp\left[ \sum_{\rm i}h_{\rm i} s_{\rm i} + {1\over 2}\sum_{\rm ij} J_{\rm ij} s_{\rm i} s_{\rm j} \right],
\label{Ising_pm1}
\end{equation}
can be rewritten as 
\begin{widetext}
\begin{equation}
P\left( \bm{s} \right) = 
{1\over Z} \left[ { {\det J} \over {(2\pi)^N} }\right]^{1/2} 
\int d^N\phi \,\exp\left[ - {1\over 2}\sum_{\rm ij} \phi_{\rm i}(J^{-1})_{\rm ij}\phi_{\rm j} + \sum_{\rm i} (h_{\rm i} + \phi_{\rm i})s_{\rm i} \right] .
\end{equation}
\end{widetext}
But because the spins appear linearly in the exponential, this can be factorized:
\begin{equation}
P\left( \bm{s} \right) = \int d^N\phi\, {\cal P}\left( \bm{\phi} \right) \prod_{{\rm i}=1}^N P_{\rm ind}(s_{\rm i}| h_{\rm i} + \phi_{\rm i}) ,\label{latent1}
\end{equation}
where the distribution of field $\bm{\phi}$ is given by
\begin{eqnarray}
{\cal P}\left( \bm{\phi} \right) &=& {1\over {2^N Z}} 
\left[ { {\det J} \over {(2\pi)^N} }\right]^{1/2} \exp\left[ - {\cal H}(\bm{\phi})\right]\\
{\cal H}(\bm{\phi}) &=& 
 {1\over 2}\sum_{\rm ij} \phi_{\rm i}(J^{-1})_{\rm ij}\phi_{\rm j} - \sum_{\rm i} \ln\cosh( \phi_{\rm i} + h_{\rm i}) ,
 \nonumber\\
 &&
\label{latent2}
\end{eqnarray}
and the conditional distribution for each neuron (or spin) responding independently is
as always
\begin{equation}
P_{\rm ind}(s | \psi  ) = {{e^{\psi s}}\over{2\cosh\psi}} ,
\end{equation}
As an aside, one might worry that the matrix $J$ is not invertible, or that it has negative eigenvalues that cause ${\cal P}\left( \bm{\phi} \right)$ to be ill--defined.  But with $s = \pm 1$ we can always add terms to the diagonal of $J$ that serve only to shift the zero of energy  but will solve these problems.

Thus, as in the textbook derivations of mean--field theory \cite{kivelson+al_24,sethna2021statistical},  we can trade interactions of neurons with one another for a picture in which they respond independently to fluctuating fields.  Models with the structure of Eq (\ref{latent1}) often are referred to as latent variable models \cite{everitt_84},  since the behavior that we observe $\{s_{\rm i}\}$ is controlled by some underlying hidden or latent variables $\{\phi_{\rm i}\}$.  Latent variable models are very popular in the neuroscience literature, where they sometimes are presented  as an alternative to the physicists' models for interacting neurons.  We see that this is a false dichotomy, since the different models are mathematically equivalent.

Ultimately we want to understand whether the latent variable description changes our interpretation of the evidence for critical behavior.  But first we should ask whether this description is a compelling alternative, independent of where real networks are in their phase diagram. The latent variable or effective field description is especially useful if it simplifies the model, and indeed advocates of this description emphasize that it is simpler than the Ising model, or more precisely that simple versions of the latent variable approach do as well as the Ising model.  One clear possibility for neural systems is that the effective fields have a direct meaning for the brain, perhaps as the variables that neural activity is encoding, or are genuinely external to the network, such as sensory inputs.

In the hippocampus, for example, we might imagine that the latent variable is the position of the animal, since we know that this is represented by the population of ``place cells''  (\S\S\ref{sec-larger} and \ref{sec-subgroups}).  But in a sufficiently long recording we can estimate the probability of each cell being active as a function of position $\mathbf x$, which is the classical place field $F_{\rm i}({\mathbf x})$ as in Eq (\ref{placefield}).  If position is the only latent variable, then cells are independent given the position, and the joint distribution   becomes 
\begin{equation}
P_{\rm ind\ place} \left( \bm{\sigma} \right) = \int dx\,{\cal P}({\mathbf x}) \prod_{{\rm i}=1}^N F_{\rm i}({\mathbf x})^{\sigma_{\rm i}} \left[ 1 - F_{\rm i}({\mathbf x})\right]^{1-\sigma_{\rm i}},
\label{indplacemodel}
\end{equation}
where $\sigma_{\rm i} = \{0, 1\}$ as in previous sections and ${\cal P}({\mathbf x})$ is the distribution of positions seen in the experiment.  In this model  all correlations are inherited from the animal's movement through the space $\mathbf x$.  Importantly this construction involves no free parameters.

\begin{figure}[t]
\includegraphics[width=\linewidth]{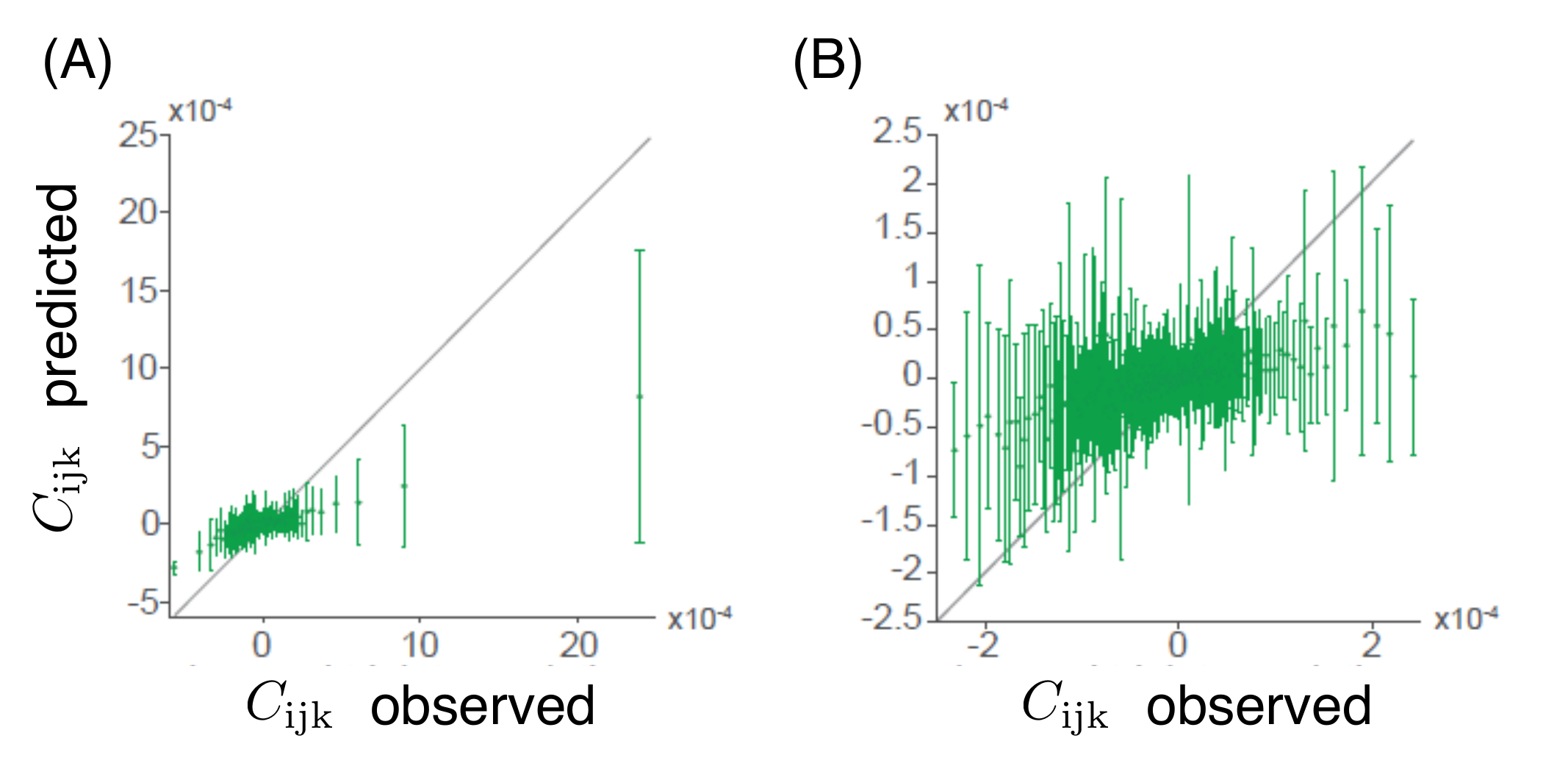}
\caption{Failure of a latent variable model for $N\sim 100$ cells in the mouse hippocampus \cite{meshulam2017collective}. Predicted vs observed triplet correlations [Eq (\ref{C3})], calculated in a model where the latent variable is position of the mouse, Eq (\ref{indplacemodel}).  (A) Full dynamic range of the data, binned along the x--axis as in Fig \ref{3pt_bin}. (B) Expanded view of the small correlations, which constitute the bulk of the data.  
\label{place_fail}}
\end{figure}

We can test the independent place cell model of Eq (\ref{indplacemodel}) in the same way that we have tested the maximum entropy models. If we compute the mean activity of each cell  it will be correct by construction.  The pairwise correlations are a nontrivial prediction, and the matrix $C_{\rm ij}$ looks roughly correct element by element but the eigenvalue spectrum is qualitatively incorrect, as noted in \S\ref{sec-larger}.  Triplet correlations are significantly underestimated (Fig \ref{place_fail}A), and in the bulk of small correlations there is essentially zero correlation between the data and the predictions of an independent place cell model (Fig \ref{place_fail}B); these results should be compared with success of the pairwise maximum entropy model in  Fig \ref{3pt_bin}.  We conclude that, in the hippocampus, an extreme version of the ``latent variable'' scheme---in which the only latent variable is position---fails dramatically \cite{meshulam2017collective}.

For the retina there has been the explicit suggestion that any successes of the maximum entropy approach should be understood in a latent variable model where the latent variables are determined by the visual stimulus itself \cite{aitchison+al_16}.   This is  the sitmulus--dependent maximum entropy model of Eqs (\ref{stimdep1}, \ref{stimdep2}) but with all interactions $J_{\rm ij} =0$, and allowing for some complicated relation between the visual input and the time dependent local fields $h_{\rm i}(t)$.  No matter how complex this relation, the  model predicts that if we show the same movie to the retina many times, then at a fixed moment in the movie there should be no correlations among the neurons, since the latent variables are fixed.  The challenge in testing this prediction is that the probability of complete silence in the network is significant, even for $N=100+$ cells, and of course in these silent moments one cannot compute the correlations.  

If we compute conditional correlations only at moments where both cells in the pair generate more than a handful of spikes, then we can indeed find examples where the correlations are near zero, but there also are many examples where the conditional correlations are {\em larger} than the overall correlations, opposite to the prediction of the latent variable model. There even are many pairs of neurons whose overall correlation is near zero, but at particular moments in a repeated movie the correlations very strong, with either sign.  These results seem to eliminate a model in which the visual inputs serve as the latent variables to explain the correlation structure of the activity in the retina \cite{tkacik2015signatures}.

Thus in two cases that have been studied carefully, we cannot find a description in which latent variables correspond to external stimuli.\footnote{To be clear, the visual inputs do generate correlations among neurons in the retina.  The point is that these are not the only source of correlations, and the separation into externally driven and internally generated correlations does not provide an immediate simplification.  It should also be emphasized that this is not a separation that is available to the brain under natural conditions. Further, the retina adapts to the distribution of its inputs, so that there is no fixed mapping from correlations in the stimulus to correlations among neurons.}  But if the latent variables are hidden from us, then it is not clear whether these variables are external to the network or emergent from the network dynamics itself.  As an example, it has been suggested that the summed activity of all the cells in a network can serve as a latent variable \cite{aitchison+al_16}, which would be like saying that the magnetization of the Ising model is a latent variable. In the mean--field limit this almost works (the natural latent variable is actually the conjugate field), but there is no doubt that the magnetization is emergent.  One also can  verify that, in the systems we have discussed, different neurons are not conditionally independent given the summed activity; see, for example, \citet{tkacik2015signatures}.

Latent variable models would be especially attractive if one could achieve an accurate description with a small number of these variables.  If the Ising model description is accurate, then a small number of latent variables requires that the rank of the coupling matrix $J_{\rm ij}$ be small.  Even better would be a case where we could identify the latent variables with measurable quantities, but we have seen that it {\em doesn't} work to identify these variables with quantities that are genuinely external, such as a sensory stimulus.  Interestingly there are popular models for the encoding of low--dimensional sensory or environmental variables in which the ``latent'' variable that represents these signals in fact emerges from network interactions \cite{zhang_96,ben-yishai+al_95,tsodyks+sejnowski_95}.  If we ask about the distribution over observable network states, then the mathematical description  is  the same no matter whether the latent variable is external or emergent.

There is a simple but compelling argument for how the seemingly mysterious linearity of entropy vs energy, and the associated signatures of criticality, can arise from fluctuating fields \cite{Schwab+al_2014}.  It is useful to place this discussion in the context of the mean--field ferromagnet \cite{sethna2021statistical,kivelson+al_24}.

The mean--field model is a collection of spins $\bm{s} \equiv \{s_{\rm i}\}$ governed by the energy function
\begin{eqnarray}
E_{\rm MF} (\bm{s}) &=& h \sum_{{\rm i}=1}^N s_{\rm i} - {J\over{2N}}\sum_{{\rm i,j}=1}^N s_{\rm i}s_{\rm j}\\
&=& N\left[ hm -(J/2) m^2\right] ,
\end{eqnarray}
where the magnetization
\begin{equation}
m = {1\over N}\sum_{{\rm i}=1}^N s_{\rm i} .
\end{equation}
This describes a system in which all spins experience the same magnetic field, and all pairs of spins interact equally; the factor of $1/N$ in the interactions insures that energy and entropy are proportional to $N$ as $N\rightarrow\infty$.  Now we can follow the same arguments that lead from the general pairwise Ising model Eq (\ref{Ising_pm1}) to the latent variable description in Eqs (\ref{latent1}, \ref{latent2}), but this case is easier because there is only one latent field.  The result is that the partition function can be written as
\begin{eqnarray}
Z_{\rm MF}   &\equiv& \sum_{\{s_{\rm i}\}} e^{-\beta E_{\rm MF} (\bm{s})}
\label{ZMF1}\\
&=& \sqrt{ {2\pi}\over J} \int d\phi\, \exp\left[ - N f_{\rm MF} (\phi, h)\right]\label{ZMF2}\\
f_{\rm MF} (\phi, h) &=& {{\phi^2}\over {2J}} - \ln\cosh(\phi + h ),
\label{ZMF3}
\end{eqnarray}
where for simplicity we choose units where the thermal energy $1/\beta = 1$.  

At large $N$ the integral in Eq (\ref{ZMF2}) is dominated by a single value of the latent field $\phi = \phi_*$ that minimizes the free energy, that is
\begin{equation}
{{\partial f_{\rm MF} (\phi, h)}\over{\partial\phi}}{\bigg |}_{\phi = \phi_*} = 0.
\end{equation}
If the second derivative 
\begin{equation}
{{\partial ^2 f_{\rm MF} (\phi, h)}\over{\partial\phi^2}}{\bigg |}_{\phi = \phi_*} = \kappa
\end{equation}
is of order unity, then fluctuations in the latent variable  will be on a scale $\delta\phi \sim 1/N^{1/2}$.  Because all spins couple equally to the latent field, these fluctuations produce correlations between spins, but because the scale of fluctuations in small these correlations also are small; the result is that covariance matrix elements $C_{\rm ij} \sim 1/N$.  The critical point is the place where the second derivative $\kappa \rightarrow 0$, and fluctuations in the latent field become anomalously large, $\delta\phi \sim 1/N^{1/4}$.  The idea of \citet{Schwab+al_2014} is to turn this around:  since criticality is marked by large fluctuations in the latent field, then if external signals drive large fluctuations in the latent variable they could also generate the signatures of criticality, generically. 

To make this idea concrete, consider a collection of Ising spins   that all couple to the same magnetic field $h$, but this field itself is drawn from a distribution $Q(h)$.  Crucially this distribution is imposed on the system by external inputs, rather than being an emergent property of the interactions.
Then the joint distribution for the state of all the spins is
\begin{eqnarray}
P_{\rm latent} (\bm{s} ) &=& \int dh\, Q(h) \prod_{{\rm i}=1}^N P(s_{\rm i}|h)\\
&=& \int dh\, Q(h) \prod_{{\rm i}=1}^N {{e^{hs_{\rm i}}}\over{2\cosh(h)}} ,
\end{eqnarray}
This becomes
\begin{eqnarray}
P_{\rm latent} (\bm{s} )  &=&  {1\over {2^N}}\int dh\, Q(h) \exp\left[ -N f(m,h)\right]
\label{schwab1}\\
f(m,h) &=&- hm + \ln\cosh(h) ,
\label{schwab2}
\end{eqnarray}
where as before the magnetization
\begin{equation}
m(\bm{s}) = {1\over N}\sum_{{\rm i}=1}^N\sigma_{\rm i} .
\end{equation}
Once again when $N$ is large the  integral over fields is dominated the value which minimizes the free energy density $f(m,h)$,
\begin{equation}
h_* (  \bm{s}  ) = h_* ( m ) = \tanh^{-1}\left(m \right) ,
\end{equation}
so long as $Q(h_*)$ is nonzero.  In making this argument it is important that the distribution of fields is externally imposed and this cannot have an $N$ dependence.  The result is that the probability of any state $ \bm{s} $ depends only on the magnetization $m( \bm{s} )$, 
\begin{eqnarray}
P_{\rm latent} ( \bm{s} ) &=& \exp\left[- E(m)\right],\\
E(m)/N &=& -h_*(m) m + \ln\cosh[h_*(m)] + \cdots ,\nonumber\\
&&\label{schwabE}
\end{eqnarray}
where we drop terms that are independent of $m$ or vanish as $N\rightarrow\infty$.
The entropy at fixed energy is then the entropy at fixed magnetization,
\begin{equation}
S(m)/N = -{{1+m}\over 2}\ln\left({{1+m}\over 2}\right) -  {{1-m}\over 2}\ln\left({{1-m}\over 2}\right) .
\label{binaryS}
\end{equation}
After some algebra, Eqs (\ref{schwabE}) and (\ref{binaryS}) can be combined to give  $S(m)/N  = E(m)/N$, as with the data in Fig \ref{SvsE}.  More generally we see that $d^2S(E)/dE^2 = 0$, which is equivalent to the divergence of the specific heat, a core signature of criticality. 

This argument generalizes beyond the case of a single fluctuating field coupled to the spins.  Not only can one have multiple fields, but they can couple to more complex functions of the system state.  What is required is that a mean--field approximation be valid, so that at large $N$ each state $\{s_{\rm i}\}$ picks a single value for all the latent variables out of some broad distribution \cite{Schwab+al_2014}.    This generality is quite striking, and it is natural to ask whether this ``explains'' the signatures of criticality that we have seen experimentally.

Although quite general, there are limits, and we need to ask whether real networks of neurons are in the regime where we expect critical phenomenology to emerge generically.  As an example, the independent place cell model discussed above is one model in the broad class considered by \citet{Schwab+al_2014}, but not an arbitrary model.  We already know that this model doesn't explain the correlation structure that we see in populations of $N\sim 100$ cells in the hippocampus, and it also is true that this model does not predict $S /N= E/N$, as shown in Fig S7 of \citet{tkacik2015signatures}.  In this sense a biologically plausible version of the latent field model evades the conditions for the generic emergence of critical behavior at reasonable $N$. 

Similarly, we can try to account for the large fluctuations in summed activity that we see in recordings from $N\sim 1500$ hippocampal neurons using models where all cells are driven by a common field, as in Eqs (\ref{schwab1}, \ref{schwab2}).  The regime where we have a generic prediction of $S/N = E/N$ is where the ``stiffness'' of the free energy restricts the fluctuations 
\begin{equation}
\delta h_{\rm f} \sim \left[ N {{\partial^2 f(m,h)}\over{\partial h^2}}\right]^{-1/2}
\end{equation}
to be much smaller than the range of fields $\delta h _Q$ spanned by the distribution $Q(h)$.  If we use this approach to look at the data analyzed in \S\ref{sec-subgroups}, we find that $\delta h_{\rm f} \sim \delta h_Q$ within a factor of two.  While not conclusive, since the correct model surely is more complex, this also suggests that this network is not in the regime where fluctuating external fields explain apparent criticality.
 
 Behind this discussion is the question of whether networks of real neurons are in a mean field limit.  We note that in the analysis of associative memories one can use mean--field theory, but the capacity of the memory is reached only when the number of latent fields is proportional to the number of neurons \cite{amit+al1987}, which is quite different from models in which the numbers of latent variables is fixed as $N\rightarrow\infty$.  We do not know of any simple test for ``mean--fieldness'' of a system, and this seems a deeper problem.
 
 The prediction of critical behavior in the maximum entropy approach emerges, with no adjustable parameters, in models that account in detail for the correlation structure among neurons.  While low dimensional latent variable models have a regime in which they can generate signatures of criticality without fine tuning, there is no example that we know of where such models account for all the observed correlation structure.  Taking seriously what the maximum entropy principle is doing---building {\em minimally structured} models---it seems that the observed correlation structure implies criticality but networks could be critical without this correlation structure.
 
 On the other hand, the models that we study also give us ways of generating surrogate data, for example a network in which every neuron has the same mean activity in the real network but the correlations are weaker or stronger, as in Eqs (\ref{Ealpha0}--\ref{Ealpha2}).  As we tune away from the real network we lose signatures of criticality such as the linearity of entropy vs energy (Fig \ref{SvsE}).  Relatedly, models that capture more of the real correlation structure have stronger signs of criticality, as for example in Fig \ref{fig-Kdynamic}. Thus while there surely are critical networks with different correlation structures, plausible changes in correlations drive the predicted behavior of the network away from criticality.  
 
Taken together, these observations suggest that the signatures of criticality that we see in networks of neurons are not a generic consequence of the system being driven by external fields.  Instead it really does seem that these systems are tuned to a special point in their parameter space.  Ordinarily such fine tuning is worrisome, but networks of neurons have an array of mechanisms for adaptation and learning that allow stabilization of non--generic behaviors.  One clear example is that the oculomotor integrator (\S\ref{sec-avalanches}) is tuned, continuously, based on visual feedback, holding it close to a bifurcation point and thereby allowing for  long, emergent time scales \cite{major+al_04a, major+al_04b}.
 
It may be useful to compare the problem of criticality in networks of neurons to the corresponding problem in flocks of birds and swarms of insects (\S\ref{app-flocks}).  In these animal groups there are good reasons to think that interactions are local, so it is tempting to think that observation of long--ranged spatial correlations would be prima facie evidence for critical behavior, but this is not quite correct. First, even in equilibrium systems the breaking of a continuous symmetry generates Goldstone modes, and fluctuations along these modes will generate long--ranged correlations.  Maximum entropy models that match local correlations provide an example of this idea, which provides a quantitative description of the directional fluctuations in flocks with no free parameters (Fig~\ref{fig-flocks}G, H).  Second, non--equilibrium effects in animal groups can generate effectively non--local interactions, and this is central to theories of active matter \cite{toner+tu_95,toner+tu_98,marchetti+al_13}.  But there are arguments that these effects are smaller than expected in real flocks \cite{mora+al_16}, so that the observed long--ranged correlations in speed fluctuations may indeed provide evidence of critical behavior.  In natural swarms one sees finite size and dynamical scaling behaviors that provide more direct evidence for criticality, independent of models \cite{attanasi+al_14b,cavagna+al_17}.  While each example must stand on its own, it is an old dream that tuning to criticality might unify our understanding of disparate living systems.

\section{Renormalization group for neurons}
\label{chapter-RG}

Physicists are known for our appreciation of simplified models, perhaps even to the point of over--simplification \cite{devine+cohen_92}.  The complexity of living systems is in obvious tension with this drive for simplification;  we can perhaps sympathize with biologists who worry that our theoretical impulses may be mismatched to the richness of life's molecular details.  A useful response is that there is nothing special about biology:  in condensed matter physics and statistical mechanics we routinely describe the macroscopic behavior of materials using models that are much simpler than the underlying microscopic mechanisms.  These simplified models succeed,  not because we are lucky but because of the renormalization group \cite{wilson_79,wilson_83}.  

The central idea of the renormalization group (RG) is to ask how our description of a system changes, systematically, as we change the scale on which we look.  The crucial qualitative result is that many different microscopic mechanisms flow toward the same macroscopic behavior as we ``zoom out'' to look at longer length scales.  This means that we can understand large scale phenomena quantitatively if we can assign them to the correct universality class, even if we can't get all the small scale details right, and this gives us license to write relatively simple models of complex systems \cite{anderson_84}.  We would like to exercise this license in the context of the brain.   To do this we need to understand how to implement the RG when many of our usual guides (locality, symmetry, ... ) are absent.  We then can ask whether there is any sign that simplification emerges from the data as we zoom out from individual neurons to more coarse--grained variables.

\subsection{Taking inspiration from the RG}
\label{sec-RG}

The development of the renormalization group is one the great chapters of theoretical physics from the second half of the twentieth century, with origins in efforts to understand matter at both short and long distances \cite{gell-mann+low_54,kadanoff_66}.  These ideas crystallized in the early 1970s and played a central role in revolutionizing our understanding of the strong interaction among elementary particles, critical phenomena at second order phase transitions, the transition to chaos, and more \cite{wilson_83}.   How can these ideas help us to think about networks of neurons?

In the standard formulation of the RG for statistical physics we start with a set of variables $\bm{z}_{\ell_0} \equiv \{z_{\rm i}(\ell_0 )\}$ defined on some microscopic length scale $\ell_0$.  Our description of these variables is given by a Hamiltonian that in turns specifies the Boltzmann distribution $P_{\ell_0}(\bm{z})$, or perhaps we will be interested in the dynamics generated by this Hamiltonian.  We then imagine ``coarse--graining'' the variables to average out the details on length scales below some $\ell > \ell_0$.  The result is a new set of variables $\bm{z}_{\ell}$, and we can ask for the effective Hamiltonian that governs these variables.  If we think of the Hamiltonian as being built from different kinds of interactions, it becomes natural to say that the effective strengths of these interactions has changed as change scale from $\ell_0$ to $\ell$, and the RG invites us to follow this flow as we change $\ell$.  Although this flow of interaction strengths or running of coupling constants often is the goal an RG analysis, it was emphasized early on by \citet{jona-lasinio_75} that we can think more generally about flow in the space of probability distributions $P_{\ell}(\bm{z})$, leaving aside any reference to Hamiltonians. 

An essential result of the renormalization group is that many different starting distributions $P_{\ell_0}(\bm{z})$ converge to the same $P_{\ell}(\bm{z})$ as $\ell$ becomes large.  Along this trajectory parameters of the distribution exhibit simple scaling behaviors as a function of $\ell$.  A familiar example is the central limit theorem, where if variables in $P_{\ell_0}(\bm{z})$ are sufficiently weakly correlated then $P_{\ell}(\bm{z})$ approaches a Gaussian as $\ell$ becomes large, and along the way the variances of the individual variables scale as $1/\ell$.  The RG predicts that more interesting starting points can flow toward stable non--Gaussian distributions, with moments scaling as non--trivial powers of $\ell$.

The renormalization group approach provides a framework to understand how we can go from discrete Ising spins on a lattice to a description of smoothly varying local magnetization,  or from the positions and momenta of individual molecules to the density of a fluid and the velocity of its flow.  In these examples, the coarse--graining operation is guided by symmetry and locality.  Perhaps the most successful development of RG ideas in a biological context has been for flocks of birds and swarms of insects, where the ideas of symmetry and locality continue to be useful (\S\ref{app-flocks}). For networks of neurons, where connections can span distances encompassing thousands of cells, the principle of locality is less of a guide, and there are no obvious symmetries.  How then do we choose a coarse--graining strategy?  

Perhaps a more serious problem in taking inspiration from the renormalization group is that the RG is formulated as an approach to understanding theories or models, taming the complexities of interactions among degrees of freedom at many scales.   These theories of course make quantitative predictions for experiment, but in the absence of a well defined model it is not clear how to proceed.   There is a recent start on renormalization group analysis of models for a network of moderately realistic spiking neurons \cite{brinkman2023}, and we hope there will be more of this.  But, keeping to the spirit of the discussion thus far, we want to ask:  How can we use the RG to guide the analysis of emerging data on large populations of real neurons?

To address these challenges we rely on two key ideas.  First, as emphasized above, modern experiments on the electrical activity in networks of neurons give us access to something analogous to the trajectory of a Monte Carlo simulation on a statistical physics model, albeit a model that we don't know how to write down.  Thus we can follow the approach used in now classical analysis of such simulations, for example by \citet{binder_81}:  We start with raw data on the most microscopic scale, construct coarse--grained variables, and follow various features of the distribution of thee variables as we change the scale of coarse--graining.

Second, we will use the measured pairwise correlations as guide to which neurons are ``neighbors,'' in the absence of locality \cite{bradde+bialek2017}.  In one version (\S\ref{sec-blockspins}), this involves averaging together the activities of the most correlated cells, building clusters of neurons  that are analogous to block spins \cite{kadanoff_66}.  In another version (\S\ref{sec-shells}), we successively filter out linear combinations of the population activity that make small contributions to the overall variance, and this is analogous to the momentum shell construction \cite{wilson_83}.  We will see that both these approaches uncover simple, precise, and reproducible scaling behaviors that now have been confirmed in multiple brain areas from multiple organisms. We then discuss the implications of these results and some future direction \S\ref{sec-RGunderstanding}.

\subsection{By analogy with real--space methods}
\label{sec-blockspins}

\begin{figure}[t]
\includegraphics[width=\linewidth]{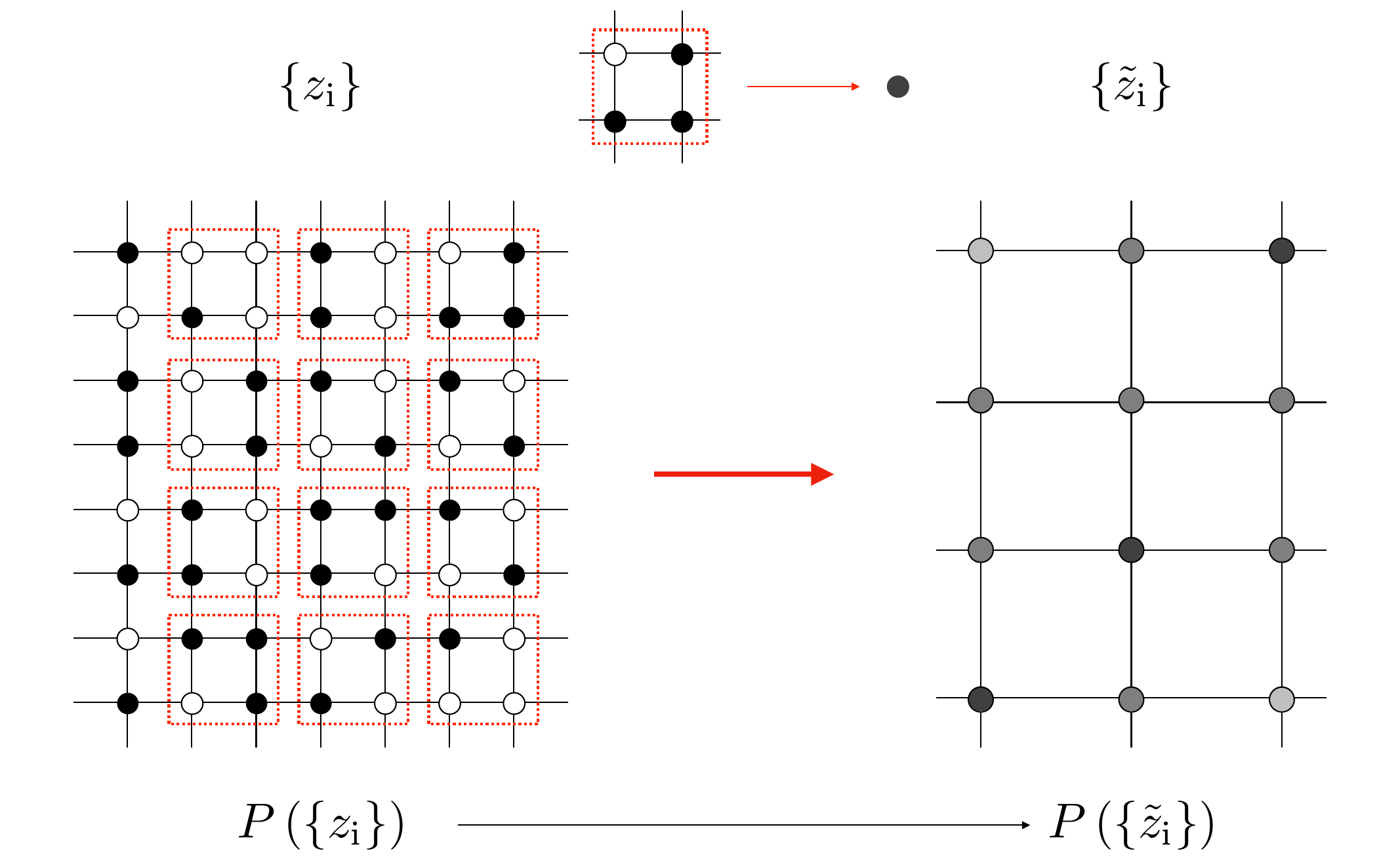}
\caption{Coarse--graining on a regular lattice. We start with binary (black/white) variables $\{z_{\rm i}\}$, and replace $2\times 2$ blocks with the average of these variables $\{\tilde{z}_{\rm i}\}$, shown as grey levels.  The interesting question is what happens to the joint distribution as we coarse--grain, not just once but iteratively.
\label{fig-blockspins}}
\end{figure}

Renormalization group methods in statistical physics rest on a notion of coarse--graining, averaging over microscopic details.  If we start with variables $\{z_{\rm i}\}$ that live on a regular lattice, the it is natural to do this by combining variables with their neighbors, as in Fig \ref{fig-blockspins}.  Formally we can write
\begin{equation}
z_{\rm i} \rightarrow {\tilde z}_{\rm i} = f\left( \sum_{{\rm j} \in {\cal N}_{\rm i}} z_{\rm j}\right ),
\end{equation}
where ${\cal N}_{\rm i}$ is a neighborhood surrounding site $\rm i$.  If the function $f(\cdot )$ is linear then we are just averaging over a neighborhood, and for example this will lead from discrete Ising--like variables to a more continuous local magnetization if we iterate.  If  $f(\cdot )$ is a threshold function  then we can implement majority rule, so that clusters of Ising--like variables are mapped into Ising--like variables on the sparser lattice, as in the original block spin construction \cite{kadanoff_66}.

\begin{figure}[b]
\includegraphics[width=\linewidth]{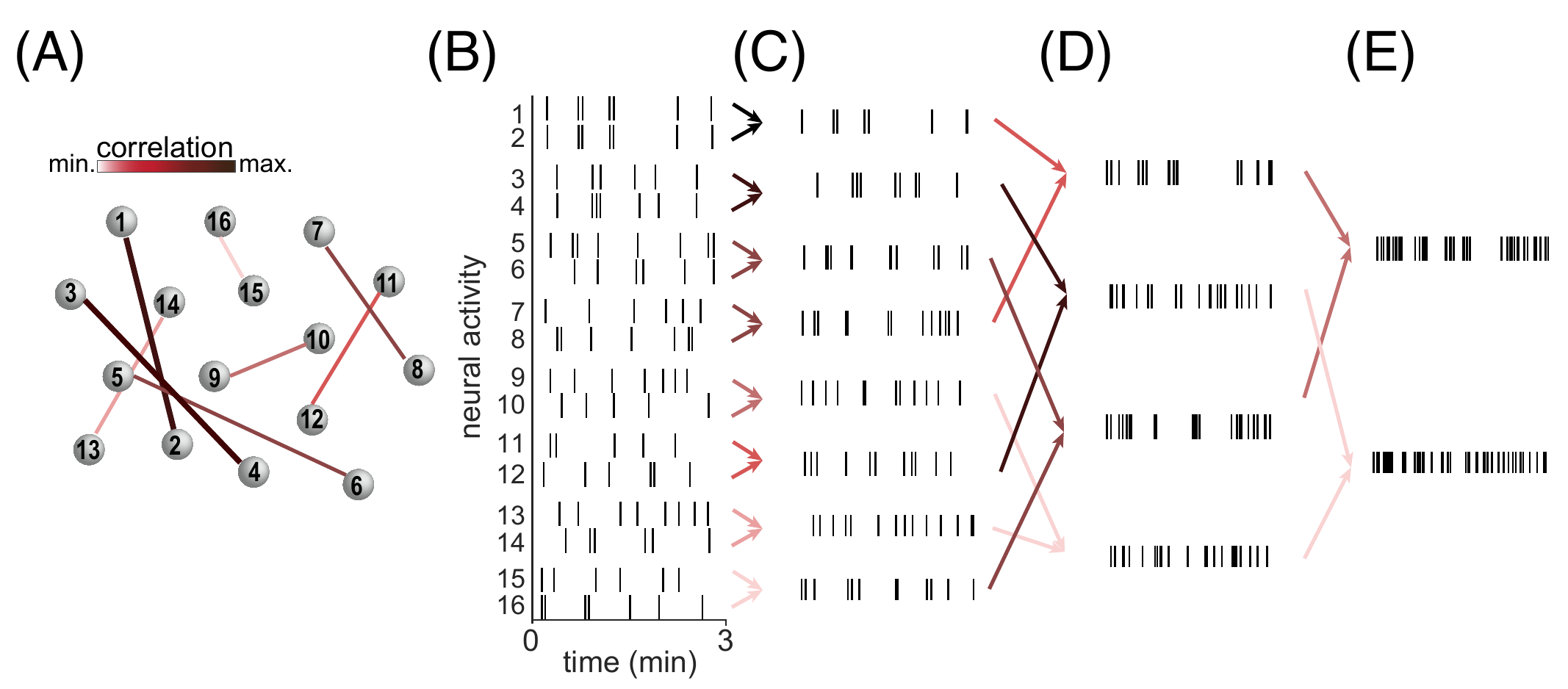}
\caption{Coarse--graining neural activity. (A) A small group of neurons with links indicating the most strongly correlated pairs, and the strength of these correlations. (B) Schematic sequence of action potentials from these cells. (C) Coarse--graining by summing the activity in highly correlated pairs.  (D) Finding the most strongly correlated pairs of coarse--grained variables in (C) and coarse--graining again by summing.  The strengths of the correlations are color coded as in (A).  (E) One more iteration of this ``real space''  coarse--graining.
\label{fig-CGschematic}}
\end{figure}

In a system with local interactions, the variables in the neighborhood typically are the most strongly correlated with one another.  This suggests that even if we don't have a notion of neighborhood, we can make progress by searching for the most correlated variables and using these to build the clusters that we use in coarse--graining.
A schematic of how this can work for neural activity is shown in Fig \ref{fig-CGschematic}.

We start with variables $\{\sigma_{\rm i}\}$, as before, describing the patterns of activity ($\sigma_{\rm i} = 1$) and silence ($\sigma_{\rm i} = 0$) across all the neurons ${\rm i} = 1,\, 2,\, \cdots ,\, N$ in a small window of time.  To emphasize that this is the most microscopic description we will write this as $\sigma_{\rm i} = \sigma^{(1)}_{\rm i}$.  Then as before we can compute the means, covariance, and correlation matrices:
\begin{eqnarray}
m_{\rm i}^{(1)} &=& \langle \sigma^{(1)}_{\rm i}\rangle \\
C_{\rm ij}^{(1)} &=& {\bigg\langle} \left[ \sigma^{(1)}_{\rm i} - m_{\rm i}^{(1)} \right] \left[ \sigma^{(1)}_{\rm j} - m_{\rm j}^{(1)} \right] {\bigg \rangle}\\
c_{\rm ij}^{(1)} &=& {{C_{\rm ij}^{(1)} }\over\sqrt{C_{\rm ii}^{(1)} C_{\rm jj}^{(1)} }} .
\end{eqnarray}
Now we search for the maximal non--diagonal element in the matrix of correlation coefficients, then zero the rows and columns associated with this pair of cells ${\rm i},{\rm j}_*({\rm i})$, and repeat.  The result is a set of maximally correlated pairs $\{{\rm i}, {\rm j}_*({\rm i})\}$, and we then define coarse--grained variables
\begin{equation}
	\sigma^{(2)}_{\rm i} = \sigma^{(1)}_{\rm i} + \sigma^{(1)}_{{\rm j}_*({\rm i})}  , \label{CG_1}
\end{equation}
where now ${\rm i} = 1,\, 2,\, \cdots ,\, N/2$.  Importantly, we can iterate this process across scales:  we compute the correlation matrix of the variables $\{\sigma^{(2)}_{\rm i} \}$ and search again for the maximally correlated pairs $\{{\rm i}, {\rm j}_*({\rm i})\}$, then define
\begin{equation}
	\sigma^{(3)}_{\rm i} = \sigma^{(2)}_{\rm i} + \sigma^{(2)}_{{\rm j}_*({\rm i})}  , \label{CG_2}
\end{equation}
and so on; at each stage we have $N_k = \lfloor N/2^{k-1}\rfloor$ variables remaining.   This coarse graining produces clusters of $K = 2,\, 4,\cdots ,\, 2^{k-1}$ neurons, and the variable $\sigma_{\rm i}^{(k)}$ is the summed activity of cluster $\rm i$. 

We emphasize that one could have different criteria for coarse--graining, and different ways of combing the variables.  We return to some of these points below (\S\ref{sec-RGunderstanding}), but for now we explore what happens when we apply this simplest scheme to a network of real neurons.  The first such example used the experiments on the activity of $1000+$ neurons described in \S\ref{sec-subgroups} \cite{meshulam2019RG,meshulam+al2018}.

We are interested in how the probability distributions transform and flow as we pass through successive scales of coarse--graining. Of course looking at the joint distribution $P(\{\sigma_{\rm i}^{(k)}\})$ is essentially impossible.  But much can be learned by looking at slices through this distribution, even the distribution of individual coarse--grained variables, as with the magnetization in the Ising model \cite{binder_81}. 

\begin{figure}
\includegraphics[width=\linewidth]{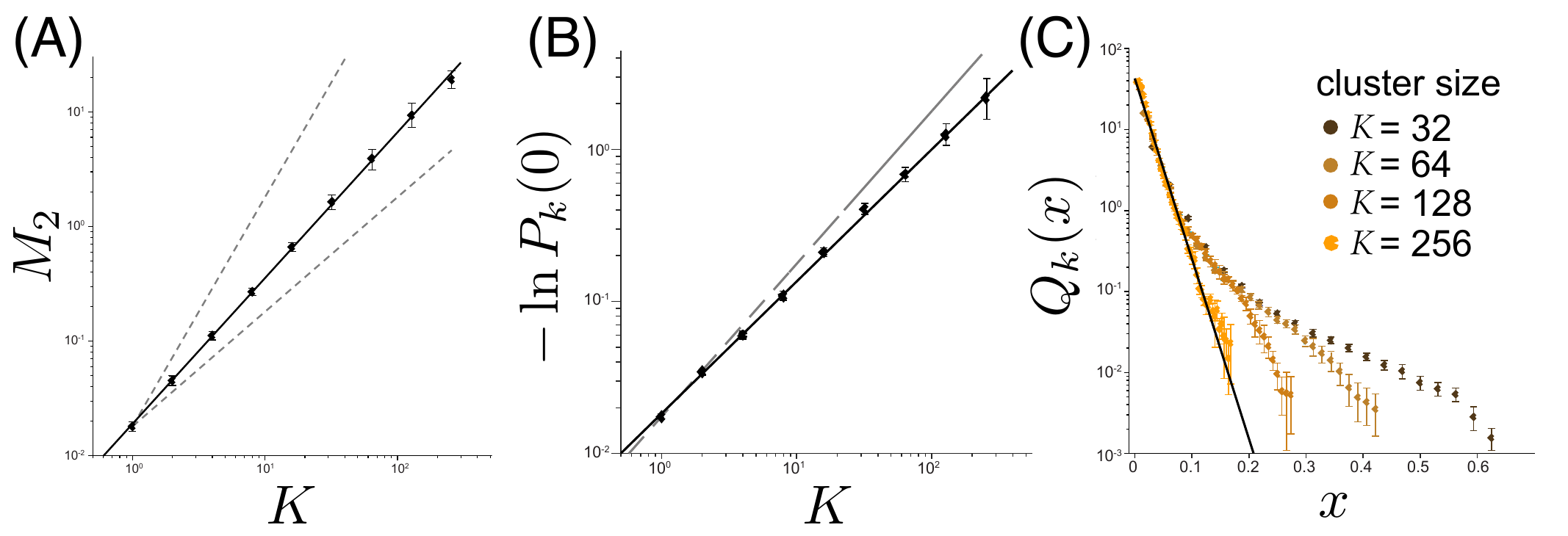}
\caption{Three slices through the distribution of coarse--grained variables \cite{meshulam+al2018, meshulam2019RG}.  (A) Variance of the activity vs.~the (real space) coarse graining scale, from Eq (\ref{eqvar}).  Solid line is $M_2\propto K^{\tilde \alpha}$, $\tilde\alpha = 1.4\pm 0.06$; dashed lines are predictions for independent ($\tilde\alpha =1$) or perfectly correlated ($\tilde\alpha =2$) neurons. (B) Probability of silence vs.~the coarse--graining scale.  Solid line is Eq (\ref{Pk0_1}) with  $\tilde\beta = 0.88\pm 0.01$; dashed line is the expectation for independent neurons, $\tilde\beta = 1$. (C) Distribution of the normalized non--zero activity, as defined in Eq (\ref{Q-def}). \label{slices1}}
\end{figure}

Since this coarse--graining  is based simply on adding the ``neighboring''  variables, the first moment of the distribution of the individual variables must scale linearly, 
\begin{equation}
	M_1(k) \equiv {1\over {N_k}} \sum_{{\rm i}=1}^{N_k} \langle \sigma_{\rm i}^{(k)}\rangle = 
	{1\over {N_k}} \sum_{{\rm i}=1}^{N_k} m_{\rm i}^{(k)} = K M_1(1)  ,
\end{equation}
where after $k$ steps we have $N_k$ clusters each involving $K = 2^{k-1}$ of the original variables.  The first non--trivial question is about the second moment, or the variance in activity,
\begin{equation}
	M_2(K) \equiv {1\over {N_k}} \sum_{{\rm i}=1}^{N_k} {\bigg \langle} \left( \sigma_{\rm i}^{(k)} - m_{\rm i}^{(k)}\right)^2 {\bigg \rangle} .
	\label{eqvar}
\end{equation}
Note that if neurons are independent we expect $M_2(K) \propto K$, and many weakly correlated populations should approach this behavior at large $K$.  If neurons are perfectly correlated, on the other hand, we expect $M_2(K) \propto K^2$.  Looking at the data, in Fig \ref{slices1}A, we see that for neurons in the hippocampus $M_2(K) \propto K^{\tilde\alpha}$, with $\tilde\alpha = 1.4\pm0.06$.  This non--trivial scaling is visible over more than two decades.

We can take another slice through the distribution by asking for the probability $P_k(0)$ that the coarse--grained variable $\sigma_{\rm i}^{(k)} = 0$.  Since we started with variables $\sigma_{\rm i} = \{0,1\}$, this is the same as asking for the probability that all of the neurons inside the cluster of size $K=2^{k-1}$ are silent.  If the neurons are independent we expect a simple scaling $P_k(0) \propto \exp(-aK)$, and once more expect to see this at large $K$ even if the cells are weakly correlated.   Experimentally we see in Fig \ref{slices1}B  that
\begin{equation}
P_k(0) = \exp( - a K^{\tilde \beta}) ,
\label{Pk0_1}
\end{equation}
with the exponent $\tilde\beta = 0.88\pm 0.01$.  Again scaling is precise over more than two decades.

If we imagine making an explicit model for the joint activity of all the neurons inside one of the clusters, perhaps in the form of the pairwise models above [Eq (\ref{E2-V})], then the probability of complete silence is dependent only on the partition function, $P_k(0) = 1/Z$.  This generalizes if we include higher--order terms, so that Fig \ref{slices1}B probes the effective free energy, which apparently behaves as $F(K) = -aK^{\tilde\beta}$.  Since $\tilde\beta <1$, the free energy is sub--extensive, and hence the free energy per neuron will vanish in the thermodynamic limit.  This is consistent with the equality of entropy and energy that we saw for retinal neurons in \S\ref{sec-thermo} (Fig \ref{SvsE}).

More generally if we define the normalized variable $x = \sigma^{(k)}/K$, then
\begin{equation}
P_k(x) = P_k(0) \delta_{x,0} + [1 - P_k(0)] Q_k(x) .
\label{Q-def}
\end{equation}
Figure \ref{slices1}C shows the evolution of $Q_k(x)$ as $K$ increases.  We see that the tail of the distribution is gradually absorbed into the bulk, which seems to approach a fixed form $Q(x) \sim e^{-x/x_0}$.   If the neurons were independent the central limit theorem would drive this distribution toward a Gaussian, but instead we see the emergence of a fixed non--Gaussian form.

In addition to looking at the distribution of single coarse--grained variables we can look at the covariance matrix of the microscopic variables within each cluster of size $K$.  The eigenvalue spectrum of this covariance matrix depends on the rank scaled by $K$, and there is a substantial region over which the spectrum is a power $\lambda \sim (K/{\rm rank})^\mu$, with $\mu = 0.71\pm 0.06$, although this is less crisp than the other examples of scaling.

Our discussion of thus far has focused on the distribution of variables at a single moment in time.  In the applications of the RG that we understand, however, we can often observe dynamic scaling  \cite{hohenberg+halperin_77}. Intuitively, fluctuations on longer length scales take longer to relax because the underlying interactions are local.  What is non--trivial is that correlation functions for variables coarse--grained to different length scales collapse to a universal form if we measure time in units of the correlation time, and this correlation time itself varies as a power of the length scale.  An elegant example of these ideas in a fully biological context is provided by dynamic scaling of the velocity fluctuations in natural swarms of insects \cite{cavagna+al_17}.

\begin{figure}[t]
\includegraphics[width=\linewidth]{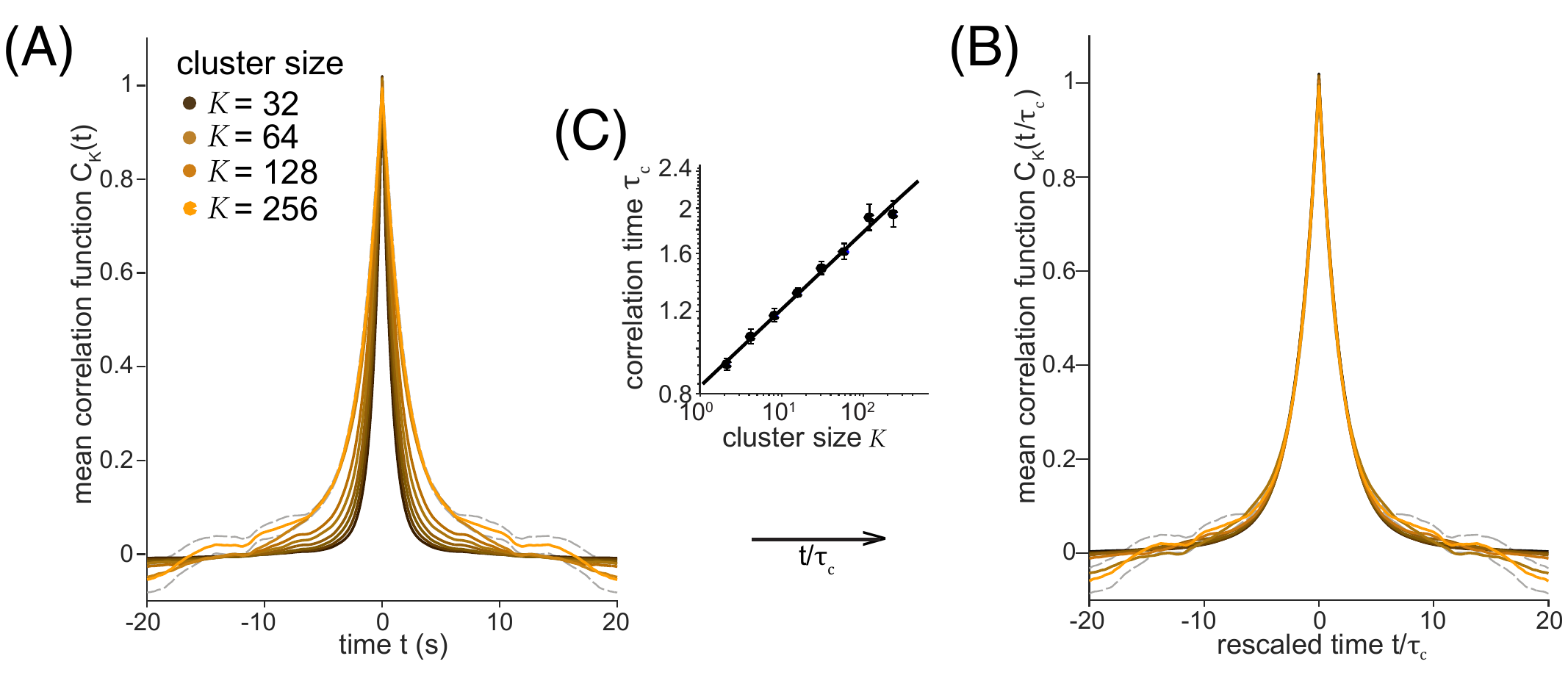}
\caption{Dynamic scaling across $1000+$ neurons in the hippocampus \cite{meshulam+al2018}.
(A) Mean correlation functions for coarse--grained variables, Eq (\ref{C_dyn1}),   in clusters of $K=2,\,, 4\, \cdots ,\, 256$ neurons (lightest orange corresponds to the largest cluster), with larger clusters exhibiting slower dynamics.  In dashed gray, $\pm$ one standard deviation across the $K = 256$ neuron clusters. 
(B) Collapse under scaling of the time axis, Eq (\ref{C_dyn2}). 
(C) Correlation time vs cluster size, fit to $\tau_c \propto K^{\tilde z}$, with $\tilde z = 0.16 \pm 0.02$.  \label{dyn-scale1}}
\end{figure}

With networks of neurons we don't expect locality to be a good guide, but it still is plausible that more strongly coarse--grained variables will have slower dynamics, and we can search for dynamic scaling.  Concretely we define the correlation function for individual variables at coarse--graining scale $k$,
\begin{equation}
\tilde C_{\rm i}^{(k)} (t) = {\bigg\langle} \left[\sigma_{\rm i}^{k}(t_0) - m_{\rm i}^{(k)}\right] \left[\sigma_{\rm i}^{k}(t_0 + t) - m_{\rm i}^{(k)}\right] {\bigg\rangle},
\end{equation}
and then we can normalize and average over the clusters to give
\begin{equation}
C^{(k)} (t) = {1\over{N_k}} \sum_{{\rm i}=1}^{N_k} {{\tilde C_{\rm i}^{(k)} (t)}\over{\tilde C_{\rm i}^{(k)} (0)}} .
\label{C_dyn1}
\end{equation}
Dynamic scaling is the hypothesis that the dependence on scale is captured by a single correlation time,
\begin{equation}
	C^{(k)}(t) = C[t/\tau_c(k)] ,
	\label{C_dyn2}
\end{equation}
with $\tau_c (k) \propto K^{\tilde z}$.   In Figure \ref{dyn-scale1} we see that all of this works for the population of hippocampal neurons.  We note that dynamic range of correlation times accessed in this experiment is limited, at short times by the dynamics of the indicator molecules and at long times by the small value of the exponent $\tilde z = 0.16 \pm 0.02$.

It is important that these scaling behavior are not somehow driven by our choice to describe neural activity with binary variables.  In these experiments, neural activity was recorded by imaging of fluorescence from  indicator molecules that provide a continuous signal as in Figs \ref{imaging} and \ref{setup}.  We can follow the same steps of coarse--graining for these continuous signals, and the results are the same \cite{meshulam2019RG}.

In the full theoretical structure of the RG, scaling exponents are signatures of universality classes.  Before we can ask about universality we have to ask about reproducibility, especially in such complex systems.  As a first step, the same analyses have been done with data from experiments on multiple mice.  Because scaling is precise across more than two decades, the error bars in determining the exponents in individual mice are small, which sets a high standard for reproducibility.\footnote{To be clear, we could see that multiple experiments ``agree''  just because the error bars on the individual experiments are large.  This of course would be much less compelling.} For example, the exponent describing the scaling of the free energy (Fig \ref{slices1}B) is $\tilde \beta = 0.87 \pm 0.014 \pm 0.015$ for the mean, the rms error in single experiments, and the standard deviation across experiments in three mice.  This holds out the hope that we have uncovered features of the emergent behavior that are reproducible in the second decimal place.

\begin{figure}[t]
	\includegraphics[width = \linewidth]{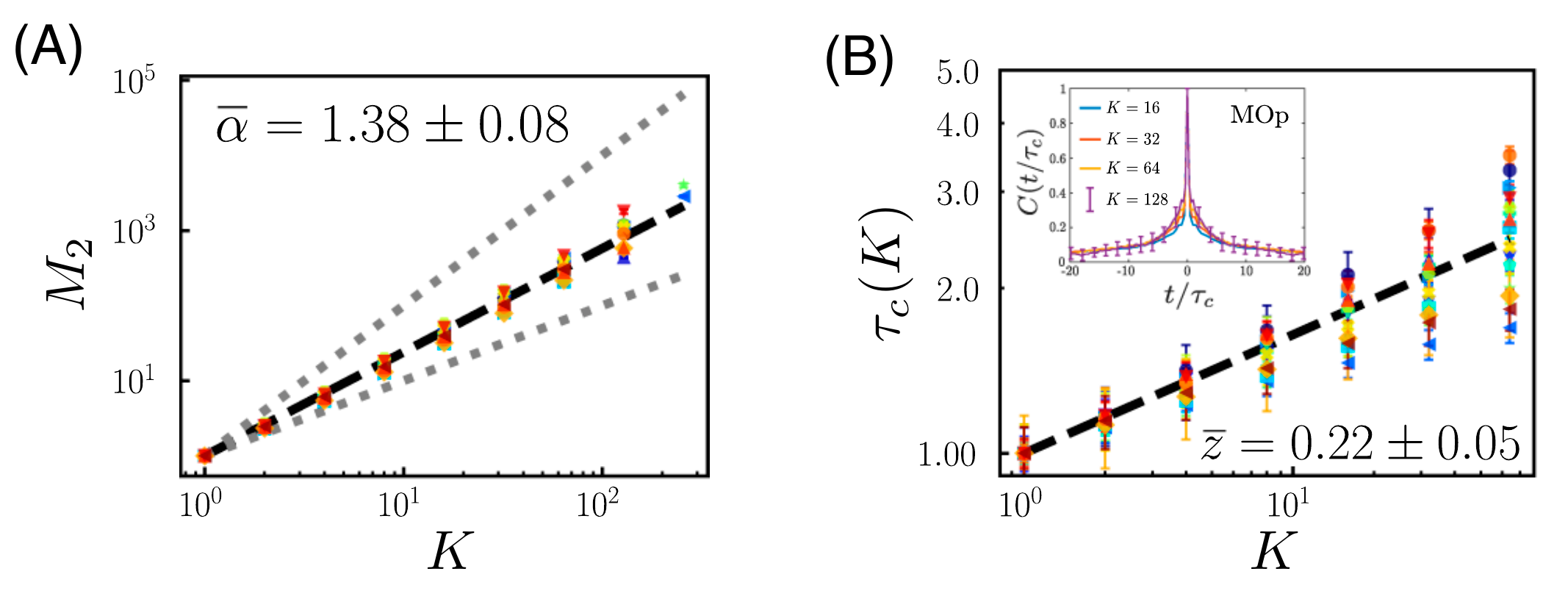}
	\caption{Scaling in mutliple distinct areas of the mouse brain \cite{morales+al2023}. neuronal data after the ``real space'' (direct correlations) coarse--graining procedure. (A) Variance of the coarse--grained activity vs cluster size for neurons in sixteen different brain regions (depicted as different markers), comparable to Fig \ref{slices1}A.  
(B) Dynamic scaling for the same brain areas. Correlation time vs cluster size, comparable to Fig \ref{dyn-scale1}.   Inset: Decay of the autocorrelation function for the neurons in one brain region (primary motor cortex) showing the collapse once time is rescaled. 
\label{scaling_real}}
\end{figure}

\begin{figure*}[t]
\includegraphics[width=\textwidth]{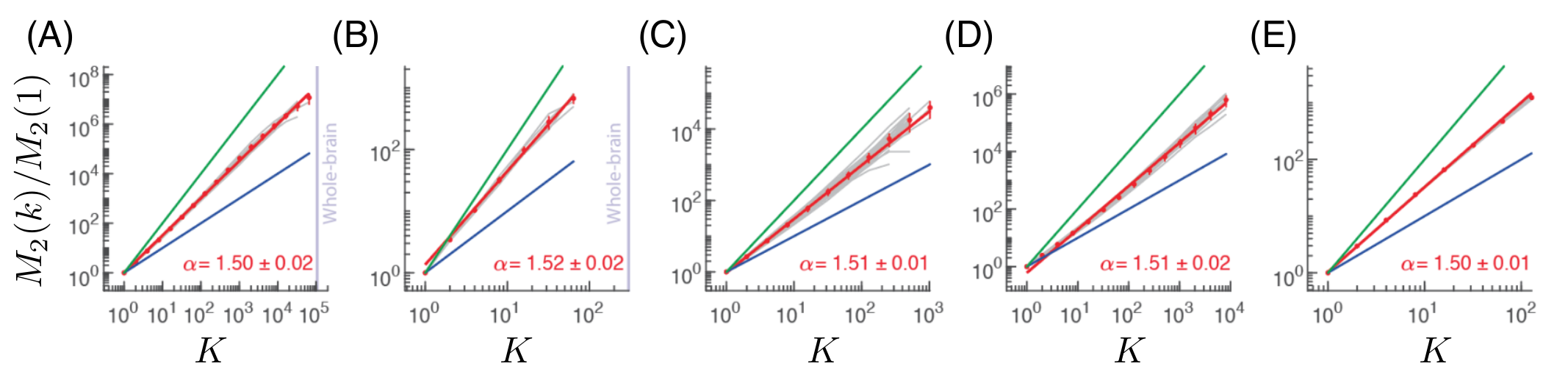}
\caption{Scaling in the variance of neural activity, Eq (\ref{eqvar}), as a function of scale across multiple species \cite{Munn+al2024}.   (A) Zebrafish. (B)  The worm {\em C.~elegans}. (C) The fruit fly {\em Drosophila melanogaster}. (D) Mouse primary visual cortex. (E)  Macaque primary visual and motor cortices.  Grey lines are results from individual animals, red points with errors are means within species, and red lines are fits to $M_2\propto K^{\tilde\alpha}$, with exponents as shown.  
Expectations for independent (blue) and completely correlated (green) populations   corresponding to the dashed lines in Fig \ref{slices1}A.  
\label{scaling_species}}
\end{figure*}

A more ambitious search for universality was undertaken by \citet{morales+al2023}.  They analyzed experiments that are part of a large effort at the Allen Institute for Brain Science, in this case using  multiple neuropixels probes (Fig \ref{neuropixels}) to record $100+$ neurons from each of many different areas of the mouse brain, simultaneously.  Note that in addition to exploring many different brain regions, the technique for recording activity is completely different than in the hippocampal imaging data analyzed in Figs \ref{slices1} and \ref{dyn-scale1}.  Nonetheless, all aspects of scaling are reproduced across all these brain areas; examples include the scaling of the variance in coarse--grained activity (Fig \ref{scaling_real}A) and dynamic scaling (Fig \ref{scaling_real}B).  

As we were completing this review a striking result was reported by \citet{Munn+al2024}.  Rather than looking at experiments across multiple brain areas in a single organism, they looked at experiments on many different organisms, from the tiny worm {\em C.~elegans} to primates much like us.  There are significant technical differences among these experiments, including differences in the calcium indicator proteins (\S\ref{sec-imagingmethods}) and differences in the sampling rate; complete resolution of individual neurons vs ``regions of interest;'' and recording from the entire brain is smaller model organisms vs.~a single sensory or motor area in larger organisms.  Many microscopic features of these networks also are very different, with the extreme being that {\em C.~elegans} neurons generate slow, graded potentials instead of discrete action potentials or spikes.  Despite these caveats, we can ask how the patterns of neural activity in these systems transform under coarse--graining across a range from two to five decades.  Results for the variance of the coarse--grained activity, $M_2(k)$ from Eq (\ref{eqvar}), are shown in Fig \ref{scaling_species}.  The apparent universality of these results is tantalizing.

\subsection{By analogy with momentum shell methods}
\label{sec-shells}

In problems where ``scale'' really is a length scale, coarse--graining is a gradual blurring out of spatial detail much as what happens when we look through a microscope and defocus.  In that analogy, the spatial pattern is Fourier transformed and then reconstructed using only a limited range of wavelengths.  Concretely, if we start with variables $\phi(\bm{\vec x})$ in a $d$--dimensional space with coordinates $\bm{\vec x}$, the coarse--graining operation becomes
\begin{eqnarray}
\phi(\bm{\vec x}) &\rightarrow& 
\phi_\Lambda (\bm{\vec x}) = 
z_\Lambda \int_{|\bm{\vec k}|< \Lambda} {{d^d k}\over{(2\pi)^d}} e^{i\bm{\vec k \cdot \vec x}} 
\tilde \phi(\bm{\vec k})\\
\tilde \phi( \bm{\vec k}) &=& \int d^d x\, e^{-i\bm{\vec k \cdot \vec x}}\phi(\bm{\vec x}) ,
\end{eqnarray}
where $\Lambda = \pi/\ell$ cuts off contributions below a length scale $\ell$ and $z_\Lambda$ serves to (re)normalize the variables; in the microscopic analogy this compensates for the loss of contrast as we defocus.  As in real space we are interested in how the probability distribution $P_\lambda [\phi_\Lambda]$ evolves as a function of the cutoff $\Lambda$.  Since the Fourier variables are continuous (in the limit of a large system) we can make infinitesimal changes $\Lambda \rightarrow \Lambda - d\Lambda$.  In quantum mechanics wave with wavevector $\bm{\vec k}$ describe particles with momentum $\bm{\vec p} = \hbar\bm{\vec k}$, so that average over the details in a range $\Lambda - d\Lambda < |\bm{\vec k}| < \Lambda$ is equivalent to integrating out a ``momentum shell'' \cite{wilson+kogut_74}.

Momentum is conserved in systems with translation invariance.  Independent of these physical principles, spatial translation invariance privileges the Fourier transform.  As an example, if variables $z_{\rm i}$ live on a lattice of points $\bm{\vec x}_{\rm i}$, translation invariance means that the covariance matrix elements $C_{\rm ij}$ can depend only on the difference in positions,
\begin{equation}
C_{\rm ij} = C(\bm{\vec x}_{\rm i} - \bm{\vec x}_{\rm j}),
\end{equation}
this matrix is diagonalized in a Fourier basis,
\begin{eqnarray}
\sum_{{\rm j}=1}^N C_{\rm ij}u_{\rm jr}  &=&  \lambda_{\rm r} u_{\rm ir} 
\label{eigen1}\\
u_{\rm jr} &\propto& \exp\left(i \bm{\vec k}_{\rm r} \cdot \bm{\vec x}_{\rm j}\right) ,
\end{eqnarray}
where we can put the modes in order by the rank of the eigenvalue $\rm r$.  

In the usual applications of the RG, large momenta correspond to small eigenvalues of the covariance matrix.  Thus suggests that we can construct coarse--grained variables by filtering out the ``modes'' that correspond to small eigenvalues, without reference to space or momenta \cite{bradde+bialek2017}.  This connects coarse--graining to a more familiar data analysis technique, principal components analysis \cite{shlens_14}. 

Concretely, if we start with microscopic variables $\{\sigma_{\rm i}\}$, we can compute the covariance matrix as usual 
\begin{equation}
C_{\rm ij} = \langle \left( \sigma_{\rm i} - \langle \sigma_{\rm i}\rangle\right)
\left( \sigma_{\rm j} - \langle \sigma_{\rm j}\rangle\right)\rangle ,
\end{equation}
and then we have eigenvalues and eigenvectors as in Eq (\ref{eigen1}).  Let's choose the rank $\rm r$ so that $\lambda_1 \geq \lambda_2 \cdots \lambda_N$.  We can define a projection onto the  $\hat K$ modes that make the largest contribution to the variance,
\begin{eqnarray}
\hat P_{\rm ij} (\hat K) &=& \sum_{{\rm r}=1}^{\hat K} u_{\rm ir} u_{\rm jr}\\
	\phi_{\hat K} ({\rm i}) &=& z_{\rm i}(\hat K) \sum_{\rm j} \hat P_{\rm ij} (\hat K) \left[ \sigma_{\rm i}  - \langle \sigma_{\rm i}  \rangle\right] ,
	\label{phiKi}
\end{eqnarray}
with the normalization $z_{\rm i}(\hat K)$ such that $\langle [\phi_{\hat K} ({\rm i})]^2\rangle = 1$.

\begin{figure}[b]
\includegraphics[width=\linewidth]{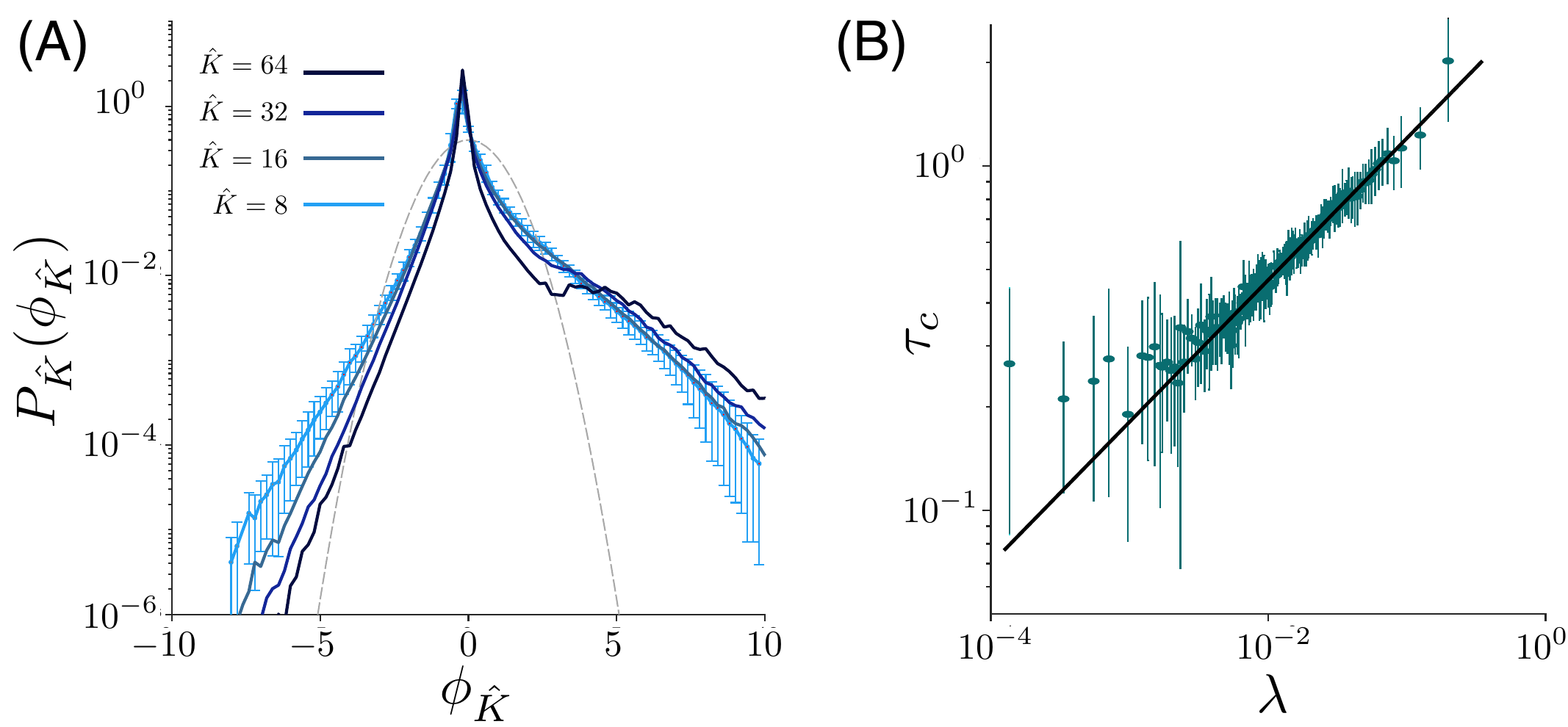}
\caption{Coarse--graining in groups of $N=128$ neurons via  ``momentum shells''  \cite{meshulam+al2018}. (A) Following the distribution of individual coarse--grained variables from Eq (\ref{phiKi}).  Different colors correspond to keeping different numbers of modes $\hat K$, as in inset; dashed line is a Gaussian for comparison.  (B) Dynamic scaling of the correlation time for fluctuations in mode $\rm r$, Eq (\ref{Cmode}), vs the associated eigenvalue of the covariance matrix, $\tau_c({\rm r}) \propto \lambda_{\rm r}^{\tilde z '}$, $\tilde z' = 0.37\pm 0.04$. \label{momentum}}
\end{figure}

As before, we want to follow the distribution of the individual coarse--grained variables, $P_{\hat K}(\phi_{\hat K})$; results are shown in Fig \ref{momentum}A.  To be sure that we have control over the full matrix $C_{\rm ij}$ we look at clusters of $N=128$ neurons identified through the real space coarse--graining above.  We can then filter out half of the modes, so that $\hat K = 64$, resulting in a distribution $P_{\hat K}(\phi_{\hat K})$ that still has some fine structure.  If we reduce to $\hat K = 32$ these wiggles disappear but the distribution remains asymmetric with long tails.   This pattern continues as we reduce to $\hat K = 16$ and then $\hat K = 8$, and in these last steps the distribution hardly changes.  This suggests that as we coarse--grain, the distribution flows toward a fixed form.  Importantly this form is {\em very} different from the Gaussian that would be guaranteed by the central limit theorem if correlations were weak.

The intuition behind dynamic scaling is that fluctuations on larger length scales relax more slowly, and we have seen that this generalizes to a network of neurons even though the meaning of ``scale'' now if more abstract (Fig \ref{dyn-scale1}).  By transforming to basis that diagonalizes the covariance matrix we have isolated the modes of fluctuation that are independent at second order, and it is natural to ask how these fluctuations along these modes relax.  Variations along mode $\rm r$ are define by
\begin{equation}
\tilde\phi_{\rm r} = \sum_{{\rm i}=1}^N \left[ \sigma_{\rm i}  - \langle \sigma_{\rm i}  \rangle\right] u_{\rm ir} ,
\end{equation}
and the correlation function is
\begin{equation}
C_{\rm r}(t) = \langle \tilde\phi_{\rm r} (t_0) \tilde\phi_{\rm r} (t_0 + t)\rangle.
\label{Cmode}
\end{equation}
Dynamic scaling is the statement that all these correlations collapse when time is scaled by a single correlation time, and that this correlation time  itself has a power--law dependence of scale.  In the usual examples this means $\tau_c \propto |\bm{\vec k}|^z$ \cite{hohenberg+halperin_77}, but near a critical point the eigenvalues of the covariance matrix also have a power--law dependence on $|\bm{\vec k}|$, so we can test directly for $\tau_c \propto \lambda^{\tilde z'}$ as shown in Fig \ref{momentum}B.  As before, the shortest correlation times are limited by the response time of the fluorescent proteins that report on electrical activity, and the longest times are limited by the magnitude of the dynamic scaling exponent; nonetheless we can observe reasonably precise scaling across two decades in $\lambda$.

The dynamic exponent $\tilde z'$ that one finds by looking at the correlation times of the modes should be related to the one we see via coarse--graining in real space, $\tilde z$ (Fig \ref{dyn-scale1}C), through the exponent $\mu$ that describes the decay of the eigenvalues of the covariance matrix, $\tilde z = \mu \tilde z'$.  This works, although error bars are large \cite{meshulam+al2018}.  More importantly, these results indicate that the network has no single characteristic time scale, but rather a continuum of time scales that can be accessed by probing on different scales.

\subsection{RG as a path to understanding}
\label{sec-RGunderstanding}

If we believe there is an underlying simplicity to be found amidst the complexity of neural network function and activity,  we might want to pause for a moment to convince ourselves that following the RG simplification can actually lead us there. This quest now feels attainable, given the explosive experimental progress in obtaining datasets with increasing number of neurons, as in the examples above. While we may not know how to manipulate ``temperature'' or ``magnetization'' in the brain, we are gaining decades in the sheer number of monitored neurons. 

The renormalization group is a powerful theoretical structure.  Because we do not have a microscopic model for neural dynamics, we are not yet able to exploit this structure.  What we have done instead is to adopt an RG--inspired approach to data analysis, which has been described as a ``phenomenological renormalization group'' \cite{nicoletti+al2020} or ``iterative coarse--graining'' \cite{Munn+al2024}.    If we apply these approaches to well understood equilibrium statistical mechanics problems, the most interesting outcome would be the flow of probability distributions toward some fixed, non--Gaussian form, and the appearance of power--law scaling along this trajectory, as would happen at a critical point.  Remarkably, this is what has been found, both in the initial application to the hippocampus and now in many other systems; scaling exponents are reproducible and perhaps even universal.  It is tempting to conclude that the underlying network dynamics must be described by a theory which is at a non--trivial fixed point of the renormalization group.

We should be cautious.  Is it possible that some of the behaviors under coarse--graining that we associate with RG fixed points could emerge, more generically, in non--equilibrium systems?  \citet{nicoletti+al2020} addressed this by analyzing simulations of the contact process, in which binary variables are turned on with a probability per unit time proportional to the density of active variables at neighboring sites, and then deactivate with a fixed probability per unit time.  This model has one parameter, the proportionality constant in the activation rate, and there is a critical value that depends on the geometry of the network \cite{marro+dickman_99}.  Below the critical point the fully inactive state is absorbing, so the question is whether the phenomenological RG can distinguish the critical point from super--critical behaviors.

\begin{figure}
\includegraphics[width = \linewidth]{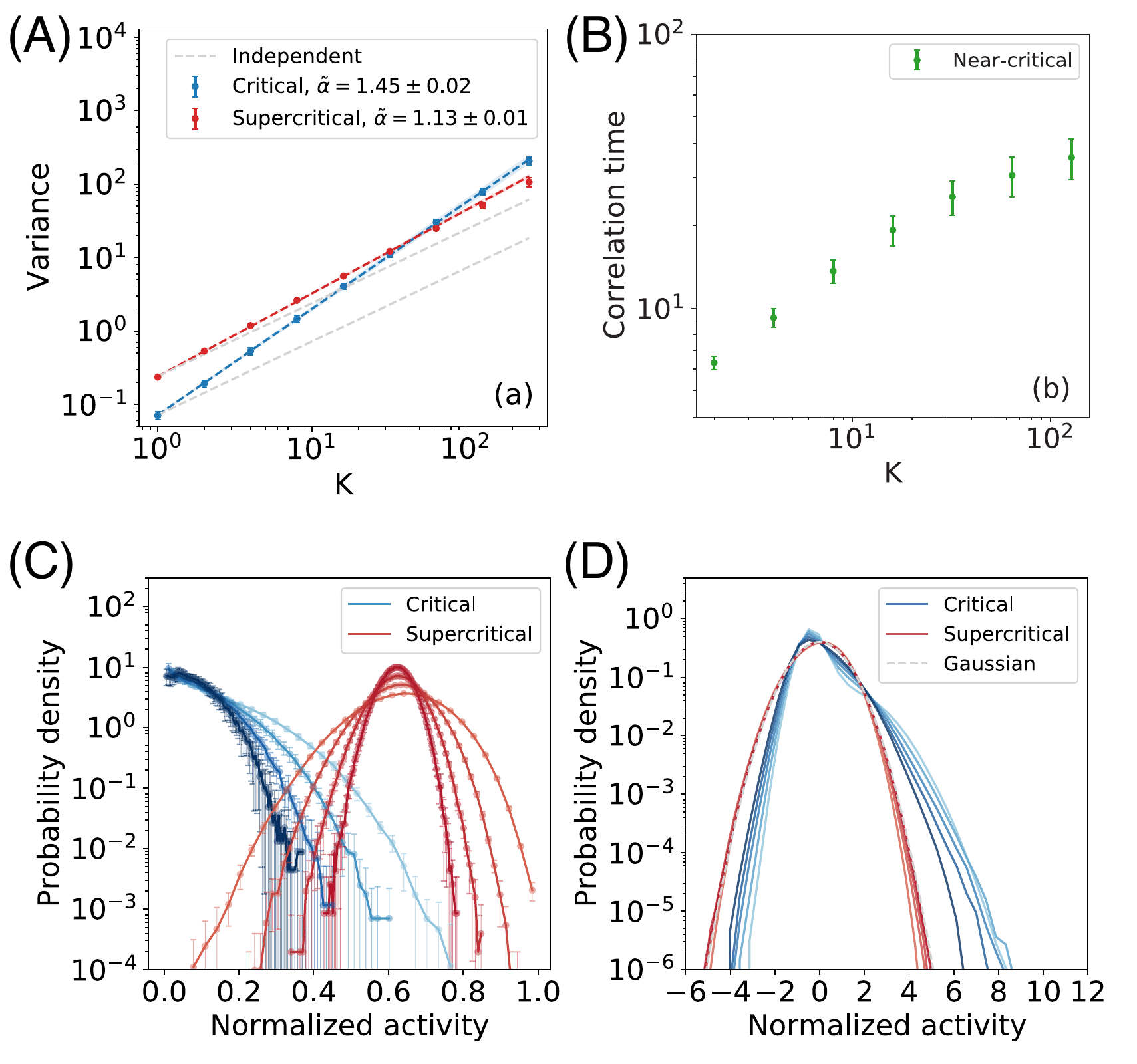}
\caption{Coarse--graining of the contact process \cite{nicoletti+al2020}.  (A) Variance of activity vs.~the scale of coarse--graining in real space, as in Figs~\ref{slices1}A and \ref{scaling_species}.  Behavior at criticality (blue) is clearly different from the super--critical case (red), which departs systematically but weakly from the expectations for independent variables (dashed lines). (B) Correlation time vs.~the scale of coarse--graining in real space, as in Fig~\ref{dyn-scale1}.  The control parameter is set close to its critical value, and we see hints of scaling at small $K$ but clear departures at large $K$.  (C) Distribution of individual coarse--grained variables for $K=32, 64, 128, 256$ at criticality (blue) and  away from criticality (red).  In both cases we see flow toward a fixed distribution, but away from criticality this is Gaussian as expected from the central limit theorem.  (D) As in (C), but with coarse--graning via momentum shells, keeping $N/8, N/16, N/32, N/64, N/128$ of the modes.   \label{fig:nicoletti}}
\end{figure}

Perhaps surprisingly, one can see (weakly) non--trivial scaling behavior in some quantities even away from the critical point, as with the variance in activity shown in Fig \ref{fig:nicoletti}A.  But other quantities show clear deviations from scaling, even very close to criticality, as with the correlation times in Fig \ref{fig:nicoletti}B.  What is unambiguous is that the probability distributions of coarse--grained variables flow toward a non--trivial fixed form at the critical point, and toward a Gaussian otherwise.  We can see this by coarse--graining in real space (Fig~\ref{fig:nicoletti}C) or via momentum shells (Fig~\ref{fig:nicoletti}D).  \citet{nicoletti+al2020} emphasize that the phenomenological RG can identify critical points unambiguously, but only if we check the full range of behaviors.

As with the (related) discussion of criticality in \S\ref{sec-alternatives}, it has been suggested that some of the phenomena uncovered by iterative coarse--graining can be reproduced in a model where neurons respond independently to latent fields \cite{morrell+al2021}.  In this view, scaling and the flow toward fixed distributions are approximate, and it is not clear why scaling exponents should be reproducible across animals; a broader notion of universality, as in Fig \ref{scaling_species}, would be even more difficult to understand.  

Certainly the suggestion that scaling behaviors emerge generically from latent variable models is incorrect. Consider models in which the effective field acting on each neuron $\rm i$ is a linear combination of $K$ latent variables drawn from a Gaussian distribution.  If the fields are weak then the covariance matrix of neural activity has the same rank as the covariance matrix of the fields.  This simple result breaks down at stronger fields, but even in the limit of infinitely strong fields there remains a gap in the eigenvalue spectrum of the covariance matrix, at least for typical choices of parameters, so that it is impossible to recover precise scaling behaviors.

We note that a concrete, biologically motivated model of latent fields---the independent place cell model discussed in \S\ref{sec-alternatives}---fails to exhibit scaling \cite{meshulam+al2018}.  This result perhaps should not be surprising. In a population of place cells, there are two length scales, the approximate width of the place fields and the mean distance between place field centers. In the one--dimensional (virtual) environment that provides the background for the hippocampal experiments analyzed here, the ratio of these lengths gives us a characteristic number of neurons, $K_c\sim 18$.  Indeed, analyses of the independent place cell model corresponding to Figs~\ref{slices1}A, B show ``breaks'' at $K\sim K_c$.  While these are approximate statements, they highlight the fact that, in the presence of such obvious scales, the observation of rather precise power--law scaling in both static and dynamic quantities really is surprising.

Faced with high--dimensional observations, a natural reaction is to search for a lower dimensional description.  In some sense the renormalization group is the opposite approach \cite{bradde+bialek2017}.  Rather than looking for the correct number of dimensions onto which to project the data, the RG invites us to examine how our description changes as we move the boundary between details that we ignore and features that we keep.   Things simplify not because we have fewer degrees of freedom but because the model describing these degrees of freedom flows toward something simpler and more universal.  The evidence thus far points toward the existence of such a simplified description.  From the theoretical side, initial efforts at an RG analysis of models for networks of more realistic neurons suggest that these are described by new universality classes \cite{brinkman2023}.

What we have not emphasized here is the connection of coarse--graining to more functional behaviors.   In the hippocampus, how is position represented in the coarse--grained variables?  More generally, do fine--grained and coarse--grained variables implement different principles for the encoding of the sensory world \cite{Munn+al2024}?  Can local networks of neurons access different scaling trajectories as the brain switches among different global states \cite{castro+al_24}?    As coarse--graining becomes a more commonly used tool for the analysis of large scale neural recordings, we expect progress on these issues over the next years.

The most detailed tests of scaling in equilibrium critical phenomena span six decades with better than one percent precision \cite{lipa+al_96}.  As described in \S\S\ref{sec-neuropixels} and \ref{sec-imagingmethods}, the experimental frontier is moving toward recording from $\sim 10^6$ neurons simultaneously.   This opens the possibility of following coarse--graining trajectories across five decades with single cell resolution, and of driving error bars down to the one percent level across more limited ranges.  The extension of existing tools to organisms with larger brains also means that we will see simultaneous recordings from more neurons in single brain areas, within which scaling seems more likely.  We already see signs that quantities which emerge from these analyses can be reproducible in the second decimal place.  One possibility is that new, larger experiments will reveal crossovers between different regimes on different scales.  Alternatively, the scaling behaviors seen thus far might prove to be essentially exact.  Whatever the outcome, it is extraordinary to think that experiments on real, functioning brains could soon reach a  precision comparable to those on equilibrium critical phenomena.  The corresponding challenge to theory should be clear.

\section{Outlook}

Statistical physics has long been a source of useful metaphors for emergent behaviors in living systems.  All the birds in a flock agreeing to fly in the same direction is like the alignment of spins in a magnet.  Recalling memories in the brain is like a spin glass settling into one of many locally stable states.  Quietly, examples have emerged that are more than metaphors.  Thus, experiments on single DNA molecules provide the most detailed tests of predictions for the random flight polymer, one of the classical models discussed in statistical mechanics courses \cite{bustamante+al_94,marko+siggia_95}. The explosion of data on networks of real neurons similarly offers the opportunity to move beyond metaphor.

We have seen that relatively simple statistical physics models---Ising models with pairwise interactions, and modest generalizations---provide detailed quantitative descriptions of real networks,  from the retina deep into the cortex and hippocampus.  Correct predictions are not limited to a few macroscopic or thermodynamic quantities, but include detailed patterns of higher--order correlations and the way in which the activity of each neuron depends on the collective state of the others.  Again it is not just some trends in these quantities that are being captured, but precise numerical values within experimental error.   The theories we are discussing may be swept away by the next generation of data, but these results set a standard for what we should demand in comparing theory with experiment.

The state of the field is such that our examples of success still are scattered, and each network that has been studied is different, being described for example by a different matrix of interactions $J_{\rm ij}$.  We can hope that as neural recordings at large $N$ become more common, and these analysis methods are applied more widely, we will learn something about the distribution from which these matrices are being drawn.  The goal is to go beyond models for particular networks toward a theory of these networks more generally.

The experimental frontier is moving rapidly, and it is reasonable to expect that $N\sim 10^3$ soon will be routine and that $N\sim 10^6$ soon will be possible with higher time resolution and higher signal to noise ratio.  We have emphasized that one cannot simply carry existing models to larger $N$ unless the duration of experiments increases in proportion.  This is not impossible, as stable recording methods allow visiting the same population of neuron day after day, not only to study non--stationary processes such as learning but to increase the volume of data from which we can estimate the correlation structures in the network.  At the same time, there are new ideas about how to build statistical physics models for networks from sparse data.  Success here means discovering some previously hidden simplicity that allows fewer measurements to characterize the global dynamics, and this will represent real theoretical insight.  

Perhaps the most fundamental question that we can sharpen by moving to larger $N$ is the construction of a thermodynamics for networks of neurons---guided by theory but built from data. Can we convince ourselves that real networks are understandable in the thermodynamic limit?   What are the relevant order parameters?  Where are real networks in the phase diagram of possible networks?  We have glimpsed possible answers to these questions at $N\sim 100$, but everything will become clearer at larger $N$, over the next few years.

The idea that networks of neurons might be poised near a critical point, or critical surface, has been a continuing source of fascination and controversy.  Much of the literature is about why this would be a good idea, or why it can't be right, rather than about the evidence, and we have tried to avoid these more ideological discussions here.  What we have seen is that populations of $N\sim 100$ neurons are described very accurately by relatively simple statistical physics models, and that if we change the temperature or the relative strength of different terms in the model then the parameter settings that describe the real system are close to criticality.   

While one can construct models that capture various aspects of critical phenomenology without the underlying structure of a critical point, it is not so easy to do this {\em and} engage with the detailed correlation structure of the real networks.  In contrast, critical behavior emerges naturally from the simplest models that are consistent with this structure. Importantly, models off criticality describe plausible networks---e.g. with slightly weaker or stronger correlations---but not the ones we see experimentally.  It should be possible to draw phase diagrams directly in a space that corresponds to these measurable quantities, perhaps ultimately without reference to more microscopic models.

In our modern understanding, saying that a system is at a critical point means that is described by a theory that is a fixed point of the renormalization group (RG).  More generally the RG invites us to ask how our description of a system changes as we include more or fewer levels of detail, and this suggests new approaches to data analysis.  It is striking that the first efforts in this direction showed that coarse--grained variables flow to non--trivial distributions, that one can see precise scaling over more than two decades, and that exponents can be reproducible in the second decimal place, with tantalizing hints of universality across brain areas and even across organisms.    Again it seems important to confront the full set of data, rather than focusing on one or two features that could by themselves be misleading.

As data collection moves to ever larger scales, coarse--graining becomes a more attractive approach to visualizing system behavior.  In conventional applications of the renormalization group, the coarse--graining step is constrained by symmetries and the associated conservation laws, as well as by locality, but these are absent in networks of neurons.  A first attempt was to average together the activity of neurons that are the most strongly correlated, and much remains to be explored using this idea.  But perhaps the example of neurons motivates a more general look at the RG itself.

Coarse--graining is an example of lossy data compression \cite{cover+thomas_91}, and in general one can choose what is preserved in making such compressions, e.g. the intelligibility of speech in the compression of acoustic waveforms.  The fundamental tradeoff is between the bits of information that coarse--grained variables carry about microscopic variables and the bit of information that they carry about ``relevant'' variables \cite{tishby+al_99}. Can we construct RG transformations with different choices for what is relevant?  We could compress the states of multiple neurons to preserve the information that the coarse--grained activity provides about other neurons in the network, about the future states of the same neurons, or about external quantities such as sensory inputs and motor outputs.   This combination of information theoretic and renormalization group ideas could give us new perspectives on classical results, but also lead to new fixed points and hence by definition new physics \cite{koch--janusz+ringel2018,gordon+al2021,kline+palmer2022}.  In a system as complex as the brain, one could even imagine that different RG flows co--exist, based on different coarse--graining schemes applied in parallel to the same population of neurons.

One basic question that the combination of information theory with the RG might help answer is how to identify order parameters.  In many contexts, once we know the order parameter we can almost immediately write down an effective field theory, and the technical apparatus of the RG tells us which terms in this theory are relevant or irrelevant.\footnote{The notion of relevance in the renormalization group is different from the notion of relevance in information theory.  But perhaps they are related.  In this spirit, see \citet{machta+al_13}.} But finding the order parameter currently relies on inspiration rather than constructive calculation.  In some contexts it seems clear that order parameters have an information theoretic interpretation, e.g. as the most compressed variable that provides information about the states of other variables at large spatial separations.  The hope is that we can turn this around, and give an information theoretic {\em definition} of order parameters that would allow their systematic discovery.

It is natural to ask what the observed scaling behaviors say about the function of neural networks in the life of the organism.  Dynamic scaling suggests that we can ``read out'' dynamics on different time scales just by averaging together the activity of different combinations of neurons, in the spirit of ideas about ``reservoir computing'' \cite{maass+al_02}.  Although attention often is focused on how to learn the correct readout scheme for a particular task, it also is essential that the reservoir be sufficiently rich.  Dynamic scaling means that there is a continuous range of available time scales, out to a longest time set only by the size of the network.  Perhaps this connects  with the ability of the brain to make predictions and drive behavior on a range of time scales.

Since the brain drives behavior, scaling in neural dynamics suggests that we might find scaling in behavioral correlations across time \cite{bialek+shaevitz_24}, although the search for these correlations is challenging.  If we can coarse--grain the activity of neurons, we should also be able to coarse--grain behavior itself, and this has been used to make notions of hierarchy in behavior more precise \cite{berman+al_16}.  There has been enormous progress in mapping high--resolutions video of animal behavior into descriptions of postural trajectories or behavioral states \cite{berman+al_14,mathis+al_18,pereira+al_19}, and perhaps we can should see this as a first step in coarse--graining, one that should be unified with subsequent analysis of the state sequences in time.  Put more simply, is there an RG for animal behavior?

If networks of neurons are described by a non--trivial fixed point of the renormalization group, the central theoretical question is of course to identify this fixed point theory (or theories, if different networks scale differently).  Even if scaling is found to break down at sufficiently large scales, the fact that we see this behavior over several decades suggests that   important aspects of network function will be controlled by the underlying fixed point.   We don't know how to fully connect the molecular events that shape the electrical dynamics of single neurons to cellular scale models of neural networks, and we can see this connection as a coarse--graining step.  There is a start on RG approaches to the cellular models \cite{brinkman2023}, and we hope to see more of this. An understanding of the fixed point theory or theories that describe the observed scaling should provide a division of more microscopic models into universality classes.

Not so long ago everything that we have said in this Outlook would have seemed like  physicists' fantasies,  disconnected from real brains (perhaps there are a few remnants of this). 
What has changed, dramatically, is that all these ideas---Ising models and correlation functions, scaling behaviors and the RG, and more---are connected to quantitative experiments on networks of real neurons.  Importantly, old worries that experiments on living systems are irreducibly messy have been overcome by demonstrating the levels of precision and reproducibility that we expect in physics.  Not all systems are equally accessible to this kind of exploration, but these results set an example for what is possible.  We can connect all the way from abstract physics concepts to the details of particular neurons in specific brain regions.  Our experimentalist friends will continue to move the frontier, combining tools from physics and biology to make more and more of the brain accessible in this way.   The outlook for theory is bright.

\section*{Acknowledgments}

Our exploration of these ideas has been shaped by the opportunity to interact with a wonderful group of experimentalist friends:  D Amodei, CD Brody, MJ Berry II, JL Gauthier, M Ioffe, SA Koay, A Leifer, O Marre, R Segev, JW Shaevitz, and DW Tank.
Our understanding of the theoretical issues has been aided enormously by discussions with many collaborators and colleagues: F Azhar, S Bradde, A Cavagna, X Chen, L Di Carlo, I Giardina, CM Holmes, K Krishnamurthy,  CW Lynn, F Mignacco, T Mora, I Nemenman, R Pang, SE Palmer, L Ramirez, C Sarra, DJ Schwab, E Schneidman, S Still, M Tikhonov, G Tka\v{c}ik, and AM Walczak.  
Special thanks also to those who gave advice about references, sometimes in extremis: ERF Aksay, EJ Chichilnisky, MS Goldman, F Rieke, DJS Schwab, and IH Stevenson.

This work was supported in part by the  National Science Foundation, through the Center for the Physics of Biological Function (PHY--1734030);   by the National Institutes of Health, through the BRAIN Initiative (R01EB026943--01); by the Burroughs Wellcome Fund through a CASI award to LM;  and by the Swartz,  Simons, and  John Simon Guggenheim Memorial Foundations.  WB also thanks the Center for Studies in Physics and Biology at Rockefeller University for its hospitality during part of this work.

\appendix

\section{Sequences, flocks, and more}
\label{sec-sequences+}

Part of what makes maximum entropy models exciting is that they can be used in a wide variety of contexts, perhaps pointing toward a more general statistical physics of biological networks.   Examples range from the evolution of protein families and its connection to protein structure to the propagation of order in flocks of birds, and more.  Exchange of methods and ideas among applications to these very different systems has been productive also for thinking about networks of neurons, so we give a brief (and hopefully not too idiosyncratic) survey here.

\subsection{Protein families}

Proteins are polymers of amino acids, and there are twenty amino acids to choose from at each site along the chain; to a large extent this sequence determines the folded structure and function of the protein.  The explosion of data on sequences has been even more dramatic than the explosion of data on networks of neurons, but thoughtful analysis of sequences began as soon as there were a handful to look at, and this played a crucial role in working out the genetic code \cite{brenner_57}.  

By the late 1970s it was clear that proteins form families with similar functions and structures  \cite{stroud_74}, and eventually the sequence data would become plentiful enough that these relationships could be detected without structural or functional measurements \cite{Pfam}.  Proteins are densely packed, and there is a strong intuition that evolutionary changes in one amino acid might need to be compensated by changes in neighboring amino acids \cite{Gobel+al_1994}; by the early 1990s there were a few families of proteins with enough sequences that one could see signatures of these correlated pairwise substitutions \cite{neher1994}.

To be concrete, define a variable $s_{\rm i}^\alpha = 1$ if the amino acid at site $\rm i$ along the chain is of type $\alpha$, and $s_{\rm i}^\alpha = 0$ otherwise; the full amino acid sequence of one protein then is  $\{s_{\rm i}^\alpha\}$ with ${\rm i} = 1,\, 2,\, \cdots ,\, N$ and $\alpha = 1,\, 2,\, \cdots ,\, 20$.  If we have $K$ proteins in a family we have a larger set of variables $\{s_{\rm i}^\alpha (n)\}$, with $n=1,\, 2,\, \cdots ,\, K$; this is called a multiple sequence alignment.\footnote{It is useful to introduce a ``blank'' state at $\alpha = 21$, allowing that one protein may have two segments that overlap strongly with others in the family but a small gap in between.} We can measure the expectation values at each  site
\begin{equation}
m_{\rm i}^\alpha = {1\over K}\sum_{n=1}^K s_{\rm i}^\alpha (n) ,
\end{equation}
which is the probability that amino acid $\alpha$ is used at site $\rm i$ in the family.  We can also define the joint probability of amino acids $\alpha$ and $\beta$ at sites ${\rm i}$ and ${\rm j}$,
\begin{equation}
C_{\rm ij}^{\alpha\beta} = {1\over K}\sum_{n=1}^K s_{\rm i}^\alpha (n) s_{\rm j}^\beta (n) .
\end{equation}
If we want to synthesize a new family of proteins $\{\tilde s_{\rm i}^\alpha(n)\}$ we could ask how similar these one-- and two--body statistics are to the original family by computing
\begin{eqnarray}
\chi^2 &=& \sum_{{\rm i}\,\alpha} W_{\rm i}^\alpha \left[ {1\over K}\sum_{n=1}^K \tilde s_{\rm i}^\alpha (n) - m_{\rm i}^\alpha\right]^2\nonumber\\
&&\,\,\,\,\,
+ {1\over 2} \sum_{{\rm i}\,\alpha}\sum_{{\rm j}\,\beta}  W_{\rm ij}^{\alpha\beta}  \left[ {1\over K}\sum_{n=1}^K \tilde s_{\rm i}^\alpha (n) \tilde s_{\rm j}^\beta (n) - C_{\rm ij}^{\alpha\beta}\right]^2 .\nonumber\\
&&
\label{chi2protein}
\end{eqnarray}
In this formulation we can give each term a different weight, perhaps in proportion to the accuracy with which we can estimate each expectation value.

In the early 2000s Ranganathan and colleagues realized that one could use the similarity measure $\chi^2$ as an energy function, and generate new families of proteins from known families by Monte Carlo simulation  \cite{socolich+al_05,russ+al_05}.   Most importantly, rather than just drawing samples out of the distribution they actually synthesized the proteins and asked whether they fold and function like the naturally occurring members of the family.  The short but compelling answer is that if one constrains only the one--body terms (i.e., set $W_{\rm ij}^{\alpha\beta} = 0$), then none of the many proteins synthesized in this way fold.  On the other hand, with the two--body terms included a reasonable fraction of all the new proteins synthesized do fold.   This was quite startling, suggesting that pairwise correlations were sufficient to capture the essence of the mapping from protein structure back to amino acid sequence.

What was missing from the original analysis was an explicit construction of the underlying probability distribution.  As it turns out, in the limit that families are large ($K\rightarrow\infty$) and the temperature of the Monte Carlo simulation is low, using $\chi^2$ in Eq (\ref{chi2protein}) as an energy function is equivalent to sampling the maximum entropy distribution consistent with one-- and two--body statistics \cite{bialek+ranganathan_07}.  This distribution has the form
\begin{eqnarray}
P\left(\{s_{\rm i}^\alpha\}\right) &=& {1\over Z} \exp\left[-E_p(\{s_{\rm i}^\alpha\})\right]\\\
E_p(\{s_{\rm i}^\alpha\}) &=& \sum_{{\rm i},\alpha} h_{\rm i}^\alpha s_{\rm i}^\alpha + {1\over 2}\sum_{{\rm i}\,\alpha}\sum_{{\rm j}\,\beta}  J_{\rm ij}^{\alpha\beta} s_{\rm i}^\alpha s_{\rm j}^\beta  ,
\label{potts}
\end{eqnarray}
where  the fields $\{h_{\rm i}^\alpha\}$ and  couplings $\{J_{\rm ij}^{\alpha\beta}\} $ are adjusted to match the means $\{m_{\rm i}^\alpha\}$ and joint probabilities $\{C_{\rm ij}^{\alpha\beta}\}$; the subscript $E_p$ reminds us that these are Potts--like models. 

\begin{figure}[t]
\includegraphics[width=\linewidth]{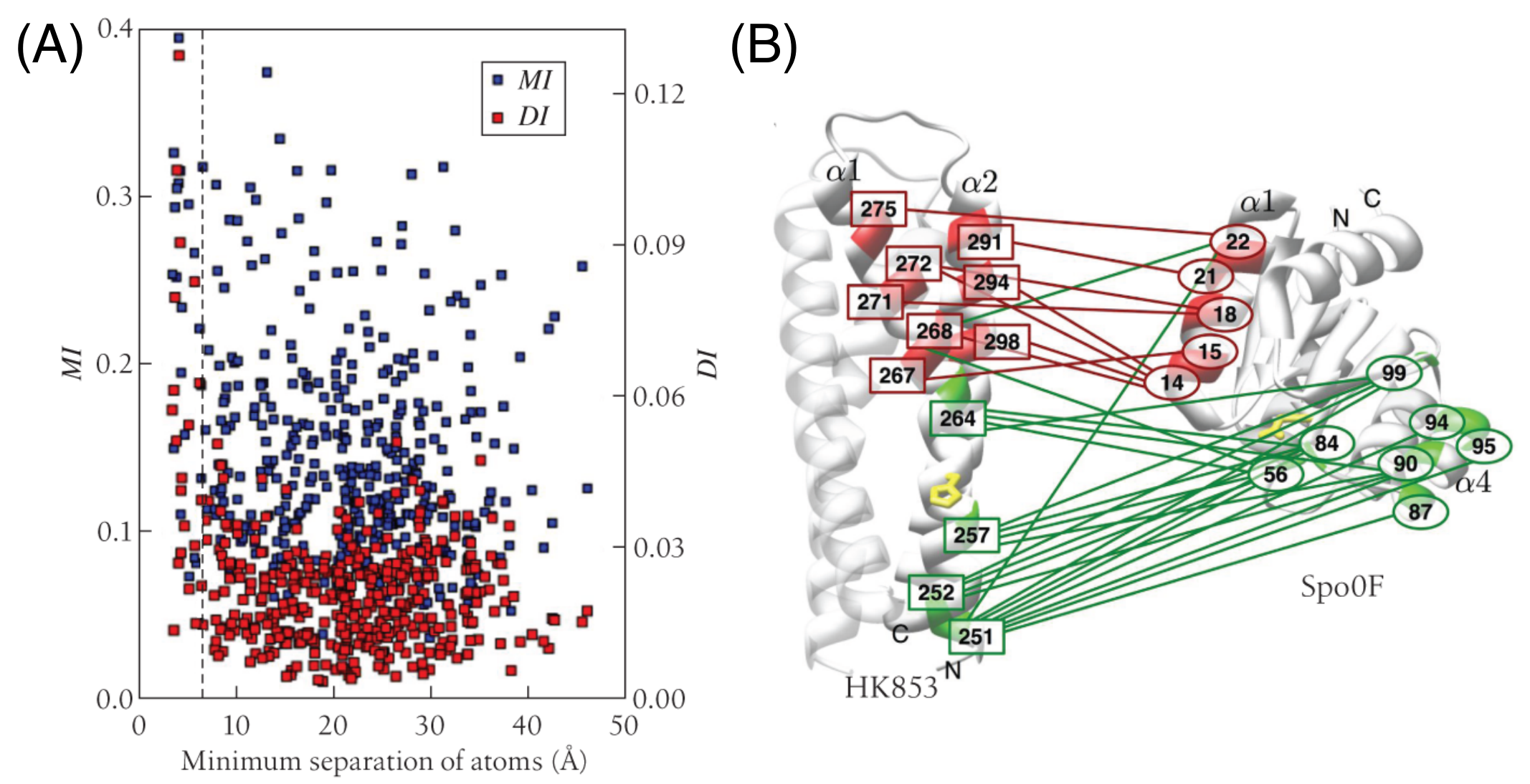}
\caption{Correlations vs.~effective interactions in protein sequence and structure, for pairs of bacterial signaling proteins. (A) Mutual information (MI) between amino acid substitutions at pairs of sites $\rm i$ and $\rm j$, compared with ``direct information'' (DI) between these sites, calculated by allowing for the interaction $J_{\rm ij}$ in Eq (\ref{potts})  but eliminating the interactions with all other sites. (B) Lines connect sites above some threshold level of MI.  Pairs in red share large DI, and are in contact.  Pairs in green have smaller DI, suggesting that correlations are indirect, and correspondingly they are not in contact.  \label{fig-weigt}}
\end{figure}

Note that in this formulation the amino acids sequences are analogous to the patterns of spiking and silence in a network of neurons.  The idea that proteins form families might correspond to these patterns forming a small number of globally structured clusters or brain states.  Synthesizing proteins with sequences drawn from some model distribution would correspond to  imposing patterns of activity onto the network, which still is a bit beyond the reach of today's experiments, though perhaps not for long.

One of the first modern applications of the maximum entropy approach was to families formed by pairs of interacting proteins that serve to convey signals across the membrane of bacterial cells  \cite{weigt+al_09}.  The crucial observation was that if one estimates the strength of correlation between amino acid choices at different sites, then this is only weakly correlated with the distance between these sites in the three dimensional structure (Fig \ref{fig-weigt}).  But we can imagine turning off all the interactions except those between sites $\rm i$ and $\rm j$, and then recomputing the mutual information; this ``direct information'' is strongly correlated with the distance between the sites, and a simple threshold allows us to identify the sites which are in contact at the interface between the two proteins. Subsequent work showed that this same principle also could be used to identify the correct interacting pairs of proteins \cite{bitbol2016inferring}.

The lesson of Fig \ref{fig-weigt} is that, as in many statistical mechanics problems, spatially extended correlations can arise from much more local interactions.  In this case the interactions are not real microscopic physical interactions, but rather effective interactions that describe the basic dependencies of amino acid substitutions on one another.   This picture was presented quite clearly well before there were large enough data sets to make inference practical \cite{lapedes+al_98}, but this seems to have been lost in conference proceedings that were not widely cited \cite{lapedes2012using}.  We note that large gaps between the range of interactions and the range of correlations, as seen here, are not generic.

If  the statistics of amino acid substitutions in protein families are described by a set of spatially local effective interactions, then we should be able to predict the three dimensional structure of these molecules from the sequence families alone.  Remarkably, this works \cite{marks2011protein,Sulkowska+al_2012}.  These successes provided a foundation for the dramatic development of AlphaFold, a deep network that achieves unprecedented accuracy in structure prediction  \cite{Jumper+al_2021}.

The emphasis on structure prediction perhaps detracted from some of the more basic questions about the use of pairwise models.  One exciting idea is that the effective energy function in Eq (\ref{potts}) might actually be related to the physical stability of the folded state.  More ambitiously if we build models for sequence variations across large populations of viruses such as HIV, the energy might predict the fitness of different sequences in the environment provided by the patient's immune system \cite{chakraborty+barton_17}.

As with the discussion of patterns of activity in networks of neurons, we'd like to know if the pairwise maximum entropy models correctly capture higher order correlations across sequence variations.  The first effort in this direction was focused just on short, highly variable sequences in antibody molecules \cite{mora2010maximum}.  This was perhaps more influential as an introduction to the idea that one could use modern sequencing data to describe the full distribution of antibody diversity, fitting into a larger stream of work on physics problems motivated by immunology \cite{altan-bonnet+al_20}.

\begin{figure}[b]
\includegraphics[width=\linewidth]{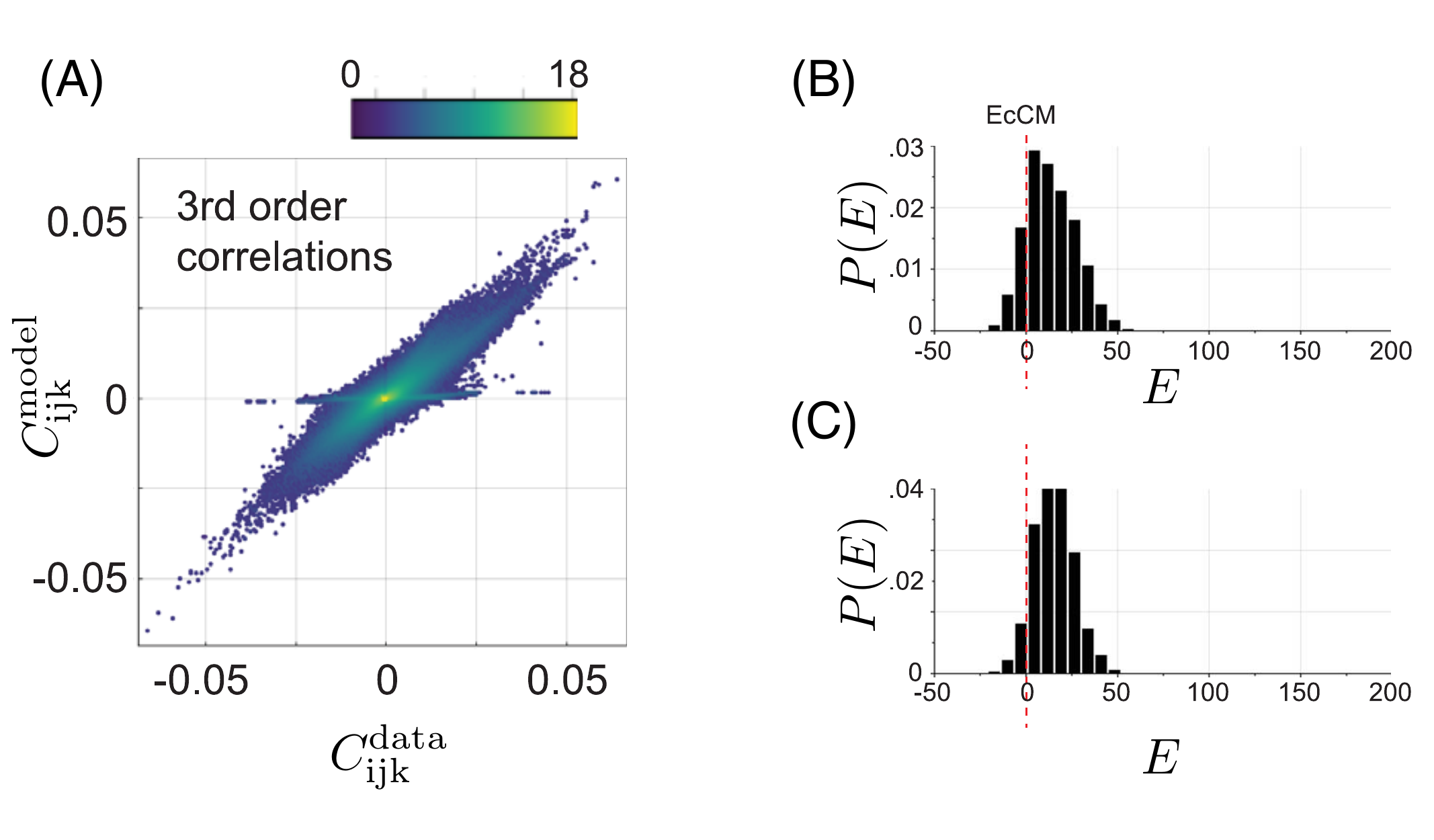}
\caption{Maximum entropy models for the ensemble of sequences in the AroQ family of chormismate mutases, enzymes involved in the synthesis of amino acids \cite{russ2020evolution}.  (A) The probability density of triplet correlations predicted by the model given the correlation observed in the data.  (B) The distribution of energies, from Eq (\ref{}), across the sequences found in the data.  Red dashed line marks the energy of a particular enzyme found in {\em E.~coli}. (C) The distribution of energies predicted by the maximum entropy model ``cooled'' to a temperature $T=0.66$.  \label{russ20}}
\end{figure}

More recent work has shown that pairwise models can capture the three--point correlations among amino acid substitutions a family of 1000+ sequences for an enzyme involved in the synthesis of amino acids, as shown in Fig \ref{russ20}A \cite{russ2020evolution}.  We emphasize that this test of the model is completely analogous to the tests in networks of neurons shown in Figs~\ref{retina_triplets} and \ref{3pt_bin}.  New sequences drawn from the effective Boltzmann distribution again are functional, and this is enhanced by ```cooling'' the distribution to lower temperature, consistent with the idea that the effective energy is a surrogate for functional behavior.  Interestingly the distribution of effective energies seen in the data (Fig~\ref{russ20}B) is closer to the distribution seen at lower temperatures (Fig~\ref{russ20}C) than at $T=1$ where the model was learned.  Even at these lower temperatures the model generates large numbers of distinct functional sequences, providing input for further design of new proteins.  

It should be noted that pairwise models for proteins are much more complicated than for neurons.  While neurons can be described well by binary variables (active/inactive, spiking/silent), each site along the amino acid chain has 20 possible amino acids, so there are $\sim 200$ elements of the matrix $J_{\rm ij}^{\alpha\beta}$ for each pair of sites $\rm ij$.  At the same time, it is not easy to find families with a number of sequences much larger than the typical number of samples in a neural recording lasting tens of minutes.  More subtly, there is a correlation structure in these samples imposed by human choices to sequences some organisms, or even particular proteins, and not others. Thus,  the literature on maximum entropy models for sequence families involves much more discussion of sampling problems than in the case of neural networks \cite{Cocco+al_2011,Cocco+al_2013a,Cocco+al_2013b}.

Perhaps the most fundamental prediction of maximum entropy models is the entropy of the sequence family itself \cite{barton+al_16}.   In addition to providing a practical guide to how many sequences will fold into structures close to some target, the entropy gives us a sense for how to locate at these particular parts of living systems on a continuum from the generic to the particular.  At one extreme we might have imagined that the interactions among amino acids are so complex that they might as well be random,\footnote{The idea that ``complex $=$ random'' was especially popular in the years immediately after the solution of the mean--field spin glass, which gave us many new tools for analyzing such random systems \cite{mezard+al_87}.} but this is wrong because random sequences typically don't fold into compact or functional proteins.  At the opposite extreme we might have imagined that every detail of the sequence matters, but this is wrong because proteins can tolerate many amino acid substitutions and remain functional.  

The explicit construction of the probability distribution for sequences in a family provides a nuanced and quantitative response to these extreme views:  proteins are not generic heteropolymers, but functionality persists in an ensemble of sequences with substantial entropy rather than being confined to particular points in sequence space.  This emphasis on the entropy of sequences associated with a single family and hence a particular structure also connects to much earlier work on global features of the sequence/structure mapping and the ``designability'' of structures
\cite{li+al_96}.  Although not usually phrased in this language, widely used descriptions of the variation in DNA sequences at protein binding sites also can be seen as maximum entropy models \cite{hippel+berg_86}.  We emphasize that writing explicit and quantitative models for the distribution of sequences is much more ambitious than the conventional use of highly simplified but tractable probabilistic models as a guide to data analysis, as in much of bioinformatics \cite{Durbin+al_1998}.

\subsection{Collective behavior}
\label{app-flocks}

At the other extreme of length scales is the use of statistical physics concepts to describe the behavior of animal groups, such as flocks of birds, schools of fish, and swarms of insects.  The qualitative phenomenology of flocks, schools, and swarms is very familiar.  These collective behaviors are dramatic, and have long been interesting to biologists because they provide a testing ground for ideas about the evolution of cooperation.  In the mid--1990s, there were efforts to write dynamical models for populations of self--propelled particles that could control their motion in relation to that of their neighbors \cite{viscek+al_95}.\footnote{Although the questions addressed were quite different, there was prior work in the computer science community on a model of ``boids'' \cite{reynolds_87}.  This in turn had precursors in the biological literature \cite{aoki_82}.} This work immediately caught the attention of the physics community in part because these models exhibited directional ordering---the emergence of a well defined direction of motion for all the ``birds'' in the flock---even in two dimensions, where this is forbidden for equilibrium systems.

The simulations of self--propelled particles have the flavor of a molecular dynamics simulation, but with microscopic entities that expend energy to keep moving on their own and with ``social'' rather than Newtonian forces.  A huge step forward was to ask whether there is a more macroscopic fluid mechanics that emerges as we coarse--grain these molecular(--ish) dynamics.  More abstractly, what is the effective field theory that describes the long distance, long time behavior of a large collection of such self--propelled particles?  The answer to this question \cite{toner+tu_95,toner+tu_98} laid the foundations for the field of active matter \cite{marchetti+al_13}.  As with the statistical mechanics of neural networks, this field now has a life independent of its origins as an effort understand real flocks and swarms.

The early models, and their field--theoretic development, captured the qualitative phenomenon of flocking.  As in other statistical physics problems, plausible local interactions lead to global ordering and the ordered state is separated from a disordered state by a phase transition.  But these theories make quantitative predictions, and at the time there were essentially no large scale, quantitative data with which to test these predictions.\footnote{There were fascinating early efforts to characterize small schools of fish in the laboratory \cite{Cullen+al_1965}.  The state of the art circa 2000 is reviewed in edited volumes \cite{Camazine+al_2001,Krause+Ruxton_2002,Parrish+Hammer_1997}.  Note that there was considerable progress in quantifying more complex collective behaviors, such as nest building by social insects, a subject as yet largely untouched by statistical physics methods.}

\begin{figure}
\includegraphics[width=\linewidth]{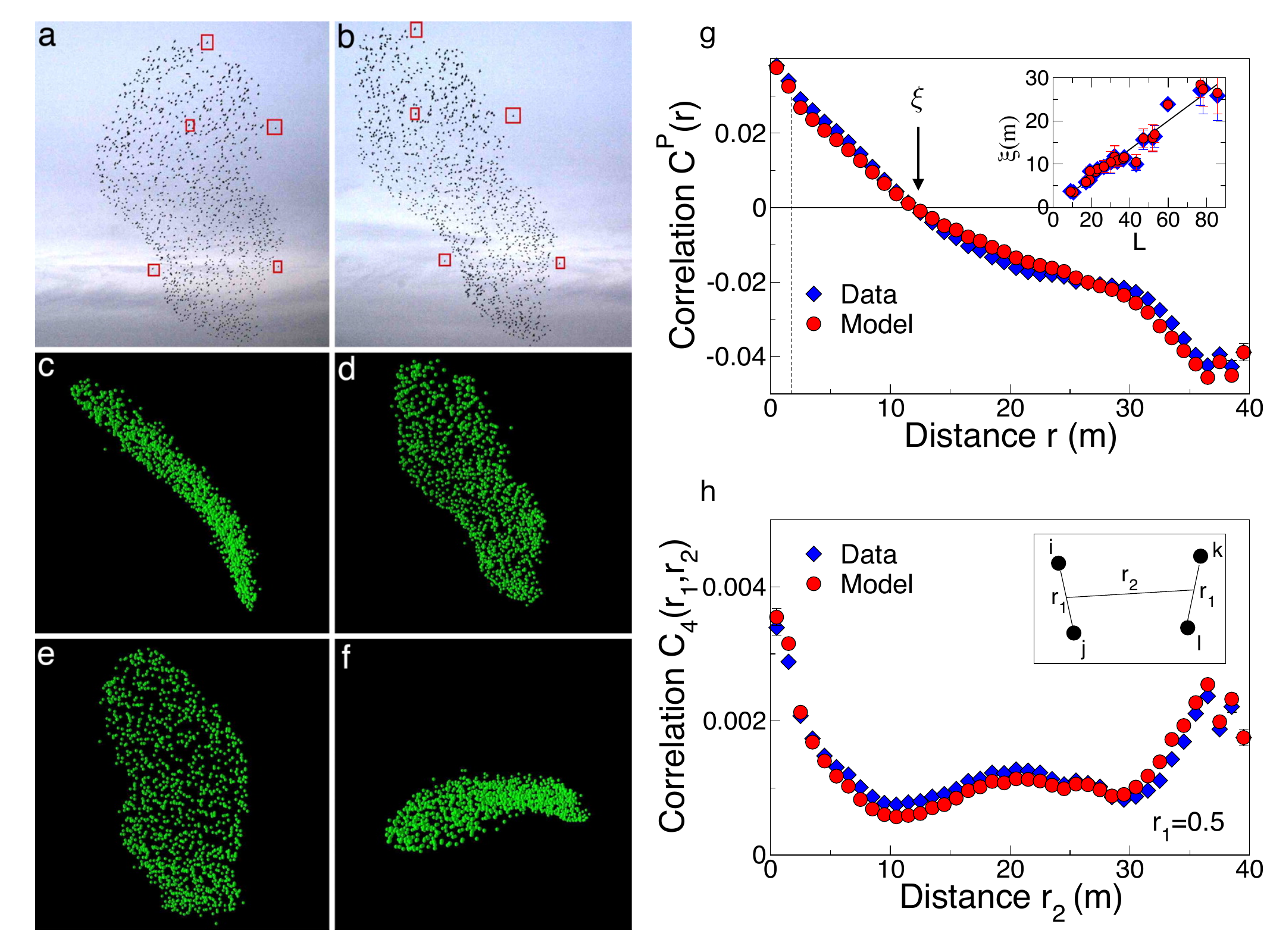}
\caption{Positions, velocities, and correlations in a flock of birds. (a, b) A flock with $N=1246$ starlings captured at a single moment in time by two cameras separated by $25\,{\rm m}$.  Red boxes highlight corresponding birds in the two images, used for reconstructing  positions in three dimensions \cite{ballerini+al2008a}. (c--f)  Three-dimensional reconstruction of the flock under four different points of view.  (g) Two--point correlations of directional fluctuations, $C^{\rm P}(r)$ from Eq (\ref{CP_def}), comparing predictions of the maximum entropy model (red) with measurements on the real flock (blue).  The typical radius of the neighborhoods ${\cal N}_{\rm i}$ is shown as a vertical dashed line.  If we define the correlation length through $C^{\rm P}(\xi) =0$, then $\xi$ is proportional to the linear size $L$ of the flock, and this also is captured by the maximum entropy models, as shown in the inset. 
(h)  Four--point correlations $C_4(r_1, r_2)$ from Eq (\ref{C4_def}), with distances defined in the inset, again comparing theory (red) and experiment (blue) \cite{bialek2012statistical}.
\label{fig-flocks}}
\end{figure}

New analysis methods made it possible to reconstruct the three--dimensional positions of every individual in large, naturally occurring animal collectives, first in flocks of thousands of birds as in Fig \ref{fig-flocks}a, b \cite{ballerini+al2008b,cavagna+al2008a,cavagna+al2008b}, and then in swarms of hundreds of insects \cite{attanasi+al_14a}.  Beyond new methods for image analysis, this work brought the conceptual framework of statistical physics to bear on the analysis of correlations in these animal groups \cite{cavagna+al_18a}.   It is particularly noteworthy that these analyses revealed precise and reproducible behaviors of animal groups out in the ``wild,'' rather than in the laboratory.

Flocks of starlings are highly polarized, but there are measurable fluctuations around this mean velocity.  Comparing flocks of different sizes and densities demonstrates that correlations among these fluctuations depend not on the physical or metric distance between birds but on the ranking of neighbors, termed ``topological distance'' \cite{ballerini+al2008a}.  Correlations between fluctuations in both direction and speed have a scale invariant form, with no characteristic length scale other than the linear dimensions of the flock itself \cite{cavagna+al_10}.  Swarms of midges are not polarized (the mean velocity is zero), but one can still see correlations in the velocity fluctuations and again these are scale invariant \cite{attanasi+al_14b}.   Analysis of events where flocks turn shows that information propagates ballistically rather than diffusively \cite{attanasi+al_14c}, and in swarms one can see dynamic scaling of the fluctuations with an exponent $z = 1.37 \pm 0.11$, far below the diffusive $z=2$ \cite{cavagna+al_17}.
None of these quantitative results agree with predictions from the original models of self--propelled particles.

Some features of the correlation structure in flocks can be captured by surprisingly simple maximum entropy models \cite{bialek2012statistical,bialek2014social}.  We start by writing the velocity of each bird $\rm i$ as $\vec{\mathbf v}_{\rm i} = v_{\rm i}\hat{\mathbf s}_{\rm i}$, where $\hat{\mathbf s}_{\rm i}$ is a unit vector pointing the flight direction and $v_{\rm i}$ is the flight speed; as a first step we focus on the flight directions and ignore the fluctuations in speed.   We expect that individual birds are orienting relative to the average of their near neighbors, and we can measure the strength of this effect through the correlation
\begin{equation}
C_{\rm local} = {1\over N}\sum_{{\rm i}=1}^N \hat{\mathbf s}_{\rm i} {\mathbf\cdot}\left( {1\over {n_c}}\sum_{{\rm j}\in {\cal N}_{\rm i}} \hat{\mathbf s}_{\rm j}\right ),
\label{eq-Clocal}
\end{equation}
where ${\cal N}_{\rm i}$ denotes the neighborhood of bird $\rm i$, and from the observations on the topological character of the correlations we take this neighborhood to include the $n_c$ nearest neighbors.  We treat all birds as equivalent,\footnote{More precisely, we treat all birds in the interior of the flock as equivalent.  The birds at the surface are special because all their neighbors are to one side of them.  In what follows we will take velocities of birds on the boundary of the flock as given, and study the response of the bulk to this boundary condition.} and so $C_{\rm local}$ is defined as an average over the flock.  Given a measured $\langle C_{\rm local}\rangle_{\rm expt}$, the maximum entropy distribution for the all the flight directions in the flock is
\begin{equation}
P\left( \{\hat{\mathbf s}_{\rm i}\}\right) = {1\over {Z(J)}} \exp\left[ J \sum_{{\rm i}=1}^N \hat{\mathbf s}_{\rm i} {\mathbf\cdot}\left( {1\over {n_c}}\sum_{{\rm j}\in {\cal N}_{\rm i}} \hat{\mathbf s}_{\rm j}\right )\right ],
\label{maxent_bird1}
\end{equation}
where the value of $J$ has to be adjusted to match $\langle C_{\rm local}\rangle$.  Because real flocks are highly polarized, all the relevant calculations can be done in a spin--wave approximation, and we can check at the end that the inferred value of $J$ is consistent with the validity of this approximation and with the measured polarization.  Finally we can find the best neighborhood size $n_c$ by maximum likelihood.  Note that once we have chosen $J$ to match the observed $\langle C_{\rm local}\rangle$, all other quantities are predicted with no free parameters.  

We can further decompose the unit vector $\hat{\mathbf s}$ into a (longitudinal) component along the mean velocity of the flock and a (two--dimensional) component $\hat{\mathbf \pi}$ perpendicular to the mean.  Then there is a natural two--point correlation function
\begin{equation}
C^{\rm P}(r) = \langle \hat{\mathbf \pi}_{\rm i} {\mathbf\cdot} \hat{\mathbf \pi}_{\rm j}\rangle_{r_{\rm ij} =r},
\label{CP_def}
\end{equation}
where the average is over all pairs of birds separated by a distance $r$.  Note that since the density of a flock is relatively uniform at a single moment in time, there is little difference between topological and metric distance in a single snapshot.  Figure \ref{fig-flocks}g compares this correlation function with the prediction from the maximum entropy model in Eq (\ref{maxent_bird1}), and we see that the agreement is excellent from the scale of the neighborhoods ${\cal N}_{\rm i}$ out to the size of the flock as a whole.  The behavior is featureless, suggesting that there is no characteristic scale; if we define a correlation length $\xi$ as the distance at which $C^{\rm P}(r)$ changes sign then $\xi$ is proportional to the size of the flock, confirming the scale invariance, and this is correctly predicted by the maximum entropy models (Fig.~\ref{fig-flocks}g, inset).  We can go even further and estimate a four--point function,  
\begin{equation}
C_4(r_1, r_2) = \langle (\hat{\mathbf \pi}_{\rm i} {\mathbf\cdot} \hat{\mathbf \pi}_{\rm j})(\hat{\mathbf \pi}_{\rm k} {\mathbf\cdot} \hat{\mathbf \pi}_{\rm l})\rangle ,
\label{C4_def}
\end{equation}
where the average is over four birds  with relative positions shown in the inset to Fig.~\ref{fig-flocks}h.  Theory and experiment again agree very well, even though these effects are quite small.

The maximum entropy model in Eq (\ref{maxent_bird1}) is equivalent to a equilibrium Heisenberg model with local interactions. Thus when $J$ is large enough to generate an ordered flock, scale--invariant fluctuations are a consequence of Goldstone's theorem.  But this theorem does not guarantee {\em quantitative} agreement with the data, as observed.  Instead of matching the average correlation of birds with their nearest $n_c$ neighbors, as in Eq (\ref{eq-Clocal}), we can try matching the correlation with the nearest neighbor, the second nearest neighbor, and so on \cite{cavagna+al_15}. Each time  we add a constraint on the ${\rm n}^{\rm th}$ neighbor we introduce a coupling $J({\rm n})$ into a generalization of Eq (\ref{maxent_bird1}), leading to
\begin{equation}
P\left( \{\hat{\mathbf s}_{\rm i}\}\right) = 
{1\over {Z(J)}} 
\exp\left[  \sum_{{\rm i}=1}^N \sum_{{\rm j}=1}^N  J(k_{\rm ij}) 
\hat{\mathbf s}_{\rm i} {\mathbf\cdot} \hat{\mathbf s}_{\rm j} 
\right ],
\label{maxent_bird2}
\end{equation}
where bird $\rm j$ is the $k_{\rm ij}^{\rm th}$ neighbor of bird $\rm i$.  The result of this exercise is that $J(n) \sim \exp(-n/n_0)$, with a range $n_0\sim 6$.  So even if we try to match the longer distance correlations explicitly, the structure of these correlations us drive the model toward short--range effective interactions.  These (few) short--range terms then are sufficient to predict the observed longer ranged and higher order correlations, quantitatively, as in Fig \ref{fig-flocks}g and h.

The equivalence to an equilibrium model might seem surprising.  The essence of the original models was that active systems generate behaviors that are not accessible in equilibrium, such as the breaking of a continuous symmetry in two dimensions.   Alternatively, in the active system dynamics generates long--ranged effective interactions in the steady state distribution. We now see explicitly that these effects are minimal in  real flocks, which we can understand because the time scales for individual birds to align with their neighbors are shorter than the time scales for neighbors to exchange places, leading to a local equilibrium \cite{mora+al_16}.  Thus in the case of flocks we not only see that the simplest maximum entropy models work, we can test explicitly that more complex models are not needed---the extra effective interactions are driven to zero by the data, and we can understand why they work.

In flocks one sees scale--invariant fluctuations not only in flight direction but also in flight speed \cite{cavagna+al_10}.  In this case there is no Goldstone theorem to help us understand the origin of this behavior.  If we try to build maximum entropy models that match the strength of local correlations, as before, the parameters of these models are driven close to a point where the correlation length diverges, and predictions match the observed long--ranged correlations \cite{bialek2014social}.  If we restrict ourselves to local models, then there is a much more general argument that the effective potential which holds individual birds' speeds near the mean must be very ``soft'' near the minimum \cite{cavgana+al_22}.  The maximum entropy approach thus suggests, strongly, that real flocks tune themselves to some non--generic point in their parameter space.  Swarms also seem to be poised at a special point in parameter space, although disordered \cite{attanasi+al_14b,cavagna+al_17}.  There is as yet no maximum entropy model for swarms, but there have been sophisticated renormalization group calculations to predict the observed dynamic scaling exponent \cite{cavagna+al_19,cavagna+al_23}.

Flocks and swarms provide a useful touchstone for thinking about networks of neurons.  In connecting theory to experiment, networks of neurons have the advantage that data sets are larger.  On the other hand, in animal groups the interactions are plausibly local and it seems reasonable to treat all individuals as equivalent;  both these considerations drive us toward a simpler set of constraints for the construction of maximum entropy models.  As with neurons, there a number of good reasons why these models of flocks might not have worked.  The detailed, quantitative successes thus encourage us to think that statistical physics approaches can provide theories of real living systems, not just metaphors that capture qualitative behaviors.

\subsection{Ecology and metabolism}

Maximum entropy methods have been used in ecology for many years, often in very simple form, searching for models that match the mean abundances of species or their energetic load on the environment \cite{harte+al_08,banavar+al_10,harte+newman_14}.  An important feature of these applications is that the chosen constraints are sums over contributions from each species, so the resulting models are non--interacting.  In the context of neural networks we have emphasized that maximum entropy provides an alternative path to connecting with statistical physics, not making assumptions about the underlying dynamics but rather pointing to particular experimental facts that we insist our models must match.  More recent work in ecology takes this point of view even further, noting that simplifying mechanistic hypotheses motivate particular quantities as being the ones that we should constrain the maximum entropy construction \cite{odwyer+al_17}. 

The project of building models for the distribution of species abundances has not yet felt the impact of dramatic improvements in the ability to measure the abundances of hundreds of species in microbial ecologies.   The most famous examples are from the bacterial communities that inhabit humans and influence our health, but there are precise measurements in marine environments \cite{ward+al_17}, in hot springs \cite{rosen+al_15,birzu+al_23}, in soils \cite{lee+al_24}, and on synthetic communities constructed in the laboratory \cite{cheng+al_22}.  In a different direction, classical ecological surveys, e.g. of trees and shrubs in a small forested region observed over several years  \cite{condit+al_14}, can be analyzed with ideas from statistical physics to discover unexpected structures \cite{villegas+al_21,villegas+al_24}.

As a final example we consider maximum entropy approaches to cellular metabolism \cite{demartino+al_18}.  There is a long tradition of abstracting from the frighteningly complex map of all the interlocking biochemical reactions  to a stochiometric matrix $S_{{\rm i}\mu}$ that connects the flux through the reaction $\rm i$ to the change in concentration of the molecule or metabolite $\mu$,
\begin{equation}
{{dc_\mu}\over{dt}} = \sum_{\rm i} S_{{\rm i}\mu}\nu_{\rm i} .
\end{equation}
If a bacterial cell is in a phase of steady state growth then ${{dc_\mu}/{dt}} =0$, defining a null space for the set of fluxes $\{\nu_{\rm i}\}$.  For example, the core bacterial metabolism of $N=86$ reactions among $M = 63$ metabolites leaves a space of $D = 23$ in which the fluxes can vary; lower and upper bounds on the fluxes mean that this space is a convex polytope.  It is sensible to take these fluxes as variables that can be controlled by the cell, since each reaction is catalyzed by an enzyme whose expression level and activity can be regulated.  In order to reproduce the cell must make a copy of itself, and this requires the synthesis of a particular combination of metabolites; plausibly then the growth rate is
\begin{equation}
\lambda \left( \{\nu_{\rm i}\}\right) = \sum_{{\rm i}=1}^N \xi_{\rm i}\nu_{\rm i} ,
\end{equation}
and the coefficients $\xi_{\rm i}$ are known, at least approximately.

\begin{figure}
\includegraphics[width=\linewidth]{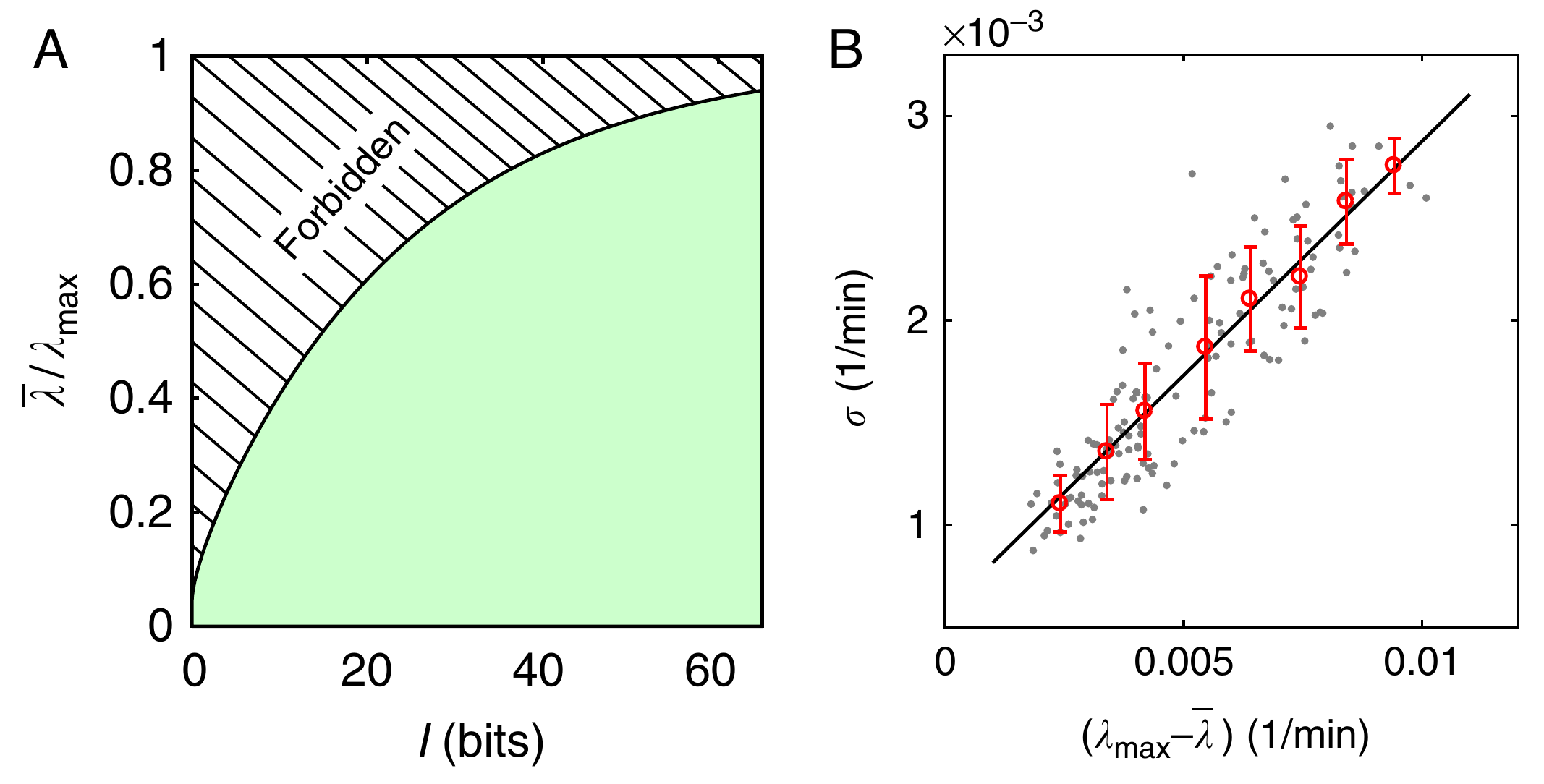}
\caption{Maximum entropy models for metabolism \cite{demartino+al_18}. (A) Achieving a particular growth rate $\lambda$ requires reducing the entropy of the joint distribution of fluxes at least by $I$ bits below the entropy of the uniform distribution (green region). Points in the hashed (forbidden) region are not achievable.
(B) Scaling of the standard deviation is growth rates with the growth rate itself. Each grey point is measured from a single lineage of cells; red points with errors are the mean and standard deviation in equally spaced bins.  The maximum entropy model predicts a linear relation, as shown by the solid line.  \label{fig-metabolism}}
\end{figure}

Being in steady state means that fluxes balance, and this ``flux balance analysis'' \cite{orth+al_10} often is supplemented by the hypothesis that fluxes are adjusted to maximize the growth rate \cite{ibarra+al_02}.  This is an extreme hypothesis, and would require infinite information to tune each flux to its optimal value.  An alternative is to ask for the ensemble of fluxes that achieves some observed mean growth rate but otherwise is as random (minimally tuned) as possible; this is the maximum entropy distribution
\begin{equation}
P\left( \{\nu_{\rm i}\}\right) = {1\over {Z(\beta )}} \exp\left[-\beta \lambda \left( \{\nu_{\rm i}\}\right)\right ].
\label{maxent-metabolism}
\end{equation}
As $\beta \rightarrow 0$ we have a completely unregulated system, allowing fluxes to vary uniformly over all allowed values.  As we increase $\beta$ the entropy in this space of fluxes is reduced and the mean growth rate increases, ultimately converging on the optimal growth rate $\lambda_{\rm max}$ as $\beta \rightarrow\infty$.  Another way of saying this is that achieving a certain mean growth rate requires specifying a certain amount of information about the fluxes relative to the unregulated, uniform distribution. This tradeoff, illustrated in Fig.~\ref{fig-metabolism}A, is an example of rate--distortion theory \cite{cover+thomas_91}.\footnote{See also the discussion in \S6.3 of \citet{bialek_12}.}

The maximum entropy model in Eq (\ref{maxent-metabolism}) has one parameter $\beta$ which must be set to match the average growth rate.  The model then predicts the mean fluxes for all the individual reactions, many of which can be measured.  Under conditions where {\em E.~coli} achieve $\sim 80\%$ of their maximal growth rate, theory and experiment agree within error bars for twenty independently measured reaction fluxes. Reaching this growth rate requires $\sim 40\,{\rm bits}$ of information about the fluxes, so that the cell must have roughly two bits of bandwidth for controlling each degree of freedom in the metabolic network. The maximum entropy model predicts that  fluxes and even the growth rate itself should be variable.  Measurements on the lineages of individual cells grown in the presence of low doses of antibiotics makes it possible to confirm the predicted dependence of the standard deviation in growth rate  on the mean, as shown in Fig.~\ref{fig-metabolism}B.  Richer behaviors are possible in models that address the spatial structure of the metabolic networks \cite{narayanankutty+al_24}.

\subsection{Coda}

In summary, maximum entropy methods have proved productive in describing emergent behaviors of biological systems on all scales, from protein molecules to ecology and from bacterial metabolism to flocks of birds.\footnote{Another natural target for this analysis is the covariation of gene expression levels in cells.  Early work used measurements averaged over many cells, but with variations across time in response to perturbations \cite{lezon+al_06}. Dramatic developments in experimental technique now make it possible to literally count almost every molecules of messenger RNA in single cells, labelled by the gene from which it was transcribed, and very recent work builds maximum entropy models to describe these data  \cite{skinner+al_24,sarra+al_24}.}  As with networks of neurons, the key idea is that these methods connect quite general statistical physics models directly to experimental data on particular systems, resulting in detailed---and often successful---quantitative predictions.   This emphasizes once more that statistical physics descriptions of living systems should not be just metaphorical. Different systems are in different regimes with respect to data set size and the complexity of the simplest plausible models, giving us the opportunity to test our algorithmic tools more extensively.  In each case we learn something about the particular system, but we also see common themes.  Notably, the parameters of the maximum entropy models that match basic facts about these systems seem to be quite non--generic. 

\section{Inference}
\label{sec-inference}

In small systems we can do an ``exact'' maximum entropy construction, but once $N$ is large we need  approximate numerical methods for solving the inverse problem.  To understand the general strategy it is useful to place the maximum entropy models into context.

From a physics point of view the maximum entropy models are special because they are the solution to a variational problem, constrained by experimental observations.  But one could also take the view that they are some interesting family of models, and we would like to fit them to the data.  Let's assume that we have made $M$ independent observations of the state $\vec{\mathbf\sigma}$, which we will index as $\vec{\mathbf\sigma}^{(n)}$, with $n = 1,\, 2,\, \cdots ,\, M$.  If our model of the probability distribution is, as in Eqs (\ref{maxent1}, \ref{maxent2}), 
\begin{equation}
P  (\vec{\mathbf\sigma}| \{\lambda_\mu\}) = {1\over Z}\exp\left[-\sum_{\mu=1}^K \lambda_\mu f_\mu(\vec{\mathbf\sigma})\right],
\end{equation}
where we note explicitly the dependence on the parameters, then the probability or likelihood of observing the data is
\begin{equation}
P \left(\{\vec{\mathbf\sigma}^{(n)}\}| \{\lambda_\mu\}\right) = {1\over{Z^M}}\exp\left[-\sum_{\mu=1}^K \lambda_\mu \sum_{n=1}^M f_\mu(\vec{\mathbf\sigma}^{(n)})\right] .
\end{equation}
A conventional strategy for estimating the parameters $\{\lambda_\mu\}$ is maximum likelihood, optimizing the probability that our model will generate the observed data.\footnote{We can also think of this as finding the model that allows us to construct the shortest code for the data \cite{cover+thomas_91,Mezard+Montanari_2009,bialek_12}.} To do this we differentiate the (log) probability with respect to each of the $\lambda_\mu$, being careful that the partition function $Z$ depends on these parameters:
\begin{equation}
{1\over M} {{\partial \ln P \left(\{\vec{\mathbf\sigma}^{(n)}\}| \{\lambda_\mu\}\right)}\over{\partial\lambda_\mu}} = - {{\partial\ln Z}\over{\partial \lambda_\mu}} - {1\over M} \sum_{n=1}^Nf_\mu(\vec{\mathbf\sigma}^{(n)}) .
\end{equation}
We have the usual identities from statistical mechanics,
\begin{equation}
{{\partial\ln Z}\over{\partial \lambda_\mu}}  = -\langle f_\mu (\vec{\mathbf\sigma}) \rangle_P ,
\end{equation}
and we recognize the average over experimental data,
\begin{equation}
{1\over M} \sum_{n=1}^Nf_\mu(\vec{\mathbf\sigma}^{(n)}) = -\langle f_\mu (\vec{\mathbf\sigma}) \rangle_{\rm expt} .
\end{equation}
Thus we have
\begin{equation}
{1\over M} {{\partial \ln P \left(\{\vec{\mathbf\sigma}^{(n)}\}| \{\lambda_\mu\}\right)}\over{\partial\lambda_\mu}} = -\left[ \langle f_\mu (\vec{\mathbf\sigma}) \rangle_{P_{\mathbf\lambda}} - \langle f_\mu (\vec{\mathbf\sigma}) \rangle_{\rm expt}\right ].
\label{gradient}
\end{equation}
This derivative vanishes, and the likelihood is maximized, when we satisfy the constraints in Eq (\ref{constraints}), matching the predicted and observed expectation values of the observable on which we choose to focus.

Equation (\ref{gradient}) tell us that the likelihood of the data is maximized when the constraints are satisfied, but it tells us more: if we adjust each $\lambda_\mu$ in proportion to the difference between the theoretical and experimental expectation values, then we are climbing the gradient in likelihood toward the point where the constraints are satisfied.   This suggests an algorithm for learning the parameters $\{\lambda_\mu\}$, or equivalently for solving the constraint Eqs (\ref{constraints}):
\begin{enumerate}
\item Choose some set of parameters $\{\lambda_\mu\}$.
\item Do a Monte Carlo simulation to generate samples from the distribution $P(\vec\sigma | \{\lambda_\mu\})$.
\item From these samples estimate the expectation values $\langle f_\mu (\vec{\mathbf\sigma}) \rangle_{P_{\mathbf\lambda}}$.
\item Update the parameters
\begin{equation}
\lambda_\mu \rightarrow \lambda_\mu - \eta\left[ \langle f_\mu (\vec{\mathbf\sigma}) \rangle_{P_{\mathbf\lambda}} - \langle f_\mu (\vec{\mathbf\sigma}) \rangle_{\rm expt}\right ],
\label{update}
\end{equation}
where $\eta$ is some small ``learning rate.''
\item Return to (2), or
\item end when constraints are satisfied within the error bars on the experimental estimates of the expectation values.
\end{enumerate}
This approach has a long history, dating back at least to \citet{ackley+al_85}.  Once the maximum entropy models for neurons were introduced, these tools were pushed quickly from $N=10$ up to $N=40$ neurons in the retina \cite{tkacik2006ising,tkacik2009spin}, and they continue to be at the core of most applications of the maximum entropy idea.

The brute force Monte Carlo methods can be improved.  One idea is to add some ``inertia'' to the updating of parameters in Eq (\ref{update}), or to allow the learning rate $\eta$ to slow with time as the algorithm gets closer to the final answer, as in simulated annealing \cite{kirkpatrick+al_83}.  More fundamentally, when the parameters change by only a small amount, one might be able to reuse the same Monte Carlo samples with new weights \cite{broderick+al_07}, as in histogram Monte Carlo \cite{ferrenberg+swendsen_88}, thus increasing efficiency. There is also some artistry involved in choosing the length of the Monte Carlo simulations to balance errors in estimating expectation values vs the efficiency of moving through parameter space, and again some sort of annealing can be useful. Many of these and other issues are described by \citet{lee+daniels_19}, who also provide Python code.

As noted above, the construction of maximum entropy models is the inverse of the usual statistical mechanics problem:  rather than being given the coupling constants and asked to compute expectation values, we are given the expectation values and asked to estimate the coupling constants.  A related  problem is to find the coupling constants at the fixed points of the renormalization group (RG), using Monte Carlo methods \cite{swendsen_84}.  The reappearance of this problem in the context of neural data analysis has led to new algorithms and the exploration of different approximations.  

If we focus on one neuron we can write the probability that it is active as a function of the state of all the other neurons; see also Eq (\ref{heff1}) below.  In the purely pairwise models, Eq (\ref{eq-E2}) with $\sigma_{\rm i} = \{0,1\}$,  this is 
\begin{equation}
P(\sigma_{\rm i} = 1 | \{\sigma_{{\rm i}\neq {\rm j}}\}) = {1\over{1 + \exp\left[- h_{\rm i}^{\rm eff}(\{\sigma_{{\rm i}\neq {\rm j}}\})\right]}},
\label{heff1}
\end{equation}
where the effective field
\begin{equation}
h_{\rm i}^{\rm eff}(\{\sigma_{{\rm i}\neq {\rm j}}\}) = h_{\rm i} + \sum_{\rm j} J_{\rm ij} \sigma_{\rm j}  ;
\end{equation}
we also have
\begin{equation}
P(\sigma_{\rm i} = 0| \{\sigma_{{\rm i}\neq {\rm j}}\}) = {1\over{1 + \exp\left[+ h_{\rm i}^{\rm eff}(\{\sigma_{{\rm i}\neq {\rm j}}\})\right]}}.
\end{equation}
We can fit  these expressions to the data in the usual way, and thus determine one row of the $J_{\rm ij}$ matrix without confronting the real difficulties of the underlying statistical mechanics problem; for this one cell the fitting problem has become a form of regression.  In the ``pseudolikelihood''' method we pretend that the fitting for each neuron is independent of all the others, so that the log probability of the data is the sum of terms from individual cells \cite{aurell+ekeberg_12}; there are interesting connections between this method and Monte Carlo RG \cite{albert+swendsen_14}.

Since the maximum entropy construction can be done exactly at $J_{\rm ij} = 0$ it is natural to ask how far we can get with perturbation theory, perhaps suitably resummed \cite{sessak+monasson_09}.   Perturbation theory is interesting both because it may provide a path to solving the inverse problem and because it can give us a sense for the strength of correlations \cite{azhar+bialek_10}.  An alternative to perturbation theory is a cluster expansion, instantiating the intuition that even in a large network interactions may be strongest among more limited groups of neurons \cite{cocco+monasson_11,cocco+monasson_12}.  For a review of these and other methods see \cite{nguyen+al_16}.

\bibliography{RMP_LM+WB.bib}

\begin{thebibliography}{339}
\expandafter\ifx\csname natexlab\endcsname\relax\def\natexlab#1{#1}\fi
\expandafter\ifx\csname bibnamefont\endcsname\relax
  \def\bibnamefont#1{#1}\fi
\expandafter\ifx\csname bibfnamefont\endcsname\relax
  \def\bibfnamefont#1{#1}\fi
\expandafter\ifx\csname citenamefont\endcsname\relax
  \def\citenamefont#1{#1}\fi
\expandafter\ifx\csname url\endcsname\relax
  \def\url#1{\texttt{#1}}\fi
\expandafter\ifx\csname urlprefix\endcsname\relax\def\urlprefix{URL }\fi
\providecommand{\bibinfo}[2]{#2}
\providecommand{\eprint}[2][]{\url{#2}}

\bibitem[{\citenamefont{Abdelfattah}
  \emph{et~al.}(2019)\citenamefont{Abdelfattah, Kawashima, Singh, Novak, Liu,
  Shuai, Huang, Campagnola, Seeman, Yu, Zheng, Grimm}
  \emph{et~al.}}]{Abdelfattah+al2019}
\bibinfo{author}{\bibnamefont{Abdelfattah}, \bibfnamefont{A.~S.}},
  \bibinfo{author}{\bibfnamefont{T.}~\bibnamefont{Kawashima}},
  \bibinfo{author}{\bibfnamefont{A.}~\bibnamefont{Singh}},
  \bibinfo{author}{\bibfnamefont{O.}~\bibnamefont{Novak}},
  \bibinfo{author}{\bibfnamefont{H.}~\bibnamefont{Liu}},
  \bibinfo{author}{\bibfnamefont{Y.}~\bibnamefont{Shuai}},
  \bibinfo{author}{\bibfnamefont{Y.-C.} \bibnamefont{Huang}},
  \bibinfo{author}{\bibfnamefont{L.}~\bibnamefont{Campagnola}},
  \bibinfo{author}{\bibfnamefont{S.~C.} \bibnamefont{Seeman}},
  \bibinfo{author}{\bibfnamefont{J.}~\bibnamefont{Yu}},
  \bibinfo{author}{\bibfnamefont{J.}~\bibnamefont{Zheng}},
  \bibinfo{author}{\bibfnamefont{J.~B.} \bibnamefont{Grimm}}, \emph{et~al.},
  \bibinfo{year}{2019}, \bibinfo{journal}{Science}
  \textbf{\bibinfo{volume}{365}}, \bibinfo{pages}{699}.

\bibitem[{\citenamefont{Ackley} \emph{et~al.}(1985)\citenamefont{Ackley,
  Hinton, and Sejnowski}}]{ackley+al_85}
\bibinfo{author}{\bibnamefont{Ackley}, \bibfnamefont{D.~H.}},
  \bibinfo{author}{\bibfnamefont{G.~E.} \bibnamefont{Hinton}}, and
  \bibinfo{author}{\bibfnamefont{T.~J.} \bibnamefont{Sejnowski}},
  \bibinfo{year}{1985}, \bibinfo{journal}{Cognitive Science}
  \textbf{\bibinfo{volume}{9}}, \bibinfo{pages}{147}.

\bibitem[{\citenamefont{Adrian}(1928)}]{Adrian1928}
\bibinfo{author}{\bibnamefont{Adrian}, \bibfnamefont{E.~D.}},
  \bibinfo{year}{1928}, \emph{\bibinfo{title}{The Basis of Sensation: The
  Action of the Sense Organs}} (\bibinfo{publisher}{W.~W. Norton, New York}).

\bibitem[{\citenamefont{Adrian and Matthews}(1934)}]{adrian+matthews_34}
\bibinfo{author}{\bibnamefont{Adrian}, \bibfnamefont{E.~D.}}, and
  \bibinfo{author}{\bibfnamefont{B.~H.~C.} \bibnamefont{Matthews}},
  \bibinfo{year}{1934}, \bibinfo{journal}{Brain}
  \textbf{\bibinfo{volume}{57.4}}, \bibinfo{pages}{355}.

\bibitem[{\citenamefont{Ahrens} \emph{et~al.}(2013)\citenamefont{Ahrens, Orger,
  Robson, Li, and Keller}}]{ahrens2013whole}
\bibinfo{author}{\bibnamefont{Ahrens}, \bibfnamefont{M.~B.}},
  \bibinfo{author}{\bibfnamefont{M.~B.} \bibnamefont{Orger}},
  \bibinfo{author}{\bibfnamefont{D.~N.} \bibnamefont{Robson}},
  \bibinfo{author}{\bibfnamefont{J.~M.} \bibnamefont{Li}}, and
  \bibinfo{author}{\bibfnamefont{P.~J.} \bibnamefont{Keller}},
  \bibinfo{year}{2013}, \bibinfo{journal}{Nature Methods}
  \textbf{\bibinfo{volume}{10}}, \bibinfo{pages}{413}.

\bibitem[{\citenamefont{Aidley}(1998)}]{Aidley1998}
\bibinfo{author}{\bibnamefont{Aidley}, \bibfnamefont{D.~J.}},
  \bibinfo{year}{1998}, \emph{\bibinfo{title}{The Physiology of Excitable
  Cells, Fourth Edition}} (\bibinfo{publisher}{Cambridge University Press,
  Cambridge}).

\bibitem[{\citenamefont{Aitchison} \emph{et~al.}(2016)\citenamefont{Aitchison,
  Corradi, and Latham}}]{aitchison+al_16}
\bibinfo{author}{\bibnamefont{Aitchison}, \bibfnamefont{L.}},
  \bibinfo{author}{\bibfnamefont{N.}~\bibnamefont{Corradi}}, and
  \bibinfo{author}{\bibfnamefont{P.~E.} \bibnamefont{Latham}},
  \bibinfo{year}{2016}, \bibinfo{journal}{PLoS Computational Biology}
  \textbf{\bibinfo{volume}{12}}, \bibinfo{pages}{e1005110}.

\bibitem[{\citenamefont{Aksay} \emph{et~al.}(2001)\citenamefont{Aksay,
  Gamkrelidze, Seung, Baker, and Tank}}]{aksay+al_01}
\bibinfo{author}{\bibnamefont{Aksay}, \bibfnamefont{E.}},
  \bibinfo{author}{\bibfnamefont{G.}~\bibnamefont{Gamkrelidze}},
  \bibinfo{author}{\bibfnamefont{H.~S.} \bibnamefont{Seung}},
  \bibinfo{author}{\bibfnamefont{R.}~\bibnamefont{Baker}}, and
  \bibinfo{author}{\bibfnamefont{D.~W.} \bibnamefont{Tank}},
  \bibinfo{year}{2001}, \bibinfo{journal}{Nature Neuroscience}
  \textbf{\bibinfo{volume}{4}}, \bibinfo{pages}{184}.

\bibitem[{\citenamefont{Albert and Swendsen}(2014)}]{albert+swendsen_14}
\bibinfo{author}{\bibnamefont{Albert}, \bibfnamefont{J.}}, and
  \bibinfo{author}{\bibfnamefont{R.~H.} \bibnamefont{Swendsen}},
  \bibinfo{year}{2014}, \bibinfo{journal}{Physics Procedia}
  \textbf{\bibinfo{volume}{57}}, \bibinfo{pages}{99}.

\bibitem[{\citenamefont{Altan-Bonnet}
  \emph{et~al.}(2020)\citenamefont{Altan-Bonnet, Mora, and
  Walczak}}]{altan-bonnet+al_20}
\bibinfo{author}{\bibnamefont{Altan-Bonnet}, \bibfnamefont{G.}},
  \bibinfo{author}{\bibfnamefont{T.}~\bibnamefont{Mora}}, and
  \bibinfo{author}{\bibfnamefont{A.~M.} \bibnamefont{Walczak}},
  \bibinfo{year}{2020}, \bibinfo{journal}{Physics Reports}
  \textbf{\bibinfo{volume}{849}}, \bibinfo{pages}{1}.

\bibitem[{\citenamefont{Amit}(1989)}]{amit_89}
\bibinfo{author}{\bibnamefont{Amit}, \bibfnamefont{D.~J.}},
  \bibinfo{year}{1989}, \emph{\bibinfo{title}{Modeling Brain Function: The
  World of Attractor Neural Networks}} (\bibinfo{publisher}{Cambridge
  University Press, Cambridge}).

\bibitem[{\citenamefont{Amit} \emph{et~al.}(1985)\citenamefont{Amit, Gutfreund,
  and Sompolinsky}}]{amit+al1985}
\bibinfo{author}{\bibnamefont{Amit}, \bibfnamefont{D.~J.}},
  \bibinfo{author}{\bibfnamefont{H.}~\bibnamefont{Gutfreund}}, and
  \bibinfo{author}{\bibfnamefont{H.}~\bibnamefont{Sompolinsky}},
  \bibinfo{year}{1985}, \bibinfo{journal}{Physical Review A}
  \textbf{\bibinfo{volume}{32}}, \bibinfo{pages}{1007}.

\bibitem[{\citenamefont{Amit} \emph{et~al.}(1987)\citenamefont{Amit, Gutfreund,
  and Sompolinsky}}]{amit+al1987}
\bibinfo{author}{\bibnamefont{Amit}, \bibfnamefont{D.~J.}},
  \bibinfo{author}{\bibfnamefont{H.}~\bibnamefont{Gutfreund}}, and
  \bibinfo{author}{\bibfnamefont{H.}~\bibnamefont{Sompolinsky}},
  \bibinfo{year}{1987}, \bibinfo{journal}{Annals of Physics}
  \textbf{\bibinfo{volume}{173}}, \bibinfo{pages}{30}.

\bibitem[{\citenamefont{Anderson}(1984)}]{anderson_84}
\bibinfo{author}{\bibnamefont{Anderson}, \bibfnamefont{P.~W.}},
  \bibinfo{year}{1984}, \emph{\bibinfo{title}{Basic Notions of Condensed Matter
  Physics}} (\bibinfo{publisher}{Benjamin/Cummings, Menlo Park CA}).

\bibitem[{\citenamefont{Aoki}(1982)}]{aoki_82}
\bibinfo{author}{\bibnamefont{Aoki}, \bibfnamefont{I.}}, \bibinfo{year}{1982},
  \bibinfo{journal}{Nippon Suisan Gakkaishi (Japanese Fisheries Academic
  Journal)} \textbf{\bibinfo{volume}{48}}, \bibinfo{pages}{1081}.

\bibitem[{\citenamefont{Attanasi}
  \emph{et~al.}(2014{\natexlab{a}})\citenamefont{Attanasi, Cavagna,
  Del~Castello, Giardina, Grigera, Jeli\'c, Melillo, Parisi, Pohl, Shen, and
  Viale}}]{attanasi+al_14c}
\bibinfo{author}{\bibnamefont{Attanasi}, \bibfnamefont{A.}},
  \bibinfo{author}{\bibfnamefont{A.}~\bibnamefont{Cavagna}},
  \bibinfo{author}{\bibfnamefont{L.}~\bibnamefont{Del~Castello}},
  \bibinfo{author}{\bibfnamefont{I.}~\bibnamefont{Giardina}},
  \bibinfo{author}{\bibfnamefont{T.~S.} \bibnamefont{Grigera}},
  \bibinfo{author}{\bibfnamefont{A.}~\bibnamefont{Jeli\'c}},
  \bibinfo{author}{\bibfnamefont{S.}~\bibnamefont{Melillo}},
  \bibinfo{author}{\bibfnamefont{L.}~\bibnamefont{Parisi}},
  \bibinfo{author}{\bibfnamefont{O.}~\bibnamefont{Pohl}},
  \bibinfo{author}{\bibfnamefont{E.}~\bibnamefont{Shen}}, and
  \bibinfo{author}{\bibfnamefont{M.}~\bibnamefont{Viale}},
  \bibinfo{year}{2014}{\natexlab{a}}, \bibinfo{journal}{Nature Physics}
  \textbf{\bibinfo{volume}{10}}, \bibinfo{pages}{691}.

\bibitem[{\citenamefont{Attanasi}
  \emph{et~al.}(2014{\natexlab{b}})\citenamefont{Attanasi, Cavagna,
  Del~Castello, Giardina, Melillo, Parisi, Pohl, Rossaro, Shen, Silvestri, and
  Viale}}]{attanasi+al_14a}
\bibinfo{author}{\bibnamefont{Attanasi}, \bibfnamefont{A.}},
  \bibinfo{author}{\bibfnamefont{A.}~\bibnamefont{Cavagna}},
  \bibinfo{author}{\bibfnamefont{L.}~\bibnamefont{Del~Castello}},
  \bibinfo{author}{\bibfnamefont{I.}~\bibnamefont{Giardina}},
  \bibinfo{author}{\bibfnamefont{S.}~\bibnamefont{Melillo}},
  \bibinfo{author}{\bibfnamefont{L.}~\bibnamefont{Parisi}},
  \bibinfo{author}{\bibfnamefont{O.}~\bibnamefont{Pohl}},
  \bibinfo{author}{\bibfnamefont{B.}~\bibnamefont{Rossaro}},
  \bibinfo{author}{\bibfnamefont{E.}~\bibnamefont{Shen}},
  \bibinfo{author}{\bibfnamefont{E.}~\bibnamefont{Silvestri}}, and
  \bibinfo{author}{\bibfnamefont{M.}~\bibnamefont{Viale}},
  \bibinfo{year}{2014}{\natexlab{b}}, \bibinfo{journal}{PLoS Computational
  Biology} \textbf{\bibinfo{volume}{10}}, \bibinfo{pages}{e1003697}.

\bibitem[{\citenamefont{Attanasi}
  \emph{et~al.}(2014{\natexlab{c}})\citenamefont{Attanasi, Cavagna,
  Del~Castello, Giardina, Melillo, Parisi, Pohl, Rossaro, Shen, Silvestri, and
  Viale}}]{attanasi+al_14b}
\bibinfo{author}{\bibnamefont{Attanasi}, \bibfnamefont{A.}},
  \bibinfo{author}{\bibfnamefont{A.}~\bibnamefont{Cavagna}},
  \bibinfo{author}{\bibfnamefont{L.}~\bibnamefont{Del~Castello}},
  \bibinfo{author}{\bibfnamefont{I.}~\bibnamefont{Giardina}},
  \bibinfo{author}{\bibfnamefont{S.}~\bibnamefont{Melillo}},
  \bibinfo{author}{\bibfnamefont{L.}~\bibnamefont{Parisi}},
  \bibinfo{author}{\bibfnamefont{O.}~\bibnamefont{Pohl}},
  \bibinfo{author}{\bibfnamefont{B.}~\bibnamefont{Rossaro}},
  \bibinfo{author}{\bibfnamefont{E.}~\bibnamefont{Shen}},
  \bibinfo{author}{\bibfnamefont{E.}~\bibnamefont{Silvestri}}, and
  \bibinfo{author}{\bibfnamefont{M.}~\bibnamefont{Viale}},
  \bibinfo{year}{2014}{\natexlab{c}}, \bibinfo{journal}{Physical Review
  Letters} \textbf{\bibinfo{volume}{113}}, \bibinfo{pages}{238102}.

\bibitem[{\citenamefont{Aurell and Ekeberg}(2012)}]{aurell+ekeberg_12}
\bibinfo{author}{\bibnamefont{Aurell}, \bibfnamefont{E.}}, and
  \bibinfo{author}{\bibfnamefont{M.}~\bibnamefont{Ekeberg}},
  \bibinfo{year}{2012}, \bibinfo{journal}{Physical Review Letters}
  \textbf{\bibinfo{volume}{108}}, \bibinfo{pages}{090201}.

\bibitem[{\citenamefont{Azhar and Bialek}(2010)}]{azhar+bialek_10}
\bibinfo{author}{\bibnamefont{Azhar}, \bibfnamefont{F.}}, and
  \bibinfo{author}{\bibfnamefont{W.}~\bibnamefont{Bialek}},
  \bibinfo{year}{2010}, \eprint{arXiv:1012.5987}.

\bibitem[{\citenamefont{Bak}(1996)}]{bak_96}
\bibinfo{author}{\bibnamefont{Bak}, \bibfnamefont{P.}}, \bibinfo{year}{1996},
  \emph{\bibinfo{title}{How Nature Works: The Science of Self--Organized
  Criticality}} (\bibinfo{publisher}{Copernicus, New York}).

\bibitem[{\citenamefont{Bak} \emph{et~al.}(1987)\citenamefont{Bak, Tang, and
  Wiesenfeld}}]{bak+al_87}
\bibinfo{author}{\bibnamefont{Bak}, \bibfnamefont{P.}},
  \bibinfo{author}{\bibfnamefont{C.}~\bibnamefont{Tang}}, and
  \bibinfo{author}{\bibfnamefont{K.}~\bibnamefont{Wiesenfeld}},
  \bibinfo{year}{1987}, \bibinfo{journal}{Physical Review Letters}
  \textbf{\bibinfo{volume}{59}}, \bibinfo{pages}{381}.

\bibitem[{\citenamefont{Ballerini}
  \emph{et~al.}(2008{\natexlab{a}})\citenamefont{Ballerini, Cabibbo, Candelier,
  Cavagna, Cisbani, Giardina, Orlandi, Parisi, Procaccini, Viale, and
  Zdravkovic}}]{ballerini+al2008b}
\bibinfo{author}{\bibnamefont{Ballerini}, \bibfnamefont{M.}},
  \bibinfo{author}{\bibfnamefont{N.}~\bibnamefont{Cabibbo}},
  \bibinfo{author}{\bibfnamefont{R.}~\bibnamefont{Candelier}},
  \bibinfo{author}{\bibfnamefont{A.}~\bibnamefont{Cavagna}},
  \bibinfo{author}{\bibfnamefont{E.}~\bibnamefont{Cisbani}},
  \bibinfo{author}{\bibfnamefont{I.}~\bibnamefont{Giardina}},
  \bibinfo{author}{\bibfnamefont{A.}~\bibnamefont{Orlandi}},
  \bibinfo{author}{\bibfnamefont{G.}~\bibnamefont{Parisi}},
  \bibinfo{author}{\bibfnamefont{A.}~\bibnamefont{Procaccini}},
  \bibinfo{author}{\bibfnamefont{M.}~\bibnamefont{Viale}}, and
  \bibinfo{author}{\bibfnamefont{V.}~\bibnamefont{Zdravkovic}},
  \bibinfo{year}{2008}{\natexlab{a}}, \bibinfo{journal}{Animal Behaviour}
  \textbf{\bibinfo{volume}{76}}, \bibinfo{pages}{201}.

\bibitem[{\citenamefont{Ballerini}
  \emph{et~al.}(2008{\natexlab{b}})\citenamefont{Ballerini, Cabibbo, Candelier,
  Cavagna, Cisbani, Giardina, Orlandi, Parisi, Procaccini, Viale, and
  Zdravkovic}}]{ballerini+al2008a}
\bibinfo{author}{\bibnamefont{Ballerini}, \bibfnamefont{M.}},
  \bibinfo{author}{\bibfnamefont{N.}~\bibnamefont{Cabibbo}},
  \bibinfo{author}{\bibfnamefont{R.}~\bibnamefont{Candelier}},
  \bibinfo{author}{\bibfnamefont{A.}~\bibnamefont{Cavagna}},
  \bibinfo{author}{\bibfnamefont{E.}~\bibnamefont{Cisbani}},
  \bibinfo{author}{\bibfnamefont{I.}~\bibnamefont{Giardina}},
  \bibinfo{author}{\bibfnamefont{A.}~\bibnamefont{Orlandi}},
  \bibinfo{author}{\bibfnamefont{G.}~\bibnamefont{Parisi}},
  \bibinfo{author}{\bibfnamefont{A.}~\bibnamefont{Procaccini}},
  \bibinfo{author}{\bibfnamefont{M.}~\bibnamefont{Viale}}, and
  \bibinfo{author}{\bibfnamefont{V.}~\bibnamefont{Zdravkovic}},
  \bibinfo{year}{2008}{\natexlab{b}}, \bibinfo{journal}{Proceedings of the
  National Academy of Sciences (USA)} \textbf{\bibinfo{volume}{105}},
  \bibinfo{pages}{1232}.

\bibitem[{\citenamefont{Banavar} \emph{et~al.}(2010)\citenamefont{Banavar,
  Maritan, and Volkov}}]{banavar+al_10}
\bibinfo{author}{\bibnamefont{Banavar}, \bibfnamefont{J.~R.}},
  \bibinfo{author}{\bibfnamefont{A.}~\bibnamefont{Maritan}}, and
  \bibinfo{author}{\bibfnamefont{I.}~\bibnamefont{Volkov}},
  \bibinfo{year}{2010}, \bibinfo{journal}{Journal of Physics: Condensed Matter}
  \textbf{\bibinfo{volume}{22}}, \bibinfo{pages}{063101}.

\bibitem[{\citenamefont{Barson} \emph{et~al.}(2020)\citenamefont{Barson,
  Hamodi, Shen, Lur, Constable, Cardin, Crair, and
  Higley}}]{barson2020simultaneous}
\bibinfo{author}{\bibnamefont{Barson}, \bibfnamefont{D.}},
  \bibinfo{author}{\bibfnamefont{A.~S.} \bibnamefont{Hamodi}},
  \bibinfo{author}{\bibfnamefont{X.}~\bibnamefont{Shen}},
  \bibinfo{author}{\bibfnamefont{G.}~\bibnamefont{Lur}},
  \bibinfo{author}{\bibfnamefont{R.~T.} \bibnamefont{Constable}},
  \bibinfo{author}{\bibfnamefont{J.~A.} \bibnamefont{Cardin}},
  \bibinfo{author}{\bibfnamefont{M.~C.} \bibnamefont{Crair}}, and
  \bibinfo{author}{\bibfnamefont{M.~J.} \bibnamefont{Higley}},
  \bibinfo{year}{2020}, \bibinfo{journal}{Nature Methods}
  \textbf{\bibinfo{volume}{17}}, \bibinfo{pages}{107}.

\bibitem[{\citenamefont{Barton} \emph{et~al.}(2016)\citenamefont{Barton,
  Chakraborty, Cocco, Jacquin, and Monasson}}]{barton+al_16}
\bibinfo{author}{\bibnamefont{Barton}, \bibfnamefont{J.~P.}},
  \bibinfo{author}{\bibfnamefont{A.~K.} \bibnamefont{Chakraborty}},
  \bibinfo{author}{\bibfnamefont{S.}~\bibnamefont{Cocco}},
  \bibinfo{author}{\bibfnamefont{H.}~\bibnamefont{Jacquin}}, and
  \bibinfo{author}{\bibfnamefont{R.}~\bibnamefont{Monasson}},
  \bibinfo{year}{2016}, \bibinfo{journal}{Journal of Statistical Physics}
  \textbf{\bibinfo{volume}{162}}, \bibinfo{pages}{1267}.

\bibitem[{\citenamefont{Beggs and Plenz}(2003)}]{beggs+plenz_03}
\bibinfo{author}{\bibnamefont{Beggs}, \bibfnamefont{J.~M.}}, and
  \bibinfo{author}{\bibfnamefont{D.}~\bibnamefont{Plenz}},
  \bibinfo{year}{2003}, \bibinfo{journal}{Journal of Neuroscience}
  \textbf{\bibinfo{volume}{23}}, \bibinfo{pages}{11167}.

\bibitem[{\citenamefont{Ben-Yishai}
  \emph{et~al.}(1995)\citenamefont{Ben-Yishai, Bar-Or, and
  Sompolinsky}}]{ben-yishai+al_95}
\bibinfo{author}{\bibnamefont{Ben-Yishai}, \bibfnamefont{R.}},
  \bibinfo{author}{\bibfnamefont{R.~L.} \bibnamefont{Bar-Or}}, and
  \bibinfo{author}{\bibfnamefont{H.}~\bibnamefont{Sompolinsky}},
  \bibinfo{year}{1995}, \bibinfo{journal}{Proceedings of the National Academy
  of Sciences (USA)} \textbf{\bibinfo{volume}{92}}, \bibinfo{pages}{3844}.

\bibitem[{\citenamefont{Berman} \emph{et~al.}(2016)\citenamefont{Berman,
  Bialek, and Shaevitz}}]{berman+al_16}
\bibinfo{author}{\bibnamefont{Berman}, \bibfnamefont{G.~J.}},
  \bibinfo{author}{\bibfnamefont{W.}~\bibnamefont{Bialek}}, and
  \bibinfo{author}{\bibfnamefont{J.~W.} \bibnamefont{Shaevitz}},
  \bibinfo{year}{2016}, \bibinfo{journal}{Proceedings of the National Academy
  of Sciences (USA)} \textbf{\bibinfo{volume}{113}}, \bibinfo{pages}{11943}.

\bibitem[{\citenamefont{Berman} \emph{et~al.}(2014)\citenamefont{Berman, Choi,
  Bialek, and Shaevitz}}]{berman+al_14}
\bibinfo{author}{\bibnamefont{Berman}, \bibfnamefont{G.~J.}},
  \bibinfo{author}{\bibfnamefont{D.~M.} \bibnamefont{Choi}},
  \bibinfo{author}{\bibfnamefont{W.}~\bibnamefont{Bialek}}, and
  \bibinfo{author}{\bibfnamefont{J.~W.} \bibnamefont{Shaevitz}},
  \bibinfo{year}{2014}, \bibinfo{journal}{Journal of The Royal Society
  Interface} \textbf{\bibinfo{volume}{11}}, \bibinfo{pages}{20140672}.

\bibitem[{\citenamefont{Bertschinger and
  Natschl\"ager}(2004)}]{bertschinger+natschlager_04}
\bibinfo{author}{\bibnamefont{Bertschinger}, \bibfnamefont{N.}}, and
  \bibinfo{author}{\bibfnamefont{T.}~\bibnamefont{Natschl\"ager}},
  \bibinfo{year}{2004}, \bibinfo{journal}{Neural Computation}
  \textbf{\bibinfo{volume}{16}}, \bibinfo{pages}{1413}.

\bibitem[{\citenamefont{Bialek}(2012)}]{bialek_12}
\bibinfo{author}{\bibnamefont{Bialek}, \bibfnamefont{W.}},
  \bibinfo{year}{2012}, \emph{\bibinfo{title}{Biophysics: Searching for
  Principles}} (\bibinfo{publisher}{Princeton University Press, Princeton}).

\bibitem[{\citenamefont{Bialek}(2024)}]{Bialek_24}
\bibinfo{author}{\bibnamefont{Bialek}, \bibfnamefont{W.}},
  \bibinfo{year}{2024}, in \emph{\bibinfo{booktitle}{Les Houches Summer School
  Lecture Notes: Theoretical Biological Physics 2023. SciPost Physics Lecture
  Notes 84}}, edited by \bibinfo{editor}{\bibfnamefont{A.-F.}
  \bibnamefont{Bitbol}},
  \bibinfo{editor}{\bibfnamefont{T.}~\bibnamefont{Mora}},
  \bibinfo{editor}{\bibfnamefont{I.}~\bibnamefont{Nemenman}}, and
  \bibinfo{editor}{\bibfnamefont{A.~M.} \bibnamefont{Walczak}}
  (\bibinfo{publisher}{SciPost Foundation, Amsterdam}).

\bibitem[{\citenamefont{Bialek} \emph{et~al.}(2014)\citenamefont{Bialek,
  Cavagna, Giardina, Mora, Pohl, Silvestri, Viale, and
  Walczak}}]{bialek2014social}
\bibinfo{author}{\bibnamefont{Bialek}, \bibfnamefont{W.}},
  \bibinfo{author}{\bibfnamefont{A.}~\bibnamefont{Cavagna}},
  \bibinfo{author}{\bibfnamefont{I.}~\bibnamefont{Giardina}},
  \bibinfo{author}{\bibfnamefont{T.}~\bibnamefont{Mora}},
  \bibinfo{author}{\bibfnamefont{O.}~\bibnamefont{Pohl}},
  \bibinfo{author}{\bibfnamefont{E.}~\bibnamefont{Silvestri}},
  \bibinfo{author}{\bibfnamefont{M.}~\bibnamefont{Viale}}, and
  \bibinfo{author}{\bibfnamefont{A.~M.} \bibnamefont{Walczak}},
  \bibinfo{year}{2014}, \bibinfo{journal}{Proceedings of the National Academy
  of Sciences (USA)} \textbf{\bibinfo{volume}{111}}, \bibinfo{pages}{7212}.

\bibitem[{\citenamefont{Bialek} \emph{et~al.}(2012)\citenamefont{Bialek,
  Cavagna, Giardina, Mora, Silvestri, Viale, and
  Walczak}}]{bialek2012statistical}
\bibinfo{author}{\bibnamefont{Bialek}, \bibfnamefont{W.}},
  \bibinfo{author}{\bibfnamefont{A.}~\bibnamefont{Cavagna}},
  \bibinfo{author}{\bibfnamefont{I.}~\bibnamefont{Giardina}},
  \bibinfo{author}{\bibfnamefont{T.}~\bibnamefont{Mora}},
  \bibinfo{author}{\bibfnamefont{E.}~\bibnamefont{Silvestri}},
  \bibinfo{author}{\bibfnamefont{M.}~\bibnamefont{Viale}}, and
  \bibinfo{author}{\bibfnamefont{A.~M.} \bibnamefont{Walczak}},
  \bibinfo{year}{2012}, \bibinfo{journal}{Proceedings of the National Academy
  of Sciences (USA)} \textbf{\bibinfo{volume}{109}}, \bibinfo{pages}{4786}.

\bibitem[{\citenamefont{Bialek} \emph{et~al.}(2020)\citenamefont{Bialek,
  Palmer, and Schwab}}]{bialek+al_20}
\bibinfo{author}{\bibnamefont{Bialek}, \bibfnamefont{W.}},
  \bibinfo{author}{\bibfnamefont{S.~E.} \bibnamefont{Palmer}}, and
  \bibinfo{author}{\bibfnamefont{D.~J.} \bibnamefont{Schwab}},
  \bibinfo{year}{2020}, \eprint{arXiv:2008.12279}.

\bibitem[{\citenamefont{Bialek and Ranganathan}(2007)}]{bialek+ranganathan_07}
\bibinfo{author}{\bibnamefont{Bialek}, \bibfnamefont{W.}}, and
  \bibinfo{author}{\bibfnamefont{R.}~\bibnamefont{Ranganathan}},
  \bibinfo{year}{2007}, \eprint{arXiv:0712.4397}.

\bibitem[{\citenamefont{Bialek and Shaevitz}(2024)}]{bialek+shaevitz_24}
\bibinfo{author}{\bibnamefont{Bialek}, \bibfnamefont{W.}}, and
  \bibinfo{author}{\bibfnamefont{J.~W.} \bibnamefont{Shaevitz}},
  \bibinfo{year}{2024}, \bibinfo{journal}{Physical Review Letters}
  \textbf{\bibinfo{volume}{132}}, \bibinfo{pages}{048401}.

\bibitem[{\citenamefont{Binder}(1981)}]{binder_81}
\bibinfo{author}{\bibnamefont{Binder}, \bibfnamefont{K.}},
  \bibinfo{year}{1981}, \bibinfo{journal}{Zeitschrift f\"ur Physik B: Condensed
  Matter} \textbf{\bibinfo{volume}{43}}, \bibinfo{pages}{119}.

\bibitem[{\citenamefont{Birzu} \emph{et~al.}(2023)\citenamefont{Birzu,
  Muralidharan, Goudeau, Malmstrom, Fisher, and Bhaya}}]{birzu+al_23}
\bibinfo{author}{\bibnamefont{Birzu}, \bibfnamefont{G.}},
  \bibinfo{author}{\bibfnamefont{S.~H.} \bibnamefont{Muralidharan}},
  \bibinfo{author}{\bibfnamefont{D.}~\bibnamefont{Goudeau}},
  \bibinfo{author}{\bibfnamefont{R.~R.} \bibnamefont{Malmstrom}},
  \bibinfo{author}{\bibfnamefont{D.~S.} \bibnamefont{Fisher}}, and
  \bibinfo{author}{\bibfnamefont{D.}~\bibnamefont{Bhaya}},
  \bibinfo{year}{2023}, \bibinfo{journal}{eLife} \textbf{\bibinfo{volume}{12}},
  \bibinfo{pages}{RP90849}.

\bibitem[{\citenamefont{Bitbol} \emph{et~al.}(2016)\citenamefont{Bitbol, Dwyer,
  Colwell, and Wingreen}}]{bitbol2016inferring}
\bibinfo{author}{\bibnamefont{Bitbol}, \bibfnamefont{A.-F.}},
  \bibinfo{author}{\bibfnamefont{R.~S.} \bibnamefont{Dwyer}},
  \bibinfo{author}{\bibfnamefont{L.~J.} \bibnamefont{Colwell}}, and
  \bibinfo{author}{\bibfnamefont{N.~S.} \bibnamefont{Wingreen}},
  \bibinfo{year}{2016}, \bibinfo{journal}{Proceedings of the National Academy
  of Sciences (USA)} \textbf{\bibinfo{volume}{113}}, \bibinfo{pages}{12180}.

\bibitem[{\citenamefont{Bliss and L{\o}mo}(1973)}]{bliss+lomo1973}
\bibinfo{author}{\bibnamefont{Bliss}, \bibfnamefont{T.~V.~P.}}, and
  \bibinfo{author}{\bibfnamefont{T.}~\bibnamefont{L{\o}mo}},
  \bibinfo{year}{1973}, \bibinfo{journal}{Journal of Physiology (London)}
  \textbf{\bibinfo{volume}{232}}, \bibinfo{pages}{331}.

\bibitem[{\citenamefont{Block}(1962)}]{block1962}
\bibinfo{author}{\bibnamefont{Block}, \bibfnamefont{H.}}, \bibinfo{year}{1962},
  \bibinfo{journal}{Reviews of Modern Physics} \textbf{\bibinfo{volume}{34}},
  \bibinfo{pages}{123}.

\bibitem[{\citenamefont{Block} \emph{et~al.}(1962)\citenamefont{Block,
  Knight~Jr, and Rosenblatt}}]{block+al1962}
\bibinfo{author}{\bibnamefont{Block}, \bibfnamefont{H.}},
  \bibinfo{author}{\bibfnamefont{B.~W.} \bibnamefont{Knight~Jr}}, and
  \bibinfo{author}{\bibfnamefont{F.}~\bibnamefont{Rosenblatt}},
  \bibinfo{year}{1962}, \bibinfo{journal}{Reviews of Modern Physics}
  \textbf{\bibinfo{volume}{34}}, \bibinfo{pages}{135}.

\bibitem[{\citenamefont{Bradde and Bialek}(2017)}]{bradde+bialek2017}
\bibinfo{author}{\bibnamefont{Bradde}, \bibfnamefont{S.}}, and
  \bibinfo{author}{\bibfnamefont{W.}~\bibnamefont{Bialek}},
  \bibinfo{year}{2017}, \bibinfo{journal}{Journal of Statistical Physics}
  \textbf{\bibinfo{volume}{167}}, \bibinfo{pages}{462}.

\bibitem[{\citenamefont{Braun} \emph{et~al.}(1984)\citenamefont{Braun, Abney,
  and Owicki}}]{braun+al1984}
\bibinfo{author}{\bibnamefont{Braun}, \bibfnamefont{J.}},
  \bibinfo{author}{\bibfnamefont{J.~R.} \bibnamefont{Abney}}, and
  \bibinfo{author}{\bibfnamefont{J.~C.} \bibnamefont{Owicki}},
  \bibinfo{year}{1984}, \bibinfo{journal}{Nature}
  \textbf{\bibinfo{volume}{310}}, \bibinfo{pages}{316}.

\bibitem[{\citenamefont{Brenner}(1957)}]{brenner_57}
\bibinfo{author}{\bibnamefont{Brenner}, \bibfnamefont{S.}},
  \bibinfo{year}{1957}, \bibinfo{journal}{Proceedings of the National Academy
  of Sciences (USA)} \textbf{\bibinfo{volume}{43}}, \bibinfo{pages}{687}.

\bibitem[{\citenamefont{Brinkman}(2023)}]{brinkman2023}
\bibinfo{author}{\bibnamefont{Brinkman}, \bibfnamefont{B.~A.~W.}},
  \bibinfo{year}{2023}, \eprint{arXiv:2301.09600}.

\bibitem[{\citenamefont{Broderick} \emph{et~al.}(2007)\citenamefont{Broderick,
  Dud\'ik, Tka\v{c}ik, Schapire, and Bialek}}]{broderick+al_07}
\bibinfo{author}{\bibnamefont{Broderick}, \bibfnamefont{T.}},
  \bibinfo{author}{\bibfnamefont{M.}~\bibnamefont{Dud\'ik}},
  \bibinfo{author}{\bibfnamefont{G.}~\bibnamefont{Tka\v{c}ik}},
  \bibinfo{author}{\bibfnamefont{R.~E.} \bibnamefont{Schapire}}, and
  \bibinfo{author}{\bibfnamefont{W.}~\bibnamefont{Bialek}},
  \bibinfo{year}{2007}, \eprint{arXiv:0712.2437}.

\bibitem[{\citenamefont{Bustamante}
  \emph{et~al.}(1994)\citenamefont{Bustamante, Marko, Siggia, and
  Smith}}]{bustamante+al_94}
\bibinfo{author}{\bibnamefont{Bustamante}, \bibfnamefont{C.}},
  \bibinfo{author}{\bibfnamefont{J.~F.} \bibnamefont{Marko}},
  \bibinfo{author}{\bibfnamefont{E.~D.} \bibnamefont{Siggia}}, and
  \bibinfo{author}{\bibfnamefont{S.}~\bibnamefont{Smith}},
  \bibinfo{year}{1994}, \bibinfo{journal}{Science}
  \textbf{\bibinfo{volume}{265}}, \bibinfo{pages}{1599}.

\bibitem[{\citenamefont{Ram\'on~y Cajal}(1893)}]{Cajal1893}
\bibinfo{author}{\bibnamefont{Ram\'on~y Cajal}, \bibfnamefont{S.}},
  \bibinfo{year}{1893}, \bibinfo{journal}{La Cellule}
  \textbf{\bibinfo{volume}{9}}, \bibinfo{pages}{17}.

\bibitem[{\citenamefont{Ram\'on~y Cajal}(1894)}]{Cajal1894}
\bibinfo{author}{\bibnamefont{Ram\'on~y Cajal}, \bibfnamefont{S.}},
  \bibinfo{year}{1894}, \bibinfo{journal}{Proceedings of the Royal Society of
  London} \textbf{\bibinfo{volume}{55}}, \bibinfo{pages}{444}.

\bibitem[{\citenamefont{Camazine} \emph{et~al.}(2001)\citenamefont{Camazine,
  Deneubourg, Francks, Sneyd, Theraulaz, and Bonabeau}}]{Camazine+al_2001}
\bibinfo{editor}{\bibnamefont{Camazine}, \bibfnamefont{S.}},
  \bibinfo{editor}{\bibfnamefont{J.-L.} \bibnamefont{Deneubourg}},
  \bibinfo{editor}{\bibfnamefont{N.~R.} \bibnamefont{Francks}},
  \bibinfo{editor}{\bibfnamefont{J.}~\bibnamefont{Sneyd}},
  \bibinfo{editor}{\bibfnamefont{G.}~\bibnamefont{Theraulaz}}, and
  \bibinfo{editor}{\bibfnamefont{E.}~\bibnamefont{Bonabeau}} (eds.),
  \bibinfo{year}{2001}, \emph{\bibinfo{title}{Self--Organization in Biological
  Systems}} (\bibinfo{publisher}{Princeton University Press, Princeton NJ}).

\bibitem[{\citenamefont{Carleo} \emph{et~al.}(2019)\citenamefont{Carleo, Cirac,
  Cranmer, Daudet, Schuld, Tishby, Vogt-Maranto, and
  Zdeborov\'a}}]{carleo+al_19}
\bibinfo{author}{\bibnamefont{Carleo}, \bibfnamefont{G.}},
  \bibinfo{author}{\bibfnamefont{I.}~\bibnamefont{Cirac}},
  \bibinfo{author}{\bibfnamefont{K.}~\bibnamefont{Cranmer}},
  \bibinfo{author}{\bibfnamefont{L.}~\bibnamefont{Daudet}},
  \bibinfo{author}{\bibfnamefont{M.}~\bibnamefont{Schuld}},
  \bibinfo{author}{\bibfnamefont{N.}~\bibnamefont{Tishby}},
  \bibinfo{author}{\bibfnamefont{L.}~\bibnamefont{Vogt-Maranto}}, and
  \bibinfo{author}{\bibfnamefont{L.}~\bibnamefont{Zdeborov\'a}},
  \bibinfo{year}{2019}, \bibinfo{journal}{Reviews of Modern Physics}
  \textbf{\bibinfo{volume}{91}}, \bibinfo{pages}{045002}.

\bibitem[{\citenamefont{Carmena} \emph{et~al.}(2003)\citenamefont{Carmena,
  Lebedev, Crist, O'Doherty, Santucci, Dimitrov, Patil, Henriquez, and
  Nicolelis}}]{carmena+al_03}
\bibinfo{author}{\bibnamefont{Carmena}, \bibfnamefont{J.~M.}},
  \bibinfo{author}{\bibfnamefont{M.~A.} \bibnamefont{Lebedev}},
  \bibinfo{author}{\bibfnamefont{R.~E.} \bibnamefont{Crist}},
  \bibinfo{author}{\bibfnamefont{J.~E.} \bibnamefont{O'Doherty}},
  \bibinfo{author}{\bibfnamefont{D.~M.} \bibnamefont{Santucci}},
  \bibinfo{author}{\bibfnamefont{D.~F.} \bibnamefont{Dimitrov}},
  \bibinfo{author}{\bibfnamefont{P.~G.} \bibnamefont{Patil}},
  \bibinfo{author}{\bibfnamefont{C.~S.} \bibnamefont{Henriquez}}, and
  \bibinfo{author}{\bibfnamefont{M.~A.~L.} \bibnamefont{Nicolelis}},
  \bibinfo{year}{2003}, \bibinfo{journal}{PLoS Biology}
  \textbf{\bibinfo{volume}{1}}, \bibinfo{pages}{e42}.

\bibitem[{\citenamefont{Castellana and Bialek}(2014)}]{Castellana+Bialek_2014}
\bibinfo{author}{\bibnamefont{Castellana}, \bibfnamefont{M.}}, and
  \bibinfo{author}{\bibfnamefont{W.}~\bibnamefont{Bialek}},
  \bibinfo{year}{2014}, \bibinfo{journal}{Physical Review Letters}
  \textbf{\bibinfo{volume}{113}}, \bibinfo{pages}{117204}.

\bibitem[{\citenamefont{Castro} \emph{et~al.}(2024)\citenamefont{Castro,
  Feliciano, de~Vasconcelos, Soares-Cunha, Coimbra, Rodrigues, Carelli, and
  Copelli}}]{castro+al_24}
\bibinfo{author}{\bibnamefont{Castro}, \bibfnamefont{D.~M.}},
  \bibinfo{author}{\bibfnamefont{T.}~\bibnamefont{Feliciano}},
  \bibinfo{author}{\bibfnamefont{N.~A.~P.} \bibnamefont{de~Vasconcelos}},
  \bibinfo{author}{\bibfnamefont{C.}~\bibnamefont{Soares-Cunha}},
  \bibinfo{author}{\bibfnamefont{B.}~\bibnamefont{Coimbra}},
  \bibinfo{author}{\bibfnamefont{A.~J.} \bibnamefont{Rodrigues}},
  \bibinfo{author}{\bibfnamefont{P.~V.} \bibnamefont{Carelli}}, and
  \bibinfo{author}{\bibfnamefont{M.}~\bibnamefont{Copelli}},
  \bibinfo{year}{2024}, \bibinfo{journal}{PRX Life}
  \textbf{\bibinfo{volume}{2}}, \bibinfo{pages}{023008}.

\bibitem[{\citenamefont{Cavagna} \emph{et~al.}(2010)\citenamefont{Cavagna,
  Cimarelli, Giardina, Parisi, Santagati, Stefanini, and
  Viale}}]{cavagna+al_10}
\bibinfo{author}{\bibnamefont{Cavagna}, \bibfnamefont{A.}},
  \bibinfo{author}{\bibfnamefont{A.}~\bibnamefont{Cimarelli}},
  \bibinfo{author}{\bibfnamefont{I.}~\bibnamefont{Giardina}},
  \bibinfo{author}{\bibfnamefont{G.}~\bibnamefont{Parisi}},
  \bibinfo{author}{\bibfnamefont{R.}~\bibnamefont{Santagati}},
  \bibinfo{author}{\bibfnamefont{F.}~\bibnamefont{Stefanini}}, and
  \bibinfo{author}{\bibfnamefont{M.}~\bibnamefont{Viale}},
  \bibinfo{year}{2010}, \bibinfo{journal}{Proceedings of the National Academy
  of Sciences (USA)} \textbf{\bibinfo{volume}{107}}, \bibinfo{pages}{11865}.

\bibitem[{\citenamefont{Cavagna} \emph{et~al.}(2017)\citenamefont{Cavagna,
  Conte, Creato, Del~Castello, Giardina, Grigera, Melillo, Parisi, , and
  Viale}}]{cavagna+al_17}
\bibinfo{author}{\bibnamefont{Cavagna}, \bibfnamefont{A.}},
  \bibinfo{author}{\bibfnamefont{D.}~\bibnamefont{Conte}},
  \bibinfo{author}{\bibfnamefont{C.}~\bibnamefont{Creato}},
  \bibinfo{author}{\bibfnamefont{L.}~\bibnamefont{Del~Castello}},
  \bibinfo{author}{\bibfnamefont{I.}~\bibnamefont{Giardina}},
  \bibinfo{author}{\bibfnamefont{T.~S.} \bibnamefont{Grigera}},
  \bibinfo{author}{\bibfnamefont{S.}~\bibnamefont{Melillo}},
  \bibinfo{author}{\bibfnamefont{L.}~\bibnamefont{Parisi}}, , and
  \bibinfo{author}{\bibfnamefont{M.}~\bibnamefont{Viale}},
  \bibinfo{year}{2017}, \bibinfo{journal}{Nature Physics}
  \textbf{\bibinfo{volume}{13}}, \bibinfo{pages}{914}.

\bibitem[{\citenamefont{Cavagna} \emph{et~al.}(2022)\citenamefont{Cavagna,
  Culla, Feng, Giardina, Grigera, Kion-Crosby, Melillo, Pisegna, Postiglione,
  and Villegas}}]{cavgana+al_22}
\bibinfo{author}{\bibnamefont{Cavagna}, \bibfnamefont{A.}},
  \bibinfo{author}{\bibfnamefont{A.}~\bibnamefont{Culla}},
  \bibinfo{author}{\bibfnamefont{X.}~\bibnamefont{Feng}},
  \bibinfo{author}{\bibfnamefont{I.}~\bibnamefont{Giardina}},
  \bibinfo{author}{\bibfnamefont{T.~S.} \bibnamefont{Grigera}},
  \bibinfo{author}{\bibfnamefont{W.}~\bibnamefont{Kion-Crosby}},
  \bibinfo{author}{\bibfnamefont{S.}~\bibnamefont{Melillo}},
  \bibinfo{author}{\bibfnamefont{G.}~\bibnamefont{Pisegna}},
  \bibinfo{author}{\bibfnamefont{L.}~\bibnamefont{Postiglione}}, and
  \bibinfo{author}{\bibfnamefont{P.}~\bibnamefont{Villegas}},
  \bibinfo{year}{2022}, \bibinfo{journal}{Nature Communications}
  \textbf{\bibinfo{volume}{13}}, \bibinfo{pages}{2315}.

\bibitem[{\citenamefont{Cavagna} \emph{et~al.}(2015)\citenamefont{Cavagna,
  Del~Castillo, Dey, Giardina, Melillo, Parisi, and Viale}}]{cavagna+al_15}
\bibinfo{author}{\bibnamefont{Cavagna}, \bibfnamefont{A.}},
  \bibinfo{author}{\bibfnamefont{L.}~\bibnamefont{Del~Castillo}},
  \bibinfo{author}{\bibfnamefont{S.}~\bibnamefont{Dey}},
  \bibinfo{author}{\bibfnamefont{I.}~\bibnamefont{Giardina}},
  \bibinfo{author}{\bibfnamefont{S.}~\bibnamefont{Melillo}},
  \bibinfo{author}{\bibfnamefont{L.}~\bibnamefont{Parisi}}, and
  \bibinfo{author}{\bibfnamefont{M.}~\bibnamefont{Viale}},
  \bibinfo{year}{2015}, \bibinfo{journal}{Physical Review E}
  \textbf{\bibinfo{volume}{92}}, \bibinfo{pages}{012705}.

\bibitem[{\citenamefont{Cavagna} \emph{et~al.}(2019)\citenamefont{Cavagna,
  Di~Carlo, Giardina, Grandinetti, Grigera, and Pisegna}}]{cavagna+al_19}
\bibinfo{author}{\bibnamefont{Cavagna}, \bibfnamefont{A.}},
  \bibinfo{author}{\bibfnamefont{L.}~\bibnamefont{Di~Carlo}},
  \bibinfo{author}{\bibfnamefont{I.}~\bibnamefont{Giardina}},
  \bibinfo{author}{\bibfnamefont{L.}~\bibnamefont{Grandinetti}},
  \bibinfo{author}{\bibfnamefont{T.~S.} \bibnamefont{Grigera}}, and
  \bibinfo{author}{\bibfnamefont{G.}~\bibnamefont{Pisegna}},
  \bibinfo{year}{2019}, \bibinfo{journal}{Physical Review Letters}
  \textbf{\bibinfo{volume}{123}}, \bibinfo{pages}{268001}.

\bibitem[{\citenamefont{Cavagna} \emph{et~al.}(2023)\citenamefont{Cavagna,
  Di~Carlo, Giardina, Grigera, Melillo, Parisi, Pisegna, and
  Scandolo}}]{cavagna+al_23}
\bibinfo{author}{\bibnamefont{Cavagna}, \bibfnamefont{A.}},
  \bibinfo{author}{\bibfnamefont{L.}~\bibnamefont{Di~Carlo}},
  \bibinfo{author}{\bibfnamefont{I.}~\bibnamefont{Giardina}},
  \bibinfo{author}{\bibfnamefont{T.~S.} \bibnamefont{Grigera}},
  \bibinfo{author}{\bibfnamefont{S.}~\bibnamefont{Melillo}},
  \bibinfo{author}{\bibfnamefont{L.}~\bibnamefont{Parisi}},
  \bibinfo{author}{\bibfnamefont{G.}~\bibnamefont{Pisegna}}, and
  \bibinfo{author}{\bibfnamefont{M.}~\bibnamefont{Scandolo}},
  \bibinfo{year}{2023}, \bibinfo{journal}{Nature Physics}
  \textbf{\bibinfo{volume}{19}}, \bibinfo{pages}{1043}.

\bibitem[{\citenamefont{Cavagna} \emph{et~al.}(2014)\citenamefont{Cavagna,
  Giardina, Ginelli, Mora, Piovani, Tavarone, and Walczak}}]{Cavagna+al_2014b}
\bibinfo{author}{\bibnamefont{Cavagna}, \bibfnamefont{A.}},
  \bibinfo{author}{\bibfnamefont{I.}~\bibnamefont{Giardina}},
  \bibinfo{author}{\bibfnamefont{F.}~\bibnamefont{Ginelli}},
  \bibinfo{author}{\bibfnamefont{T.}~\bibnamefont{Mora}},
  \bibinfo{author}{\bibfnamefont{D.}~\bibnamefont{Piovani}},
  \bibinfo{author}{\bibfnamefont{R.}~\bibnamefont{Tavarone}}, and
  \bibinfo{author}{\bibfnamefont{A.~M.} \bibnamefont{Walczak}},
  \bibinfo{year}{2014}, \bibinfo{journal}{Physical Review E}
  \textbf{\bibinfo{volume}{89}}, \bibinfo{pages}{042707}.

\bibitem[{\citenamefont{Cavagna} \emph{et~al.}(2018)\citenamefont{Cavagna,
  Giardina, and Grigera}}]{cavagna+al_18a}
\bibinfo{author}{\bibnamefont{Cavagna}, \bibfnamefont{A.}},
  \bibinfo{author}{\bibfnamefont{I.}~\bibnamefont{Giardina}}, and
  \bibinfo{author}{\bibfnamefont{T.~A.} \bibnamefont{Grigera}},
  \bibinfo{year}{2018}, \bibinfo{journal}{Physics Reports}
  \textbf{\bibinfo{volume}{728}}, \bibinfo{pages}{1}.

\bibitem[{\citenamefont{Cavagna}
  \emph{et~al.}(2008{\natexlab{a}})\citenamefont{Cavagna, Giardina, Orlandi,
  Parisi, and Procaccini}}]{cavagna+al2008b}
\bibinfo{author}{\bibnamefont{Cavagna}, \bibfnamefont{A.}},
  \bibinfo{author}{\bibfnamefont{I.}~\bibnamefont{Giardina}},
  \bibinfo{author}{\bibfnamefont{A.}~\bibnamefont{Orlandi}},
  \bibinfo{author}{\bibfnamefont{G.}~\bibnamefont{Parisi}}, and
  \bibinfo{author}{\bibfnamefont{A.}~\bibnamefont{Procaccini}},
  \bibinfo{year}{2008}{\natexlab{a}}, \bibinfo{journal}{Animal Behaviour}
  \textbf{\bibinfo{volume}{76}}, \bibinfo{pages}{237}.

\bibitem[{\citenamefont{Cavagna}
  \emph{et~al.}(2008{\natexlab{b}})\citenamefont{Cavagna, Giardina, Orlandi,
  Parisi, Procaccini, Viale, and Zdravkovic}}]{cavagna+al2008a}
\bibinfo{author}{\bibnamefont{Cavagna}, \bibfnamefont{A.}},
  \bibinfo{author}{\bibfnamefont{I.}~\bibnamefont{Giardina}},
  \bibinfo{author}{\bibfnamefont{A.}~\bibnamefont{Orlandi}},
  \bibinfo{author}{\bibfnamefont{G.}~\bibnamefont{Parisi}},
  \bibinfo{author}{\bibfnamefont{A.}~\bibnamefont{Procaccini}},
  \bibinfo{author}{\bibfnamefont{M.}~\bibnamefont{Viale}}, and
  \bibinfo{author}{\bibfnamefont{V.}~\bibnamefont{Zdravkovic}},
  \bibinfo{year}{2008}{\natexlab{b}}, \bibinfo{journal}{Animal Behaviour}
  \textbf{\bibinfo{volume}{76}}, \bibinfo{pages}{217}.

\bibitem[{\citenamefont{Chakraborty and Barton}(2017)}]{chakraborty+barton_17}
\bibinfo{author}{\bibnamefont{Chakraborty}, \bibfnamefont{A.~K.}}, and
  \bibinfo{author}{\bibfnamefont{J.~P.} \bibnamefont{Barton}},
  \bibinfo{year}{2017}, \bibinfo{journal}{Reports on Progress in Physics}
  \textbf{\bibinfo{volume}{80}}, \bibinfo{pages}{032601}.

\bibitem[{\citenamefont{Chalfie} \emph{et~al.}(1994)\citenamefont{Chalfie, Tu,
  Euskirchen, Ward, and Prasher}}]{chalfie+al_94}
\bibinfo{author}{\bibnamefont{Chalfie}, \bibfnamefont{M.}},
  \bibinfo{author}{\bibfnamefont{Y.}~\bibnamefont{Tu}},
  \bibinfo{author}{\bibfnamefont{G.}~\bibnamefont{Euskirchen}},
  \bibinfo{author}{\bibfnamefont{W.~W.} \bibnamefont{Ward}}, and
  \bibinfo{author}{\bibfnamefont{D.~C.} \bibnamefont{Prasher}},
  \bibinfo{year}{1994}, \bibinfo{journal}{Science}
  \textbf{\bibinfo{volume}{263}}, \bibinfo{pages}{802}.

\bibitem[{\citenamefont{Chayes} \emph{et~al.}(1984)\citenamefont{Chayes,
  Chayes, and Lieb}}]{chayes+al_84}
\bibinfo{author}{\bibnamefont{Chayes}, \bibfnamefont{J.~T.}},
  \bibinfo{author}{\bibfnamefont{L.}~\bibnamefont{Chayes}}, and
  \bibinfo{author}{\bibfnamefont{E.~H.} \bibnamefont{Lieb}},
  \bibinfo{year}{1984}, \bibinfo{journal}{Communications in Mathematical
  Physics} \textbf{\bibinfo{volume}{93}}, \bibinfo{pages}{57}.

\bibitem[{\citenamefont{Chen} \emph{et~al.}(2013)\citenamefont{Chen, Wardill,
  Sun, Pulver, Renninger, Baohan, Schreiter, Kerr, Orger, Jayaraman, Looger,
  Svoboda} \emph{et~al.}}]{Chen+al_2013}
\bibinfo{author}{\bibnamefont{Chen}, \bibfnamefont{T.-W.}},
  \bibinfo{author}{\bibfnamefont{T.~J.} \bibnamefont{Wardill}},
  \bibinfo{author}{\bibfnamefont{Y.}~\bibnamefont{Sun}},
  \bibinfo{author}{\bibfnamefont{S.~R.} \bibnamefont{Pulver}},
  \bibinfo{author}{\bibfnamefont{S.~L.} \bibnamefont{Renninger}},
  \bibinfo{author}{\bibfnamefont{A.}~\bibnamefont{Baohan}},
  \bibinfo{author}{\bibfnamefont{E.~R.} \bibnamefont{Schreiter}},
  \bibinfo{author}{\bibfnamefont{R.~A.} \bibnamefont{Kerr}},
  \bibinfo{author}{\bibfnamefont{M.~B.} \bibnamefont{Orger}},
  \bibinfo{author}{\bibfnamefont{V.}~\bibnamefont{Jayaraman}},
  \bibinfo{author}{\bibfnamefont{L.~L.} \bibnamefont{Looger}},
  \bibinfo{author}{\bibfnamefont{K.}~\bibnamefont{Svoboda}}, \emph{et~al.},
  \bibinfo{year}{2013}, \bibinfo{journal}{Nature}
  \textbf{\bibinfo{volume}{499}}, \bibinfo{pages}{295}.

\bibitem[{\citenamefont{Chen and Dzakpasu}(2010)}]{chen+dzakpasu_10}
\bibinfo{author}{\bibnamefont{Chen}, \bibfnamefont{X.}}, and
  \bibinfo{author}{\bibfnamefont{R.}~\bibnamefont{Dzakpasu}},
  \bibinfo{year}{2010}, \bibinfo{journal}{Physical Review E}
  \textbf{\bibinfo{volume}{82}}, \bibinfo{pages}{031907}.

\bibitem[{\citenamefont{Chen} \emph{et~al.}(2019)\citenamefont{Chen, Randi,
  Leifer, and Bialek}}]{chen+al_19}
\bibinfo{author}{\bibnamefont{Chen}, \bibfnamefont{X.}},
  \bibinfo{author}{\bibfnamefont{F.}~\bibnamefont{Randi}},
  \bibinfo{author}{\bibfnamefont{A.~M.} \bibnamefont{Leifer}}, and
  \bibinfo{author}{\bibfnamefont{W.}~\bibnamefont{Bialek}},
  \bibinfo{year}{2019}, \bibinfo{journal}{Physical Review E}
  \textbf{\bibinfo{volume}{99}}, \bibinfo{pages}{052418}.

\bibitem[{\citenamefont{Chen} \emph{et~al.}(2023)\citenamefont{Chen, Winiarski,
  Pu\'scian, Knapska, Walczak, and Mora}}]{chen+al_23}
\bibinfo{author}{\bibnamefont{Chen}, \bibfnamefont{X.}},
  \bibinfo{author}{\bibfnamefont{M.}~\bibnamefont{Winiarski}},
  \bibinfo{author}{\bibfnamefont{A.}~\bibnamefont{Pu\'scian}},
  \bibinfo{author}{\bibfnamefont{E.}~\bibnamefont{Knapska}},
  \bibinfo{author}{\bibfnamefont{A.~M.} \bibnamefont{Walczak}}, and
  \bibinfo{author}{\bibfnamefont{T.}~\bibnamefont{Mora}}, \bibinfo{year}{2023},
  \bibinfo{journal}{Physical Review X} \textbf{\bibinfo{volume}{13}},
  \bibinfo{pages}{041053}.

\bibitem[{\citenamefont{Cheng} \emph{et~al.}(2022)\citenamefont{Cheng, Ho,
  Aranda-D\'iaz, Jain, Yu, Meng, Wang, Iakiviak, Nagashima, Zhao, Murugkar,
  Patil} \emph{et~al.}}]{cheng+al_22}
\bibinfo{author}{\bibnamefont{Cheng}, \bibfnamefont{A.~G.}},
  \bibinfo{author}{\bibfnamefont{P.-Y.} \bibnamefont{Ho}},
  \bibinfo{author}{\bibfnamefont{A.}~\bibnamefont{Aranda-D\'iaz}},
  \bibinfo{author}{\bibfnamefont{S.}~\bibnamefont{Jain}},
  \bibinfo{author}{\bibfnamefont{F.~B.} \bibnamefont{Yu}},
  \bibinfo{author}{\bibfnamefont{X.}~\bibnamefont{Meng}},
  \bibinfo{author}{\bibfnamefont{M.}~\bibnamefont{Wang}},
  \bibinfo{author}{\bibfnamefont{M.}~\bibnamefont{Iakiviak}},
  \bibinfo{author}{\bibfnamefont{K.}~\bibnamefont{Nagashima}},
  \bibinfo{author}{\bibfnamefont{A.}~\bibnamefont{Zhao}},
  \bibinfo{author}{\bibfnamefont{P.}~\bibnamefont{Murugkar}},
  \bibinfo{author}{\bibfnamefont{A.}~\bibnamefont{Patil}}, \emph{et~al.},
  \bibinfo{year}{2022}, \bibinfo{journal}{Cell} \textbf{\bibinfo{volume}{185}},
  \bibinfo{pages}{3617}.

\bibitem[{\citenamefont{Chung} \emph{et~al.}(2019)\citenamefont{Chung, Joo,
  Fan, Liu, Barnett, Chen, Geaghan-Breiner, Karlsson, Karlsson, Lee}
  \emph{et~al.}}]{chung2019high}
\bibinfo{author}{\bibnamefont{Chung}, \bibfnamefont{J.~E.}},
  \bibinfo{author}{\bibfnamefont{H.~R.} \bibnamefont{Joo}},
  \bibinfo{author}{\bibfnamefont{J.~L.} \bibnamefont{Fan}},
  \bibinfo{author}{\bibfnamefont{D.~F.} \bibnamefont{Liu}},
  \bibinfo{author}{\bibfnamefont{A.~H.} \bibnamefont{Barnett}},
  \bibinfo{author}{\bibfnamefont{S.}~\bibnamefont{Chen}},
  \bibinfo{author}{\bibfnamefont{C.}~\bibnamefont{Geaghan-Breiner}},
  \bibinfo{author}{\bibfnamefont{M.~P.} \bibnamefont{Karlsson}},
  \bibinfo{author}{\bibfnamefont{M.}~\bibnamefont{Karlsson}},
  \bibinfo{author}{\bibfnamefont{K.~Y.} \bibnamefont{Lee}}, \emph{et~al.},
  \bibinfo{year}{2019}, \bibinfo{journal}{Neuron}
  \textbf{\bibinfo{volume}{101}}, \bibinfo{pages}{21}.

\bibitem[{\citenamefont{Chung} \emph{et~al.}(2017)\citenamefont{Chung, Magland,
  Barnett, Tolosa, Tooker, Lee, Shah, Felix, Frank, and
  Greengard}}]{chung+al_17}
\bibinfo{author}{\bibnamefont{Chung}, \bibfnamefont{J.~E.}},
  \bibinfo{author}{\bibfnamefont{J.~F.} \bibnamefont{Magland}},
  \bibinfo{author}{\bibfnamefont{A.~H.} \bibnamefont{Barnett}},
  \bibinfo{author}{\bibfnamefont{V.~M.} \bibnamefont{Tolosa}},
  \bibinfo{author}{\bibfnamefont{A.~C.} \bibnamefont{Tooker}},
  \bibinfo{author}{\bibfnamefont{K.~Y.} \bibnamefont{Lee}},
  \bibinfo{author}{\bibfnamefont{K.~G.} \bibnamefont{Shah}},
  \bibinfo{author}{\bibfnamefont{S.~H.} \bibnamefont{Felix}},
  \bibinfo{author}{\bibfnamefont{L.~M.} \bibnamefont{Frank}}, and
  \bibinfo{author}{\bibfnamefont{L.~F.} \bibnamefont{Greengard}},
  \bibinfo{year}{2017}, \bibinfo{journal}{Neuron}
  \textbf{\bibinfo{volume}{95}}, \bibinfo{pages}{1381}.

\bibitem[{\citenamefont{Cocco and Monasson}(2011)}]{cocco+monasson_11}
\bibinfo{author}{\bibnamefont{Cocco}, \bibfnamefont{S.}}, and
  \bibinfo{author}{\bibfnamefont{R.}~\bibnamefont{Monasson}},
  \bibinfo{year}{2011}, \bibinfo{journal}{Physical Review Letters}
  \textbf{\bibinfo{volume}{106}}, \bibinfo{pages}{090601}.

\bibitem[{\citenamefont{Cocco and Monasson}(2012)}]{cocco+monasson_12}
\bibinfo{author}{\bibnamefont{Cocco}, \bibfnamefont{S.}}, and
  \bibinfo{author}{\bibfnamefont{R.}~\bibnamefont{Monasson}},
  \bibinfo{year}{2012}, \bibinfo{journal}{Journal of Statistical Physics}
  \textbf{\bibinfo{volume}{147}}, \bibinfo{pages}{252}.

\bibitem[{\citenamefont{Cocco} \emph{et~al.}(2011)\citenamefont{Cocco,
  Monasson, and Sessak}}]{Cocco+al_2011}
\bibinfo{author}{\bibnamefont{Cocco}, \bibfnamefont{S.}},
  \bibinfo{author}{\bibfnamefont{R.}~\bibnamefont{Monasson}}, and
  \bibinfo{author}{\bibfnamefont{V.}~\bibnamefont{Sessak}},
  \bibinfo{year}{2011}, \bibinfo{journal}{Physical Review E}
  \textbf{\bibinfo{volume}{83}}, \bibinfo{pages}{051123}.

\bibitem[{\citenamefont{Cocco}
  \emph{et~al.}(2013{\natexlab{a}})\citenamefont{Cocco, Monasson, and
  Weigt}}]{Cocco+al_2013a}
\bibinfo{author}{\bibnamefont{Cocco}, \bibfnamefont{S.}},
  \bibinfo{author}{\bibfnamefont{R.}~\bibnamefont{Monasson}}, and
  \bibinfo{author}{\bibfnamefont{M.}~\bibnamefont{Weigt}},
  \bibinfo{year}{2013}{\natexlab{a}}, \bibinfo{journal}{PLoS Computational
  Biology} \textbf{\bibinfo{volume}{9}}, \bibinfo{pages}{e1003176}.

\bibitem[{\citenamefont{Cocco}
  \emph{et~al.}(2013{\natexlab{b}})\citenamefont{Cocco, Monasson, and
  Weigt}}]{Cocco+al_2013b}
\bibinfo{author}{\bibnamefont{Cocco}, \bibfnamefont{S.}},
  \bibinfo{author}{\bibfnamefont{R.}~\bibnamefont{Monasson}}, and
  \bibinfo{author}{\bibfnamefont{M.}~\bibnamefont{Weigt}},
  \bibinfo{year}{2013}{\natexlab{b}}, \bibinfo{journal}{Journal of Physics:
  Conference Series} \textbf{\bibinfo{volume}{473}}, \bibinfo{pages}{012010}.

\bibitem[{\citenamefont{Cohen and Salzberg}(1978)}]{cohen+salzberg_78}
\bibinfo{author}{\bibnamefont{Cohen}, \bibfnamefont{L.~B.}}, and
  \bibinfo{author}{\bibfnamefont{B.~M.} \bibnamefont{Salzberg}},
  \bibinfo{year}{1978}, \bibinfo{journal}{Reviews of Physiology, Biochemistry
  and Pharmacology} \textbf{\bibinfo{volume}{83}}, \bibinfo{pages}{35}.

\bibitem[{\citenamefont{Condit} \emph{et~al.}(2014)\citenamefont{Condit, Lao,
  Singh, Esufali, and Dolins}}]{condit+al_14}
\bibinfo{author}{\bibnamefont{Condit}, \bibfnamefont{R.}},
  \bibinfo{author}{\bibfnamefont{S.}~\bibnamefont{Lao}},
  \bibinfo{author}{\bibfnamefont{A.}~\bibnamefont{Singh}},
  \bibinfo{author}{\bibfnamefont{S.}~\bibnamefont{Esufali}}, and
  \bibinfo{author}{\bibfnamefont{S.}~\bibnamefont{Dolins}},
  \bibinfo{year}{2014}, \bibinfo{journal}{Forest Ecology and Management}
  \textbf{\bibinfo{volume}{316}}, \bibinfo{pages}{21}.

\bibitem[{\citenamefont{Cook} \emph{et~al.}(2019)\citenamefont{Cook, Jarrell,
  Brittin, Wang, Bloniarz, Yakovlev, Nguyen, Tang, Bayer, Duerr, B\"ulow,
  Hobert} \emph{et~al.}}]{cook+al_19}
\bibinfo{author}{\bibnamefont{Cook}, \bibfnamefont{S.~J.}},
  \bibinfo{author}{\bibfnamefont{T.~A.} \bibnamefont{Jarrell}},
  \bibinfo{author}{\bibfnamefont{C.~A.} \bibnamefont{Brittin}},
  \bibinfo{author}{\bibfnamefont{Y.}~\bibnamefont{Wang}},
  \bibinfo{author}{\bibfnamefont{A.~E.} \bibnamefont{Bloniarz}},
  \bibinfo{author}{\bibfnamefont{M.~A.} \bibnamefont{Yakovlev}},
  \bibinfo{author}{\bibfnamefont{K.~C.~Q.} \bibnamefont{Nguyen}},
  \bibinfo{author}{\bibfnamefont{L.~T.-H.} \bibnamefont{Tang}},
  \bibinfo{author}{\bibfnamefont{E.~A.} \bibnamefont{Bayer}},
  \bibinfo{author}{\bibfnamefont{J.~S.} \bibnamefont{Duerr}},
  \bibinfo{author}{\bibfnamefont{H.~E.} \bibnamefont{B\"ulow}},
  \bibinfo{author}{\bibfnamefont{O.}~\bibnamefont{Hobert}}, \emph{et~al.},
  \bibinfo{year}{2019}, \bibinfo{journal}{Nature}
  \textbf{\bibinfo{volume}{571}}, \bibinfo{pages}{63}.

\bibitem[{\citenamefont{Cooper}(1973)}]{cooper_73}
\bibinfo{author}{\bibnamefont{Cooper}, \bibfnamefont{L.~N.}},
  \bibinfo{year}{1973}, in \emph{\bibinfo{booktitle}{Collective Properties of
  Physical Systems: Proceedings of Nobel Symposium 24}}, edited by
  \bibinfo{editor}{\bibfnamefont{B.}~\bibnamefont{Lundqvist}} and
  \bibinfo{editor}{\bibfnamefont{S.}~\bibnamefont{Lundqvist}}
  (\bibinfo{publisher}{Academic Press, New York}), pp.
  \bibinfo{pages}{252--264}.

\bibitem[{\citenamefont{Cover and Thomas}(1991)}]{cover+thomas_91}
\bibinfo{author}{\bibnamefont{Cover}, \bibfnamefont{T.~M.}}, and
  \bibinfo{author}{\bibfnamefont{J.~A.} \bibnamefont{Thomas}},
  \bibinfo{year}{1991}, \emph{\bibinfo{title}{Elements of Information Theory}}
  (\bibinfo{publisher}{Wiley and Sons, New York}).

\bibitem[{\citenamefont{Cragg and Temperley}(1954)}]{cragg+temperley_54}
\bibinfo{author}{\bibnamefont{Cragg}, \bibfnamefont{B.~G.}}, and
  \bibinfo{author}{\bibfnamefont{H.~N.~V.} \bibnamefont{Temperley}},
  \bibinfo{year}{1954}, \bibinfo{journal}{Electroencephalography and Clinical
  Neurophysiology} \textbf{\bibinfo{volume}{6}}, \bibinfo{pages}{85}.

\bibitem[{\citenamefont{Cullen} \emph{et~al.}(1965)\citenamefont{Cullen, Shaw,
  and Baldwin}}]{Cullen+al_1965}
\bibinfo{author}{\bibnamefont{Cullen}, \bibfnamefont{J.~M.}},
  \bibinfo{author}{\bibfnamefont{E.}~\bibnamefont{Shaw}}, and
  \bibinfo{author}{\bibfnamefont{H.~A.} \bibnamefont{Baldwin}},
  \bibinfo{year}{1965}, \bibinfo{journal}{Animal Behaviour}
  \textbf{\bibinfo{volume}{13}}, \bibinfo{pages}{534}.

\bibitem[{\citenamefont{Dayan and Abbott}(2001)}]{Dayan+Abbott_2001}
\bibinfo{author}{\bibnamefont{Dayan}, \bibfnamefont{P.}}, and
  \bibinfo{author}{\bibfnamefont{L.~F.} \bibnamefont{Abbott}},
  \bibinfo{year}{2001}, \emph{\bibinfo{title}{Theoretical Neuroscience:
  Computational and Mathematical Modeling of Neural Systems}}
  (\bibinfo{publisher}{MIT Press, Cambridge MA}).

\bibitem[{\citenamefont{De~Martino}
  \emph{et~al.}(2018)\citenamefont{De~Martino, Andersson, Bergmiller, Guet, and
  Tka\v{c}ik}}]{demartino+al_18}
\bibinfo{author}{\bibnamefont{De~Martino}, \bibfnamefont{D.}},
  \bibinfo{author}{\bibfnamefont{A.~M.~C.} \bibnamefont{Andersson}},
  \bibinfo{author}{\bibfnamefont{T.}~\bibnamefont{Bergmiller}},
  \bibinfo{author}{\bibfnamefont{C.~C.} \bibnamefont{Guet}}, and
  \bibinfo{author}{\bibfnamefont{G.}~\bibnamefont{Tka\v{c}ik}},
  \bibinfo{year}{2018}, \bibinfo{journal}{Nature Communications}
  \textbf{\bibinfo{volume}{9}}, \bibinfo{pages}{2988}.

\bibitem[{\citenamefont{Demas} \emph{et~al.}(2021)\citenamefont{Demas, Manley,
  Tejera, Barber, Kim, Traub, Chen, and Vaziri}}]{demas+al2021}
\bibinfo{author}{\bibnamefont{Demas}, \bibfnamefont{J.}},
  \bibinfo{author}{\bibfnamefont{J.}~\bibnamefont{Manley}},
  \bibinfo{author}{\bibfnamefont{F.}~\bibnamefont{Tejera}},
  \bibinfo{author}{\bibfnamefont{K.}~\bibnamefont{Barber}},
  \bibinfo{author}{\bibfnamefont{H.}~\bibnamefont{Kim}},
  \bibinfo{author}{\bibfnamefont{F.~M.} \bibnamefont{Traub}},
  \bibinfo{author}{\bibfnamefont{B.}~\bibnamefont{Chen}}, and
  \bibinfo{author}{\bibfnamefont{A.}~\bibnamefont{Vaziri}},
  \bibinfo{year}{2021}, \bibinfo{journal}{Nature Methods}
  \textbf{\bibinfo{volume}{18}}, \bibinfo{pages}{1103}.

\bibitem[{\citenamefont{Derrida}(1981)}]{derrida_81}
\bibinfo{author}{\bibnamefont{Derrida}, \bibfnamefont{B.}},
  \bibinfo{year}{1981}, \bibinfo{journal}{Physical Review B}
  \textbf{\bibinfo{volume}{24}}, \bibinfo{pages}{2613}.

\bibitem[{\citenamefont{Devine and Cohen}(1992)}]{devine+cohen_92}
\bibinfo{author}{\bibnamefont{Devine}, \bibfnamefont{B.}}, and
  \bibinfo{author}{\bibfnamefont{J.~E.} \bibnamefont{Cohen}},
  \bibinfo{year}{1992}, \emph{\bibinfo{title}{Absolute Zero Gravity: Science
  Jokes, Quotes and Anecdotes}} (\bibinfo{publisher}{Simon \& Schuster, New
  York}).

\bibitem[{\citenamefont{Dombeck} \emph{et~al.}(2010)\citenamefont{Dombeck,
  Harvey, Tian, Looger, and Tank}}]{dombeck2010functional}
\bibinfo{author}{\bibnamefont{Dombeck}, \bibfnamefont{D.~A.}},
  \bibinfo{author}{\bibfnamefont{C.~D.} \bibnamefont{Harvey}},
  \bibinfo{author}{\bibfnamefont{L.}~\bibnamefont{Tian}},
  \bibinfo{author}{\bibfnamefont{L.~L.} \bibnamefont{Looger}}, and
  \bibinfo{author}{\bibfnamefont{D.~W.} \bibnamefont{Tank}},
  \bibinfo{year}{2010}, \bibinfo{journal}{Nature Neuroscience}
  \textbf{\bibinfo{volume}{13}}, \bibinfo{pages}{1433}.

\bibitem[{\citenamefont{Durbin} \emph{et~al.}(1998)\citenamefont{Durbin, Eddy,
  Krogh, and Mitchison}}]{Durbin+al_1998}
\bibinfo{author}{\bibnamefont{Durbin}, \bibfnamefont{R.}},
  \bibinfo{author}{\bibfnamefont{S.~R.} \bibnamefont{Eddy}},
  \bibinfo{author}{\bibfnamefont{A.}~\bibnamefont{Krogh}}, and
  \bibinfo{author}{\bibfnamefont{G.}~\bibnamefont{Mitchison}},
  \bibinfo{year}{1998}, \emph{\bibinfo{title}{Biological Sequence Analysis:
  Probabilistic Models of Proteins and Nucleic Acids}}
  (\bibinfo{publisher}{Cambridge University Press, Cambridge}).

\bibitem[{\citenamefont{Everitt}(1984)}]{everitt_84}
\bibinfo{author}{\bibnamefont{Everitt}, \bibfnamefont{B.}},
  \bibinfo{year}{1984}, \emph{\bibinfo{title}{An Introduction to Latent
  Variable Models}} (\bibinfo{publisher}{Chapman and Hall, London and New
  York}).

\bibitem[{\citenamefont{Ferrari} \emph{et~al.}(2018)\citenamefont{Ferrari,
  Deny, Chalk, Tka\v{c}ik, Marre, and Mora}}]{ferrari+al2018}
\bibinfo{author}{\bibnamefont{Ferrari}, \bibfnamefont{U.}},
  \bibinfo{author}{\bibfnamefont{S.}~\bibnamefont{Deny}},
  \bibinfo{author}{\bibfnamefont{M.}~\bibnamefont{Chalk}},
  \bibinfo{author}{\bibfnamefont{G.}~\bibnamefont{Tka\v{c}ik}},
  \bibinfo{author}{\bibfnamefont{O.}~\bibnamefont{Marre}}, and
  \bibinfo{author}{\bibfnamefont{T.}~\bibnamefont{Mora}}, \bibinfo{year}{2018},
  \bibinfo{journal}{Physical Review E} \textbf{\bibinfo{volume}{98}},
  \bibinfo{pages}{042410}.

\bibitem[{\citenamefont{Ferrari} \emph{et~al.}(2017)\citenamefont{Ferrari,
  Obuchi, and Mora}}]{ferrari+al_17}
\bibinfo{author}{\bibnamefont{Ferrari}, \bibfnamefont{U.}},
  \bibinfo{author}{\bibfnamefont{T.}~\bibnamefont{Obuchi}}, and
  \bibinfo{author}{\bibfnamefont{T.}~\bibnamefont{Mora}}, \bibinfo{year}{2017},
  \bibinfo{journal}{Physical Review E} \textbf{\bibinfo{volume}{95}},
  \bibinfo{pages}{042321}.

\bibitem[{\citenamefont{Ferrenberg and
  Swendsen}(1988)}]{ferrenberg+swendsen_88}
\bibinfo{author}{\bibnamefont{Ferrenberg}, \bibfnamefont{A.~M.}}, and
  \bibinfo{author}{\bibfnamefont{R.~H.} \bibnamefont{Swendsen}},
  \bibinfo{year}{1988}, \bibinfo{journal}{Physical Review Letters}
  \textbf{\bibinfo{volume}{61}}, \bibinfo{pages}{2635}.

\bibitem[{\citenamefont{Field and Chichilnisky}(2007)}]{field+chichilnisky_07}
\bibinfo{author}{\bibnamefont{Field}, \bibfnamefont{G.~D.}}, and
  \bibinfo{author}{\bibfnamefont{E.~J.} \bibnamefont{Chichilnisky}},
  \bibinfo{year}{2007}, \bibinfo{journal}{Annual Review of Neuroscience}
  \textbf{\bibinfo{volume}{30}}, \bibinfo{pages}{1}.

\bibitem[{\citenamefont{Finn} \emph{et~al.}(2014)\citenamefont{Finn, Bateman,
  Clements, Coggill, Eberhardt, Eddy, Heger, Hetherington, Hol, Mistry,
  Sonnhammer, Tate} \emph{et~al.}}]{Pfam}
\bibinfo{author}{\bibnamefont{Finn}, \bibfnamefont{R.~D.}},
  \bibinfo{author}{\bibfnamefont{A.}~\bibnamefont{Bateman}},
  \bibinfo{author}{\bibfnamefont{J.}~\bibnamefont{Clements}},
  \bibinfo{author}{\bibfnamefont{P.}~\bibnamefont{Coggill}},
  \bibinfo{author}{\bibfnamefont{R.~Y.} \bibnamefont{Eberhardt}},
  \bibinfo{author}{\bibfnamefont{S.~R.} \bibnamefont{Eddy}},
  \bibinfo{author}{\bibfnamefont{A.}~\bibnamefont{Heger}},
  \bibinfo{author}{\bibfnamefont{K.}~\bibnamefont{Hetherington}},
  \bibinfo{author}{\bibfnamefont{L.}~\bibnamefont{Hol}},
  \bibinfo{author}{\bibfnamefont{J.}~\bibnamefont{Mistry}},
  \bibinfo{author}{\bibfnamefont{E.~L.~L.} \bibnamefont{Sonnhammer}},
  \bibinfo{author}{\bibfnamefont{J.}~\bibnamefont{Tate}}, \emph{et~al.},
  \bibinfo{year}{2014}, \bibinfo{journal}{Nucleic Acids Research}
  \textbf{\bibinfo{volume}{42}}, \bibinfo{pages}{D222}.

\bibitem[{\citenamefont{Fontenele} \emph{et~al.}(2024)\citenamefont{Fontenele,
  Sooter, Norman, Gautam, and Shew}}]{fontenele+al_24}
\bibinfo{author}{\bibnamefont{Fontenele}, \bibfnamefont{A.~J.}},
  \bibinfo{author}{\bibfnamefont{J.~S.} \bibnamefont{Sooter}},
  \bibinfo{author}{\bibfnamefont{V.~K.} \bibnamefont{Norman}},
  \bibinfo{author}{\bibfnamefont{S.~H.} \bibnamefont{Gautam}}, and
  \bibinfo{author}{\bibfnamefont{W.~L.} \bibnamefont{Shew}},
  \bibinfo{year}{2024}, \bibinfo{journal}{Science Advances}
  \textbf{\bibinfo{volume}{17}}, \bibinfo{pages}{eadj9303}.

\bibitem[{\citenamefont{Fontenele} \emph{et~al.}(2019)\citenamefont{Fontenele,
  de~Vasconcelos, Feliciano, Aguiar, Soares-Cunha, Coimbra, Dalla~Porta,
  Ribeiro, Rodrigues, Sousa, Carelli, and Copelli}}]{fontenele+al_19}
\bibinfo{author}{\bibnamefont{Fontenele}, \bibfnamefont{A.~J.}},
  \bibinfo{author}{\bibfnamefont{N.~A.~P.} \bibnamefont{de~Vasconcelos}},
  \bibinfo{author}{\bibfnamefont{T.}~\bibnamefont{Feliciano}},
  \bibinfo{author}{\bibfnamefont{L.~A.~A.} \bibnamefont{Aguiar}},
  \bibinfo{author}{\bibfnamefont{C.}~\bibnamefont{Soares-Cunha}},
  \bibinfo{author}{\bibfnamefont{B.}~\bibnamefont{Coimbra}},
  \bibinfo{author}{\bibfnamefont{L.}~\bibnamefont{Dalla~Porta}},
  \bibinfo{author}{\bibfnamefont{S.}~\bibnamefont{Ribeiro}},
  \bibinfo{author}{\bibfnamefont{A.~J.} \bibnamefont{Rodrigues}},
  \bibinfo{author}{\bibfnamefont{N.}~\bibnamefont{Sousa}},
  \bibinfo{author}{\bibfnamefont{P.~V.} \bibnamefont{Carelli}}, and
  \bibinfo{author}{\bibfnamefont{M.}~\bibnamefont{Copelli}},
  \bibinfo{year}{2019}, \bibinfo{journal}{Physical Review Letters}
  \textbf{\bibinfo{volume}{122}}, \bibinfo{pages}{208101}.

\bibitem[{\citenamefont{Friedman} \emph{et~al.}(2012)\citenamefont{Friedman,
  Ito, Brinkman, Shimono, DeVille, Dahmen, Beggs, and Butler}}]{friedman+al_12}
\bibinfo{author}{\bibnamefont{Friedman}, \bibfnamefont{N.}},
  \bibinfo{author}{\bibfnamefont{S.}~\bibnamefont{Ito}},
  \bibinfo{author}{\bibfnamefont{B.~A.~W.} \bibnamefont{Brinkman}},
  \bibinfo{author}{\bibfnamefont{M.}~\bibnamefont{Shimono}},
  \bibinfo{author}{\bibfnamefont{R.~E.~L.} \bibnamefont{DeVille}},
  \bibinfo{author}{\bibfnamefont{K.~A.} \bibnamefont{Dahmen}},
  \bibinfo{author}{\bibfnamefont{J.~M.} \bibnamefont{Beggs}}, and
  \bibinfo{author}{\bibfnamefont{T.~C.} \bibnamefont{Butler}},
  \bibinfo{year}{2012}, \bibinfo{journal}{Physical Review Letters}
  \textbf{\bibinfo{volume}{108}}, \bibinfo{pages}{208102}.

\bibitem[{\citenamefont{Gardner}(1988)}]{gardner_88}
\bibinfo{author}{\bibnamefont{Gardner}, \bibfnamefont{E.}},
  \bibinfo{year}{1988}, \bibinfo{journal}{Journal of Physics A: Mathematical
  and General} \textbf{\bibinfo{volume}{21}}, \bibinfo{pages}{257}.

\bibitem[{\citenamefont{Gardner and Derrida}(1988)}]{gardner+derrida_88}
\bibinfo{author}{\bibnamefont{Gardner}, \bibfnamefont{E.}}, and
  \bibinfo{author}{\bibfnamefont{B.}~\bibnamefont{Derrida}},
  \bibinfo{year}{1988}, \bibinfo{journal}{Journal of Physics A: Mathematical
  and General} \textbf{\bibinfo{volume}{21}}, \bibinfo{pages}{271}.

\bibitem[{\citenamefont{Gauthier} \emph{et~al.}(2009)\citenamefont{Gauthier,
  Field, Sher, Greschner, Shlens, Litke, and Chichilnisky}}]{gauthier+al_09}
\bibinfo{author}{\bibnamefont{Gauthier}, \bibfnamefont{J.~L.}},
  \bibinfo{author}{\bibfnamefont{G.~D.} \bibnamefont{Field}},
  \bibinfo{author}{\bibfnamefont{A.}~\bibnamefont{Sher}},
  \bibinfo{author}{\bibfnamefont{M.}~\bibnamefont{Greschner}},
  \bibinfo{author}{\bibfnamefont{J.}~\bibnamefont{Shlens}},
  \bibinfo{author}{\bibfnamefont{A.~M.} \bibnamefont{Litke}}, and
  \bibinfo{author}{\bibfnamefont{E.~J.} \bibnamefont{Chichilnisky}},
  \bibinfo{year}{2009}, \bibinfo{journal}{PLoS Biology}
  \textbf{\bibinfo{volume}{7}}, \bibinfo{pages}{e1000063}.

\bibitem[{\citenamefont{Gell-Mann and Low}(1954)}]{gell-mann+low_54}
\bibinfo{author}{\bibnamefont{Gell-Mann}, \bibfnamefont{M.}}, and
  \bibinfo{author}{\bibfnamefont{F.~E.} \bibnamefont{Low}},
  \bibinfo{year}{1954}, \bibinfo{journal}{Physical Review}
  \textbf{\bibinfo{volume}{95}}, \bibinfo{pages}{1300}.

\bibitem[{\citenamefont{Ghosh} \emph{et~al.}(2020)\citenamefont{Ghosh, Dixit,
  Agozzino, and Dill}}]{ghosh+al_20}
\bibinfo{author}{\bibnamefont{Ghosh}, \bibfnamefont{K.}},
  \bibinfo{author}{\bibfnamefont{P.~D.} \bibnamefont{Dixit}},
  \bibinfo{author}{\bibfnamefont{L.}~\bibnamefont{Agozzino}}, and
  \bibinfo{author}{\bibfnamefont{K.~A.} \bibnamefont{Dill}},
  \bibinfo{year}{2020}, \bibinfo{journal}{Annual Review of Physical Chemistry}
  \textbf{\bibinfo{volume}{71}}, \bibinfo{pages}{213}.

\bibitem[{\citenamefont{G\"obel} \emph{et~al.}(1994)\citenamefont{G\"obel,
  Sander, Scheiner, and Valencia}}]{Gobel+al_1994}
\bibinfo{author}{\bibnamefont{G\"obel}, \bibfnamefont{U.}},
  \bibinfo{author}{\bibfnamefont{C.}~\bibnamefont{Sander}},
  \bibinfo{author}{\bibfnamefont{R.}~\bibnamefont{Scheiner}}, and
  \bibinfo{author}{\bibfnamefont{A.}~\bibnamefont{Valencia}},
  \bibinfo{year}{1994}, \bibinfo{journal}{Proteins}
  \textbf{\bibinfo{volume}{18}}, \bibinfo{pages}{309}.

\bibitem[{\citenamefont{Gordon} \emph{et~al.}(2021)\citenamefont{Gordon,
  Banerjee, Koch-Janusz, and Ringel}}]{gordon+al2021}
\bibinfo{author}{\bibnamefont{Gordon}, \bibfnamefont{A.}},
  \bibinfo{author}{\bibfnamefont{A.}~\bibnamefont{Banerjee}},
  \bibinfo{author}{\bibfnamefont{M.}~\bibnamefont{Koch-Janusz}}, and
  \bibinfo{author}{\bibfnamefont{Z.}~\bibnamefont{Ringel}},
  \bibinfo{year}{2021}, \bibinfo{journal}{Physical Review Letters}
  \textbf{\bibinfo{volume}{126}}, \bibinfo{pages}{240601}.

\bibitem[{\citenamefont{Granot-Atedgi}
  \emph{et~al.}(2013)\citenamefont{Granot-Atedgi, Tka\v{c}ik, Segev, and
  Schneidman}}]{granot-atedgi+al2013}
\bibinfo{author}{\bibnamefont{Granot-Atedgi}, \bibfnamefont{E.}},
  \bibinfo{author}{\bibfnamefont{G.}~\bibnamefont{Tka\v{c}ik}},
  \bibinfo{author}{\bibfnamefont{R.}~\bibnamefont{Segev}}, and
  \bibinfo{author}{\bibfnamefont{E.}~\bibnamefont{Schneidman}},
  \bibinfo{year}{2013}, \bibinfo{journal}{PLoS Computational Biology}
  \textbf{\bibinfo{volume}{9}}, \bibinfo{pages}{e1002922}.

\bibitem[{\citenamefont{Hampson} \emph{et~al.}(1999)\citenamefont{Hampson,
  Simeral, and Deadwyler}}]{hampson+al_99}
\bibinfo{author}{\bibnamefont{Hampson}, \bibfnamefont{R.~E.}},
  \bibinfo{author}{\bibfnamefont{J.~D.} \bibnamefont{Simeral}}, and
  \bibinfo{author}{\bibfnamefont{S.~A.} \bibnamefont{Deadwyler}},
  \bibinfo{year}{1999}, \bibinfo{journal}{Nature}
  \textbf{\bibinfo{volume}{402}}, \bibinfo{pages}{610}.

\bibitem[{\citenamefont{Harte and Newman}(2014)}]{harte+newman_14}
\bibinfo{author}{\bibnamefont{Harte}, \bibfnamefont{J.}}, and
  \bibinfo{author}{\bibfnamefont{E.~A.} \bibnamefont{Newman}},
  \bibinfo{year}{2014}, \bibinfo{journal}{Trends in Ecology and Evolution}
  \textbf{\bibinfo{volume}{29}}, \bibinfo{pages}{384}.

\bibitem[{\citenamefont{Harte} \emph{et~al.}(2008)\citenamefont{Harte, Zillio,
  Conlisk, and Smith}}]{harte+al_08}
\bibinfo{author}{\bibnamefont{Harte}, \bibfnamefont{J.}},
  \bibinfo{author}{\bibfnamefont{T.}~\bibnamefont{Zillio}},
  \bibinfo{author}{\bibfnamefont{E.}~\bibnamefont{Conlisk}}, and
  \bibinfo{author}{\bibfnamefont{A.~B.} \bibnamefont{Smith}},
  \bibinfo{year}{2008}, \bibinfo{journal}{Ecology}
  \textbf{\bibinfo{volume}{89}}, \bibinfo{pages}{2700}.

\bibitem[{\citenamefont{Harvey} \emph{et~al.}(2009)\citenamefont{Harvey,
  Collman, Dombeck, and Tank}}]{harvey2009intracellular}
\bibinfo{author}{\bibnamefont{Harvey}, \bibfnamefont{C.~D.}},
  \bibinfo{author}{\bibfnamefont{F.}~\bibnamefont{Collman}},
  \bibinfo{author}{\bibfnamefont{D.~A.} \bibnamefont{Dombeck}}, and
  \bibinfo{author}{\bibfnamefont{D.~W.} \bibnamefont{Tank}},
  \bibinfo{year}{2009}, \bibinfo{journal}{Nature}
  \textbf{\bibinfo{volume}{461}}, \bibinfo{pages}{941}.

\bibitem[{\citenamefont{Hebb}(1949)}]{Hebb1949}
\bibinfo{author}{\bibnamefont{Hebb}, \bibfnamefont{D.~O.}},
  \bibinfo{year}{1949}, \emph{\bibinfo{title}{The Organization of Behavior: A
  Neuropsychological Theory}} (\bibinfo{publisher}{John Wiley and Sons, New
  York}).

\bibitem[{\citenamefont{Helmchen and Denk}(2005)}]{helmchen2005deep}
\bibinfo{author}{\bibnamefont{Helmchen}, \bibfnamefont{F.}}, and
  \bibinfo{author}{\bibfnamefont{W.}~\bibnamefont{Denk}}, \bibinfo{year}{2005},
  \bibinfo{journal}{Nature Methods} \textbf{\bibinfo{volume}{2}},
  \bibinfo{pages}{932}.

\bibitem[{\citenamefont{Hertz} \emph{et~al.}(1991)\citenamefont{Hertz, Krogh,
  and Palmer}}]{hertz+al_91}
\bibinfo{author}{\bibnamefont{Hertz}, \bibfnamefont{J.}},
  \bibinfo{author}{\bibfnamefont{A.}~\bibnamefont{Krogh}}, and
  \bibinfo{author}{\bibfnamefont{R.~G.} \bibnamefont{Palmer}},
  \bibinfo{year}{1991}, \emph{\bibinfo{title}{Introduction to the Theory of
  Neural Computation}} (\bibinfo{publisher}{Addison--Wesley, Redwood City}).

\bibitem[{\citenamefont{von Hippel and Berg}(1986)}]{hippel+berg_86}
\bibinfo{author}{\bibnamefont{von Hippel}, \bibfnamefont{P.~H.}}, and
  \bibinfo{author}{\bibfnamefont{O.~G.} \bibnamefont{Berg}},
  \bibinfo{year}{1986}, \bibinfo{journal}{Proceedings of the National Academy
  of Sciences (USA)} \textbf{\bibinfo{volume}{83}}, \bibinfo{pages}{1608}.

\bibitem[{\citenamefont{Hires} \emph{et~al.}(2015)\citenamefont{Hires,
  Gutnisky, Yu, O'Connor, and Svoboda}}]{hires+al_15}
\bibinfo{author}{\bibnamefont{Hires}, \bibfnamefont{S.~A.}},
  \bibinfo{author}{\bibfnamefont{D.~A.} \bibnamefont{Gutnisky}},
  \bibinfo{author}{\bibfnamefont{J.}~\bibnamefont{Yu}},
  \bibinfo{author}{\bibfnamefont{D.~H.} \bibnamefont{O'Connor}}, and
  \bibinfo{author}{\bibfnamefont{K.}~\bibnamefont{Svoboda}},
  \bibinfo{year}{2015}, \bibinfo{journal}{eLife} \textbf{\bibinfo{volume}{6}},
  \bibinfo{pages}{e00619}.

\bibitem[{\citenamefont{Hochberg} \emph{et~al.}(2012)\citenamefont{Hochberg,
  Bacher, Jarosiewicz, Masse, Simeral, Vogel, Haddadin, Liu, Cash, van~der
  Smagt, and Donoghue}}]{hochberg+al_12}
\bibinfo{author}{\bibnamefont{Hochberg}, \bibfnamefont{L.~R.}},
  \bibinfo{author}{\bibfnamefont{D.}~\bibnamefont{Bacher}},
  \bibinfo{author}{\bibfnamefont{B.}~\bibnamefont{Jarosiewicz}},
  \bibinfo{author}{\bibfnamefont{N.~Y.} \bibnamefont{Masse}},
  \bibinfo{author}{\bibfnamefont{J.~D.} \bibnamefont{Simeral}},
  \bibinfo{author}{\bibfnamefont{J.}~\bibnamefont{Vogel}},
  \bibinfo{author}{\bibfnamefont{S.}~\bibnamefont{Haddadin}},
  \bibinfo{author}{\bibfnamefont{J.}~\bibnamefont{Liu}},
  \bibinfo{author}{\bibfnamefont{S.~S.} \bibnamefont{Cash}},
  \bibinfo{author}{\bibfnamefont{P.}~\bibnamefont{van~der Smagt}}, and
  \bibinfo{author}{\bibfnamefont{J.~P.} \bibnamefont{Donoghue}},
  \bibinfo{year}{2012}, \bibinfo{journal}{Nature}
  \textbf{\bibinfo{volume}{485}}, \bibinfo{pages}{372}.

\bibitem[{\citenamefont{Hodgkin and Huxley}(1952)}]{Hodgkin+Huxley_1952}
\bibinfo{author}{\bibnamefont{Hodgkin}, \bibfnamefont{A.~L.}}, and
  \bibinfo{author}{\bibfnamefont{A.~F.} \bibnamefont{Huxley}},
  \bibinfo{year}{1952}, \bibinfo{journal}{Journal of Physiology (London)}
  \textbf{\bibinfo{volume}{117}}, \bibinfo{pages}{500}.

\bibitem[{\citenamefont{Hohenberg and Halperin}(1977)}]{hohenberg+halperin_77}
\bibinfo{author}{\bibnamefont{Hohenberg}, \bibfnamefont{P.~C.}}, and
  \bibinfo{author}{\bibfnamefont{B.~I.} \bibnamefont{Halperin}},
  \bibinfo{year}{1977}, \bibinfo{journal}{Reviews of Modern Physics}
  \textbf{\bibinfo{volume}{49}}, \bibinfo{pages}{435}.

\bibitem[{\citenamefont{Hopfield}(1982)}]{hopfield1982}
\bibinfo{author}{\bibnamefont{Hopfield}, \bibfnamefont{J.~J.}},
  \bibinfo{year}{1982}, \bibinfo{journal}{Proceedings of the National Academy
  of Sciences (USA)} \textbf{\bibinfo{volume}{79}}, \bibinfo{pages}{2554}.

\bibitem[{\citenamefont{Hopfield}(1984)}]{hopfield1984}
\bibinfo{author}{\bibnamefont{Hopfield}, \bibfnamefont{J.~J.}},
  \bibinfo{year}{1984}, \bibinfo{journal}{Proceedings of the National Academy
  of Sciences (USA)} \textbf{\bibinfo{volume}{81}}, \bibinfo{pages}{3088}.

\bibitem[{\citenamefont{Hopfield and Tank}(1985)}]{hopfield+tank1985}
\bibinfo{author}{\bibnamefont{Hopfield}, \bibfnamefont{J.~J.}}, and
  \bibinfo{author}{\bibfnamefont{D.~W.} \bibnamefont{Tank}},
  \bibinfo{year}{1985}, \bibinfo{journal}{Biological Cybernetics}
  \textbf{\bibinfo{volume}{52}}, \bibinfo{pages}{141}.

\bibitem[{\citenamefont{Hopfield and Tank}(1986)}]{hopfield+tank1986}
\bibinfo{author}{\bibnamefont{Hopfield}, \bibfnamefont{J.~J.}}, and
  \bibinfo{author}{\bibfnamefont{D.~W.} \bibnamefont{Tank}},
  \bibinfo{year}{1986}, \bibinfo{journal}{Science}
  \textbf{\bibinfo{volume}{233}}, \bibinfo{pages}{625}.

\bibitem[{\citenamefont{Hornik} \emph{et~al.}(1989)\citenamefont{Hornik,
  Stinchcombe, and White}}]{hornik+al_89}
\bibinfo{author}{\bibnamefont{Hornik}, \bibfnamefont{K.}},
  \bibinfo{author}{\bibfnamefont{M.}~\bibnamefont{Stinchcombe}}, and
  \bibinfo{author}{\bibfnamefont{H.}~\bibnamefont{White}},
  \bibinfo{year}{1989}, \bibinfo{journal}{Neural Networks}
  \textbf{\bibinfo{volume}{2}}, \bibinfo{pages}{359}.

\bibitem[{\citenamefont{Hoshal} \emph{et~al.}(2023)\citenamefont{Hoshal,
  Holmes, Bojanek, Salisbury, Berry~II, Marre, and Palmer}}]{hoshal+al_23}
\bibinfo{author}{\bibnamefont{Hoshal}, \bibfnamefont{B.~D.}},
  \bibinfo{author}{\bibfnamefont{C.~M.} \bibnamefont{Holmes}},
  \bibinfo{author}{\bibfnamefont{K.}~\bibnamefont{Bojanek}},
  \bibinfo{author}{\bibfnamefont{J.}~\bibnamefont{Salisbury}},
  \bibinfo{author}{\bibfnamefont{M.~J.} \bibnamefont{Berry~II}},
  \bibinfo{author}{\bibfnamefont{O.}~\bibnamefont{Marre}}, and
  \bibinfo{author}{\bibfnamefont{S.~E.} \bibnamefont{Palmer}},
  \bibinfo{year}{2023}, \eprint{bioRxiv:2023.08.08.552526}.

\bibitem[{\citenamefont{Ibarra} \emph{et~al.}(2002)\citenamefont{Ibarra,
  Edwards, and Palsson}}]{ibarra+al_02}
\bibinfo{author}{\bibnamefont{Ibarra}, \bibfnamefont{R.~U.}},
  \bibinfo{author}{\bibfnamefont{J.~S.} \bibnamefont{Edwards}}, and
  \bibinfo{author}{\bibfnamefont{B.~O.} \bibnamefont{Palsson}},
  \bibinfo{year}{2002}, \bibinfo{journal}{Nature}
  \textbf{\bibinfo{volume}{420}}, \bibinfo{pages}{186}.

\bibitem[{\citenamefont{James}(1904)}]{James1904}
\bibinfo{author}{\bibnamefont{James}, \bibfnamefont{W.}}, \bibinfo{year}{1904},
  \emph{\bibinfo{title}{Psychology: Briefer Course}} (\bibinfo{publisher}{Henry
  Holt, New York}).

\bibitem[{\citenamefont{Jaynes}(1957)}]{jaynes1957information}
\bibinfo{author}{\bibnamefont{Jaynes}, \bibfnamefont{E.~T.}},
  \bibinfo{year}{1957}, \bibinfo{journal}{Physical Review}
  \textbf{\bibinfo{volume}{106}}, \bibinfo{pages}{620}.

\bibitem[{\citenamefont{Jaynes}(1982)}]{jaynes1982rationale}
\bibinfo{author}{\bibnamefont{Jaynes}, \bibfnamefont{E.~T.}},
  \bibinfo{year}{1982}, \bibinfo{journal}{Proceedings of the IEEE}
  \textbf{\bibinfo{volume}{70}}, \bibinfo{pages}{939}.

\bibitem[{\citenamefont{Jin} \emph{et~al.}(2012)\citenamefont{Jin, Han,
  Platisa, Wooltorton, Cohen, and Pieribone}}]{Jin+al2012}
\bibinfo{author}{\bibnamefont{Jin}, \bibfnamefont{L.}},
  \bibinfo{author}{\bibfnamefont{Z.}~\bibnamefont{Han}},
  \bibinfo{author}{\bibfnamefont{J.}~\bibnamefont{Platisa}},
  \bibinfo{author}{\bibfnamefont{J.~R.~A.} \bibnamefont{Wooltorton}},
  \bibinfo{author}{\bibfnamefont{L.~B.} \bibnamefont{Cohen}}, and
  \bibinfo{author}{\bibfnamefont{V.~A.} \bibnamefont{Pieribone}},
  \bibinfo{year}{2012}, \bibinfo{journal}{Neuron}
  \textbf{\bibinfo{volume}{75}}, \bibinfo{pages}{779}.

\bibitem[{\citenamefont{Johnson} \emph{et~al.}(1962)\citenamefont{Johnson,
  Shimomura, Saiga, Gershman, Reynolds, and Waters~Jr.}}]{johnson+al_62}
\bibinfo{author}{\bibnamefont{Johnson}, \bibfnamefont{F.~H.}},
  \bibinfo{author}{\bibfnamefont{O.}~\bibnamefont{Shimomura}},
  \bibinfo{author}{\bibfnamefont{Y.}~\bibnamefont{Saiga}},
  \bibinfo{author}{\bibfnamefont{L.~C.} \bibnamefont{Gershman}},
  \bibinfo{author}{\bibfnamefont{G.~T.} \bibnamefont{Reynolds}}, and
  \bibinfo{author}{\bibfnamefont{J.~R.} \bibnamefont{Waters~Jr.}},
  \bibinfo{year}{1962}, \bibinfo{journal}{Journal of Cellular and Comparative
  Physiology} \textbf{\bibinfo{volume}{60}}, \bibinfo{pages}{85}.

\bibitem[{\citenamefont{Jona-Lasinio}(1975)}]{jona-lasinio_75}
\bibinfo{author}{\bibnamefont{Jona-Lasinio}, \bibfnamefont{G.}},
  \bibinfo{year}{1975}, \bibinfo{journal}{Il Nuovo Cimento}
  \textbf{\bibinfo{volume}{26B}}, \bibinfo{pages}{99}.

\bibitem[{\citenamefont{Jones} \emph{et~al.}(1992)\citenamefont{Jones,
  Campbell, and Normann}}]{jones+al_92}
\bibinfo{author}{\bibnamefont{Jones}, \bibfnamefont{K.~E.}},
  \bibinfo{author}{\bibfnamefont{P.~K.} \bibnamefont{Campbell}}, and
  \bibinfo{author}{\bibfnamefont{R.~A.} \bibnamefont{Normann}},
  \bibinfo{year}{1992}, \bibinfo{journal}{Annals of Biomedical Engineering}
  \textbf{\bibinfo{volume}{20}}, \bibinfo{pages}{423}.

\bibitem[{\citenamefont{Joshua and Lisberger}(2015)}]{joshua+lisberger_15}
\bibinfo{author}{\bibnamefont{Joshua}, \bibfnamefont{M.}}, and
  \bibinfo{author}{\bibfnamefont{S.~G.} \bibnamefont{Lisberger}},
  \bibinfo{year}{2015}, \bibinfo{journal}{Neuroscience}
  \textbf{\bibinfo{volume}{296}}, \bibinfo{pages}{80}.

\bibitem[{\citenamefont{Jumper} \emph{et~al.}(2021)\citenamefont{Jumper, Evans,
  Pritzel, Green, Figurnov, Ronneberger, Tunyasuvunakool, Bates, \v{Z}ídek,
  Potapenko, Bridgland, Meyer} \emph{et~al.}}]{Jumper+al_2021}
\bibinfo{author}{\bibnamefont{Jumper}, \bibfnamefont{J.}},
  \bibinfo{author}{\bibfnamefont{R.}~\bibnamefont{Evans}},
  \bibinfo{author}{\bibfnamefont{A.}~\bibnamefont{Pritzel}},
  \bibinfo{author}{\bibfnamefont{T.}~\bibnamefont{Green}},
  \bibinfo{author}{\bibfnamefont{M.}~\bibnamefont{Figurnov}},
  \bibinfo{author}{\bibfnamefont{O.}~\bibnamefont{Ronneberger}},
  \bibinfo{author}{\bibfnamefont{K.}~\bibnamefont{Tunyasuvunakool}},
  \bibinfo{author}{\bibfnamefont{R.}~\bibnamefont{Bates}},
  \bibinfo{author}{\bibfnamefont{A.}~\bibnamefont{\v{Z}ídek}},
  \bibinfo{author}{\bibfnamefont{A.}~\bibnamefont{Potapenko}},
  \bibinfo{author}{\bibfnamefont{A.}~\bibnamefont{Bridgland}},
  \bibinfo{author}{\bibfnamefont{C.}~\bibnamefont{Meyer}}, \emph{et~al.},
  \bibinfo{year}{2021}, \bibinfo{journal}{Nature}
  \textbf{\bibinfo{volume}{596}}, \bibinfo{pages}{583}.

\bibitem[{\citenamefont{Jun} \emph{et~al.}(2017)\citenamefont{Jun, Steinmetz,
  Siegle, Denman, Bauza, Barbarits, Lee, Anastassiou, Andrei, Aydın, Barbic,
  Blanche} \emph{et~al.}}]{jun+al_17}
\bibinfo{author}{\bibnamefont{Jun}, \bibfnamefont{J.~J.}},
  \bibinfo{author}{\bibfnamefont{N.~A.} \bibnamefont{Steinmetz}},
  \bibinfo{author}{\bibfnamefont{J.~H.} \bibnamefont{Siegle}},
  \bibinfo{author}{\bibfnamefont{D.~J.} \bibnamefont{Denman}},
  \bibinfo{author}{\bibfnamefont{M.}~\bibnamefont{Bauza}},
  \bibinfo{author}{\bibfnamefont{B.}~\bibnamefont{Barbarits}},
  \bibinfo{author}{\bibfnamefont{A.~K.} \bibnamefont{Lee}},
  \bibinfo{author}{\bibfnamefont{C.~A.} \bibnamefont{Anastassiou}},
  \bibinfo{author}{\bibfnamefont{A.}~\bibnamefont{Andrei}},
  \bibinfo{author}{\bibfnamefont{C.}~\bibnamefont{Aydın}},
  \bibinfo{author}{\bibfnamefont{M.}~\bibnamefont{Barbic}},
  \bibinfo{author}{\bibfnamefont{T.~J.} \bibnamefont{Blanche}}, \emph{et~al.},
  \bibinfo{year}{2017}, \bibinfo{journal}{Nature}
  \textbf{\bibinfo{volume}{551}}, \bibinfo{pages}{232}.

\bibitem[{\citenamefont{Kadanoff}(1966)}]{kadanoff_66}
\bibinfo{author}{\bibnamefont{Kadanoff}, \bibfnamefont{L.~P.}},
  \bibinfo{year}{1966}, \bibinfo{journal}{Physics}
  \textbf{\bibinfo{volume}{2}}, \bibinfo{pages}{263}.

\bibitem[{\citenamefont{Kaifosh} \emph{et~al.}(2014)\citenamefont{Kaifosh,
  Zaremba, Danielson, and Losonczy}}]{Kaifosh2014}
\bibinfo{author}{\bibnamefont{Kaifosh}, \bibfnamefont{P.}},
  \bibinfo{author}{\bibfnamefont{J.~D.} \bibnamefont{Zaremba}},
  \bibinfo{author}{\bibfnamefont{N.~B.} \bibnamefont{Danielson}}, and
  \bibinfo{author}{\bibfnamefont{A.}~\bibnamefont{Losonczy}},
  \bibinfo{year}{2014}, \bibinfo{journal}{Frontiers in Neuroinformatics}
  \textbf{\bibinfo{volume}{8}}, \bibinfo{pages}{80}.

\bibitem[{\citenamefont{Kandel} \emph{et~al.}(2012)\citenamefont{Kandel,
  Schwarts, Jessell, Siegelbaum, and Hudspeth}}]{Kandel+al_2001}
\bibinfo{author}{\bibnamefont{Kandel}, \bibfnamefont{E.~R.}},
  \bibinfo{author}{\bibfnamefont{J.~H.} \bibnamefont{Schwarts}},
  \bibinfo{author}{\bibfnamefont{T.~M.} \bibnamefont{Jessell}},
  \bibinfo{author}{\bibfnamefont{S.~A.} \bibnamefont{Siegelbaum}}, and
  \bibinfo{author}{\bibfnamefont{A.~J.} \bibnamefont{Hudspeth}},
  \bibinfo{year}{2012}, \emph{\bibinfo{title}{Principles of Neural Science,
  Fifth Edition}} (\bibinfo{publisher}{McGraw--Hill, New York}).

\bibitem[{\citenamefont{Keller and Zumino}(1959)}]{keller+zumino1959}
\bibinfo{author}{\bibnamefont{Keller}, \bibfnamefont{J.~B.}}, and
  \bibinfo{author}{\bibfnamefont{B.}~\bibnamefont{Zumino}},
  \bibinfo{year}{1959}, \bibinfo{journal}{Journal of Chemical Physics}
  \textbf{\bibinfo{volume}{30}}, \bibinfo{pages}{1351}.

\bibitem[{\citenamefont{Kirkpatrick}
  \emph{et~al.}(1983)\citenamefont{Kirkpatrick, Gelatt~Jr., and
  Vecchi}}]{kirkpatrick+al_83}
\bibinfo{author}{\bibnamefont{Kirkpatrick}, \bibfnamefont{S.}},
  \bibinfo{author}{\bibfnamefont{C.~D.} \bibnamefont{Gelatt~Jr.}}, and
  \bibinfo{author}{\bibfnamefont{M.~P.} \bibnamefont{Vecchi}},
  \bibinfo{year}{1983}, \bibinfo{journal}{Science}
  \textbf{\bibinfo{volume}{220}}, \bibinfo{pages}{671}.

\bibitem[{\citenamefont{Kirkpatrick and Selman}(1994)}]{kirkpatrick+selman_94}
\bibinfo{author}{\bibnamefont{Kirkpatrick}, \bibfnamefont{S.}}, and
  \bibinfo{author}{\bibfnamefont{B.}~\bibnamefont{Selman}},
  \bibinfo{year}{1994}, \bibinfo{journal}{Science}
  \textbf{\bibinfo{volume}{264}}, \bibinfo{pages}{1297}.

\bibitem[{\citenamefont{Kivelson} \emph{et~al.}(2024)\citenamefont{Kivelson,
  Jiang, and Chang}}]{kivelson+al_24}
\bibinfo{author}{\bibnamefont{Kivelson}, \bibfnamefont{S.~A.}},
  \bibinfo{author}{\bibfnamefont{J.~M.} \bibnamefont{Jiang}}, and
  \bibinfo{author}{\bibfnamefont{J.}~\bibnamefont{Chang}},
  \bibinfo{year}{2024}, \emph{\bibinfo{title}{Statistical Mechanics of Phases
  and Phase Transitions}} (\bibinfo{publisher}{Princeton University Press,
  Princeton}).

\bibitem[{\citenamefont{Kline and Palmer}(2022)}]{kline+palmer2022}
\bibinfo{author}{\bibnamefont{Kline}, \bibfnamefont{A.~G.}}, and
  \bibinfo{author}{\bibfnamefont{S.~E.} \bibnamefont{Palmer}},
  \bibinfo{year}{2022}, \bibinfo{journal}{New Journal of Physics}
  \textbf{\bibinfo{volume}{24}}(\bibinfo{number}{3}), \bibinfo{pages}{033007}.

\bibitem[{\citenamefont{Koch-Janusz and
  Ringel}(2018)}]{koch--janusz+ringel2018}
\bibinfo{author}{\bibnamefont{Koch-Janusz}, \bibfnamefont{M.}}, and
  \bibinfo{author}{\bibfnamefont{Z.}~\bibnamefont{Ringel}},
  \bibinfo{year}{2018}, \bibinfo{journal}{Nature Physics}
  \textbf{\bibinfo{volume}{14}}, \bibinfo{pages}{578}.

\bibitem[{\citenamefont{Krause and Ruxton}(2002)}]{Krause+Ruxton_2002}
\bibinfo{author}{\bibnamefont{Krause}, \bibfnamefont{J.}}, and
  \bibinfo{author}{\bibfnamefont{G.~D.} \bibnamefont{Ruxton}},
  \bibinfo{year}{2002}, \emph{\bibinfo{title}{Living in Groups}}
  (\bibinfo{publisher}{Oxford University Press, Oxford}).

\bibitem[{\citenamefont{Krishnamurthy}
  \emph{et~al.}(2022)\citenamefont{Krishnamurthy, Can, and
  Schwab}}]{krishnamurthy+al_20}
\bibinfo{author}{\bibnamefont{Krishnamurthy}, \bibfnamefont{K.}},
  \bibinfo{author}{\bibfnamefont{T.}~\bibnamefont{Can}}, and
  \bibinfo{author}{\bibfnamefont{D.~J.} \bibnamefont{Schwab}},
  \bibinfo{year}{2022}, \bibinfo{journal}{Physical Review X}
  \textbf{\bibinfo{volume}{12}}, \bibinfo{pages}{011011}.

\bibitem[{\citenamefont{Kunkin and Firsch}(1969)}]{kunkin+frisch_69}
\bibinfo{author}{\bibnamefont{Kunkin}, \bibfnamefont{W.}}, and
  \bibinfo{author}{\bibfnamefont{H.~W.} \bibnamefont{Firsch}},
  \bibinfo{year}{1969}, \bibinfo{journal}{Physical Review}
  \textbf{\bibinfo{volume}{177}}, \bibinfo{pages}{282}.

\bibitem[{\citenamefont{Landau and Lifshitz}(1977)}]{landau+lifshitz}
\bibinfo{author}{\bibnamefont{Landau}, \bibfnamefont{L.~D.}}, and
  \bibinfo{author}{\bibfnamefont{E.~M.} \bibnamefont{Lifshitz}},
  \bibinfo{year}{1977}, \emph{\bibinfo{title}{Statistical Physics}}
  (\bibinfo{publisher}{Pergamon Press, Oxford}).

\bibitem[{\citenamefont{Lapedes and Farber}(1988)}]{LapedesFarber1988}
\bibinfo{author}{\bibnamefont{Lapedes}, \bibfnamefont{A.}}, and
  \bibinfo{author}{\bibfnamefont{R.}~\bibnamefont{Farber}},
  \bibinfo{year}{1988}, in \emph{\bibinfo{booktitle}{Neural Information
  Processing Systems}}, edited by \bibinfo{editor}{\bibfnamefont{D.~Z.}
  \bibnamefont{Anderson}} (\bibinfo{publisher}{American Institute of Physics,
  New York}), pp. \bibinfo{pages}{442--456}.

\bibitem[{\citenamefont{Lapedes} \emph{et~al.}(2012)\citenamefont{Lapedes,
  Giraud, and Jarzynski}}]{lapedes2012using}
\bibinfo{author}{\bibnamefont{Lapedes}, \bibfnamefont{A.}},
  \bibinfo{author}{\bibfnamefont{B.}~\bibnamefont{Giraud}}, and
  \bibinfo{author}{\bibfnamefont{C.}~\bibnamefont{Jarzynski}},
  \bibinfo{year}{2012}, \eprint{arXiv:1207.2484}.

\bibitem[{\citenamefont{Lapedes} \emph{et~al.}(1998)\citenamefont{Lapedes,
  Giraud, Liu, and Stormo}}]{lapedes+al_98}
\bibinfo{author}{\bibnamefont{Lapedes}, \bibfnamefont{A.~S.}},
  \bibinfo{author}{\bibfnamefont{B.~G.} \bibnamefont{Giraud}},
  \bibinfo{author}{\bibfnamefont{L.~C.} \bibnamefont{Liu}}, and
  \bibinfo{author}{\bibfnamefont{G.~D.} \bibnamefont{Stormo}},
  \bibinfo{year}{1998}, \bibinfo{journal}{Los Alamos National Laboratory Report
  LA--UR--98--1094} .

\bibitem[{\citenamefont{LeCun}(1987)}]{LeCun1987}
\bibinfo{author}{\bibnamefont{LeCun}, \bibfnamefont{Y.}}, \bibinfo{year}{1987},
  \emph{\bibinfo{title}{Mod\`eles Connexionnistes de l’Apprentissage}}, Ph.D.
  thesis, \bibinfo{school}{Universit\'e Pierre et Marie Curie}.

\bibitem[{\citenamefont{LeCun} \emph{et~al.}(2015)\citenamefont{LeCun, Bengio,
  and Hinton}}]{lecun2015deep}
\bibinfo{author}{\bibnamefont{LeCun}, \bibfnamefont{Y.}},
  \bibinfo{author}{\bibfnamefont{Y.}~\bibnamefont{Bengio}}, and
  \bibinfo{author}{\bibfnamefont{G.}~\bibnamefont{Hinton}},
  \bibinfo{year}{2015}, \bibinfo{journal}{Nature}
  \textbf{\bibinfo{volume}{521}}, \bibinfo{pages}{436}.

\bibitem[{\citenamefont{Lee and Daniels}(2019)}]{lee+daniels_19}
\bibinfo{author}{\bibnamefont{Lee}, \bibfnamefont{E.~D.}}, and
  \bibinfo{author}{\bibfnamefont{B.~C.} \bibnamefont{Daniels}},
  \bibinfo{year}{2019}, \bibinfo{journal}{Journal of Open Research Software}
  \textbf{\bibinfo{volume}{7}}, \bibinfo{pages}{3}.

\bibitem[{\citenamefont{Lee} \emph{et~al.}(2024)\citenamefont{Lee, Liu, Croker,
  Huggins, Tikhonov, Mani, and Kuehn}}]{lee+al_24}
\bibinfo{author}{\bibnamefont{Lee}, \bibfnamefont{K.~K.}},
  \bibinfo{author}{\bibfnamefont{S.}~\bibnamefont{Liu}},
  \bibinfo{author}{\bibfnamefont{K.}~\bibnamefont{Croker}},
  \bibinfo{author}{\bibfnamefont{D.~R.} \bibnamefont{Huggins}},
  \bibinfo{author}{\bibfnamefont{M.}~\bibnamefont{Tikhonov}},
  \bibinfo{author}{\bibfnamefont{M.}~\bibnamefont{Mani}}, and
  \bibinfo{author}{\bibfnamefont{S.}~\bibnamefont{Kuehn}},
  \bibinfo{year}{2024}, \eprint{bioRxiv:2024.03.15.584851}.

\bibitem[{\citenamefont{Leifer} \emph{et~al.}(2011)\citenamefont{Leifer,
  Fang-Yen, Gershow, Alema, and Samuel}}]{leifer+al_11}
\bibinfo{author}{\bibnamefont{Leifer}, \bibfnamefont{A.~M.}},
  \bibinfo{author}{\bibfnamefont{C.}~\bibnamefont{Fang-Yen}},
  \bibinfo{author}{\bibfnamefont{M.}~\bibnamefont{Gershow}},
  \bibinfo{author}{\bibfnamefont{M.~J.} \bibnamefont{Alema}}, and
  \bibinfo{author}{\bibfnamefont{A.~D.~T.} \bibnamefont{Samuel}},
  \bibinfo{year}{2011}, \bibinfo{journal}{Nature Methods}
  \textbf{\bibinfo{volume}{8}}, \bibinfo{pages}{147}.

\bibitem[{\citenamefont{Levin} \emph{et~al.}(1990)\citenamefont{Levin, Tishby,
  and Solla}}]{levin+al_90}
\bibinfo{author}{\bibnamefont{Levin}, \bibfnamefont{E.}},
  \bibinfo{author}{\bibfnamefont{N.}~\bibnamefont{Tishby}}, and
  \bibinfo{author}{\bibfnamefont{S.~A.} \bibnamefont{Solla}},
  \bibinfo{year}{1990}, \bibinfo{journal}{Proceedings of the IEEE}
  \textbf{\bibinfo{volume}{78}}, \bibinfo{pages}{1568}.

\bibitem[{\citenamefont{Lezon} \emph{et~al.}(2006)\citenamefont{Lezon, Banavar,
  Cieplak, Maritan, and Fedoroff}}]{lezon+al_06}
\bibinfo{author}{\bibnamefont{Lezon}, \bibfnamefont{T.~R.}},
  \bibinfo{author}{\bibfnamefont{J.~R.} \bibnamefont{Banavar}},
  \bibinfo{author}{\bibfnamefont{M.}~\bibnamefont{Cieplak}},
  \bibinfo{author}{\bibfnamefont{A.}~\bibnamefont{Maritan}}, and
  \bibinfo{author}{\bibfnamefont{N.~V.} \bibnamefont{Fedoroff}},
  \bibinfo{year}{2006}, \bibinfo{journal}{Proceedings of the National Academy
  of Sciences (USA)} \textbf{\bibinfo{volume}{103}}, \bibinfo{pages}{19033}.

\bibitem[{\citenamefont{Li} \emph{et~al.}(1996)\citenamefont{Li, Helling, Tang,
  and Wingreen}}]{li+al_96}
\bibinfo{author}{\bibnamefont{Li}, \bibfnamefont{H.}},
  \bibinfo{author}{\bibfnamefont{R.}~\bibnamefont{Helling}},
  \bibinfo{author}{\bibfnamefont{C.}~\bibnamefont{Tang}}, and
  \bibinfo{author}{\bibfnamefont{N.~S.} \bibnamefont{Wingreen}},
  \bibinfo{year}{1996}, \bibinfo{journal}{Science}
  \textbf{\bibinfo{volume}{273}}, \bibinfo{pages}{666}.

\bibitem[{\citenamefont{Lipa} \emph{et~al.}(1996)\citenamefont{Lipa, Swanson,
  Nissen, Chui, and Israelsson}}]{lipa+al_96}
\bibinfo{author}{\bibnamefont{Lipa}, \bibfnamefont{J.~A.}},
  \bibinfo{author}{\bibfnamefont{D.~R.} \bibnamefont{Swanson}},
  \bibinfo{author}{\bibfnamefont{J.~A.} \bibnamefont{Nissen}},
  \bibinfo{author}{\bibfnamefont{T.~C.~P.} \bibnamefont{Chui}}, and
  \bibinfo{author}{\bibfnamefont{U.~E.} \bibnamefont{Israelsson}},
  \bibinfo{year}{1996}, \bibinfo{journal}{Physical Review Letters}
  \textbf{\bibinfo{volume}{76}}, \bibinfo{pages}{944}.

\bibitem[{\citenamefont{Litke} \emph{et~al.}(2004)\citenamefont{Litke,
  Bezayiff, Chichilnisky, Cunningham, Dabrowski, Grillo, Grivich, Grybos,
  Hottowy, Kachiguine, Kalmar, Mathieson} \emph{et~al.}}]{litke+al2004}
\bibinfo{author}{\bibnamefont{Litke}, \bibfnamefont{A.~M.}},
  \bibinfo{author}{\bibfnamefont{N.}~\bibnamefont{Bezayiff}},
  \bibinfo{author}{\bibfnamefont{E.~J.} \bibnamefont{Chichilnisky}},
  \bibinfo{author}{\bibfnamefont{W.}~\bibnamefont{Cunningham}},
  \bibinfo{author}{\bibfnamefont{W.}~\bibnamefont{Dabrowski}},
  \bibinfo{author}{\bibfnamefont{A.~A.} \bibnamefont{Grillo}},
  \bibinfo{author}{\bibfnamefont{M.}~\bibnamefont{Grivich}},
  \bibinfo{author}{\bibfnamefont{P.}~\bibnamefont{Grybos}},
  \bibinfo{author}{\bibfnamefont{P.}~\bibnamefont{Hottowy}},
  \bibinfo{author}{\bibfnamefont{S.}~\bibnamefont{Kachiguine}},
  \bibinfo{author}{\bibfnamefont{R.~S.} \bibnamefont{Kalmar}},
  \bibinfo{author}{\bibfnamefont{K.}~\bibnamefont{Mathieson}}, \emph{et~al.},
  \bibinfo{year}{2004}, \bibinfo{journal}{IEEE Transactions on Nuclear Science}
  \textbf{\bibinfo{volume}{51}}, \bibinfo{pages}{1434}.

\bibitem[{\citenamefont{Little}(1974)}]{little_74}
\bibinfo{author}{\bibnamefont{Little}, \bibfnamefont{W.~A.}},
  \bibinfo{year}{1974}, \bibinfo{journal}{Mathematical Biosciences}
  \textbf{\bibinfo{volume}{19}}, \bibinfo{pages}{101}.

\bibitem[{\citenamefont{Little and Shaw}(1975)}]{little+shaw_75}
\bibinfo{author}{\bibnamefont{Little}, \bibfnamefont{W.~A.}}, and
  \bibinfo{author}{\bibfnamefont{G.~L.} \bibnamefont{Shaw}},
  \bibinfo{year}{1975}, \bibinfo{journal}{Behavioural Biology}
  \textbf{\bibinfo{volume}{14}}, \bibinfo{pages}{115}.

\bibitem[{\citenamefont{Little and Shaw}(1978)}]{little+shaw_78}
\bibinfo{author}{\bibnamefont{Little}, \bibfnamefont{W.~A.}}, and
  \bibinfo{author}{\bibfnamefont{G.~L.} \bibnamefont{Shaw}},
  \bibinfo{year}{1978}, \bibinfo{journal}{Mathematical Biosciences}
  \textbf{\bibinfo{volume}{39}}, \bibinfo{pages}{281}.

\bibitem[{\citenamefont{Liu} \emph{et~al.}(2009)\citenamefont{Liu, Stevens, and
  Sharpee}}]{liu+al_09}
\bibinfo{author}{\bibnamefont{Liu}, \bibfnamefont{Y.~S.}},
  \bibinfo{author}{\bibfnamefont{C.~F.} \bibnamefont{Stevens}}, and
  \bibinfo{author}{\bibfnamefont{T.~O.} \bibnamefont{Sharpee}},
  \bibinfo{year}{2009}, \bibinfo{journal}{Proceedings of the National Academy
  of Sciences (USA)} \textbf{\bibinfo{volume}{106}}, \bibinfo{pages}{16499}.

\bibitem[{\citenamefont{Loback} \emph{et~al.}(2017)\citenamefont{Loback,
  Prentice, Ioffe, and Berry~II}}]{loback+al_17}
\bibinfo{author}{\bibnamefont{Loback}, \bibfnamefont{A.}},
  \bibinfo{author}{\bibfnamefont{J.}~\bibnamefont{Prentice}},
  \bibinfo{author}{\bibfnamefont{M.}~\bibnamefont{Ioffe}}, and
  \bibinfo{author}{\bibfnamefont{M.~J.} \bibnamefont{Berry~II}},
  \bibinfo{year}{2017}, \bibinfo{journal}{Neural Computation}
  \textbf{\bibinfo{volume}{29}}, \bibinfo{pages}{3119}.

\bibitem[{\citenamefont{Lynn}
  \emph{et~al.}(2022{\natexlab{a}})\citenamefont{Lynn, Holmes, Bialek, and
  Schwab}}]{lynn+al_22a}
\bibinfo{author}{\bibnamefont{Lynn}, \bibfnamefont{C.~W.}},
  \bibinfo{author}{\bibfnamefont{C.~M.} \bibnamefont{Holmes}},
  \bibinfo{author}{\bibfnamefont{W.}~\bibnamefont{Bialek}}, and
  \bibinfo{author}{\bibfnamefont{D.~J.} \bibnamefont{Schwab}},
  \bibinfo{year}{2022}{\natexlab{a}}, \bibinfo{journal}{Physical Review
  Letters} \textbf{\bibinfo{volume}{129}}, \bibinfo{pages}{118101}.

\bibitem[{\citenamefont{Lynn}
  \emph{et~al.}(2022{\natexlab{b}})\citenamefont{Lynn, Holmes, Bialek, and
  Schwab}}]{lynn+al_22b}
\bibinfo{author}{\bibnamefont{Lynn}, \bibfnamefont{C.~W.}},
  \bibinfo{author}{\bibfnamefont{C.~M.} \bibnamefont{Holmes}},
  \bibinfo{author}{\bibfnamefont{W.}~\bibnamefont{Bialek}}, and
  \bibinfo{author}{\bibfnamefont{D.~J.} \bibnamefont{Schwab}},
  \bibinfo{year}{2022}{\natexlab{b}}, \bibinfo{journal}{Physical Review E}
  \textbf{\bibinfo{volume}{106}}, \bibinfo{pages}{034102}.

\bibitem[{\citenamefont{Lynn} \emph{et~al.}(2023)\citenamefont{Lynn, Yu, Pang,
  Bialek, and Palmer}}]{lynn+al_23a}
\bibinfo{author}{\bibnamefont{Lynn}, \bibfnamefont{C.~W.}},
  \bibinfo{author}{\bibfnamefont{Q.}~\bibnamefont{Yu}},
  \bibinfo{author}{\bibfnamefont{R.}~\bibnamefont{Pang}},
  \bibinfo{author}{\bibfnamefont{W.}~\bibnamefont{Bialek}}, and
  \bibinfo{author}{\bibfnamefont{S.~E.} \bibnamefont{Palmer}},
  \bibinfo{year}{2023}, \eprint{arXiv:2310.10860}.

\bibitem[{\citenamefont{Lynn} \emph{et~al.}(2024)\citenamefont{Lynn, Yu, Pang,
  E., and Bialek}}]{lynn+al_24}
\bibinfo{author}{\bibnamefont{Lynn}, \bibfnamefont{C.~W.}},
  \bibinfo{author}{\bibfnamefont{Q.}~\bibnamefont{Yu}},
  \bibinfo{author}{\bibfnamefont{R.}~\bibnamefont{Pang}},
  \bibinfo{author}{\bibfnamefont{S.}~\bibnamefont{E.}}, and
  \bibinfo{author}{\bibfnamefont{W.}~\bibnamefont{Bialek}},
  \bibinfo{year}{2024}, \eprint{arXiv:2402.00007}.

\bibitem[{\citenamefont{Maass} \emph{et~al.}(2002)\citenamefont{Maass,
  Natschl\"ager, and Markram}}]{maass+al_02}
\bibinfo{author}{\bibnamefont{Maass}, \bibfnamefont{W.}},
  \bibinfo{author}{\bibfnamefont{T.}~\bibnamefont{Natschl\"ager}}, and
  \bibinfo{author}{\bibfnamefont{H.}~\bibnamefont{Markram}},
  \bibinfo{year}{2002}, \bibinfo{journal}{Neural Computation}
  \textbf{\bibinfo{volume}{14}}, \bibinfo{pages}{2531}.

\bibitem[{\citenamefont{Machta} \emph{et~al.}(2013)\citenamefont{Machta,
  Chachra, Transtrum, and Sethna}}]{machta+al_13}
\bibinfo{author}{\bibnamefont{Machta}, \bibfnamefont{B.~B.}},
  \bibinfo{author}{\bibfnamefont{R.}~\bibnamefont{Chachra}},
  \bibinfo{author}{\bibfnamefont{M.~K.} \bibnamefont{Transtrum}}, and
  \bibinfo{author}{\bibfnamefont{J.~P.} \bibnamefont{Sethna}},
  \bibinfo{year}{2013}, \bibinfo{journal}{Science}
  \textbf{\bibinfo{volume}{342}}, \bibinfo{pages}{604}.

\bibitem[{\citenamefont{Macke}
  \emph{et~al.}(2011{\natexlab{a}})\citenamefont{Macke, Murray, and
  Latham}}]{macke+al_11}
\bibinfo{author}{\bibnamefont{Macke}, \bibfnamefont{J.~H.}},
  \bibinfo{author}{\bibfnamefont{I.}~\bibnamefont{Murray}}, and
  \bibinfo{author}{\bibfnamefont{P.}~\bibnamefont{Latham}},
  \bibinfo{year}{2011}{\natexlab{a}}, in \emph{\bibinfo{booktitle}{Advances in
  Neural Information Processing Systems}}, edited by
  \bibinfo{editor}{\bibfnamefont{J.}~\bibnamefont{Shawe-Taylor}},
  \bibinfo{editor}{\bibfnamefont{R.}~\bibnamefont{Zemel}},
  \bibinfo{editor}{\bibfnamefont{P.}~\bibnamefont{Bartlett}},
  \bibinfo{editor}{\bibfnamefont{F.}~\bibnamefont{Pereira}}, and
  \bibinfo{editor}{\bibfnamefont{K.}~\bibnamefont{Weinberger}}
  (\bibinfo{publisher}{Curran Associates, Inc.}), volume~\bibinfo{volume}{24}.

\bibitem[{\citenamefont{Macke}
  \emph{et~al.}(2011{\natexlab{b}})\citenamefont{Macke, Opper, and
  Bethge}}]{macke2011common}
\bibinfo{author}{\bibnamefont{Macke}, \bibfnamefont{J.~H.}},
  \bibinfo{author}{\bibfnamefont{M.}~\bibnamefont{Opper}}, and
  \bibinfo{author}{\bibfnamefont{M.}~\bibnamefont{Bethge}},
  \bibinfo{year}{2011}{\natexlab{b}}, \bibinfo{journal}{Physical Review
  Letters} \textbf{\bibinfo{volume}{106}}, \bibinfo{pages}{208102}.

\bibitem[{\citenamefont{Maheswaranathan}
  \emph{et~al.}(2023)\citenamefont{Maheswaranathan, McIntosh, Tanaka, Grant,
  Kastner, Melander, Nayebi, Brezovec, Wang, Ganguli, and
  Baccus}}]{maheswaranathan+al_23}
\bibinfo{author}{\bibnamefont{Maheswaranathan}, \bibfnamefont{N.}},
  \bibinfo{author}{\bibfnamefont{L.~T.} \bibnamefont{McIntosh}},
  \bibinfo{author}{\bibfnamefont{H.}~\bibnamefont{Tanaka}},
  \bibinfo{author}{\bibfnamefont{S.}~\bibnamefont{Grant}},
  \bibinfo{author}{\bibfnamefont{D.~B.} \bibnamefont{Kastner}},
  \bibinfo{author}{\bibfnamefont{J.~B.} \bibnamefont{Melander}},
  \bibinfo{author}{\bibfnamefont{A.}~\bibnamefont{Nayebi}},
  \bibinfo{author}{\bibfnamefont{L.~E.} \bibnamefont{Brezovec}},
  \bibinfo{author}{\bibfnamefont{J.~H.} \bibnamefont{Wang}},
  \bibinfo{author}{\bibfnamefont{S.}~\bibnamefont{Ganguli}}, and
  \bibinfo{author}{\bibfnamefont{S.~A.} \bibnamefont{Baccus}},
  \bibinfo{year}{2023}, \bibinfo{journal}{Neuron}
  \textbf{\bibinfo{volume}{111}}, \bibinfo{pages}{2742}.

\bibitem[{\citenamefont{Major}
  \emph{et~al.}(2004{\natexlab{a}})\citenamefont{Major, Baker, Aksay, Mensh,
  Seung, and Tank}}]{major+al_04a}
\bibinfo{author}{\bibnamefont{Major}, \bibfnamefont{G.}},
  \bibinfo{author}{\bibfnamefont{R.}~\bibnamefont{Baker}},
  \bibinfo{author}{\bibfnamefont{E.}~\bibnamefont{Aksay}},
  \bibinfo{author}{\bibfnamefont{B.}~\bibnamefont{Mensh}},
  \bibinfo{author}{\bibfnamefont{H.~S.} \bibnamefont{Seung}}, and
  \bibinfo{author}{\bibfnamefont{D.~W.} \bibnamefont{Tank}},
  \bibinfo{year}{2004}{\natexlab{a}}, \bibinfo{journal}{Proceedings of the
  National Academy of Sciences (USA)} \textbf{\bibinfo{volume}{101}},
  \bibinfo{pages}{7739}.

\bibitem[{\citenamefont{Major}
  \emph{et~al.}(2004{\natexlab{b}})\citenamefont{Major, Baker, Aksay, Seung,
  and Tank}}]{major+al_04b}
\bibinfo{author}{\bibnamefont{Major}, \bibfnamefont{G.}},
  \bibinfo{author}{\bibfnamefont{R.}~\bibnamefont{Baker}},
  \bibinfo{author}{\bibfnamefont{E.}~\bibnamefont{Aksay}},
  \bibinfo{author}{\bibfnamefont{H.~S.} \bibnamefont{Seung}}, and
  \bibinfo{author}{\bibfnamefont{D.~W.} \bibnamefont{Tank}},
  \bibinfo{year}{2004}{\natexlab{b}}, \bibinfo{journal}{Proceedings of the
  National Academy of Sciences (USA)} \textbf{\bibinfo{volume}{101}},
  \bibinfo{pages}{7745}.

\bibitem[{\citenamefont{Manley} \emph{et~al.}(2024)\citenamefont{Manley, Lu,
  Barber, Demas, Kim, Meyer, Traub, and Vaziri}}]{manley2024simultaneous}
\bibinfo{author}{\bibnamefont{Manley}, \bibfnamefont{J.}},
  \bibinfo{author}{\bibfnamefont{S.}~\bibnamefont{Lu}},
  \bibinfo{author}{\bibfnamefont{K.}~\bibnamefont{Barber}},
  \bibinfo{author}{\bibfnamefont{J.}~\bibnamefont{Demas}},
  \bibinfo{author}{\bibfnamefont{H.}~\bibnamefont{Kim}},
  \bibinfo{author}{\bibfnamefont{D.}~\bibnamefont{Meyer}},
  \bibinfo{author}{\bibfnamefont{F.~M.} \bibnamefont{Traub}}, and
  \bibinfo{author}{\bibfnamefont{A.}~\bibnamefont{Vaziri}},
  \bibinfo{year}{2024}, \bibinfo{journal}{Neuron}
  \textbf{\bibinfo{volume}{112}}, \bibinfo{pages}{1694}.

\bibitem[{\citenamefont{Maoz} \emph{et~al.}(2020)\citenamefont{Maoz,
  Tka\v{c}ik, Esteki, Kiani, and Schneidman}}]{maoz+al_21}
\bibinfo{author}{\bibnamefont{Maoz}, \bibfnamefont{O.}},
  \bibinfo{author}{\bibfnamefont{G.}~\bibnamefont{Tka\v{c}ik}},
  \bibinfo{author}{\bibfnamefont{M.~S.} \bibnamefont{Esteki}},
  \bibinfo{author}{\bibfnamefont{R.}~\bibnamefont{Kiani}}, and
  \bibinfo{author}{\bibfnamefont{E.}~\bibnamefont{Schneidman}},
  \bibinfo{year}{2020}, \bibinfo{journal}{Proceedings of the National Academy
  of Sciences (USA)} \textbf{\bibinfo{volume}{117}}, \bibinfo{pages}{25066}.

\bibitem[{\citenamefont{Marchetti} \emph{et~al.}(2013)\citenamefont{Marchetti,
  Joanny, Ramaswamy, Liverpool, Prost, Rao, and Simha}}]{marchetti+al_13}
\bibinfo{author}{\bibnamefont{Marchetti}, \bibfnamefont{M.~C.}},
  \bibinfo{author}{\bibfnamefont{J.~F.} \bibnamefont{Joanny}},
  \bibinfo{author}{\bibfnamefont{S.}~\bibnamefont{Ramaswamy}},
  \bibinfo{author}{\bibfnamefont{T.~B.} \bibnamefont{Liverpool}},
  \bibinfo{author}{\bibfnamefont{J.}~\bibnamefont{Prost}},
  \bibinfo{author}{\bibfnamefont{M.}~\bibnamefont{Rao}}, and
  \bibinfo{author}{\bibfnamefont{R.~A.} \bibnamefont{Simha}},
  \bibinfo{year}{2013}, \bibinfo{journal}{Reviews of Modern Physics}
  \textbf{\bibinfo{volume}{85}}, \bibinfo{pages}{1143}.

\bibitem[{\citenamefont{Marko and Siggia}(1995)}]{marko+siggia_95}
\bibinfo{author}{\bibnamefont{Marko}, \bibfnamefont{J.~F.}}, and
  \bibinfo{author}{\bibfnamefont{E.~D.} \bibnamefont{Siggia}},
  \bibinfo{year}{1995}, \bibinfo{journal}{Macromolecules}
  \textbf{\bibinfo{volume}{28}}, \bibinfo{pages}{8759}.

\bibitem[{\citenamefont{Marks} \emph{et~al.}(2011)\citenamefont{Marks, Colwell,
  Sheridan, Hopf, Pagnani, Zecchina, and Sander}}]{marks2011protein}
\bibinfo{author}{\bibnamefont{Marks}, \bibfnamefont{D.~S.}},
  \bibinfo{author}{\bibfnamefont{L.~J.} \bibnamefont{Colwell}},
  \bibinfo{author}{\bibfnamefont{R.}~\bibnamefont{Sheridan}},
  \bibinfo{author}{\bibfnamefont{T.~A.} \bibnamefont{Hopf}},
  \bibinfo{author}{\bibfnamefont{A.}~\bibnamefont{Pagnani}},
  \bibinfo{author}{\bibfnamefont{R.}~\bibnamefont{Zecchina}}, and
  \bibinfo{author}{\bibfnamefont{C.}~\bibnamefont{Sander}},
  \bibinfo{year}{2011}, \bibinfo{journal}{PLoS One}
  \textbf{\bibinfo{volume}{6}}, \bibinfo{pages}{e28766}.

\bibitem[{\citenamefont{Marre} \emph{et~al.}(2012)\citenamefont{Marre, Amodei,
  Sadeghi, Soo, Holy, and Berry~II}}]{marre+al2012}
\bibinfo{author}{\bibnamefont{Marre}, \bibfnamefont{O.}},
  \bibinfo{author}{\bibfnamefont{D.}~\bibnamefont{Amodei}},
  \bibinfo{author}{\bibfnamefont{K.}~\bibnamefont{Sadeghi}},
  \bibinfo{author}{\bibfnamefont{F.}~\bibnamefont{Soo}},
  \bibinfo{author}{\bibfnamefont{T.~E.} \bibnamefont{Holy}}, and
  \bibinfo{author}{\bibfnamefont{M.~J.} \bibnamefont{Berry~II}},
  \bibinfo{year}{2012}, \bibinfo{journal}{Journal of Neuroscience}
  \textbf{\bibinfo{volume}{32}}, \bibinfo{pages}{14859}.

\bibitem[{\citenamefont{Marro and Dickman}(1999)}]{marro+dickman_99}
\bibinfo{author}{\bibnamefont{Marro}, \bibfnamefont{J.}}, and
  \bibinfo{author}{\bibfnamefont{R.}~\bibnamefont{Dickman}},
  \bibinfo{year}{1999}, \emph{\bibinfo{title}{Nonequilibrium Phase Transitions
  in Lattice Models}} (\bibinfo{publisher}{Cambridge University Press,
  Cambridge}).

\bibitem[{\citenamefont{Martignon} \emph{et~al.}(2000)\citenamefont{Martignon,
  Deco, Laskey, Diamond, Freiwald, and Vaadia}}]{martignon+al2000}
\bibinfo{author}{\bibnamefont{Martignon}, \bibfnamefont{L.}},
  \bibinfo{author}{\bibfnamefont{G.}~\bibnamefont{Deco}},
  \bibinfo{author}{\bibfnamefont{K.}~\bibnamefont{Laskey}},
  \bibinfo{author}{\bibfnamefont{M.}~\bibnamefont{Diamond}},
  \bibinfo{author}{\bibfnamefont{W.}~\bibnamefont{Freiwald}}, and
  \bibinfo{author}{\bibfnamefont{E.}~\bibnamefont{Vaadia}},
  \bibinfo{year}{2000}, \bibinfo{journal}{Neural Computation}
  \textbf{\bibinfo{volume}{12}}, \bibinfo{pages}{2621}.

\bibitem[{\citenamefont{Martin} \emph{et~al.}(1973)\citenamefont{Martin,
  Siggia, and Rose}}]{martin+al_73}
\bibinfo{author}{\bibnamefont{Martin}, \bibfnamefont{P.~C.}},
  \bibinfo{author}{\bibfnamefont{E.~D.} \bibnamefont{Siggia}}, and
  \bibinfo{author}{\bibfnamefont{H.~A.} \bibnamefont{Rose}},
  \bibinfo{year}{1973}, \bibinfo{journal}{Physical Review A}
  \textbf{\bibinfo{volume}{8}}, \bibinfo{pages}{423}.

\bibitem[{\citenamefont{Maruyama} \emph{et~al.}(2014)\citenamefont{Maruyama,
  Maeda, Moroda, Kato, Inoue, Miyakawa, and Aonishi}}]{Maruyama2014}
\bibinfo{author}{\bibnamefont{Maruyama}, \bibfnamefont{R.}},
  \bibinfo{author}{\bibfnamefont{K.}~\bibnamefont{Maeda}},
  \bibinfo{author}{\bibfnamefont{H.}~\bibnamefont{Moroda}},
  \bibinfo{author}{\bibfnamefont{I.}~\bibnamefont{Kato}},
  \bibinfo{author}{\bibfnamefont{M.}~\bibnamefont{Inoue}},
  \bibinfo{author}{\bibfnamefont{H.}~\bibnamefont{Miyakawa}}, and
  \bibinfo{author}{\bibfnamefont{T.}~\bibnamefont{Aonishi}},
  \bibinfo{year}{2014}, \bibinfo{journal}{Neural Networks}
  \textbf{\bibinfo{volume}{55}}, \bibinfo{pages}{11}.

\bibitem[{\citenamefont{Mathis} \emph{et~al.}(2018)\citenamefont{Mathis,
  Mamidanna, Cury, Abe, Murthy, Mathis, and Bethge}}]{mathis+al_18}
\bibinfo{author}{\bibnamefont{Mathis}, \bibfnamefont{A.}},
  \bibinfo{author}{\bibfnamefont{P.}~\bibnamefont{Mamidanna}},
  \bibinfo{author}{\bibfnamefont{K.~M.} \bibnamefont{Cury}},
  \bibinfo{author}{\bibfnamefont{T.}~\bibnamefont{Abe}},
  \bibinfo{author}{\bibfnamefont{V.~N.} \bibnamefont{Murthy}},
  \bibinfo{author}{\bibfnamefont{M.~W.} \bibnamefont{Mathis}}, and
  \bibinfo{author}{\bibfnamefont{M.}~\bibnamefont{Bethge}},
  \bibinfo{year}{2018}, \bibinfo{journal}{Nature Neuroscience}
  \textbf{\bibinfo{volume}{21}}, \bibinfo{pages}{1281}.

\bibitem[{\citenamefont{McCulloch and Pitts}(1943)}]{mcculloch+pitts_43}
\bibinfo{author}{\bibnamefont{McCulloch}, \bibfnamefont{W.~S.}}, and
  \bibinfo{author}{\bibfnamefont{W.}~\bibnamefont{Pitts}},
  \bibinfo{year}{1943}, \bibinfo{journal}{Bulletin of Mathematical Biology}
  \textbf{\bibinfo{volume}{5}}, \bibinfo{pages}{115}.

\bibitem[{\citenamefont{McNaughton}
  \emph{et~al.}(1983)\citenamefont{McNaughton, O'Keefe, and
  Barnes}}]{mcnaughton+al_83}
\bibinfo{author}{\bibnamefont{McNaughton}, \bibfnamefont{B.~L.}},
  \bibinfo{author}{\bibfnamefont{J.}~\bibnamefont{O'Keefe}}, and
  \bibinfo{author}{\bibfnamefont{C.~A.} \bibnamefont{Barnes}},
  \bibinfo{year}{1983}, \bibinfo{journal}{Journal of Neuroscience Methods}
  \textbf{\bibinfo{volume}{8}}, \bibinfo{pages}{391}.

\bibitem[{\citenamefont{Mead}(1989)}]{Mead1989}
\bibinfo{author}{\bibnamefont{Mead}, \bibfnamefont{C.~A.}},
  \bibinfo{year}{1989}, \emph{\bibinfo{title}{Analog VLSI and Neural Systems}}
  (\bibinfo{publisher}{Addison--Wesley, Redwood City CA}).

\bibitem[{\citenamefont{Mehta} \emph{et~al.}(2019)\citenamefont{Mehta, Bukov,
  Wang, Day, Richardson, Fisher, and Schwab}}]{mehta+al_19}
\bibinfo{author}{\bibnamefont{Mehta}, \bibfnamefont{P.}},
  \bibinfo{author}{\bibfnamefont{M.}~\bibnamefont{Bukov}},
  \bibinfo{author}{\bibfnamefont{C.-H.} \bibnamefont{Wang}},
  \bibinfo{author}{\bibfnamefont{A.~G.~R.} \bibnamefont{Day}},
  \bibinfo{author}{\bibfnamefont{C.}~\bibnamefont{Richardson}},
  \bibinfo{author}{\bibfnamefont{C.~K.} \bibnamefont{Fisher}}, and
  \bibinfo{author}{\bibfnamefont{D.~J.} \bibnamefont{Schwab}},
  \bibinfo{year}{2019}, \bibinfo{journal}{Physics Reports}
  \textbf{\bibinfo{volume}{810}}, \bibinfo{pages}{1}.

\bibitem[{\citenamefont{Meister} \emph{et~al.}(1994)\citenamefont{Meister,
  Pine, and Baylor}}]{meister+al1994}
\bibinfo{author}{\bibnamefont{Meister}, \bibfnamefont{M.}},
  \bibinfo{author}{\bibfnamefont{J.}~\bibnamefont{Pine}}, and
  \bibinfo{author}{\bibfnamefont{D.~A.} \bibnamefont{Baylor}},
  \bibinfo{year}{1994}, \bibinfo{journal}{Journal of Neuroscience Methods}
  \textbf{\bibinfo{volume}{51}}, \bibinfo{pages}{95}.

\bibitem[{\citenamefont{Merchan and Nemenman}(2016)}]{Merchan+Nemenman_2016}
\bibinfo{author}{\bibnamefont{Merchan}, \bibfnamefont{L.}}, and
  \bibinfo{author}{\bibfnamefont{I.}~\bibnamefont{Nemenman}},
  \bibinfo{year}{2016}, \bibinfo{journal}{Journal of Statistical Physics}
  \textbf{\bibinfo{volume}{162}}, \bibinfo{pages}{1294}.

\bibitem[{\citenamefont{Meshulam} \emph{et~al.}(2017)\citenamefont{Meshulam,
  Gauthier, Brody, Tank, and Bialek}}]{meshulam2017collective}
\bibinfo{author}{\bibnamefont{Meshulam}, \bibfnamefont{L.}},
  \bibinfo{author}{\bibfnamefont{J.~L.} \bibnamefont{Gauthier}},
  \bibinfo{author}{\bibfnamefont{C.~D.} \bibnamefont{Brody}},
  \bibinfo{author}{\bibfnamefont{D.~W.} \bibnamefont{Tank}}, and
  \bibinfo{author}{\bibfnamefont{W.}~\bibnamefont{Bialek}},
  \bibinfo{year}{2017}, \bibinfo{journal}{Neuron}
  \textbf{\bibinfo{volume}{96}}, \bibinfo{pages}{1178}.

\bibitem[{\citenamefont{Meshulam} \emph{et~al.}(2018)\citenamefont{Meshulam,
  Gauthier, Brody, Tank, and Bialek}}]{meshulam+al2018}
\bibinfo{author}{\bibnamefont{Meshulam}, \bibfnamefont{L.}},
  \bibinfo{author}{\bibfnamefont{J.~L.} \bibnamefont{Gauthier}},
  \bibinfo{author}{\bibfnamefont{C.~D.} \bibnamefont{Brody}},
  \bibinfo{author}{\bibfnamefont{D.~W.} \bibnamefont{Tank}}, and
  \bibinfo{author}{\bibfnamefont{W.}~\bibnamefont{Bialek}},
  \bibinfo{year}{2018}, \eprint{arXiv:1812.11904}.

\bibitem[{\citenamefont{Meshulam} \emph{et~al.}(2019)\citenamefont{Meshulam,
  Gauthier, Brody, Tank, and Bialek}}]{meshulam2019RG}
\bibinfo{author}{\bibnamefont{Meshulam}, \bibfnamefont{L.}},
  \bibinfo{author}{\bibfnamefont{J.~L.} \bibnamefont{Gauthier}},
  \bibinfo{author}{\bibfnamefont{C.~D.} \bibnamefont{Brody}},
  \bibinfo{author}{\bibfnamefont{D.~W.} \bibnamefont{Tank}}, and
  \bibinfo{author}{\bibfnamefont{W.}~\bibnamefont{Bialek}},
  \bibinfo{year}{2019}, \bibinfo{journal}{Physical Review Letters}
  \textbf{\bibinfo{volume}{123}}, \bibinfo{pages}{178103}.

\bibitem[{\citenamefont{Meshulam} \emph{et~al.}(2021)\citenamefont{Meshulam,
  Gauthier, Brody, Tank, and Bialek}}]{Meshulam+al_2021}
\bibinfo{author}{\bibnamefont{Meshulam}, \bibfnamefont{L.}},
  \bibinfo{author}{\bibfnamefont{J.~L.} \bibnamefont{Gauthier}},
  \bibinfo{author}{\bibfnamefont{C.~D.} \bibnamefont{Brody}},
  \bibinfo{author}{\bibfnamefont{D.~W.} \bibnamefont{Tank}}, and
  \bibinfo{author}{\bibfnamefont{W.}~\bibnamefont{Bialek}},
  \bibinfo{year}{2021}, \eprint{arXiv:2112.14735}.

\bibitem[{\citenamefont{M\'ezard and Montanari}(2009)}]{Mezard+Montanari_2009}
\bibinfo{author}{\bibnamefont{M\'ezard}, \bibfnamefont{M.}}, and
  \bibinfo{author}{\bibfnamefont{A.}~\bibnamefont{Montanari}},
  \bibinfo{year}{2009}, \emph{\bibinfo{title}{Information, Physics, and
  Computation}} (\bibinfo{publisher}{Oxford University Press, Oxford and New
  York}).

\bibitem[{\citenamefont{M\'ezard} \emph{et~al.}(1987)\citenamefont{M\'ezard,
  Parisi, and Virasoro}}]{mezard+al_87}
\bibinfo{author}{\bibnamefont{M\'ezard}, \bibfnamefont{M.}},
  \bibinfo{author}{\bibfnamefont{G.}~\bibnamefont{Parisi}}, and
  \bibinfo{author}{\bibfnamefont{M.~A.} \bibnamefont{Virasoro}},
  \bibinfo{year}{1987}, \emph{\bibinfo{title}{Spin Glass Theory and Beyond}}
  (\bibinfo{publisher}{World Scientific, Singapore}).

\bibitem[{\citenamefont{Minaee} \emph{et~al.}(2024)\citenamefont{Minaee,
  Mikolov, Nikzad, Chenaghlu, Socher, Amatriain, and Gao}}]{minae+al_24}
\bibinfo{author}{\bibnamefont{Minaee}, \bibfnamefont{S.}},
  \bibinfo{author}{\bibfnamefont{T.}~\bibnamefont{Mikolov}},
  \bibinfo{author}{\bibfnamefont{N.}~\bibnamefont{Nikzad}},
  \bibinfo{author}{\bibfnamefont{M.}~\bibnamefont{Chenaghlu}},
  \bibinfo{author}{\bibfnamefont{R.}~\bibnamefont{Socher}},
  \bibinfo{author}{\bibfnamefont{X.}~\bibnamefont{Amatriain}}, and
  \bibinfo{author}{\bibfnamefont{J.}~\bibnamefont{Gao}}, \bibinfo{year}{2024},
  \eprint{arXiv:2402.06196}.

\bibitem[{\citenamefont{Minsky and Papert}(1969)}]{MinksyPapert1969}
\bibinfo{author}{\bibnamefont{Minsky}, \bibfnamefont{M.}}, and
  \bibinfo{author}{\bibfnamefont{S.}~\bibnamefont{Papert}},
  \bibinfo{year}{1969}, \emph{\bibinfo{title}{Perceptrons}}
  (\bibinfo{publisher}{MIT Press, Cambridge MA}).

\bibitem[{\citenamefont{Monasson} \emph{et~al.}(1999)\citenamefont{Monasson,
  Zecchina, Kirkpatrick, Selman, and Troyansky}}]{monasson+al_99}
\bibinfo{author}{\bibnamefont{Monasson}, \bibfnamefont{R.}},
  \bibinfo{author}{\bibfnamefont{R.}~\bibnamefont{Zecchina}},
  \bibinfo{author}{\bibfnamefont{S.}~\bibnamefont{Kirkpatrick}},
  \bibinfo{author}{\bibfnamefont{B.}~\bibnamefont{Selman}}, and
  \bibinfo{author}{\bibfnamefont{L.}~\bibnamefont{Troyansky}},
  \bibinfo{year}{1999}, \bibinfo{journal}{Nature}
  \textbf{\bibinfo{volume}{400}}, \bibinfo{pages}{133}.

\bibitem[{\citenamefont{Mora and Bialek}(2011)}]{mora+bialek_11}
\bibinfo{author}{\bibnamefont{Mora}, \bibfnamefont{T.}}, and
  \bibinfo{author}{\bibfnamefont{W.}~\bibnamefont{Bialek}},
  \bibinfo{year}{2011}, \bibinfo{journal}{Journal of Statistical Physics}
  \textbf{\bibinfo{volume}{144}}, \bibinfo{pages}{268}.

\bibitem[{\citenamefont{Mora} \emph{et~al.}(2015)\citenamefont{Mora, Deny, and
  Marre}}]{mora+al2015}
\bibinfo{author}{\bibnamefont{Mora}, \bibfnamefont{T.}},
  \bibinfo{author}{\bibfnamefont{S.}~\bibnamefont{Deny}}, and
  \bibinfo{author}{\bibfnamefont{O.}~\bibnamefont{Marre}},
  \bibinfo{year}{2015}, \bibinfo{journal}{Physical Review Letters}
  \textbf{\bibinfo{volume}{114}}, \bibinfo{pages}{078105}.

\bibitem[{\citenamefont{Mora} \emph{et~al.}(2010)\citenamefont{Mora, Walczak,
  Bialek, and Callan~Jr.}}]{mora2010maximum}
\bibinfo{author}{\bibnamefont{Mora}, \bibfnamefont{T.}},
  \bibinfo{author}{\bibfnamefont{A.~M.} \bibnamefont{Walczak}},
  \bibinfo{author}{\bibfnamefont{W.}~\bibnamefont{Bialek}}, and
  \bibinfo{author}{\bibfnamefont{C.~G.} \bibnamefont{Callan~Jr.}},
  \bibinfo{year}{2010}, \bibinfo{journal}{Proceedings of the National Academy
  of Sciences (USA)} \textbf{\bibinfo{volume}{107}}, \bibinfo{pages}{5405}.

\bibitem[{\citenamefont{Mora} \emph{et~al.}(2016)\citenamefont{Mora, Walczak,
  Del~Castello, Ginelli, Melillo, Parisi, Viale, Cavagna, and
  Giardina}}]{mora+al_16}
\bibinfo{author}{\bibnamefont{Mora}, \bibfnamefont{T.}},
  \bibinfo{author}{\bibfnamefont{A.~M.} \bibnamefont{Walczak}},
  \bibinfo{author}{\bibfnamefont{L.}~\bibnamefont{Del~Castello}},
  \bibinfo{author}{\bibfnamefont{F.}~\bibnamefont{Ginelli}},
  \bibinfo{author}{\bibfnamefont{S.}~\bibnamefont{Melillo}},
  \bibinfo{author}{\bibfnamefont{L.}~\bibnamefont{Parisi}},
  \bibinfo{author}{\bibfnamefont{M.}~\bibnamefont{Viale}},
  \bibinfo{author}{\bibfnamefont{A.}~\bibnamefont{Cavagna}}, and
  \bibinfo{author}{\bibfnamefont{I.}~\bibnamefont{Giardina}},
  \bibinfo{year}{2016}, \bibinfo{journal}{Nature Physics}
  \textbf{\bibinfo{volume}{12}}, \bibinfo{pages}{1153}.

\bibitem[{\citenamefont{Morales} \emph{et~al.}(2023)\citenamefont{Morales,
  di~Santo, and Mu\~noz}}]{morales+al2023}
\bibinfo{author}{\bibnamefont{Morales}, \bibfnamefont{G.~B.}},
  \bibinfo{author}{\bibfnamefont{S.}~\bibnamefont{di~Santo}}, and
  \bibinfo{author}{\bibfnamefont{M.~A.} \bibnamefont{Mu\~noz}},
  \bibinfo{year}{2023}, \bibinfo{journal}{Proceedings of the National Academy
  of Sciences (USA)} \textbf{\bibinfo{volume}{120}},
  \bibinfo{pages}{e2208998120}.

\bibitem[{\citenamefont{Morrell} \emph{et~al.}(2021)\citenamefont{Morrell,
  Sederberg, and Nemenman}}]{morrell+al2021}
\bibinfo{author}{\bibnamefont{Morrell}, \bibfnamefont{M.~C.}},
  \bibinfo{author}{\bibfnamefont{A.~J.} \bibnamefont{Sederberg}}, and
  \bibinfo{author}{\bibfnamefont{I.}~\bibnamefont{Nemenman}},
  \bibinfo{year}{2021}, \bibinfo{journal}{Physical Review Letters}
  \textbf{\bibinfo{volume}{126}}, \bibinfo{pages}{118302}.

\bibitem[{\citenamefont{Mu\~noz}(2018)}]{munoz2018}
\bibinfo{author}{\bibnamefont{Mu\~noz}, \bibfnamefont{M.~A.}},
  \bibinfo{year}{2018}, \bibinfo{journal}{Reviews of Modern Physics}
  \textbf{\bibinfo{volume}{90}}, \bibinfo{pages}{031001}.

\bibitem[{\citenamefont{Mukamel} \emph{et~al.}(2009)\citenamefont{Mukamel,
  Nimmerjahn, and Schnitzer}}]{Mukamel2009}
\bibinfo{author}{\bibnamefont{Mukamel}, \bibfnamefont{E.~A.}},
  \bibinfo{author}{\bibfnamefont{A.}~\bibnamefont{Nimmerjahn}}, and
  \bibinfo{author}{\bibfnamefont{M.~J.} \bibnamefont{Schnitzer}},
  \bibinfo{year}{2009}, \bibinfo{journal}{Neuron}
  \textbf{\bibinfo{volume}{63}}, \bibinfo{pages}{747}.

\bibitem[{\citenamefont{Munn} \emph{et~al.}(2024)\citenamefont{Munn, M\"uller,
  Favre-Bulle, Scott, Breakspear, and Shine}}]{Munn+al2024}
\bibinfo{author}{\bibnamefont{Munn}, \bibfnamefont{B.~R.}},
  \bibinfo{author}{\bibfnamefont{E.}~\bibnamefont{M\"uller}},
  \bibinfo{author}{\bibfnamefont{I.}~\bibnamefont{Favre-Bulle}},
  \bibinfo{author}{\bibfnamefont{E.}~\bibnamefont{Scott}},
  \bibinfo{author}{\bibfnamefont{M.}~\bibnamefont{Breakspear}}, and
  \bibinfo{author}{\bibfnamefont{J.~M.} \bibnamefont{Shine}},
  \bibinfo{year}{2024}, \eprint{bioRxiv:2024.06.22.600219}.

\bibitem[{\citenamefont{Musallam} \emph{et~al.}(2004)\citenamefont{Musallam,
  Corneil, Greger, Scherberger, and Andersen}}]{musallam+al_04}
\bibinfo{author}{\bibnamefont{Musallam}, \bibfnamefont{S.}},
  \bibinfo{author}{\bibfnamefont{B.~D.} \bibnamefont{Corneil}},
  \bibinfo{author}{\bibfnamefont{B.}~\bibnamefont{Greger}},
  \bibinfo{author}{\bibfnamefont{H.}~\bibnamefont{Scherberger}}, and
  \bibinfo{author}{\bibfnamefont{R.~A.} \bibnamefont{Andersen}},
  \bibinfo{year}{2004}, \bibinfo{journal}{Science}
  \textbf{\bibinfo{volume}{305}}, \bibinfo{pages}{258}.

\bibitem[{\citenamefont{Narayanankutty}
  \emph{et~al.}(2024)\citenamefont{Narayanankutty, Pereiro-Morejon, Ferrero,
  Onesto, Forciniti, delMercato, Mulet, DeMartino, Tourigny, and
  DeMartino}}]{narayanankutty+al_24}
\bibinfo{author}{\bibnamefont{Narayanankutty}, \bibfnamefont{K.}},
  \bibinfo{author}{\bibfnamefont{J.~A.} \bibnamefont{Pereiro-Morejon}},
  \bibinfo{author}{\bibfnamefont{A.}~\bibnamefont{Ferrero}},
  \bibinfo{author}{\bibfnamefont{V.}~\bibnamefont{Onesto}},
  \bibinfo{author}{\bibfnamefont{S.}~\bibnamefont{Forciniti}},
  \bibinfo{author}{\bibfnamefont{L.~L.} \bibnamefont{delMercato}},
  \bibinfo{author}{\bibfnamefont{R.}~\bibnamefont{Mulet}},
  \bibinfo{author}{\bibfnamefont{A.}~\bibnamefont{DeMartino}},
  \bibinfo{author}{\bibfnamefont{D.~S.} \bibnamefont{Tourigny}}, and
  \bibinfo{author}{\bibfnamefont{D.}~\bibnamefont{DeMartino}},
  \bibinfo{year}{2024}, \eprint{arXiv:0712.4397}.

\bibitem[{\citenamefont{Neher}(1994)}]{neher1994}
\bibinfo{author}{\bibnamefont{Neher}, \bibfnamefont{E.}}, \bibinfo{year}{1994},
  \bibinfo{journal}{Proceedings of the National Academy of Sciences (USA)}
  \textbf{\bibinfo{volume}{91}}, \bibinfo{pages}{98}.

\bibitem[{\citenamefont{Nemenman} \emph{et~al.}(2008)\citenamefont{Nemenman,
  Lewen, Bialek, and de~Ruyter~van Steveninck}}]{nemenman+al_08}
\bibinfo{author}{\bibnamefont{Nemenman}, \bibfnamefont{I.}},
  \bibinfo{author}{\bibfnamefont{G.~D.} \bibnamefont{Lewen}},
  \bibinfo{author}{\bibfnamefont{W.}~\bibnamefont{Bialek}}, and
  \bibinfo{author}{\bibfnamefont{R.~R.} \bibnamefont{de~Ruyter~van
  Steveninck}}, \bibinfo{year}{2008}, \bibinfo{journal}{PLoS Computational
  Biology} \textbf{\bibinfo{volume}{4}}, \bibinfo{pages}{e1000025}.

\bibitem[{\citenamefont{Ngampruetikorn and
  Schwab}(2023)}]{ngampruetikorn+schwab_23}
\bibinfo{author}{\bibnamefont{Ngampruetikorn}, \bibfnamefont{V.}}, and
  \bibinfo{author}{\bibfnamefont{D.~J.} \bibnamefont{Schwab}},
  \bibinfo{year}{2023}, \eprint{arXiv:2309.14047}.

\bibitem[{\citenamefont{Nguyen}
  \emph{et~al.}(2016{\natexlab{a}})\citenamefont{Nguyen, Zecchina, and
  Berg}}]{nguyen+al_16}
\bibinfo{author}{\bibnamefont{Nguyen}, \bibfnamefont{H.~C.}},
  \bibinfo{author}{\bibfnamefont{R.}~\bibnamefont{Zecchina}}, and
  \bibinfo{author}{\bibfnamefont{J.}~\bibnamefont{Berg}},
  \bibinfo{year}{2016}{\natexlab{a}}, \bibinfo{journal}{Advances in Physics}
  \textbf{\bibinfo{volume}{66}}, \bibinfo{pages}{197}.

\bibitem[{\citenamefont{Nguyen}
  \emph{et~al.}(2016{\natexlab{b}})\citenamefont{Nguyen, Shipley, Linder,
  Plummer, Shaevitz, and Leifer}}]{nguyen+al2016}
\bibinfo{author}{\bibnamefont{Nguyen}, \bibfnamefont{J.~P.}},
  \bibinfo{author}{\bibfnamefont{F.~B.} \bibnamefont{Shipley}},
  \bibinfo{author}{\bibfnamefont{A.~N.} \bibnamefont{Linder}},
  \bibinfo{author}{\bibfnamefont{G.~S.} \bibnamefont{Plummer}},
  \bibinfo{author}{\bibfnamefont{J.~W.} \bibnamefont{Shaevitz}}, and
  \bibinfo{author}{\bibfnamefont{A.~M.} \bibnamefont{Leifer}},
  \bibinfo{year}{2016}{\natexlab{b}}, \bibinfo{journal}{Proceedings of the
  National Academy of Sciences (USA)} \textbf{\bibinfo{volume}{113}},
  \bibinfo{pages}{E1074}.

\bibitem[{\citenamefont{Nicoletti} \emph{et~al.}(2020)\citenamefont{Nicoletti,
  Suweis, and Maritan}}]{nicoletti+al2020}
\bibinfo{author}{\bibnamefont{Nicoletti}, \bibfnamefont{G.}},
  \bibinfo{author}{\bibfnamefont{S.}~\bibnamefont{Suweis}}, and
  \bibinfo{author}{\bibfnamefont{A.}~\bibnamefont{Maritan}},
  \bibinfo{year}{2020}, \bibinfo{journal}{Physical Review Research}
  \textbf{\bibinfo{volume}{2}}, \bibinfo{pages}{023144}.

\bibitem[{\citenamefont{Obuchi} \emph{et~al.}(2015)\citenamefont{Obuchi, Cocco,
  and Monasson}}]{Obuchi+al_2015}
\bibinfo{author}{\bibnamefont{Obuchi}, \bibfnamefont{T.}},
  \bibinfo{author}{\bibfnamefont{S.}~\bibnamefont{Cocco}}, and
  \bibinfo{author}{\bibfnamefont{R.}~\bibnamefont{Monasson}},
  \bibinfo{year}{2015}, \bibinfo{journal}{Journal of Statistical Physics}
  \textbf{\bibinfo{volume}{161}}, \bibinfo{pages}{598}.

\bibitem[{\citenamefont{O'Dwyer} \emph{et~al.}(2017)\citenamefont{O'Dwyer,
  Rominger, and Xiao}}]{odwyer+al_17}
\bibinfo{author}{\bibnamefont{O'Dwyer}, \bibfnamefont{J.~P.}},
  \bibinfo{author}{\bibfnamefont{A.}~\bibnamefont{Rominger}}, and
  \bibinfo{author}{\bibfnamefont{X.}~\bibnamefont{Xiao}}, \bibinfo{year}{2017},
  \bibinfo{journal}{Ecology Letters} \textbf{\bibinfo{volume}{20}},
  \bibinfo{pages}{832}.

\bibitem[{\citenamefont{O'Keefe and Dostrovsky}(1971)}]{OKeefe1971}
\bibinfo{author}{\bibnamefont{O'Keefe}, \bibfnamefont{J.}}, and
  \bibinfo{author}{\bibfnamefont{J.}~\bibnamefont{Dostrovsky}},
  \bibinfo{year}{1971}, \bibinfo{journal}{Brain Research}
  \textbf{\bibinfo{volume}{34}}, \bibinfo{pages}{171}.

\bibitem[{\citenamefont{O'Keefe and Nadel}(1978)}]{okeefe1978hippocampus}
\bibinfo{author}{\bibnamefont{O'Keefe}, \bibfnamefont{J.}}, and
  \bibinfo{author}{\bibfnamefont{L.}~\bibnamefont{Nadel}},
  \bibinfo{year}{1978}, \emph{\bibinfo{title}{The Hippocampus as a Cognitive
  Map}} (\bibinfo{publisher}{Oxford University Press, Oxford}).

\bibitem[{\citenamefont{Orth} \emph{et~al.}(2010)\citenamefont{Orth, Thiele,
  and Palsson}}]{orth+al_10}
\bibinfo{author}{\bibnamefont{Orth}, \bibfnamefont{J.}},
  \bibinfo{author}{\bibfnamefont{I.}~\bibnamefont{Thiele}}, and
  \bibinfo{author}{\bibfnamefont{B.~O.} \bibnamefont{Palsson}},
  \bibinfo{year}{2010}, \bibinfo{journal}{Nature Biotechnolgy}
  \textbf{\bibinfo{volume}{28,}}, \bibinfo{pages}{245}.

\bibitem[{\citenamefont{Pachitariu}
  \emph{et~al.}(2013)\citenamefont{Pachitariu, Packer, Pettit, Dalgleish,
  Hausser, and Sahani}}]{Pachitariu2013}
\bibinfo{author}{\bibnamefont{Pachitariu}, \bibfnamefont{M.}},
  \bibinfo{author}{\bibfnamefont{A.~M.} \bibnamefont{Packer}},
  \bibinfo{author}{\bibfnamefont{N.}~\bibnamefont{Pettit}},
  \bibinfo{author}{\bibfnamefont{H.}~\bibnamefont{Dalgleish}},
  \bibinfo{author}{\bibfnamefont{M.}~\bibnamefont{Hausser}}, and
  \bibinfo{author}{\bibfnamefont{M.}~\bibnamefont{Sahani}},
  \bibinfo{year}{2013}, in \emph{\bibinfo{booktitle}{Advances in Neural
  Information Processing Systems}}, edited by
  \bibinfo{editor}{\bibfnamefont{C.}~\bibnamefont{Burges}},
  \bibinfo{editor}{\bibfnamefont{L.}~\bibnamefont{Bottou}},
  \bibinfo{editor}{\bibfnamefont{M.}~\bibnamefont{Welling}},
  \bibinfo{editor}{\bibfnamefont{Z.}~\bibnamefont{Ghahramani}}, and
  \bibinfo{editor}{\bibfnamefont{K.}~\bibnamefont{Weinberger}}
  (\bibinfo{publisher}{Curran Associates, Inc.}), volume~\bibinfo{volume}{26}.

\bibitem[{\citenamefont{Packer} \emph{et~al.}(2015)\citenamefont{Packer,
  Russell, Dalgleish, and H\"ausser}}]{packer+al_15}
\bibinfo{author}{\bibnamefont{Packer}, \bibfnamefont{A.~M.}},
  \bibinfo{author}{\bibfnamefont{L.~E.} \bibnamefont{Russell}},
  \bibinfo{author}{\bibfnamefont{H.~W.~P.} \bibnamefont{Dalgleish}}, and
  \bibinfo{author}{\bibfnamefont{M.}~\bibnamefont{H\"ausser}},
  \bibinfo{year}{2015}, \bibinfo{journal}{Nature Methods}
  \textbf{\bibinfo{volume}{12}}, \bibinfo{pages}{140}.

\bibitem[{\citenamefont{Parrish and Hammer}(1997)}]{Parrish+Hammer_1997}
\bibinfo{editor}{\bibnamefont{Parrish}, \bibfnamefont{J.~K.}}, and
  \bibinfo{editor}{\bibfnamefont{W.~M.} \bibnamefont{Hammer}} (eds.),
  \bibinfo{year}{1997}, \emph{\bibinfo{title}{Animal Groups in Three
  Dimensions}} (\bibinfo{publisher}{Cambridge University Press, Cambridge}).

\bibitem[{\citenamefont{Pascanu} \emph{et~al.}(2013)\citenamefont{Pascanu,
  Mikolov, and Bengio}}]{pascanu+al_13}
\bibinfo{author}{\bibnamefont{Pascanu}, \bibfnamefont{R.}},
  \bibinfo{author}{\bibfnamefont{T.}~\bibnamefont{Mikolov}}, and
  \bibinfo{author}{\bibfnamefont{Y.}~\bibnamefont{Bengio}},
  \bibinfo{year}{2013}, \bibinfo{journal}{Proceedings of Machine Learning
  Research} \textbf{\bibinfo{volume}{28}}, \bibinfo{pages}{1310}.

\bibitem[{\citenamefont{Pereira} \emph{et~al.}(2019)\citenamefont{Pereira,
  Aldarondo, Willmore, Kislin, Wang, Murthy, and Shaevitz}}]{pereira+al_19}
\bibinfo{author}{\bibnamefont{Pereira}, \bibfnamefont{T.~D.}},
  \bibinfo{author}{\bibfnamefont{D.~E.} \bibnamefont{Aldarondo}},
  \bibinfo{author}{\bibfnamefont{L.}~\bibnamefont{Willmore}},
  \bibinfo{author}{\bibfnamefont{M.}~\bibnamefont{Kislin}},
  \bibinfo{author}{\bibfnamefont{S.~S.-H.} \bibnamefont{Wang}},
  \bibinfo{author}{\bibfnamefont{M.}~\bibnamefont{Murthy}}, and
  \bibinfo{author}{\bibfnamefont{J.~W.} \bibnamefont{Shaevitz}},
  \bibinfo{year}{2019}, \bibinfo{journal}{Nature Methods}
  \textbf{\bibinfo{volume}{16}}, \bibinfo{pages}{117}.

\bibitem[{\citenamefont{Pine and Gilbert}(1982)}]{pine+gilbert1982}
\bibinfo{author}{\bibnamefont{Pine}, \bibfnamefont{J.}}, and
  \bibinfo{author}{\bibfnamefont{J.}~\bibnamefont{Gilbert}},
  \bibinfo{year}{1982}, \bibinfo{journal}{Soc Neurosci Abs}
  \textbf{\bibinfo{volume}{8}}, \bibinfo{pages}{670}.

\bibitem[{\citenamefont{Platisa} \emph{et~al.}(2023)\citenamefont{Platisa, Ye,
  Ahrens, Liu, Chen, Davison, Tian, Pieribone, and Chen}}]{platisa+al2023}
\bibinfo{author}{\bibnamefont{Platisa}, \bibfnamefont{J.}},
  \bibinfo{author}{\bibfnamefont{X.}~\bibnamefont{Ye}},
  \bibinfo{author}{\bibfnamefont{A.~M.} \bibnamefont{Ahrens}},
  \bibinfo{author}{\bibfnamefont{C.}~\bibnamefont{Liu}},
  \bibinfo{author}{\bibfnamefont{I.~A.} \bibnamefont{Chen}},
  \bibinfo{author}{\bibfnamefont{I.~G.} \bibnamefont{Davison}},
  \bibinfo{author}{\bibfnamefont{L.}~\bibnamefont{Tian}},
  \bibinfo{author}{\bibfnamefont{V.~A.} \bibnamefont{Pieribone}}, and
  \bibinfo{author}{\bibfnamefont{J.~L.} \bibnamefont{Chen}},
  \bibinfo{year}{2023}, \bibinfo{journal}{Nature Methods}
  \textbf{\bibinfo{volume}{20}}, \bibinfo{pages}{1095}.

\bibitem[{\citenamefont{Pontryagin}(1987)}]{pontrjagin}
\bibinfo{author}{\bibnamefont{Pontryagin}, \bibfnamefont{L.~S.}},
  \bibinfo{year}{1987}, \emph{\bibinfo{title}{L. S. Pontryagin Collected Works,
  Volume Four: Mathematical Theory of Optimal Processes.}}
  (\bibinfo{publisher}{Routledge, London}).

\bibitem[{\citenamefont{Prasher} \emph{et~al.}(1992)\citenamefont{Prasher,
  Eckenrode, Ward, Prendergast, and Cormier}}]{prasher+al_92}
\bibinfo{author}{\bibnamefont{Prasher}, \bibfnamefont{D.~C.}},
  \bibinfo{author}{\bibfnamefont{V.~K.} \bibnamefont{Eckenrode}},
  \bibinfo{author}{\bibfnamefont{W.~W.} \bibnamefont{Ward}},
  \bibinfo{author}{\bibfnamefont{F.~G.} \bibnamefont{Prendergast}}, and
  \bibinfo{author}{\bibfnamefont{M.~J.} \bibnamefont{Cormier}},
  \bibinfo{year}{1992}, \bibinfo{journal}{Gene} \textbf{\bibinfo{volume}{111}},
  \bibinfo{pages}{229}.

\bibitem[{\citenamefont{Prentice} \emph{et~al.}(2011)\citenamefont{Prentice,
  Homann, Simmons, Tka\v{c}ik, Balasubramanian, and Nelson}}]{prentice+al_11}
\bibinfo{author}{\bibnamefont{Prentice}, \bibfnamefont{J.~S.}},
  \bibinfo{author}{\bibfnamefont{J.}~\bibnamefont{Homann}},
  \bibinfo{author}{\bibfnamefont{K.~D.} \bibnamefont{Simmons}},
  \bibinfo{author}{\bibfnamefont{G.}~\bibnamefont{Tka\v{c}ik}},
  \bibinfo{author}{\bibfnamefont{V.}~\bibnamefont{Balasubramanian}}, and
  \bibinfo{author}{\bibfnamefont{P.~C.} \bibnamefont{Nelson}},
  \bibinfo{year}{2011}, \bibinfo{journal}{PLoS One}
  \textbf{\bibinfo{volume}{6}}, \bibinfo{pages}{e19884}.

\bibitem[{\citenamefont{Press\'e} \emph{et~al.}(2013)\citenamefont{Press\'e,
  Ghosh, Lee, and Dill}}]{presse+al2013}
\bibinfo{author}{\bibnamefont{Press\'e}, \bibfnamefont{S.}},
  \bibinfo{author}{\bibfnamefont{K.}~\bibnamefont{Ghosh}},
  \bibinfo{author}{\bibfnamefont{J.}~\bibnamefont{Lee}}, and
  \bibinfo{author}{\bibfnamefont{K.~A.} \bibnamefont{Dill}},
  \bibinfo{year}{2013}, \bibinfo{journal}{Reviews of Modern Physics}
  \textbf{\bibinfo{volume}{85}}, \bibinfo{pages}{1115}.

\bibitem[{\citenamefont{Radvansky and Dombeck}(2018)}]{radvansky2018olfactory}
\bibinfo{author}{\bibnamefont{Radvansky}, \bibfnamefont{B.~A.}}, and
  \bibinfo{author}{\bibfnamefont{D.~A.} \bibnamefont{Dombeck}},
  \bibinfo{year}{2018}, \bibinfo{journal}{Nature Communications}
  \textbf{\bibinfo{volume}{9}}, \bibinfo{pages}{1}.

\bibitem[{\citenamefont{Ramirez and Bialek}(2021)}]{ramirez+bialek_21}
\bibinfo{author}{\bibnamefont{Ramirez}, \bibfnamefont{L.}}, and
  \bibinfo{author}{\bibfnamefont{W.}~\bibnamefont{Bialek}},
  \bibinfo{year}{2021}, \eprint{arXiv:2112.14334}.

\bibitem[{\citenamefont{Randi} \emph{et~al.}(2023)\citenamefont{Randi, Sharma,
  Dvali, and Leifer}}]{randi+al_23}
\bibinfo{author}{\bibnamefont{Randi}, \bibfnamefont{F.}},
  \bibinfo{author}{\bibfnamefont{A.~K.} \bibnamefont{Sharma}},
  \bibinfo{author}{\bibfnamefont{S.}~\bibnamefont{Dvali}}, and
  \bibinfo{author}{\bibfnamefont{A.~M.} \bibnamefont{Leifer}},
  \bibinfo{year}{2023}, \bibinfo{journal}{Nature}
  \textbf{\bibinfo{volume}{623}}, \bibinfo{pages}{406}.

\bibitem[{\citenamefont{Reynolds}(1987)}]{reynolds_87}
\bibinfo{author}{\bibnamefont{Reynolds}, \bibfnamefont{C.}},
  \bibinfo{year}{1987}, in \emph{\bibinfo{booktitle}{Proceedings of the 14th
  annual conference on Computer graphics and interactive techniques}}
  (\bibinfo{publisher}{Association for Computing Machinery}), pp.
  \bibinfo{pages}{25--34}.

\bibitem[{\citenamefont{Rickgauer} \emph{et~al.}(2014)\citenamefont{Rickgauer,
  Deisseroth, and Tank}}]{rickgauer2014simultaneous}
\bibinfo{author}{\bibnamefont{Rickgauer}, \bibfnamefont{J.~P.}},
  \bibinfo{author}{\bibfnamefont{K.}~\bibnamefont{Deisseroth}}, and
  \bibinfo{author}{\bibfnamefont{D.~W.} \bibnamefont{Tank}},
  \bibinfo{year}{2014}, \bibinfo{journal}{Nature Neuroscience}
  \textbf{\bibinfo{volume}{17}}, \bibinfo{pages}{1816}.

\bibitem[{\citenamefont{Rieke} \emph{et~al.}(1997)\citenamefont{Rieke, Warland,
  de~Ruyter~van Steveninck, and Bialek}}]{spikesbook}
\bibinfo{author}{\bibnamefont{Rieke}, \bibfnamefont{F.}},
  \bibinfo{author}{\bibfnamefont{D.}~\bibnamefont{Warland}},
  \bibinfo{author}{\bibfnamefont{R.}~\bibnamefont{de~Ruyter~van Steveninck}},
  and \bibinfo{author}{\bibfnamefont{W.}~\bibnamefont{Bialek}},
  \bibinfo{year}{1997}, \emph{\bibinfo{title}{Spikes: Exploring the Neural
  Code}} (\bibinfo{publisher}{MIT Press, Cambridge}).

\bibitem[{\citenamefont{Roberts}(2021)}]{roberts_21}
\bibinfo{author}{\bibnamefont{Roberts}, \bibfnamefont{D.~A.}},
  \bibinfo{year}{2021}, \eprint{arXiv:2104.00008}.

\bibitem[{\citenamefont{Roberts and Yaida}(2022)}]{roberts+yaida_22}
\bibinfo{author}{\bibnamefont{Roberts}, \bibfnamefont{D.~A.}}, and
  \bibinfo{author}{\bibfnamefont{S.}~\bibnamefont{Yaida}},
  \bibinfo{year}{2022}, \emph{\bibinfo{title}{The Principles of Deep Learning
  Theory: An Effective Theory Approach to Understanding Neural Networks}}
  (\bibinfo{publisher}{Cambridge University Press, Cambridge}).

\bibitem[{\citenamefont{Rosen} \emph{et~al.}(2015)\citenamefont{Rosen, Davison,
  Bhaya, and Fisher}}]{rosen+al_15}
\bibinfo{author}{\bibnamefont{Rosen}, \bibfnamefont{M.~J.}},
  \bibinfo{author}{\bibfnamefont{M.}~\bibnamefont{Davison}},
  \bibinfo{author}{\bibfnamefont{D.}~\bibnamefont{Bhaya}}, and
  \bibinfo{author}{\bibfnamefont{D.~S.} \bibnamefont{Fisher}},
  \bibinfo{year}{2015}, \bibinfo{journal}{Science}
  \textbf{\bibinfo{volume}{348}}, \bibinfo{pages}{1019}.

\bibitem[{\citenamefont{Rosenblatt}(1961)}]{Rosenblatt1961}
\bibinfo{author}{\bibnamefont{Rosenblatt}, \bibfnamefont{F.}},
  \bibinfo{year}{1961}, \emph{\bibinfo{title}{Principles of Neurodynamics:
  Perceptrons and the Theory of Brain Mechanisms}} (\bibinfo{publisher}{Spartan
  Books, Washington DC}).

\bibitem[{\citenamefont{Roudi} \emph{et~al.}(2009)\citenamefont{Roudi,
  Nirenberg, and Latham}}]{roudi2009pairwise}
\bibinfo{author}{\bibnamefont{Roudi}, \bibfnamefont{Y.}},
  \bibinfo{author}{\bibfnamefont{S.}~\bibnamefont{Nirenberg}}, and
  \bibinfo{author}{\bibfnamefont{P.~E.} \bibnamefont{Latham}},
  \bibinfo{year}{2009}, \bibinfo{journal}{PLoS Computational Biology}
  \textbf{\bibinfo{volume}{5}}, \bibinfo{pages}{e1000380}.

\bibitem[{\citenamefont{Roy} \emph{et~al.}(2021)\citenamefont{Roy, Jun, Davis,
  Pearson, and Field}}]{roy+al_21}
\bibinfo{author}{\bibnamefont{Roy}, \bibfnamefont{S.}},
  \bibinfo{author}{\bibfnamefont{N.~Y.} \bibnamefont{Jun}},
  \bibinfo{author}{\bibfnamefont{E.~L.} \bibnamefont{Davis}},
  \bibinfo{author}{\bibfnamefont{J.}~\bibnamefont{Pearson}}, and
  \bibinfo{author}{\bibfnamefont{G.~D.} \bibnamefont{Field}},
  \bibinfo{year}{2021}, \bibinfo{journal}{Nature}
  \textbf{\bibinfo{volume}{592}}, \bibinfo{pages}{409}.

\bibitem[{\citenamefont{Rumelhart} \emph{et~al.}(1986)\citenamefont{Rumelhart,
  Hinton, and Williams}}]{rumelhart+al_1986}
\bibinfo{author}{\bibnamefont{Rumelhart}, \bibfnamefont{D.~E.}},
  \bibinfo{author}{\bibfnamefont{G.~E.} \bibnamefont{Hinton}}, and
  \bibinfo{author}{\bibfnamefont{R.~J.} \bibnamefont{Williams}},
  \bibinfo{year}{1986}, \bibinfo{journal}{Nature}
  \textbf{\bibinfo{volume}{323}}, \bibinfo{pages}{533}.

\bibitem[{\citenamefont{Russ} \emph{et~al.}(2005)\citenamefont{Russ, Lowery,
  Mishra, Yaffe, and Ranganathan}}]{russ+al_05}
\bibinfo{author}{\bibnamefont{Russ}, \bibfnamefont{W.}},
  \bibinfo{author}{\bibfnamefont{D.~M.} \bibnamefont{Lowery}},
  \bibinfo{author}{\bibfnamefont{P.}~\bibnamefont{Mishra}},
  \bibinfo{author}{\bibfnamefont{M.~B.} \bibnamefont{Yaffe}}, and
  \bibinfo{author}{\bibfnamefont{R.}~\bibnamefont{Ranganathan}},
  \bibinfo{year}{2005}, \bibinfo{journal}{Nature}
  \textbf{\bibinfo{volume}{437}}, \bibinfo{pages}{579}.

\bibitem[{\citenamefont{Russ} \emph{et~al.}(2020)\citenamefont{Russ, Figliuzzi,
  Stocker, Barrat-Charlaix, Socolich, Kast, Hilvert, Monasson, Cocco, Weigt}
  \emph{et~al.}}]{russ2020evolution}
\bibinfo{author}{\bibnamefont{Russ}, \bibfnamefont{W.~P.}},
  \bibinfo{author}{\bibfnamefont{M.}~\bibnamefont{Figliuzzi}},
  \bibinfo{author}{\bibfnamefont{C.}~\bibnamefont{Stocker}},
  \bibinfo{author}{\bibfnamefont{P.}~\bibnamefont{Barrat-Charlaix}},
  \bibinfo{author}{\bibfnamefont{M.}~\bibnamefont{Socolich}},
  \bibinfo{author}{\bibfnamefont{P.}~\bibnamefont{Kast}},
  \bibinfo{author}{\bibfnamefont{D.}~\bibnamefont{Hilvert}},
  \bibinfo{author}{\bibfnamefont{R.}~\bibnamefont{Monasson}},
  \bibinfo{author}{\bibfnamefont{S.}~\bibnamefont{Cocco}},
  \bibinfo{author}{\bibfnamefont{M.}~\bibnamefont{Weigt}}, \emph{et~al.},
  \bibinfo{year}{2020}, \bibinfo{journal}{Science}
  \textbf{\bibinfo{volume}{369}}, \bibinfo{pages}{440}.

\bibitem[{\citenamefont{Sampaio~Filho}
  \emph{et~al.}(2024)\citenamefont{Sampaio~Filho, de~Arcangelis, Herrmann,
  Plenz, Kells, Ribeiro, and Andrade~Jr.}}]{SampaioFilho+al2024}
\bibinfo{author}{\bibnamefont{Sampaio~Filho}, \bibfnamefont{C.~I.~N.}},
  \bibinfo{author}{\bibfnamefont{L.}~\bibnamefont{de~Arcangelis}},
  \bibinfo{author}{\bibfnamefont{H.~J.} \bibnamefont{Herrmann}},
  \bibinfo{author}{\bibfnamefont{D.}~\bibnamefont{Plenz}},
  \bibinfo{author}{\bibfnamefont{P.}~\bibnamefont{Kells}},
  \bibinfo{author}{\bibfnamefont{T.~L.} \bibnamefont{Ribeiro}}, and
  \bibinfo{author}{\bibfnamefont{J.~S.} \bibnamefont{Andrade~Jr.}},
  \bibinfo{year}{2024}, \bibinfo{journal}{Scientific Reports}
  \textbf{\bibinfo{volume}{14}}, \bibinfo{pages}{7002}.

\bibitem[{\citenamefont{Sarra} \emph{et~al.}(2024)\citenamefont{Sarra, Sarra,
  Di~Carlo, GrandPre, Zhang, Callan~Jr., and Bialek}}]{sarra+al_24}
\bibinfo{author}{\bibnamefont{Sarra}, \bibfnamefont{C.}},
  \bibinfo{author}{\bibfnamefont{L.}~\bibnamefont{Sarra}},
  \bibinfo{author}{\bibfnamefont{L.}~\bibnamefont{Di~Carlo}},
  \bibinfo{author}{\bibfnamefont{T.}~\bibnamefont{GrandPre}},
  \bibinfo{author}{\bibfnamefont{Y.}~\bibnamefont{Zhang}},
  \bibinfo{author}{\bibfnamefont{C.~G.} \bibnamefont{Callan~Jr.}}, and
  \bibinfo{author}{\bibfnamefont{W.}~\bibnamefont{Bialek}},
  \bibinfo{year}{2024}, \eprint{arXiv:2408.08037}.

\bibitem[{\citenamefont{Schneidman}
  \emph{et~al.}(2006)\citenamefont{Schneidman, Berry~II, Segev, and
  Bialek}}]{schneidman2006weak}
\bibinfo{author}{\bibnamefont{Schneidman}, \bibfnamefont{E.}},
  \bibinfo{author}{\bibfnamefont{M.~J.} \bibnamefont{Berry~II}},
  \bibinfo{author}{\bibfnamefont{R.}~\bibnamefont{Segev}}, and
  \bibinfo{author}{\bibfnamefont{W.}~\bibnamefont{Bialek}},
  \bibinfo{year}{2006}, \bibinfo{journal}{Nature}
  \textbf{\bibinfo{volume}{440}}, \bibinfo{pages}{1007}.

\bibitem[{\citenamefont{Schneidman}
  \emph{et~al.}(2003)\citenamefont{Schneidman, Still, Berry~II, and
  Bialek}}]{schneidman+al2003}
\bibinfo{author}{\bibnamefont{Schneidman}, \bibfnamefont{E.}},
  \bibinfo{author}{\bibfnamefont{S.}~\bibnamefont{Still}},
  \bibinfo{author}{\bibfnamefont{M.~J.} \bibnamefont{Berry~II}}, and
  \bibinfo{author}{\bibfnamefont{W.}~\bibnamefont{Bialek}},
  \bibinfo{year}{2003}, \bibinfo{journal}{Physical Review Letters}
  \textbf{\bibinfo{volume}{91}}, \bibinfo{pages}{238701}.

\bibitem[{\citenamefont{Schnitzer and Meister}(2003)}]{schnitzer+meister_03}
\bibinfo{author}{\bibnamefont{Schnitzer}, \bibfnamefont{M.~J.}}, and
  \bibinfo{author}{\bibfnamefont{M.}~\bibnamefont{Meister}},
  \bibinfo{year}{2003}, \bibinfo{journal}{Neuron}
  \textbf{\bibinfo{volume}{37}}, \bibinfo{pages}{499}.

\bibitem[{\citenamefont{Schwab} \emph{et~al.}(2014)\citenamefont{Schwab,
  Nemenman, and Mehta}}]{Schwab+al_2014}
\bibinfo{author}{\bibnamefont{Schwab}, \bibfnamefont{D.~J.}},
  \bibinfo{author}{\bibfnamefont{I.}~\bibnamefont{Nemenman}}, and
  \bibinfo{author}{\bibfnamefont{P.}~\bibnamefont{Mehta}},
  \bibinfo{year}{2014}, \bibinfo{journal}{Physical Review Letters}
  \textbf{\bibinfo{volume}{113}}, \bibinfo{pages}{068102}.

\bibitem[{\citenamefont{Segev} \emph{et~al.}(2004)\citenamefont{Segev,
  Goodhouse, Puchalla, and Berry~II}}]{segev+al2004}
\bibinfo{author}{\bibnamefont{Segev}, \bibfnamefont{R.}},
  \bibinfo{author}{\bibfnamefont{J.}~\bibnamefont{Goodhouse}},
  \bibinfo{author}{\bibfnamefont{J.}~\bibnamefont{Puchalla}}, and
  \bibinfo{author}{\bibfnamefont{M.~J.} \bibnamefont{Berry~II}},
  \bibinfo{year}{2004}, \bibinfo{journal}{Nature Neuroscience}
  \textbf{\bibinfo{volume}{7}}, \bibinfo{pages}{1155}.

\bibitem[{\citenamefont{Serruya} \emph{et~al.}(2002)\citenamefont{Serruya,
  Hatsopoulos, Paninski, Fellows, and Donoghue}}]{serruya+al_02}
\bibinfo{author}{\bibnamefont{Serruya}, \bibfnamefont{M.~D.}},
  \bibinfo{author}{\bibfnamefont{N.~G.} \bibnamefont{Hatsopoulos}},
  \bibinfo{author}{\bibfnamefont{L.}~\bibnamefont{Paninski}},
  \bibinfo{author}{\bibfnamefont{M.~R.} \bibnamefont{Fellows}}, and
  \bibinfo{author}{\bibfnamefont{J.~P.} \bibnamefont{Donoghue}},
  \bibinfo{year}{2002}, \bibinfo{journal}{Nature}
  \textbf{\bibinfo{volume}{416}}, \bibinfo{pages}{141}.

\bibitem[{\citenamefont{Sessak and Monasson}(2009)}]{sessak+monasson_09}
\bibinfo{author}{\bibnamefont{Sessak}, \bibfnamefont{V.}}, and
  \bibinfo{author}{\bibfnamefont{R.}~\bibnamefont{Monasson}},
  \bibinfo{year}{2009}, \bibinfo{journal}{Journal of Physics A}
  \textbf{\bibinfo{volume}{42}}, \bibinfo{pages}{055001}.

\bibitem[{\citenamefont{Sethna}(2021)}]{sethna2021statistical}
\bibinfo{author}{\bibnamefont{Sethna}, \bibfnamefont{J.}},
  \bibinfo{year}{2021}, \emph{\bibinfo{title}{Statistical Mechanics: Entropy,
  Order Parameters, and Complexity}} (\bibinfo{publisher}{Oxford University
  Press, Oxford}).

\bibitem[{\citenamefont{Seung}(1996)}]{seung_96}
\bibinfo{author}{\bibnamefont{Seung}, \bibfnamefont{H.~S.}},
  \bibinfo{year}{1996}, \bibinfo{journal}{Proceedings of the National Academy
  of Sciences (USA)} \textbf{\bibinfo{volume}{93}}, \bibinfo{pages}{13339}.

\bibitem[{\citenamefont{Shannon}(1948)}]{shannon1948mathematical}
\bibinfo{author}{\bibnamefont{Shannon}, \bibfnamefont{C.~E.}},
  \bibinfo{year}{1948}, \bibinfo{journal}{The Bell System Technical Journal}
  \textbf{\bibinfo{volume}{27}}, \bibinfo{pages}{379}.

\bibitem[{\citenamefont{Shimomura} \emph{et~al.}(1962)\citenamefont{Shimomura,
  Johnson, and Saiga}}]{shinomura+al_62}
\bibinfo{author}{\bibnamefont{Shimomura}, \bibfnamefont{O.}},
  \bibinfo{author}{\bibfnamefont{F.~H.} \bibnamefont{Johnson}}, and
  \bibinfo{author}{\bibfnamefont{Y.}~\bibnamefont{Saiga}},
  \bibinfo{year}{1962}, \bibinfo{journal}{Journal of Cellular and Comparative
  Physiology} \textbf{\bibinfo{volume}{59}}, \bibinfo{pages}{223}.

\bibitem[{\citenamefont{Shlens}(2014)}]{shlens_14}
\bibinfo{author}{\bibnamefont{Shlens}, \bibfnamefont{J.}},
  \bibinfo{year}{2014}, \eprint{arXiv:1404.1100}.

\bibitem[{\citenamefont{Shlens} \emph{et~al.}(2006)\citenamefont{Shlens, Field,
  Gauthier, Grivich, Petrusca, Sher, Litke, and Chichilnisky}}]{shlens+al2006}
\bibinfo{author}{\bibnamefont{Shlens}, \bibfnamefont{J.}},
  \bibinfo{author}{\bibfnamefont{G.~D.} \bibnamefont{Field}},
  \bibinfo{author}{\bibfnamefont{J.~L.} \bibnamefont{Gauthier}},
  \bibinfo{author}{\bibfnamefont{M.~I.} \bibnamefont{Grivich}},
  \bibinfo{author}{\bibfnamefont{D.}~\bibnamefont{Petrusca}},
  \bibinfo{author}{\bibfnamefont{A.}~\bibnamefont{Sher}},
  \bibinfo{author}{\bibfnamefont{A.~M.} \bibnamefont{Litke}}, and
  \bibinfo{author}{\bibfnamefont{E.~J.} \bibnamefont{Chichilnisky}},
  \bibinfo{year}{2006}, \bibinfo{journal}{Journal of Neuroscience}
  \textbf{\bibinfo{volume}{26}}, \bibinfo{pages}{8254}.

\bibitem[{\citenamefont{Skinner} \emph{et~al.}(2024)\citenamefont{Skinner,
  Lamaire, and Mani}}]{skinner+al_24}
\bibinfo{author}{\bibnamefont{Skinner}, \bibfnamefont{D.~J.}},
  \bibinfo{author}{\bibfnamefont{P.}~\bibnamefont{Lamaire}}, and
  \bibinfo{author}{\bibfnamefont{M.}~\bibnamefont{Mani}}, \bibinfo{year}{2024},
  \eprint{bioRxiv:2024.07.26.605398}.

\bibitem[{\citenamefont{Smith and H\"ausser}(2010)}]{SmithHäusser2010}
\bibinfo{author}{\bibnamefont{Smith}, \bibfnamefont{S.~L.}}, and
  \bibinfo{author}{\bibfnamefont{M.}~\bibnamefont{H\"ausser}},
  \bibinfo{year}{2010}, \bibinfo{journal}{Nature Neuroscience}
  \textbf{\bibinfo{volume}{13}}, \bibinfo{pages}{1144}.

\bibitem[{\citenamefont{Socolich} \emph{et~al.}(2005)\citenamefont{Socolich,
  Lockless, Russ, Lee, Gardner, and Ranganathan}}]{socolich+al_05}
\bibinfo{author}{\bibnamefont{Socolich}, \bibfnamefont{M.}},
  \bibinfo{author}{\bibfnamefont{S.~W.} \bibnamefont{Lockless}},
  \bibinfo{author}{\bibfnamefont{W.~P.} \bibnamefont{Russ}},
  \bibinfo{author}{\bibfnamefont{H.}~\bibnamefont{Lee}},
  \bibinfo{author}{\bibfnamefont{K.~H.} \bibnamefont{Gardner}}, and
  \bibinfo{author}{\bibfnamefont{R.}~\bibnamefont{Ranganathan}},
  \bibinfo{year}{2005}, \bibinfo{journal}{Nature}
  \textbf{\bibinfo{volume}{437}}, \bibinfo{pages}{512}.

\bibitem[{\citenamefont{Sofroniew} \emph{et~al.}(2016)\citenamefont{Sofroniew,
  Flickinger, King, and Svoboda}}]{sofroniew2016large}
\bibinfo{author}{\bibnamefont{Sofroniew}, \bibfnamefont{N.~J.}},
  \bibinfo{author}{\bibfnamefont{D.}~\bibnamefont{Flickinger}},
  \bibinfo{author}{\bibfnamefont{J.}~\bibnamefont{King}}, and
  \bibinfo{author}{\bibfnamefont{K.}~\bibnamefont{Svoboda}},
  \bibinfo{year}{2016}, \bibinfo{journal}{eLife} \textbf{\bibinfo{volume}{5}},
  \bibinfo{pages}{e14472}.

\bibitem[{\citenamefont{Solovey} \emph{et~al.}(2015)\citenamefont{Solovey,
  Alonso, Yanagawa, Fujii, Magnasco, Cecchi, and Proekt}}]{solovey+al_15}
\bibinfo{author}{\bibnamefont{Solovey}, \bibfnamefont{G.}},
  \bibinfo{author}{\bibfnamefont{L.~M.} \bibnamefont{Alonso}},
  \bibinfo{author}{\bibfnamefont{T.}~\bibnamefont{Yanagawa}},
  \bibinfo{author}{\bibfnamefont{N.}~\bibnamefont{Fujii}},
  \bibinfo{author}{\bibfnamefont{M.~O.} \bibnamefont{Magnasco}},
  \bibinfo{author}{\bibfnamefont{G.~A.} \bibnamefont{Cecchi}}, and
  \bibinfo{author}{\bibfnamefont{A.}~\bibnamefont{Proekt}},
  \bibinfo{year}{2015}, \bibinfo{journal}{Journal of Neuroscience}
  \textbf{\bibinfo{volume}{35}}, \bibinfo{pages}{10866}.

\bibitem[{\citenamefont{Srivastava}
  \emph{et~al.}(2017)\citenamefont{Srivastava, Holmes, Vellema, Pack, Elemans,
  Nemenman, and Sober}}]{srivastava+al_17}
\bibinfo{author}{\bibnamefont{Srivastava}, \bibfnamefont{K.~H.}},
  \bibinfo{author}{\bibfnamefont{C.~M.} \bibnamefont{Holmes}},
  \bibinfo{author}{\bibfnamefont{M.}~\bibnamefont{Vellema}},
  \bibinfo{author}{\bibfnamefont{A.~R.} \bibnamefont{Pack}},
  \bibinfo{author}{\bibfnamefont{C.~P.~H.} \bibnamefont{Elemans}},
  \bibinfo{author}{\bibfnamefont{I.}~\bibnamefont{Nemenman}}, and
  \bibinfo{author}{\bibfnamefont{S.~J.} \bibnamefont{Sober}},
  \bibinfo{year}{2017}, \bibinfo{journal}{(5) 1171-1176}
  \textbf{\bibinfo{volume}{114}}, \bibinfo{pages}{1171}.

\bibitem[{\citenamefont{Steinmetz} \emph{et~al.}(2021)\citenamefont{Steinmetz,
  Aydin, Lebedeva, Okun, Pachitariu, Bauza, Beau, Bhagat, B{\"o}hm, Broux}
  \emph{et~al.}}]{steinmetz2021neuropixels}
\bibinfo{author}{\bibnamefont{Steinmetz}, \bibfnamefont{N.~A.}},
  \bibinfo{author}{\bibfnamefont{C.}~\bibnamefont{Aydin}},
  \bibinfo{author}{\bibfnamefont{A.}~\bibnamefont{Lebedeva}},
  \bibinfo{author}{\bibfnamefont{M.}~\bibnamefont{Okun}},
  \bibinfo{author}{\bibfnamefont{M.}~\bibnamefont{Pachitariu}},
  \bibinfo{author}{\bibfnamefont{M.}~\bibnamefont{Bauza}},
  \bibinfo{author}{\bibfnamefont{M.}~\bibnamefont{Beau}},
  \bibinfo{author}{\bibfnamefont{J.}~\bibnamefont{Bhagat}},
  \bibinfo{author}{\bibfnamefont{C.}~\bibnamefont{B{\"o}hm}},
  \bibinfo{author}{\bibfnamefont{M.}~\bibnamefont{Broux}}, \emph{et~al.},
  \bibinfo{year}{2021}, \bibinfo{journal}{Science}
  \textbf{\bibinfo{volume}{372}}, \bibinfo{pages}{eabf4588}.

\bibitem[{\citenamefont{Steinmetz} \emph{et~al.}(2018)\citenamefont{Steinmetz,
  Koch, Harris, and Carandini}}]{steinmetz2018challenges}
\bibinfo{author}{\bibnamefont{Steinmetz}, \bibfnamefont{N.~A.}},
  \bibinfo{author}{\bibfnamefont{C.}~\bibnamefont{Koch}},
  \bibinfo{author}{\bibfnamefont{K.~D.} \bibnamefont{Harris}}, and
  \bibinfo{author}{\bibfnamefont{M.}~\bibnamefont{Carandini}},
  \bibinfo{year}{2018}, \bibinfo{journal}{Current Opinion in Neurobiology}
  \textbf{\bibinfo{volume}{50}}, \bibinfo{pages}{92}.

\bibitem[{\citenamefont{Stevenson and Kording}(2011)}]{stevenson+kording2011}
\bibinfo{author}{\bibnamefont{Stevenson}, \bibfnamefont{I.~H.}}, and
  \bibinfo{author}{\bibfnamefont{K.~P.} \bibnamefont{Kording}},
  \bibinfo{year}{2011}, \bibinfo{journal}{Nature Neuroscience}
  \textbf{\bibinfo{volume}{14}}, \bibinfo{pages}{139}.

\bibitem[{\citenamefont{Stroud}(1974)}]{stroud_74}
\bibinfo{author}{\bibnamefont{Stroud}, \bibfnamefont{R.~M.}},
  \bibinfo{year}{1974}, \bibinfo{journal}{Scientific American}
  \textbf{\bibinfo{volume}{231}}, \bibinfo{pages}{74}.

\bibitem[{\citenamefont{Su\l{}kowska}
  \emph{et~al.}(2012)\citenamefont{Su\l{}kowska, Morcos, Weigt, Hwa, and
  Onuchic}}]{Sulkowska+al_2012}
\bibinfo{author}{\bibnamefont{Su\l{}kowska}, \bibfnamefont{J.~I.}},
  \bibinfo{author}{\bibfnamefont{F.}~\bibnamefont{Morcos}},
  \bibinfo{author}{\bibfnamefont{M.}~\bibnamefont{Weigt}},
  \bibinfo{author}{\bibfnamefont{T.}~\bibnamefont{Hwa}}, and
  \bibinfo{author}{\bibfnamefont{J.~N.} \bibnamefont{Onuchic}},
  \bibinfo{year}{2012}, \bibinfo{journal}{Proceedings of the National Academy
  of Sciences (USA)} \textbf{\bibinfo{volume}{109}}, \bibinfo{pages}{10340}.

\bibitem[{\citenamefont{Swendsen}(1984)}]{swendsen_84}
\bibinfo{author}{\bibnamefont{Swendsen}, \bibfnamefont{R.~H.}},
  \bibinfo{year}{1984}, \bibinfo{journal}{Physical Review Letters}
  \textbf{\bibinfo{volume}{52}}, \bibinfo{pages}{1165}.

\bibitem[{\citenamefont{Tang} \emph{et~al.}(2008)\citenamefont{Tang, Jackson,
  Hobbs, Chen, Smith, Patel, Prieto, Petrusca, Grivich, Sher, Hottowy,
  Dabrowski} \emph{et~al.}}]{tang+al2008}
\bibinfo{author}{\bibnamefont{Tang}, \bibfnamefont{A.}},
  \bibinfo{author}{\bibfnamefont{D.}~\bibnamefont{Jackson}},
  \bibinfo{author}{\bibfnamefont{J.}~\bibnamefont{Hobbs}},
  \bibinfo{author}{\bibfnamefont{W.}~\bibnamefont{Chen}},
  \bibinfo{author}{\bibfnamefont{J.~L.} \bibnamefont{Smith}},
  \bibinfo{author}{\bibfnamefont{H.}~\bibnamefont{Patel}},
  \bibinfo{author}{\bibfnamefont{A.}~\bibnamefont{Prieto}},
  \bibinfo{author}{\bibfnamefont{D.}~\bibnamefont{Petrusca}},
  \bibinfo{author}{\bibfnamefont{M.~I.} \bibnamefont{Grivich}},
  \bibinfo{author}{\bibfnamefont{A.}~\bibnamefont{Sher}},
  \bibinfo{author}{\bibfnamefont{P.}~\bibnamefont{Hottowy}},
  \bibinfo{author}{\bibfnamefont{W.}~\bibnamefont{Dabrowski}}, \emph{et~al.},
  \bibinfo{year}{2008}, \bibinfo{journal}{Journal of Neuroscience}
  \textbf{\bibinfo{volume}{28}}, \bibinfo{pages}{505}.

\bibitem[{\citenamefont{Tang} \emph{et~al.}(1987)\citenamefont{Tang,
  Wiesenfeld, Bak, Coppersmith, and Littlewood}}]{tang+al_87}
\bibinfo{author}{\bibnamefont{Tang}, \bibfnamefont{C.}},
  \bibinfo{author}{\bibfnamefont{K.}~\bibnamefont{Wiesenfeld}},
  \bibinfo{author}{\bibfnamefont{P.}~\bibnamefont{Bak}},
  \bibinfo{author}{\bibfnamefont{S.}~\bibnamefont{Coppersmith}}, and
  \bibinfo{author}{\bibfnamefont{P.}~\bibnamefont{Littlewood}},
  \bibinfo{year}{1987}, \bibinfo{journal}{Physical Review Letters}
  \textbf{\bibinfo{volume}{58}}, \bibinfo{pages}{1161}.

\bibitem[{\citenamefont{Tavoni} \emph{et~al.}(2016)\citenamefont{Tavoni, Cocco,
  and Monasson}}]{tavoni+al_16}
\bibinfo{author}{\bibnamefont{Tavoni}, \bibfnamefont{G.}},
  \bibinfo{author}{\bibfnamefont{S.}~\bibnamefont{Cocco}}, and
  \bibinfo{author}{\bibfnamefont{R.}~\bibnamefont{Monasson}},
  \bibinfo{year}{2016}, \bibinfo{journal}{Journal of Computational
  Neuroscience} \textbf{\bibinfo{volume}{41}}, \bibinfo{pages}{269}.

\bibitem[{\citenamefont{Tavoni} \emph{et~al.}(2017)\citenamefont{Tavoni,
  Ferrari, Battaglia, Cocco, and Monasson}}]{tavoni2017functional}
\bibinfo{author}{\bibnamefont{Tavoni}, \bibfnamefont{G.}},
  \bibinfo{author}{\bibfnamefont{U.}~\bibnamefont{Ferrari}},
  \bibinfo{author}{\bibfnamefont{F.~P.} \bibnamefont{Battaglia}},
  \bibinfo{author}{\bibfnamefont{S.}~\bibnamefont{Cocco}}, and
  \bibinfo{author}{\bibfnamefont{R.}~\bibnamefont{Monasson}},
  \bibinfo{year}{2017}, \bibinfo{journal}{Network Neuroscience}
  \textbf{\bibinfo{volume}{1}}, \bibinfo{pages}{275}.

\bibitem[{\citenamefont{Taylor} \emph{et~al.}(2002)\citenamefont{Taylor,
  Tillery, and Schwartz}}]{taylor+al_02}
\bibinfo{author}{\bibnamefont{Taylor}, \bibfnamefont{D.~M.}},
  \bibinfo{author}{\bibfnamefont{S.~I.~H.} \bibnamefont{Tillery}}, and
  \bibinfo{author}{\bibfnamefont{A.~B.} \bibnamefont{Schwartz}},
  \bibinfo{year}{2002}, \bibinfo{journal}{Science}
  \textbf{\bibinfo{volume}{296}}, \bibinfo{pages}{1829}.

\bibitem[{\citenamefont{Tian} \emph{et~al.}(2012)\citenamefont{Tian, Akerboom,
  Schreiter, and Looger}}]{tian2012neural}
\bibinfo{author}{\bibnamefont{Tian}, \bibfnamefont{L.}},
  \bibinfo{author}{\bibfnamefont{J.}~\bibnamefont{Akerboom}},
  \bibinfo{author}{\bibfnamefont{E.~R.} \bibnamefont{Schreiter}}, and
  \bibinfo{author}{\bibfnamefont{L.~L.} \bibnamefont{Looger}},
  \bibinfo{year}{2012}, \bibinfo{journal}{Progress in Brain Research}
  \textbf{\bibinfo{volume}{196}}, \bibinfo{pages}{79}.

\bibitem[{\citenamefont{Tishby} \emph{et~al.}(1999)\citenamefont{Tishby,
  Pereira, and Bialek}}]{tishby+al_99}
\bibinfo{author}{\bibnamefont{Tishby}, \bibfnamefont{N.}},
  \bibinfo{author}{\bibfnamefont{F.~C.} \bibnamefont{Pereira}}, and
  \bibinfo{author}{\bibfnamefont{W.}~\bibnamefont{Bialek}},
  \bibinfo{year}{1999}, in \emph{\bibinfo{booktitle}{Proceedings of the 37th
  Annual Allerton Conference on Communication, Control and Computing}}, edited
  by \bibinfo{editor}{\bibfnamefont{B.}~\bibnamefont{Hajek}} and
  \bibinfo{editor}{\bibfnamefont{R.~S.} \bibnamefont{Sreenivas}}
  (\bibinfo{publisher}{University of Illinois}), pp. \bibinfo{pages}{368--377},
  \eprint{arXiv:physics/0004057}.

\bibitem[{\citenamefont{Tka{\v{c}}ik}
  \emph{et~al.}(2014)\citenamefont{Tka{\v{c}}ik, Marre, Amodei, Schneidman,
  Bialek, and Berry}}]{tkavcik2014searching}
\bibinfo{author}{\bibnamefont{Tka{\v{c}}ik}, \bibfnamefont{G.}},
  \bibinfo{author}{\bibfnamefont{O.}~\bibnamefont{Marre}},
  \bibinfo{author}{\bibfnamefont{D.}~\bibnamefont{Amodei}},
  \bibinfo{author}{\bibfnamefont{E.}~\bibnamefont{Schneidman}},
  \bibinfo{author}{\bibfnamefont{W.}~\bibnamefont{Bialek}}, and
  \bibinfo{author}{\bibfnamefont{M.~J.} \bibnamefont{Berry}},
  \bibinfo{year}{2014}, \bibinfo{journal}{PLoS Computational Biology}
  \textbf{\bibinfo{volume}{10}}, \bibinfo{pages}{e1003408}.

\bibitem[{\citenamefont{Tka{\v{c}}ik}
  \emph{et~al.}(2013)\citenamefont{Tka{\v{c}}ik, Marre, Mora, Amodei, Berry~II,
  and Bialek}}]{tkavcik2013simplest}
\bibinfo{author}{\bibnamefont{Tka{\v{c}}ik}, \bibfnamefont{G.}},
  \bibinfo{author}{\bibfnamefont{O.}~\bibnamefont{Marre}},
  \bibinfo{author}{\bibfnamefont{T.}~\bibnamefont{Mora}},
  \bibinfo{author}{\bibfnamefont{D.}~\bibnamefont{Amodei}},
  \bibinfo{author}{\bibfnamefont{M.~J.} \bibnamefont{Berry~II}}, and
  \bibinfo{author}{\bibfnamefont{W.}~\bibnamefont{Bialek}},
  \bibinfo{year}{2013}, \bibinfo{journal}{Journal of Statistical Mechanics:
  Theory and Experiment} \textbf{\bibinfo{volume}{2013}},
  \bibinfo{pages}{P03011}.

\bibitem[{\citenamefont{Tka\v{c}ik}
  \emph{et~al.}(2015)\citenamefont{Tka\v{c}ik, Mora, Marre, Amodei, Palmer,
  Berry~II, and Bialek}}]{tkacik2015signatures}
\bibinfo{author}{\bibnamefont{Tka\v{c}ik}, \bibfnamefont{G.}},
  \bibinfo{author}{\bibfnamefont{T.}~\bibnamefont{Mora}},
  \bibinfo{author}{\bibfnamefont{O.}~\bibnamefont{Marre}},
  \bibinfo{author}{\bibfnamefont{D.}~\bibnamefont{Amodei}},
  \bibinfo{author}{\bibfnamefont{S.~E.} \bibnamefont{Palmer}},
  \bibinfo{author}{\bibfnamefont{M.~J.} \bibnamefont{Berry~II}}, and
  \bibinfo{author}{\bibfnamefont{W.}~\bibnamefont{Bialek}},
  \bibinfo{year}{2015}, \bibinfo{journal}{Proceedings of the National Academy
  of Sciences (USA)} \textbf{\bibinfo{volume}{112}}, \bibinfo{pages}{11508}.

\bibitem[{\citenamefont{Tka\v{c}ik}
  \emph{et~al.}(2010)\citenamefont{Tka\v{c}ik, Prentice, Balasubramanian, and
  Schneidman}}]{tkacik+al_10}
\bibinfo{author}{\bibnamefont{Tka\v{c}ik}, \bibfnamefont{G.}},
  \bibinfo{author}{\bibfnamefont{J.~S.} \bibnamefont{Prentice}},
  \bibinfo{author}{\bibfnamefont{V.}~\bibnamefont{Balasubramanian}}, and
  \bibinfo{author}{\bibfnamefont{E.}~\bibnamefont{Schneidman}},
  \bibinfo{year}{2010}, \bibinfo{journal}{Proceedings of the National Academy
  of Sciences (USA)} \textbf{\bibinfo{volume}{107}}, \bibinfo{pages}{14419}.

\bibitem[{\citenamefont{Tka\v{c}ik}
  \emph{et~al.}(2006)\citenamefont{Tka\v{c}ik, Schneidman, Berry~II, and
  Bialek}}]{tkacik2006ising}
\bibinfo{author}{\bibnamefont{Tka\v{c}ik}, \bibfnamefont{G.}},
  \bibinfo{author}{\bibfnamefont{E.}~\bibnamefont{Schneidman}},
  \bibinfo{author}{\bibfnamefont{M.~J.} \bibnamefont{Berry~II}}, and
  \bibinfo{author}{\bibfnamefont{W.}~\bibnamefont{Bialek}},
  \bibinfo{year}{2006}, \eprint{arXiv:q-bio/0611072}.

\bibitem[{\citenamefont{Tka\v{c}ik}
  \emph{et~al.}(2009)\citenamefont{Tka\v{c}ik, Schneidman, Berry~II, and
  Bialek}}]{tkacik2009spin}
\bibinfo{author}{\bibnamefont{Tka\v{c}ik}, \bibfnamefont{G.}},
  \bibinfo{author}{\bibfnamefont{E.}~\bibnamefont{Schneidman}},
  \bibinfo{author}{\bibfnamefont{M.~J.} \bibnamefont{Berry~II}}, and
  \bibinfo{author}{\bibfnamefont{W.}~\bibnamefont{Bialek}},
  \bibinfo{year}{2009}, \eprint{arXiv:0912.5409}.

\bibitem[{\citenamefont{Toner and Tu}(1995)}]{toner+tu_95}
\bibinfo{author}{\bibnamefont{Toner}, \bibfnamefont{J.}}, and
  \bibinfo{author}{\bibfnamefont{Y.}~\bibnamefont{Tu}}, \bibinfo{year}{1995},
  \bibinfo{journal}{Physical Review Letters} \textbf{\bibinfo{volume}{75}},
  \bibinfo{pages}{4326}.

\bibitem[{\citenamefont{Toner and Tu}(1998)}]{toner+tu_98}
\bibinfo{author}{\bibnamefont{Toner}, \bibfnamefont{J.}}, and
  \bibinfo{author}{\bibfnamefont{Y.}~\bibnamefont{Tu}}, \bibinfo{year}{1998},
  \bibinfo{journal}{Physical Review E} \textbf{\bibinfo{volume}{58}},
  \bibinfo{pages}{4828}.

\bibitem[{\citenamefont{Trautmann} \emph{et~al.}(2023)\citenamefont{Trautmann,
  Hesse, Stine, Xia, Zhu, O'Shea, Karsh, Colonell, Lanfranchi, Vyas, Zimnik,
  Stenmann} \emph{et~al.}}]{trautmann+al2023}
\bibinfo{author}{\bibnamefont{Trautmann}, \bibfnamefont{E.~M.}},
  \bibinfo{author}{\bibfnamefont{J.~K.} \bibnamefont{Hesse}},
  \bibinfo{author}{\bibfnamefont{G.~M.} \bibnamefont{Stine}},
  \bibinfo{author}{\bibfnamefont{R.}~\bibnamefont{Xia}},
  \bibinfo{author}{\bibfnamefont{S.}~\bibnamefont{Zhu}},
  \bibinfo{author}{\bibfnamefont{D.~J.} \bibnamefont{O'Shea}},
  \bibinfo{author}{\bibfnamefont{B.}~\bibnamefont{Karsh}},
  \bibinfo{author}{\bibfnamefont{J.}~\bibnamefont{Colonell}},
  \bibinfo{author}{\bibfnamefont{F.~F.} \bibnamefont{Lanfranchi}},
  \bibinfo{author}{\bibfnamefont{S.}~\bibnamefont{Vyas}},
  \bibinfo{author}{\bibfnamefont{A.}~\bibnamefont{Zimnik}},
  \bibinfo{author}{\bibfnamefont{N.~A.} \bibnamefont{Stenmann}}, \emph{et~al.},
  \bibinfo{year}{2023}, \eprint{bioRxiv:2023.02.01.526664}.

\bibitem[{\citenamefont{Treves} \emph{et~al.}(1992)\citenamefont{Treves,
  Miglino, and Parisi}}]{treves+al_92}
\bibinfo{author}{\bibnamefont{Treves}, \bibfnamefont{A.}},
  \bibinfo{author}{\bibfnamefont{O.}~\bibnamefont{Miglino}}, and
  \bibinfo{author}{\bibfnamefont{D.}~\bibnamefont{Parisi}},
  \bibinfo{year}{1992}, \bibinfo{journal}{Pyschobiology}
  \textbf{\bibinfo{volume}{20}}, \bibinfo{pages}{1}.

\bibitem[{\citenamefont{Tsai} \emph{et~al.}(2017)\citenamefont{Tsai, Sawyer,
  Bradd, Yuste, and Shepard}}]{tsai2017very}
\bibinfo{author}{\bibnamefont{Tsai}, \bibfnamefont{D.}},
  \bibinfo{author}{\bibfnamefont{D.}~\bibnamefont{Sawyer}},
  \bibinfo{author}{\bibfnamefont{A.}~\bibnamefont{Bradd}},
  \bibinfo{author}{\bibfnamefont{R.}~\bibnamefont{Yuste}}, and
  \bibinfo{author}{\bibfnamefont{K.~L.} \bibnamefont{Shepard}},
  \bibinfo{year}{2017}, \bibinfo{journal}{Nature Communications}
  \textbf{\bibinfo{volume}{8}}, \bibinfo{pages}{1}.

\bibitem[{\citenamefont{Tsien}(2009)}]{tsien_09}
\bibinfo{author}{\bibnamefont{Tsien}, \bibfnamefont{R.~Y.}},
  \bibinfo{year}{2009}, \bibinfo{journal}{Angewandte Chemie International
  Edition} \textbf{\bibinfo{volume}{48}}, \bibinfo{pages}{5612}.

\bibitem[{\citenamefont{Tsoar} \emph{et~al.}(2011)\citenamefont{Tsoar, Nathan,
  Bartan, Vyssotski, Dell'Omo, and Ulanovsky}}]{tsoar+al_11}
\bibinfo{author}{\bibnamefont{Tsoar}, \bibfnamefont{A.}},
  \bibinfo{author}{\bibfnamefont{R.}~\bibnamefont{Nathan}},
  \bibinfo{author}{\bibfnamefont{Y.}~\bibnamefont{Bartan}},
  \bibinfo{author}{\bibfnamefont{A.}~\bibnamefont{Vyssotski}},
  \bibinfo{author}{\bibfnamefont{G.}~\bibnamefont{Dell'Omo}}, and
  \bibinfo{author}{\bibfnamefont{N.}~\bibnamefont{Ulanovsky}},
  \bibinfo{year}{2011}, \bibinfo{journal}{Proceedings of the National Academy
  of Sciences (USA)} \textbf{\bibinfo{volume}{108}}, \bibinfo{pages}{E718}.

\bibitem[{\citenamefont{Tsodyks and Sejnowski}(1995)}]{tsodyks+sejnowski_95}
\bibinfo{author}{\bibnamefont{Tsodyks}, \bibfnamefont{M.}}, and
  \bibinfo{author}{\bibfnamefont{T.}~\bibnamefont{Sejnowski}},
  \bibinfo{year}{1995}, \bibinfo{journal}{International Journal of Neural
  Systems} \textbf{\bibinfo{volume}{6}}, \bibinfo{pages}{81}.

\bibitem[{\citenamefont{Turing}(1937)}]{turing_37}
\bibinfo{author}{\bibnamefont{Turing}, \bibfnamefont{A.~M.}},
  \bibinfo{year}{1937}, \bibinfo{journal}{Proceedings of the London
  Mathematical Society} \textbf{\bibinfo{volume}{s2--42}},
  \bibinfo{pages}{230}.

\bibitem[{\citenamefont{Urai} \emph{et~al.}(2022)\citenamefont{Urai, Doiron,
  Leifer, and Churchland}}]{urai+al_22}
\bibinfo{author}{\bibnamefont{Urai}, \bibfnamefont{A.~E.}},
  \bibinfo{author}{\bibfnamefont{B.}~\bibnamefont{Doiron}},
  \bibinfo{author}{\bibfnamefont{A.~M.} \bibnamefont{Leifer}}, and
  \bibinfo{author}{\bibfnamefont{A.~K.} \bibnamefont{Churchland}},
  \bibinfo{year}{2022}, \bibinfo{journal}{Nature Neuroscience}
  \textbf{\bibinfo{volume}{25}}, \bibinfo{pages}{11}.

\bibitem[{\citenamefont{Varshney} \emph{et~al.}(2011)\citenamefont{Varshney,
  Chen, Paniagua, Hall, and Chklovskii}}]{varshney+al_11}
\bibinfo{author}{\bibnamefont{Varshney}, \bibfnamefont{L.~R.}},
  \bibinfo{author}{\bibfnamefont{B.~L.} \bibnamefont{Chen}},
  \bibinfo{author}{\bibfnamefont{E.}~\bibnamefont{Paniagua}},
  \bibinfo{author}{\bibfnamefont{D.~H.} \bibnamefont{Hall}}, and
  \bibinfo{author}{\bibfnamefont{D.~B.} \bibnamefont{Chklovskii}},
  \bibinfo{year}{2011}, \bibinfo{journal}{PLoS Computational Biology}
  \textbf{\bibinfo{volume}{7}}, \bibinfo{pages}{e1001066}.

\bibitem[{\citenamefont{Vicsek} \emph{et~al.}(1995)\citenamefont{Vicsek,
  Czir\'ok, Ben-Jacob, Cohen, and Shochet}}]{viscek+al_95}
\bibinfo{author}{\bibnamefont{Vicsek}, \bibfnamefont{T.}},
  \bibinfo{author}{\bibfnamefont{A.}~\bibnamefont{Czir\'ok}},
  \bibinfo{author}{\bibfnamefont{E.}~\bibnamefont{Ben-Jacob}},
  \bibinfo{author}{\bibfnamefont{I.}~\bibnamefont{Cohen}}, and
  \bibinfo{author}{\bibfnamefont{O.}~\bibnamefont{Shochet}},
  \bibinfo{year}{1995}, \bibinfo{journal}{Physical Review Letters}
  \textbf{\bibinfo{volume}{75}}, \bibinfo{pages}{1226}.

\bibitem[{\citenamefont{Villegas} \emph{et~al.}(2021)\citenamefont{Villegas,
  Cavagna, Cencini, Fort, and Grigera}}]{villegas+al_21}
\bibinfo{author}{\bibnamefont{Villegas}, \bibfnamefont{P.}},
  \bibinfo{author}{\bibfnamefont{A.}~\bibnamefont{Cavagna}},
  \bibinfo{author}{\bibfnamefont{M.}~\bibnamefont{Cencini}},
  \bibinfo{author}{\bibfnamefont{H.}~\bibnamefont{Fort}}, and
  \bibinfo{author}{\bibfnamefont{T.~S.} \bibnamefont{Grigera}},
  \bibinfo{year}{2021}, \bibinfo{journal}{Royal Society Open Science}
  \textbf{\bibinfo{volume}{8}}, \bibinfo{pages}{202200}.

\bibitem[{\citenamefont{Villegas} \emph{et~al.}(2024)\citenamefont{Villegas,
  Gili, Caldarelli, and Gabrielli}}]{villegas+al_24}
\bibinfo{author}{\bibnamefont{Villegas}, \bibfnamefont{P.}},
  \bibinfo{author}{\bibfnamefont{T.}~\bibnamefont{Gili}},
  \bibinfo{author}{\bibfnamefont{G.}~\bibnamefont{Caldarelli}}, and
  \bibinfo{author}{\bibfnamefont{A.}~\bibnamefont{Gabrielli}},
  \bibinfo{year}{2024}, \bibinfo{journal}{Physical Review E}
  \textbf{\bibinfo{volume}{109}}, \bibinfo{pages}{L042402}.

\bibitem[{\citenamefont{Villette} \emph{et~al.}(2019)\citenamefont{Villette,
  Chavarha, Dimov, Bradley, Pradhan, Mathieu, Evans, Chamberland, Shi, Yang}
  \emph{et~al.}}]{villette2019}
\bibinfo{author}{\bibnamefont{Villette}, \bibfnamefont{V.}},
  \bibinfo{author}{\bibfnamefont{M.}~\bibnamefont{Chavarha}},
  \bibinfo{author}{\bibfnamefont{I.~K.} \bibnamefont{Dimov}},
  \bibinfo{author}{\bibfnamefont{J.}~\bibnamefont{Bradley}},
  \bibinfo{author}{\bibfnamefont{L.}~\bibnamefont{Pradhan}},
  \bibinfo{author}{\bibfnamefont{B.}~\bibnamefont{Mathieu}},
  \bibinfo{author}{\bibfnamefont{S.~W.} \bibnamefont{Evans}},
  \bibinfo{author}{\bibfnamefont{S.}~\bibnamefont{Chamberland}},
  \bibinfo{author}{\bibfnamefont{D.}~\bibnamefont{Shi}},
  \bibinfo{author}{\bibfnamefont{R.}~\bibnamefont{Yang}}, \emph{et~al.},
  \bibinfo{year}{2019}, \bibinfo{journal}{Cell}
  \textbf{\bibinfo{volume}{179}}(\bibinfo{number}{7}), \bibinfo{pages}{1590}.

\bibitem[{\citenamefont{Vishwanathan}
  \emph{et~al.}(2024)\citenamefont{Vishwanathan, Sood, Wu, Ramirez, Yang,
  Kemnitz, Ih, Turner, Lee, Tartavull, Silversmith, Jordan}
  \emph{et~al.}}]{vishwanathan+al_24}
\bibinfo{author}{\bibnamefont{Vishwanathan}, \bibfnamefont{A.}},
  \bibinfo{author}{\bibfnamefont{A.}~\bibnamefont{Sood}},
  \bibinfo{author}{\bibfnamefont{J.}~\bibnamefont{Wu}},
  \bibinfo{author}{\bibfnamefont{A.~D.} \bibnamefont{Ramirez}},
  \bibinfo{author}{\bibfnamefont{R.}~\bibnamefont{Yang}},
  \bibinfo{author}{\bibfnamefont{N.}~\bibnamefont{Kemnitz}},
  \bibinfo{author}{\bibfnamefont{D.}~\bibnamefont{Ih}},
  \bibinfo{author}{\bibfnamefont{N.}~\bibnamefont{Turner}},
  \bibinfo{author}{\bibfnamefont{K.}~\bibnamefont{Lee}},
  \bibinfo{author}{\bibfnamefont{I.}~\bibnamefont{Tartavull}},
  \bibinfo{author}{\bibfnamefont{W.~M.} \bibnamefont{Silversmith}},
  \bibinfo{author}{\bibfnamefont{C.~S.} \bibnamefont{Jordan}}, \emph{et~al.},
  \bibinfo{year}{2024}, \eprint{bioRxiv:2020.10.28.359620}.

\bibitem[{\citenamefont{Vogels} \emph{et~al.}(2005)\citenamefont{Vogels, Rajan,
  and Abbott}}]{vogels+al_05}
\bibinfo{author}{\bibnamefont{Vogels}, \bibfnamefont{T.~P.}},
  \bibinfo{author}{\bibfnamefont{K.}~\bibnamefont{Rajan}}, and
  \bibinfo{author}{\bibfnamefont{L.~F.} \bibnamefont{Abbott}},
  \bibinfo{year}{2005}, \bibinfo{journal}{Annual Review of Neuroscience}
  \textbf{\bibinfo{volume}{28}}, \bibinfo{pages}{357}.

\bibitem[{\citenamefont{Vorontsov} \emph{et~al.}(2017)\citenamefont{Vorontsov,
  Trabelsi, Kadoury, and Pal}}]{vorontsov+al_17}
\bibinfo{author}{\bibnamefont{Vorontsov}, \bibfnamefont{E.}},
  \bibinfo{author}{\bibfnamefont{C.}~\bibnamefont{Trabelsi}},
  \bibinfo{author}{\bibfnamefont{S.}~\bibnamefont{Kadoury}}, and
  \bibinfo{author}{\bibfnamefont{C.}~\bibnamefont{Pal}}, \bibinfo{year}{2017},
  \bibinfo{journal}{Proceedings of Machine Learning Research}
  \textbf{\bibinfo{volume}{70}}, \bibinfo{pages}{3570}.

\bibitem[{\citenamefont{van Vreeswijk and
  Sompolinsky}(1998)}]{vreeswijk+sompolinsky_98}
\bibinfo{author}{\bibnamefont{van Vreeswijk}, \bibfnamefont{C.}}, and
  \bibinfo{author}{\bibfnamefont{H.}~\bibnamefont{Sompolinsky}},
  \bibinfo{year}{1998}, \bibinfo{journal}{Neural Computation}
  \textbf{\bibinfo{volume}{10}}, \bibinfo{pages}{1321–1371}.

\bibitem[{\citenamefont{Ward} \emph{et~al.}(2017)\citenamefont{Ward, Yung,
  Davis, Blinebry, Williams, Johnson, and E.}}]{ward+al_17}
\bibinfo{author}{\bibnamefont{Ward}, \bibfnamefont{C.~S.}},
  \bibinfo{author}{\bibfnamefont{C.-M.} \bibnamefont{Yung}},
  \bibinfo{author}{\bibfnamefont{K.~M.} \bibnamefont{Davis}},
  \bibinfo{author}{\bibfnamefont{S.~K.} \bibnamefont{Blinebry}},
  \bibinfo{author}{\bibfnamefont{T.~C.} \bibnamefont{Williams}},
  \bibinfo{author}{\bibfnamefont{Z.~I.} \bibnamefont{Johnson}}, and
  \bibinfo{author}{\bibfnamefont{H.~D.} \bibnamefont{E.}},
  \bibinfo{year}{2017}, \bibinfo{journal}{The ISME Journal: Multidisciplinary
  Journal of Microbial Ecology} \textbf{\bibinfo{volume}{11}},
  \bibinfo{pages}{1412}.

\bibitem[{\citenamefont{Watkin} \emph{et~al.}(1993)\citenamefont{Watkin, Rau,
  and Biehl}}]{watkin+al_93}
\bibinfo{author}{\bibnamefont{Watkin}, \bibfnamefont{T.~L.~H.}},
  \bibinfo{author}{\bibfnamefont{A.}~\bibnamefont{Rau}}, and
  \bibinfo{author}{\bibfnamefont{M.}~\bibnamefont{Biehl}},
  \bibinfo{year}{1993}, \bibinfo{journal}{Reviews of Modern Physics}
  \textbf{\bibinfo{volume}{65}}, \bibinfo{pages}{499}.

\bibitem[{\citenamefont{Weigt} \emph{et~al.}(2009)\citenamefont{Weigt, White,
  Szurmant, Hoch, and Hwa}}]{weigt+al_09}
\bibinfo{author}{\bibnamefont{Weigt}, \bibfnamefont{M.}},
  \bibinfo{author}{\bibfnamefont{R.}~\bibnamefont{White}},
  \bibinfo{author}{\bibfnamefont{H.}~\bibnamefont{Szurmant}},
  \bibinfo{author}{\bibfnamefont{J.}~\bibnamefont{Hoch}}, and
  \bibinfo{author}{\bibfnamefont{T.}~\bibnamefont{Hwa}}, \bibinfo{year}{2009},
  \bibinfo{journal}{Proceedings of the National Academy of Sciences (USA)}
  \textbf{\bibinfo{volume}{106}}, \bibinfo{pages}{67}.

\bibitem[{\citenamefont{Weisenburger}
  \emph{et~al.}(2019)\citenamefont{Weisenburger, Tejera, Demas, Chen, Manley,
  Sparks, Traub, Daigle, Zeng, Losonczy}
  \emph{et~al.}}]{weisenburger2019volumetric}
\bibinfo{author}{\bibnamefont{Weisenburger}, \bibfnamefont{S.}},
  \bibinfo{author}{\bibfnamefont{F.}~\bibnamefont{Tejera}},
  \bibinfo{author}{\bibfnamefont{J.}~\bibnamefont{Demas}},
  \bibinfo{author}{\bibfnamefont{B.}~\bibnamefont{Chen}},
  \bibinfo{author}{\bibfnamefont{J.}~\bibnamefont{Manley}},
  \bibinfo{author}{\bibfnamefont{F.~T.} \bibnamefont{Sparks}},
  \bibinfo{author}{\bibfnamefont{F.~M.} \bibnamefont{Traub}},
  \bibinfo{author}{\bibfnamefont{T.}~\bibnamefont{Daigle}},
  \bibinfo{author}{\bibfnamefont{H.}~\bibnamefont{Zeng}},
  \bibinfo{author}{\bibfnamefont{A.}~\bibnamefont{Losonczy}}, \emph{et~al.},
  \bibinfo{year}{2019}, \bibinfo{journal}{Cell} \textbf{\bibinfo{volume}{177}},
  \bibinfo{pages}{1050}.

\bibitem[{\citenamefont{White} \emph{et~al.}(1986)\citenamefont{White,
  Southgate, Thomson, and Brenner}}]{white+al_86}
\bibinfo{author}{\bibnamefont{White}, \bibfnamefont{J.~G.}},
  \bibinfo{author}{\bibfnamefont{E.}~\bibnamefont{Southgate}},
  \bibinfo{author}{\bibfnamefont{J.~N.} \bibnamefont{Thomson}}, and
  \bibinfo{author}{\bibfnamefont{S.}~\bibnamefont{Brenner}},
  \bibinfo{year}{1986}, \textbf{\bibinfo{volume}{314}}, \bibinfo{pages}{1}.

\bibitem[{\citenamefont{Wiener}(1958)}]{wiener_58}
\bibinfo{author}{\bibnamefont{Wiener}, \bibfnamefont{N.}},
  \bibinfo{year}{1958}, \emph{\bibinfo{title}{Nonlinear Problems in Random
  Theory}} (\bibinfo{publisher}{MIT Press, Cambridge MA}).

\bibitem[{\citenamefont{Wiener} \emph{et~al.}(1989)\citenamefont{Wiener, Paul,
  and Eichenbaum}}]{wiener+al_89}
\bibinfo{author}{\bibnamefont{Wiener}, \bibfnamefont{S.~I.}},
  \bibinfo{author}{\bibfnamefont{C.~A.} \bibnamefont{Paul}}, and
  \bibinfo{author}{\bibfnamefont{H.}~\bibnamefont{Eichenbaum}},
  \bibinfo{year}{1989}, \bibinfo{journal}{Journal of Neuroscience}
  \textbf{\bibinfo{volume}{9}}, \bibinfo{pages}{2737}.

\bibitem[{\citenamefont{Willett} \emph{et~al.}(2021)\citenamefont{Willett,
  Avansino, Hochberg, Henderson, and Shenoy}}]{willett+al_21}
\bibinfo{author}{\bibnamefont{Willett}, \bibfnamefont{F.~R.}},
  \bibinfo{author}{\bibfnamefont{D.~T.} \bibnamefont{Avansino}},
  \bibinfo{author}{\bibfnamefont{L.~R.} \bibnamefont{Hochberg}},
  \bibinfo{author}{\bibfnamefont{J.~M.} \bibnamefont{Henderson}}, and
  \bibinfo{author}{\bibfnamefont{K.~V.} \bibnamefont{Shenoy}},
  \bibinfo{year}{2021}, \bibinfo{journal}{Nature}
  \textbf{\bibinfo{volume}{593}}, \bibinfo{pages}{249}.

\bibitem[{\citenamefont{Wilson}(1979)}]{wilson_79}
\bibinfo{author}{\bibnamefont{Wilson}, \bibfnamefont{K.~G.}},
  \bibinfo{year}{1979}, \bibinfo{journal}{Scientific American}
  \textbf{\bibinfo{volume}{241}}, \bibinfo{pages}{158}.

\bibitem[{\citenamefont{Wilson}(1983)}]{wilson_83}
\bibinfo{author}{\bibnamefont{Wilson}, \bibfnamefont{K.~G.}},
  \bibinfo{year}{1983}, \bibinfo{journal}{Reviews of Modern Physics}
  \textbf{\bibinfo{volume}{55}}, \bibinfo{pages}{583}.

\bibitem[{\citenamefont{Wilson and Kogut}(1974)}]{wilson+kogut_74}
\bibinfo{author}{\bibnamefont{Wilson}, \bibfnamefont{K.~G.}}, and
  \bibinfo{author}{\bibfnamefont{J.}~\bibnamefont{Kogut}},
  \bibinfo{year}{1974}, \bibinfo{journal}{Physics Reports}
  \textbf{\bibinfo{volume}{12}}, \bibinfo{pages}{75}.

\bibitem[{\citenamefont{Wilson and McNaughton}(1993)}]{wilson+mcnaughton_93}
\bibinfo{author}{\bibnamefont{Wilson}, \bibfnamefont{M.~A.}}, and
  \bibinfo{author}{\bibfnamefont{B.~L.} \bibnamefont{McNaughton}},
  \bibinfo{year}{1993}, \bibinfo{journal}{Science}
  \textbf{\bibinfo{volume}{261}}, \bibinfo{pages}{1055}.

\bibitem[{\citenamefont{Wu} \emph{et~al.}(2023)\citenamefont{Wu, Rudzite,
  Bohlen, Li, Kling, Cooler, Rhoades, Brackbill, Gogliettino, Shah, Madugula,
  Sher} \emph{et~al.}}]{wu+al_23}
\bibinfo{author}{\bibnamefont{Wu}, \bibfnamefont{E.~G.}},
  \bibinfo{author}{\bibfnamefont{A.~M.} \bibnamefont{Rudzite}},
  \bibinfo{author}{\bibfnamefont{M.~O.} \bibnamefont{Bohlen}},
  \bibinfo{author}{\bibfnamefont{P.~H.} \bibnamefont{Li}},
  \bibinfo{author}{\bibfnamefont{A.}~\bibnamefont{Kling}},
  \bibinfo{author}{\bibfnamefont{S.}~\bibnamefont{Cooler}},
  \bibinfo{author}{\bibfnamefont{C.}~\bibnamefont{Rhoades}},
  \bibinfo{author}{\bibfnamefont{N.}~\bibnamefont{Brackbill}},
  \bibinfo{author}{\bibfnamefont{A.~R.} \bibnamefont{Gogliettino}},
  \bibinfo{author}{\bibfnamefont{N.~P.} \bibnamefont{Shah}},
  \bibinfo{author}{\bibfnamefont{S.~S.} \bibnamefont{Madugula}},
  \bibinfo{author}{\bibfnamefont{A.}~\bibnamefont{Sher}}, \emph{et~al.},
  \bibinfo{year}{2023}, \eprint{bioRxiv:2023.11.06.565889}.

\bibitem[{\citenamefont{Yartsev and Ulanovsky}(2013)}]{yartsev+ulanovsky_13}
\bibinfo{author}{\bibnamefont{Yartsev}, \bibfnamefont{M.~M.}}, and
  \bibinfo{author}{\bibfnamefont{N.}~\bibnamefont{Ulanovsky}},
  \bibinfo{year}{2013}, \bibinfo{journal}{Science}
  \textbf{\bibinfo{volume}{108}}, \bibinfo{pages}{E718}.

\bibitem[{\citenamefont{Zhang}(1996)}]{zhang_96}
\bibinfo{author}{\bibnamefont{Zhang}, \bibfnamefont{K.}}, \bibinfo{year}{1996},
  \bibinfo{journal}{Journal of Neuroscience} \textbf{\bibinfo{volume}{16}},
  \bibinfo{pages}{2112}.

\bibitem[{\citenamefont{Zhang} \emph{et~al.}(2023)\citenamefont{Zhang,
  R{\'o}zsa, Liang, Bushey, Wei, Zheng, Reep, Broussard, Tsang, Tsegaye}
  \emph{et~al.}}]{ZhangLooger2023GCamp8}
\bibinfo{author}{\bibnamefont{Zhang}, \bibfnamefont{Y.}},
  \bibinfo{author}{\bibfnamefont{M.}~\bibnamefont{R{\'o}zsa}},
  \bibinfo{author}{\bibfnamefont{Y.}~\bibnamefont{Liang}},
  \bibinfo{author}{\bibfnamefont{D.}~\bibnamefont{Bushey}},
  \bibinfo{author}{\bibfnamefont{Z.}~\bibnamefont{Wei}},
  \bibinfo{author}{\bibfnamefont{J.}~\bibnamefont{Zheng}},
  \bibinfo{author}{\bibfnamefont{D.}~\bibnamefont{Reep}},
  \bibinfo{author}{\bibfnamefont{G.~J.} \bibnamefont{Broussard}},
  \bibinfo{author}{\bibfnamefont{A.}~\bibnamefont{Tsang}},
  \bibinfo{author}{\bibfnamefont{G.}~\bibnamefont{Tsegaye}}, \emph{et~al.},
  \bibinfo{year}{2023}, \bibinfo{journal}{Nature}
  \textbf{\bibinfo{volume}{615}}, \bibinfo{pages}{884}.

\bibitem[{\citenamefont{Zhang} \emph{et~al.}(2021)\citenamefont{Zhang, Bai,
  Cong, Yu, Zhang, Shi, Li, Du, and Wang}}]{Zhang+al2021}
\bibinfo{author}{\bibnamefont{Zhang}, \bibfnamefont{Z.}},
  \bibinfo{author}{\bibfnamefont{L.}~\bibnamefont{Bai}},
  \bibinfo{author}{\bibfnamefont{L.}~\bibnamefont{Cong}},
  \bibinfo{author}{\bibfnamefont{P.}~\bibnamefont{Yu}},
  \bibinfo{author}{\bibfnamefont{T.}~\bibnamefont{Zhang}},
  \bibinfo{author}{\bibfnamefont{W.}~\bibnamefont{Shi}},
  \bibinfo{author}{\bibfnamefont{F.}~\bibnamefont{Li}},
  \bibinfo{author}{\bibfnamefont{J.}~\bibnamefont{Du}}, and
  \bibinfo{author}{\bibfnamefont{K.}~\bibnamefont{Wang}}, \bibinfo{year}{2021},
  \bibinfo{journal}{Nature Biotechnology} \textbf{\bibinfo{volume}{39}},
  \bibinfo{pages}{74}.

\bibitem[{\citenamefont{Zhao} \emph{et~al.}(2023)\citenamefont{Zhao, Zhu, Li,
  Sun, He, Chung, Liu, Frank, Luan, and Xie}}]{zhao+al2023}
\bibinfo{author}{\bibnamefont{Zhao}, \bibfnamefont{Z.}},
  \bibinfo{author}{\bibfnamefont{H.}~\bibnamefont{Zhu}},
  \bibinfo{author}{\bibfnamefont{X.}~\bibnamefont{Li}},
  \bibinfo{author}{\bibfnamefont{L.}~\bibnamefont{Sun}},
  \bibinfo{author}{\bibfnamefont{F.}~\bibnamefont{He}},
  \bibinfo{author}{\bibfnamefont{J.~E.} \bibnamefont{Chung}},
  \bibinfo{author}{\bibfnamefont{D.~F.} \bibnamefont{Liu}},
  \bibinfo{author}{\bibfnamefont{L.}~\bibnamefont{Frank}},
  \bibinfo{author}{\bibfnamefont{L.}~\bibnamefont{Luan}}, and
  \bibinfo{author}{\bibfnamefont{C.}~\bibnamefont{Xie}}, \bibinfo{year}{2023},
  \bibinfo{journal}{Nature Biomedical Engineering}
  \textbf{\bibinfo{volume}{7}}, \bibinfo{pages}{520}.

\bibitem[{\citenamefont{Zhu} \emph{et~al.}(1997)\citenamefont{Zhu, Wu, and
  Mumford}}]{zhu+al_97}
\bibinfo{author}{\bibnamefont{Zhu}, \bibfnamefont{S.~C.}},
  \bibinfo{author}{\bibfnamefont{Y.~N.} \bibnamefont{Wu}}, and
  \bibinfo{author}{\bibfnamefont{D.}~\bibnamefont{Mumford}},
  \bibinfo{year}{1997}, \bibinfo{journal}{Neural Computation}
  \textbf{\bibinfo{volume}{9}}, \bibinfo{pages}{1627}.

\bibitem[{\citenamefont{Ziv} \emph{et~al.}(2013)\citenamefont{Ziv, Burns,
  Cocker, Hamel, Ghosh, Kitch, El~Gamal, and Schnitzer}}]{ziv+al2013}
\bibinfo{author}{\bibnamefont{Ziv}, \bibfnamefont{Y.}},
  \bibinfo{author}{\bibfnamefont{L.~D.} \bibnamefont{Burns}},
  \bibinfo{author}{\bibfnamefont{W.~D.} \bibnamefont{Cocker}},
  \bibinfo{author}{\bibfnamefont{W.~O.} \bibnamefont{Hamel}},
  \bibinfo{author}{\bibfnamefont{K.~K.} \bibnamefont{Ghosh}},
  \bibinfo{author}{\bibfnamefont{L.~J.} \bibnamefont{Kitch}},
  \bibinfo{author}{\bibfnamefont{A.}~\bibnamefont{El~Gamal}}, and
  \bibinfo{author}{\bibfnamefont{M.~J.} \bibnamefont{Schnitzer}},
  \bibinfo{year}{2013}, \bibinfo{journal}{Nature Neuroscience}
  \textbf{\bibinfo{volume}{16}}, \bibinfo{pages}{264}.

\bibitem[{\citenamefont{Zong} \emph{et~al.}(2017)\citenamefont{Zong, Wu, Li,
  Hu, Li, Li, Rong, Wu, Xu, Lu} \emph{et~al.}}]{zong2017fast}
\bibinfo{author}{\bibnamefont{Zong}, \bibfnamefont{W.}},
  \bibinfo{author}{\bibfnamefont{R.}~\bibnamefont{Wu}},
  \bibinfo{author}{\bibfnamefont{M.}~\bibnamefont{Li}},
  \bibinfo{author}{\bibfnamefont{Y.}~\bibnamefont{Hu}},
  \bibinfo{author}{\bibfnamefont{Y.}~\bibnamefont{Li}},
  \bibinfo{author}{\bibfnamefont{J.}~\bibnamefont{Li}},
  \bibinfo{author}{\bibfnamefont{H.}~\bibnamefont{Rong}},
  \bibinfo{author}{\bibfnamefont{H.}~\bibnamefont{Wu}},
  \bibinfo{author}{\bibfnamefont{Y.}~\bibnamefont{Xu}},
  \bibinfo{author}{\bibfnamefont{Y.}~\bibnamefont{Lu}}, \emph{et~al.},
  \bibinfo{year}{2017}, \bibinfo{journal}{Nature Methods}
  \textbf{\bibinfo{volume}{14}}, \bibinfo{pages}{713}.

\end{thebibliography}

\end{document}